%% file: hor.tex
\documentclass[secnum,seceqn,secthm]{elsart}

\usepackage{graphicx}

\usepackage{amssymb}

\usepackage{bm} 

\usepackage[usenames,dvipsnames]{color}

\makeatletter
 \@addtoreset{figure}{section}
 
 \@addtoreset{table}{section}
 
\makeatother

\newcommand{\encadre}[1]{\fbox{$\displaystyle #1$}}
\newcommand{\der}[2]{\frac{\partial #1}{\partial #2}}
\newcommand{\dert}[2]{{\partial #1 / \partial #2}}

\newcommand{\w}[1]{\bm{#1}}
\newcommand{\be}{\begin{equation}}
\newcommand{\ee}{\end{equation}}
\newcommand{\bea}{\begin{eqnarray}}
\newcommand{\eea}{\end{eqnarray}}
\newcommand{\nn}{\nonumber}

\newcommand{\Hor}{{\mathcal H}}
\newcommand{\M}{{\mathcal M}}
\newcommand{\Sp}{{\mathcal S}}
\newcommand{\T}{{\mathcal T}}
\newcommand{\el}{\w{\ell}}
\newcommand{\uel}{\underline{\el}}
\newcommand{\uk}{\underline{\w{k}}}
\newcommand{\dd}{\bm{\mathrm{d}}}
\newcommand{\Lie}[1]{\bm{\mathcal L}_{\w{#1}}\,}
\newcommand{\Liec}[1]{{\mathcal L}_{\w{#1}}\,}
\newcommand{\LieS}[1]{{}^{\Sp}\!\Lie{#1}}
\newcommand{\LieH}[1]{{}^{\Hor}\!\Lie{#1}}
\newcommand{\LieHc}[1]{{}^{\Hor}\!\Liec{#1}}
\newcommand{\Kil}[2]{\,{\rm Kil}(#1,#2)}
\newcommand{\DS}{{}^{2}\!D}
\newcommand{\tDS}{{}^{2}\!\tilde D}
\newcommand{\tv}{\w{t}}
\newcommand{\tvc}{t}

\newcommand{\hajicek}{H\'a\'\j i\v{c}ek}
\newcommand{\we}{\w{e}}
\newcommand{\wo}{\w{\omega}}
\newcommand{\tD}{\tilde D}
\newcommand{\equalH}{\stackrel{\Hor}{=}}

\begin{document}

\begin{frontmatter}



\title{A 3+1 perspective on null hypersurfaces and isolated horizons}


\author{Eric Gourgoulhon}
\ead{eric.gourgoulhon@obspm.fr} and
\author{Jos\'e Luis Jaramillo}
\ead{jose-luis.jaramillo@obspm.fr}
\address{Laboratoire de l'Univers et de ses Th\'eories,
UMR 8102 du C.N.R.S., Observatoire de Paris, F-92195 Meudon Cedex,  France}

\begin{abstract}
The isolated horizon formalism recently introduced by Ashtekar et
al. aims at providing a quasi-local concept of a black hole in 
equilibrium in an otherwise possibly dynamical spacetime. In this formalism, a
hierarchy of geometrical structures is constructed on a null 
hypersurface. On the other side, the 3+1 formulation of general
relativity provides a powerful setting for studying the spacetime
dynamics, in particular gravitational radiation from black hole systems. Here
we revisit the kinematics and dynamics of null hypersurfaces  by making use
of some 3+1 slicing of spacetime. In particular, the additional structures 
induced on null hypersurfaces by the 3+1 slicing permit a natural
extension to the full spacetime  of geometrical quantities defined on
the null hypersurface. This 4-dimensional point of view facilitates the
link between the null and spatial geometries. We proceed by
reformulating the isolated horizon  structure in this framework. We also
reformulate previous works, such as Damour's black hole mechanics,
and make the link with a previous 3+1 approach of black hole horizon,
namely the {\em membrane paradigm}. We
explicit all geometrical objects  in terms of 3+1 quantities, putting a
special emphasis on the conformal 3+1  formulation. This is in
particular relevant for the initial data problem  of black hole
spacetimes for numerical relativity. Illustrative examples  are provided
by considering various slicings of Schwarzschild and  Kerr spacetimes.
\end{abstract}

\begin{keyword}


\end{keyword}

\end{frontmatter}

\tableofcontents


\input{intro} 

\input{geom_null} 

\input{foliat} 

\input{indfoliat} 

\input{kinema} 

\input{dynam} 

\input{neh} 

\input{isolhor} 

\input{isolhorII} 

\input{express3p1} 

\input{applinitdata.tex} 

\input{concl} 

\appendix

\input{lieflow} 

\input{cartan} 

\input{symplectic} 

\input{kerr} 

\input{notsummary} 

\input{ref}

\end{document}

%% file: intro.tex
%
%
\section{Introduction}
\label{s:INTRO}

\subsection{Scope of this article}

Black holes are currently the subject of intense research,
both from the observational and theoretical points of view. 
Numerous observations of black holes in X-ray binaries and in the center
of most galaxies, including ours, have firmly established black holes 
as `standard' objects in the astronomical field
\cite{McCliR04,Rees03,Naray05}. Moreover, black holes are one of the main
targets of gravitational wave observatories which are currently
starting to acquire data: LIGO \cite{Gonza04}, VIRGO \cite{Acern_al04}, 
GEO600 \cite{Smith_al04}, and TAMA \cite{Ando_al01}, 
or are scheduled for the next decade: advanced ground-based interferometers and the 
space antenna LISA \cite{LISA05}. 
These vigorous observational activities constitute 
one of the main motivations for new theoretical developments on
black holes, ranging from new quasi-local formalisms \cite{AshteK05,Booth05}
to numerical relativity \cite{Alcub05,BaumgS03}, going through
perturbative techniques \cite{Poiss05} and post-Newtonian ones
\cite{Blanc02b}. In particular, special emphasis is devoted to the
computation of the merger of inspiralling binary black holes, a not yet solved cornerstone
which is \emph{par excellence} the 2-body problem of general relativity,
and constitutes one of the most promising sources for the 
interferometric gravitational wave detectors \cite{BaumgS03}.

In this article, we concentrate on the geometrical description of the
black hole horizon as a {\em null hypersurface} embedded in spacetime, mainly
aiming at numerical relativity applications. Let us recall that 
a null hypersurface is a 3-dimensional surface ruled by null 
geodesics (i.e. light rays), like the light cone in Minkowski spacetime,
and that it is always a ``one-way membrane'': it divides locally spacetime
in two regions, $A$ and $B$ let's say, such that any future directed 
causal (i.e. null or timelike) 
curve can move from region $A$ to region $B$, but not in the reverse way. 
Regarding black holes, null hypersurfaces are relevant in
two contexts. Firstly, wherever it is smooth, 
the black hole {\em event horizon} is a null hypersurface of 
spacetime\footnote{More generally the event horizon is an 
achronal set \cite{HawkiE73}.} \cite{Carte79b,Carte87,Chrus02} .
Let us stress that the event horizon constitutes an intrinsically {\em global}
concept, in the sense that its definition requires the knowledge of the 
whole spacetime (to determine whether null geodesics can reach null infinity). 
Secondly, a systematic attempt to provide a 
{\em quasi-local} description\footnote{By {\em quasi-local} 
we mean an analysis restricted to a submanifold of spacetime
(typically a 3-dimensional hypersurface with compact sections, but also 
a single compact 2-dimensional surface).} of black holes
has been initiated in the recent years by Hayward \cite{Haywa94,Haywa94b}
(concept of {\em future trapping horizons}) 
and Ashtekar and collaborators 
\cite{AshteBF99,AshteBDFKLW00,AshteBL01,AshteBL02,AshteEPB04,AshteFK00,%
AshteK02,AshteK03}
(see Refs.~\cite{AshteK05,Booth05} for a review)
(concepts of {\em isolated} and {\em dynamical horizons}).  
Restricted to the quasi-equilibrium case (isolated horizons), 
the quasi-local description amounts to model the black hole horizon 
by a null hypersurface. 
This line of research finds its
motivations and subsequent applications in a variety of fields of gravitational 
physics such as black hole mechanics, mathematical relativity, quantum gravity 
and, due to its quasi-local character, 
numerical relativity \cite{DreyeKSS03,Kr02,JaramGM04,DJK05,BaiotHMLRSFS05}.


The geometry of a null hypersurface $\Hor$ is usually  described in terms of
objects that are intrinsic to $\Hor$. At  least some of them admit no natural 
extension outside the hypersurface $\Hor$. In the case of the isolated horizon
formalism  this leads, in a natural way, to a discussion which is eminently 
intrinsic to $\Hor$ in a twofold manner. On one hand, the derived
expressions are generically valid only on $\Hor$ without canonical
extensions to a neighbourhood of the surrounding spacetime. 
On the other hand, since in this setting the hypersurface $\Hor$ can be seen as 
representing the 
history of a spacelike 2-sphere (a {\it world-tube} in spacetime),
the study of $\Hor$'s geometry from a strictly intrinsic point of view
leads to a strategy in which one firstly discuss {\it evolution}
concepts, and {\it then} one considers the initial
conditions on the 2-sphere which are compatible with such an evolution. 
We may call this an {\it up-down} strategy.

On the contrary, the dynamics of black hole spacetimes is 
mostly studied within the 3+1 formalism (see e.g. \cite{BaumgS03,York79}
for a review), 
which amounts in the foliation of spacetime by a family of 
spacelike hypersurfaces. 
In this case, one deals with a Cauchy problem, starting from some initial 
spacelike hypersurface $\Sigma_0$
and evolving it in order to construct
the proper spacetime objects. In particular, this applies to
the construction of the horizon $\Hor$ as a worldtube. We may call this 
a {\it down-up} strategy.

In this article, we analyse the dynamics of null hypersurfaces
from the 3+1 point of view. As a methodological strategy, we adopt 
a complete 4-dimensional description, 
even when considering objects which are actually intrinsic 
to a given (hyper)surface. This facilitates the link between the null horizon 
hypersurface and the spatial hypersurfaces of the 3+1 slicing.
Moreover, the 3+1 foliation of spacetime induces an
additional structure on $\Hor$ which allows to normalize
unambiguously the null normal to $\Hor$ and to define a projector
onto $\Hor$. Let us recall that a distinctive feature of null 
hypersurfaces is the lack of such canonical constructions, contrary
to the spacelike or timelike case where one can unambiguously define
the unit normal vector and the orthogonal projector onto the hypersurface.


Null hypersurfaces have been extensively studied in the literature
in connection with black hole horizons, from many different points
of view. In the seventies, \hajicek\ conducted geometrical
studies of non-expanding null hypersurfaces to model stationary
black hole horizons \cite{Hajic73,Hajic74,Hajic75}. 
By studying the response of the (null) event horizon to external perturbations, 
Hawking and Hartle \cite{HawkiH72,Hartl73,Hartl74} 
introduced the concept of black hole viscosity.
This hydrodynamical analogy was extended by
Damour \cite{Damou78,Damou79,Damou82}. Electromagnetic aspects were
studied by Damour \cite{Damou78,Damou79,Damou82}
and Znajek \cite{Znaje77,Znaje78}. 
These studies led to the famous {\em membrane paradigm}
for the description of black holes (\cite{PriceT86,ThornPM86}
and references therein).
In particular, this paradigm represents the first systematic 3+1 approach 
to black hole physics. 
Whereas these studies all dealt with the event horizon
(the global aspect of a black hole),
a quasi-local approach, based on the notion of trapped surface \cite{Penro65},
has been initiated in the nineties by 
Hayward \cite{Haywa94,Haywa94b}, in the framework of the 2+2 formalism. 
Closely related to these ideas, a systematic  
quasi-local treatment has been developed these last years
by Ashtekar and collaborators
\cite{AshteBF99,AshteBDFKLW00,AshteBL01,AshteBL02,AshteEPB04,AshteFK00,%
AshteK02,AshteK03}
(see Refs.~\cite{AshteK05,Booth05} for a review), giving rise to the notion
of {\em isolated horizons} and more recently to that of 
{\em dynamical horizons}, the latter not being constructed on 
a null hypersurface, but on a spacelike one. 

One purpose of this article is to fill the gap existing 
between the mathematical techniques used in null geometry and 
the standard expertise in the
numerical relativity community.
Consequently, an important effort will be devoted to the derivation
of explicit expressions of null-geometry quantities 
in terms of 3+1 objects. 
More generally, the article is relatively self-contained, and requires
only an elementary knowledge of differential geometry, at the level
of introductory textbooks in general relativity 
\cite{HawkiE73,MisneTW73,Wald84}. 
We have tried to be quite pedagogical, by providing concrete examples
and detailed derivations of the main results. In fact, 
these explicit developments permit to access directly to 
intermediate steps, which might be useful in actual numerical 
implementation. 
We rederive the basic properties of null hypersurfaces, taking 
advantage of our 3+1 perspective, namely the unambiguous
definition of the null normal and transverse projector provided
by the 3+1 spacelike slicing. 
Therefore the present article 
should not be considered as a substitute for comprehensive formal 
presentations of the intrinsic geometry of null surfaces, 
as Refs.~\cite{Gallow04,Gallow00,Kupel87,Jezie04,JezieKC02}. 
Likewise, it is not the aim here to review the isolated horizon
formalism and its applications, something already carried out in a full 
extent in Ref.~\cite{AshteK05}. 

Despite the length of the article, some important topics are not 
treated here, namely electromagnetic properties of black holes
or black hole thermodynamics. In particular, we will not develop
the Hamiltonian description of black hole mechanics in the
isolated horizon scheme, except for the minimum required to 
discuss the physical parameters
associated with the black hole. 
We do not comment either on the application of the isolated horizon 
framework beyond Einstein-Maxwell theory to include, e.g. Yang-Mills fields.
Even though these fields are not expected to be significant in an astrophysical 
setting, their inclusion involves a major conceptual and structural interest; 
we refer the reader to chapter 6 in Ref. \cite{AshteK05} for a review on 
the achievements in this line of research, namely on the mass of solitonic solutions.


The plan of the article is as follows. 
After setting the notations in the next subsection, 
we start by reviewing the basic properties of null hypersurfaces
in Sec.~\ref{s:NH:geom_null}. Then the spacelike slicing of the 3+1 formalism
widely used in numerical relativity is introduced in Sec.~\ref{s:FO}. 
The additional structures induced by this slicing on a given
null hypersurface $\Hor$ are discussed in Sec.~\ref{s:IN}; in particular, 
this involves a privileged null normal, a null transverse vector and 
the associated projector onto $\Hor$. 
Equipped with these tools, we proceed in Sec.~\ref{s:KI}
to describe the {\em kinematics} of null hypersurfaces, namely relations involving the
first ``time'' derivative of their degenerate metric. 
The next logical step corresponds to {\em dynamics}, namely the second order
derivatives of the metric, which is explored in Sec.~\ref{s:DY}. 
The Einstein equation naturally enters the scene at this level.
In particular, we recover in Sec.~\ref{s:DY} previous results
from the membrane paradigm, like Damour's Navier-Stokes equation
or the tidal-force equation. Then in Sec.~\ref{s:NE} we move to the 
quasi-local approach of black holes by restricting to null hypersurfaces 
with vanishing expansion, which are the ``perfect horizons'' of \hajicek\ and
constitute the first step in Ashtekar et al. hierarchy
leading to isolated horizons. The next levels in the hierarchy 
are studied in Secs.~\ref{s:IH} and \ref{s:IHII}, where 
we discuss the weakly and strongly isolated horizon structures. 
Due to the extension of the material, these two sections rely 
more explicitly on the existing literature and, 
as a consequence, the intrinsic point of 
view of the geometry of $\Hor$ (the up-down strategy referred above)
acquires there a more important role
than in the rest of the article.
In Sec.~\ref{s:TP}, we express basic objects of null geometry
in terms of the 3+1 quantities, including the standard conformal
decompositions of 3+1 objects. This allows to translate in Sec.~\ref{s:BC} the
isolated horizon prescriptions into boundary conditions
for the relevant 3+1 fields on some excised sphere,
making the link with numerical relativity.  
Some technical details are treated in appendices: 
the relationship between different derivatives along the null normal is
given Appendix~\ref{s:LF}; Appendix~\ref{s:CA} is devoted to the
complete computation of the spacetime Riemann tensor. 
In contrast with some works on null 
hypersurfaces, we do not make use of the Newman-Penrose formalism
but rely instead on Cartan's structure equations. Appendix~\ref{s:SP}
briefly presents, with the aid of examples, the basics of the
Hamiltonian description. Appendix~\ref{s:KE} provides the concrete
example of the horizon of a Kerr black hole, while simpler examples, 
based on Minkowski or 
Schwarzschild spacetimes, are provided throughout the main text.
Finally Appendix~\ref{s:SY}
gathers the different symbols used throughout the article.


\subsection{Notations and conventions} \label{s:notations}

For the benefit of the reader, we give here a somewhat detailed
exposure of the notations used throughout the article. This is
also the occasion to recall some concepts from elementary
differential geometry employed in the article.  

We consider a spacetime $(\M,\w{g})$ 
where $\M$ is a real smooth (i.e. $\mathcal C^\infty$) manifold
of dimension 4 and $\w{g}$ 
a Lorentzian metric on $\M$, of signature $(-,+,+,+)$.
We denote by $\w{\nabla}$ the affine connection associated
with $\w{g}$, and call it the {\em spacetime connection}
to distinguish it from other connections introduced in the text.

At a given point $p\in \M$, we denote by $\T_p(\M)$ the {\em tangent
space}, i.e. the (4-dimensional) space of vectors at $p$.
Its dual space (also called {\em cotangent space})
is denoted by $\T_p^*(\M)$ and is constituted 
by all linear forms at $p$. 
We denote by $\T(\M)$ (resp. $\T^*(\M)$) the space of smooth
vector fields (resp. 1-forms) on $\M$.  
The experienced reader is warned that $\T(\M)$ does not stand 
for the tangent bundle of $\M$ (it rather corresponds to the 
space of smooth cross-sections of that bundle). No confusion may arise since 
we shall not use the notion of bundle in this article.

\subsubsection{Tensors: `index' notation versus `intrinsic' notation}
\label{s:IN:index_intrinsic}

Since we will manipulate geometrical quantities 
which are not well suited to the index notation
(like Lie derivatives or exterior derivatives),
we will use quite often an index-free notation. 
When dealing with indices, we adopt the following conventions:
all Greek indices run in $\{0,1,2,3\}$.
We will use letters from the
beginning of the alphabet ($\alpha$, $\beta$, $\gamma$, ...) for free indices,
and letters starting from $\mu$ ($\mu$, $\nu$, $\rho$, ...) as dumb indices
for contraction (in this way the tensorial degree (valence) of any 
equation is immediately apparent). 
All capital Latin indices ($A$, $B$, $C$, ...) run in $\{0,2,3\}$ and 
lower case Latin indices starting from the letter $i$ ($i$, $j$, $k$, ...) 
run in $\{1,2,3\}$, while those starting from the
beginning of the alphabet ($a$, $b$, $c$, ...) run in $\{2,3\}$ only. 

For the sake of clarity, let us recall that if $(\w{e}_\alpha)$
is a vector basis of the tangent space $\T_p(\M)$ and 
$(\w{e}^\alpha)$ is the associate
dual basis, i.e. the basis of $\T_p^*(\M)$ such that 
$\w{e}^\alpha(\w{e}_\beta)=\delta^\alpha_{\ \, \beta}$,
the components 
$T^{\alpha_1\ldots\alpha_p}_{\qquad\ \; \beta_1\ldots\beta_q}$
of a tensor $\w{T}$ of type $\left({p \atop q}\right)$ with 
respect to the bases $(\w{e}_\alpha)$ and $(\w{e}^\alpha)$ 
are given by the expansion
\be \label{e:IN:comp_tens}
	\w{T} = T^{\alpha_1\ldots\alpha_p}_{\qquad\ \; \beta_1\ldots\beta_q}
		\; \w{e}_{\alpha_1} \otimes \ldots \otimes \w{e}_{\alpha_p} 
                \otimes
		\w{e}^{\beta_1} \otimes \ldots \otimes \w{e}^{\beta_q} .
\ee
The components
$\nabla_{\gamma}  T^{\alpha_1\ldots\alpha_p}_{\qquad\ \; \beta_1\ldots\beta_q}$
of the covariant derivative $\w{\nabla}\w{T}$ are defined by the expansion
\be \label{e:IN:cov_der_comp}
	\w{\nabla}\w{T} = 
	\nabla_{\gamma} \, 
        T^{\alpha_1\ldots\alpha_p}_{\qquad\ \; \beta_1\ldots\beta_q}
		\; \w{e}_{\alpha_1} \otimes \ldots \otimes \w{e}_{\alpha_p} 
                \otimes
		\w{e}^{\beta_1} \otimes \ldots \otimes \w{e}^{\beta_q} 
		\otimes \w{e}^\gamma  .
\ee
Note the position of the ``derivative index'' $\gamma$ : 
$\w{e}^\gamma$ is the
{\em last} 1-form of the tensorial product on the right-hand side. In this
respect, the notation 
$T^{\alpha_1\ldots\alpha_p}_{\qquad\ \; \beta_1\ldots\beta_q;\gamma}$ instead of 
$\nabla_{\gamma} \, 
T^{\alpha_1\ldots\alpha_p}_{\qquad\ \; \beta_1\ldots\beta_q}$
would have been more appropriate .
This index convention agrees with that 
of MTW \cite{MisneTW73} [cf. their Eq.~(10.17)].
As a result, the covariant derivative of the tensor $\w{T}$ along any
vector field $\w{u}$ is related to $\w{\nabla}\w{T}$ by
\be \label{e:IN:directional_der}
    \w{\nabla}_{\w{u}}\w{T} = \w{\nabla}\w{T}
        (\underbrace{.,\ldots,.}_{p+q\ {\rm slots}},\w{u}) . 
\ee
The components of $\w{\nabla}_{\w{u}}\w{T}$ are then 
$u^\mu \nabla_{\mu} 
T^{\alpha_1\ldots\alpha_p}_{\qquad\ \; \beta_1\ldots\beta_q}$. 

Given a vector field $\w{v}$ on $\M$, the infinitesimal change of 
any tensor field $\w{T}$
along the flow of $\w{v}$, is given by the Lie derivative of $\w{T}$
with respect to $\w{v}$, denoted by $\Lie{\w{v}}\w{T}$ and whose components are
\bea
\label{e:IN:Lie_derivative}
(\Liec{v} T)^{\alpha_1\ldots\alpha_p}_{\qquad\ \; \beta_1\ldots\beta_q}&=&
v^\mu \nabla_{\mu} \, 
        T^{\alpha_1\ldots\alpha_p}_{\qquad\ \; \beta_1\ldots\beta_q} \\
&-& \sum_{i=1}^p T^{\alpha_1\ldots\mu\ldots\alpha_p}_{\qquad\ \ \ \  \; \beta_1\ldots\beta_q}
 \nabla_{\mu}v^{\alpha_i}
+  \sum_{i=1}^q T^{\alpha_1\ldots\alpha_p}_{\qquad\ \; \beta_1\ldots\mu\ldots\beta_q} 
\nabla_{\beta_i}v^{\mu} \nn \ ,
\eea 
where the connection $\w{\nabla}$ can be substituted by any other torsion-free
connection. Actually let us recall that the Lie derivative depends only
upon the differentiable structure of the manifold $\M$ and not upon the metric $\w{g}$ nor
a particular affine connection. 
In this article, extensive use will be made of expression 
(\ref{e:IN:Lie_derivative}),
as well as of its straightforward analogues on submanifolds of
$\M$ (see also Appendix~\ref{s:LF}). 

We denote the scalar product of two vectors with respect to the
metric $\w{g}$ by a dot:
\be
  \forall (\w{u},\w{v}) \in \T_p(\M)\times\T_p(\M),\quad 
  \w{u}\cdot\w{v} := \w{g}(\w{u},\w{v}) .
\ee
We also use a dot for the contraction of two tensors $\w{A}$
and $\w{B}$ on the last index of $\w{A}$ and the first index of 
$\w{B}$ (provided of course that these indices are of opposite types). 
For instance if $\w{A}$ is a bilinear form and $\w{B}$ a vector, 
$\w{A}\cdot\w{B}$ is the linear form which components are
\be
    (A\cdot B)_\alpha = A_{\alpha\mu} B^\mu . 
\ee
However, to denote the action of linear forms on vectors, we will use 
brackets instead of a dot: 
\be \label{e:IN:brackets}
	\forall (\w{\omega},\w{v}) \in \T_p^*(\M)\times\T_p(\M),\quad 
        \langle \w{\omega},\w{v} \rangle = 
        \w{\omega} \cdot \w{v} = \omega_\mu \, v^\mu .
\ee
Given a 1-form $\w{\omega}$ and a vector field $\w{u}$, the directional
covariant derivative $\w{\nabla}_{\w{u}} \, \w{\omega}$ is a 1-form and
we have [combining the notations (\ref{e:IN:brackets}) and 
(\ref{e:IN:directional_der})]
\be \label{e:IN:direc_deriv_1form}
	\forall (\w{\omega},\w{u},\w{v}) \in \T^*(\M)\times\T(\M)\times\T(\M),\quad 
        \langle \w{\nabla}_{\w{u}} \, \w{\omega},\w{v} \rangle = 
        \w{\nabla}\w{\omega} (\w{v},\w{u}).
\ee
Again, notice the ordering in the arguments of the bilinear form $\w{\nabla}\w{\omega}$.
Taking the risk of insisting outrageously, let us stress that this is equivalent
to say that the components 
$(\nabla\omega)_{\alpha\beta}$ of $\w{\nabla}\w{\omega}$ with respect to 
a given basis $(\w{e}^\alpha\otimes\w{e}^\beta)$ of $\T^*(\M)\otimes\T^*(\M)$ are
$\nabla_\beta\omega_\alpha$:
\be \label{e:IN:grad_1form}
    \w{\nabla}\w{\omega} = \nabla_\beta\omega_\alpha \; 
    \w{e}^\alpha\otimes\w{e}^\beta ,
\ee
this relation constituting a particular case of Eq.~(\ref{e:IN:cov_der_comp}).

The metric $\w{g}$ induces an isomorphism between
$\T_p(\M)$ (vectors) and $\T_p^*(\M)$ (linear forms) which, in the index notation, 
corresponds to the lowering or raising of the index by contraction
with $g_{\alpha\beta}$ or $g^{\alpha\beta}$. 
In the present article, an index-free symbol will always denote
a tensor with a fixed covariance type (e.g. a vector, a 1-form,
a bilinear form, etc...). We will therefore use a different symbol
to denote its image under the metric isomorphism. 
In particular, we denote by an underbar the 
isomorphism $\T_p(\M) \rightarrow \T_p^*(\M)$
and by an arrow the reverse isomorphism $\T_p^*(\M) \rightarrow \T_p(\M)$:
\begin{enumerate}
\item for any vector $\w{u}$ in $\T_p(\M)$, $\underline{\w{u}}$ stands for 
the unique linear form such that 
\be \label{e:IN:underbar}
	\forall \w{v} \in \T_p(\M),\quad \langle \underline{\w{u}}, \w{v}
		\rangle = \w{g}(\w{u},\w{v}) .
\ee
However, we will omit the underlining on the components
of $\underline{\w{u}}$, since
the position of the index allows to distinguish between vectors
and  linear forms, following the standard usage:
if the components of 
$\w{u}$ in a given basis $(\w{e}_\alpha)$ are denoted by $u^\alpha$,
the components of $\underline{\w{u}}$ in the dual basis $(\w{e}^\alpha)$
are then denoted by $u_\alpha$
[in agreement with Eq.~(\ref{e:IN:comp_tens})].
\item for any linear form $\w{\omega}$ in $\T_p^*(\M)$, $\vec{\w{\omega}}$
stands for the unique vector of $\T_p(\M)$ such that
\be \label{e:IN:arrow_form}
	\forall \w{v} \in \T_p(\M),\quad 
        \w{g}(\vec{\w{\omega}},\w{v}) = 
        \langle \w{\omega}, \w{v} \rangle .
\ee
As for the underbar, we will omit the arrow over the components
of $\vec{\w{\omega}}$ by denoting them $\omega^\alpha$. 
\item we extend the arrow notation to {\em bilinear} forms on $\T_p(\M)$:
for any bilinear form $\w{T}\, : \, \T_p(\M)\times\T_p(\M) \rightarrow \mathbb{R}$,
we denote by $\vec{\w{T}}$ the (unique) endomorphism 
$T(\M) \rightarrow T(\M)$ which satisfies 
\be \label{e:IN:arrow_endo}
    \forall (\w{u},\w{v}) \in \T_p(\M)\times\T_p(\M), \quad 
    \w{T}(\w{u},\w{v}) = \w{u} \cdot \vec{\w{T}}(\w{v}) . 
\ee
If $T_{\alpha\beta}$ are the components of the bilinear form $\w{T}$ 
in some basis $\w{e}^\alpha\otimes\w{e}^\beta$, the matrix of
the endomorphism $\vec{\w{T}}$ with respect to the vector basis 
$\w{e}_\alpha$ (dual to $\w{e}^\alpha$) is $T^\alpha_{\ \, \beta}$.
\end{enumerate}

\subsubsection{Curvature tensor} \label{s:IN:curvat}

We follow the MTW convention \cite{MisneTW73} and define the
{\em Riemann curvature tensor} of the spacetime connection 
$\w{\nabla}$ by
\be \label{e:IN:def_Riemann}
	 \begin{array}{cccc}
	\mathrm{\bf Riem} \ : & \T^*(\M)\times\T(\M)^3 & 
	\longrightarrow & \mathcal{C}^\infty(\M,\mathbb{R}) \\
		& (\w{\omega},\w{w},\w{u},\w{v}) 
		& \longmapsto & \bigg\langle \w{\omega} , \ 
                \w{\nabla}_{\w{u}} \w{\nabla}_{\w{v}} \w{w}
		-  \w{\nabla}_{\w{v}} \w{\nabla}_{\w{u}} \w{w} \\
                & & &
		- \w{\nabla}_{[\w{u},\w{v}]} \w{w} \bigg\rangle ,
	\end{array}  
\ee
where $\mathcal{C}^\infty(\M,\mathbb{R})$ denotes the space of
smooth scalar fields on $\M$. As it is well known, the above
formula does define a tensor field on $\M$, i.e. the value
of $\mathrm{\bf Riem}(\w{\omega},\w{w},\w{u},\w{v})$ at a given
point $p\in\M$ depends only upon the values of the fields 
$\w{\omega}$, $\w{w}$, $\w{u}$ and $\w{v}$ at $p$ and not
upon their behaviors away from $p$, as the gradients in 
Eq.~(\ref{e:IN:def_Riemann}) might suggest. 
We denote the components of this tensor in 
a given basis $(\w{e}_\alpha)$, not by 
${\rm Riem}^\gamma_{\ \, \delta \alpha\beta}$, but by
$R^\gamma_{\ \, \delta \alpha\beta}$. 
The definition (\ref{e:IN:def_Riemann}) leads then to the
following writing 
(called {\em Ricci identity}):
\be \label{e:IN:Ricci_ident}
    \forall\w{w}\in\T(\M),\quad 
        \left(\nabla_\alpha\nabla_\beta  
        - \nabla_\beta\nabla_\alpha\right) w^\gamma
        = R^\gamma_{\ \, \mu \alpha\beta} \, w^\mu ,  
\ee
From the definition (\ref{e:IN:def_Riemann}), the Riemann tensor is
clearly antisymmetric with respect to its last two arguments $(\w{u},\w{v})$.
The fact that the connection $\w{\nabla}$ is associated with a metric
(i.e. $\w{g}$) implies the additional well-known 
antisymmetry:
\be \label{e:IN:Riemann_antisym12}
    \forall (\w{\omega},\w{w})\in \T^*(\M)\times\T(\M),\
    \mathrm{\bf Riem}(\w{\omega},\w{w},\cdot,\cdot)
    = - \mathrm{\bf Riem}(\underline{\w{w}},\vec{\w{\omega}},\cdot,\cdot) .
\ee
In addition, the Riemann tensor satisfies the cyclic property
\bea   
& & \forall (\w{u},\w{v},\w{w})\in \T(\M)^3, \nonumber \\
& & \ \quad
\mathrm{\bf Riem}(\cdot,\w{u},\w{v},\w{w}) 
+\mathrm{\bf Riem}(\cdot,\w{w},\w{u},\w{v})
+\mathrm{\bf Riem}(\cdot,\w{v},\w{w},\w{u}) = 0 \ . \label{e:IN:Riemann_cyclic}
\eea

The {\em Ricci tensor} of the spacetime connection $\w{\nabla}$ is
the bilinear form $\w{R}$ defined by 
\be \label{e:IN:def_Ricci}
	 \begin{array}{cccc}
	\w{R} \ : & \T(\M)\times\T(\M) & 
	\longrightarrow & \mathcal{C}^\infty(\M,\mathbb{R}) \\
		& (\w{u},\w{v}) 
		& \longmapsto & 
                \mathrm{\bf Riem}(\w{e}^\mu,\w{u},\w{e}_\mu,\w{v}) .
	\end{array}  
\ee
This definition is independent of the choice of the basis $(\w{e}_\alpha)$
and its dual counterpart $(\w{e}^\alpha)$. Moreover the bilinear form
$\w{R}$ is symmetric. 
In terms of components:
\be \label{e:IN:def_Ricci_comp}
    R_{\alpha\beta} = R^\mu_{\ \, \alpha\mu\beta}.
\ee
Note that, following the standard usage, we are denoting the components
of both the Riemann and Ricci tensors by the same letter $R$, the 
number of indices allowing to distinguish between the two tensors.
On the contrary we are using different symbols, $\mathrm{\bf Riem}$ and
$\w{R}$, when dealing with the `intrinsic' notation.

Finally, the Riemann tensor can 
be split into (i) a ``trace-trace'' part, represented
by the Ricci scalar $R:=g^{\mu\nu} R_{\mu\nu}$, 
(ii) a ``trace'' part, 
represented by the Ricci tensor $\w{R}$
[cf. Eq.~(\ref{e:IN:def_Ricci_comp})], and (iii) a ``traceless'' part,
which is constituted by the {\em Weyl conformal curvature tensor}, $\w{C}$:
\bea
	R^\gamma_{\ \; \delta\alpha\beta}   & = & 
        C^\gamma_{\ \; \delta\alpha\beta}
	+ \frac{1}{2} \left( R^\gamma_{\ \, \alpha} \, g_{\delta\beta}
	   - R^\gamma_{\ \, \beta}\,  g_{\delta\alpha}
	   + R_{\delta\beta} \, \delta^\gamma_{\ \, \alpha}
	   - R_{\delta\alpha} \, \delta^\gamma_{\ \, \beta} \right) 
                            \nonumber \\
	 &&   + \frac{1}{6} R \left( g_{\delta\alpha} \, 
         \delta^\gamma_{\ \, \beta}
	   - g_{\delta\beta} \, \delta^\gamma_{\ \, \alpha} \right) . \label{e:IN:Weyl}
\eea
The above relation can be taken as the definition of $\w{C}$. 
It implies that $\w{C}$ is traceless: 
\bea
\label{e:IN:Weyl_traceless}
C^\mu_{\ \, \alpha\mu\beta}=0 \ .
\eea
The other possible traces are zero thanks to the symmetry properties of 
the Riemann tensor. 
It is well known that the $20$ independent components
of the Riemann tensor distribute in the $10$ components in the 
Ricci tensor, that are fixed by Einstein equation, and $10$
independent components in the Weyl tensor.

\subsubsection{Differential forms and exterior calculus}  \label{s:IN:exterior}

In this article, we will make use of p-forms, mostly 1-forms and 2-forms.
Let us recall that a {\em p-form} is a type $\left({0 \atop p}\right)$
tensor field which is antisymmetric with respect to all its $p$ arguments. 
In other words, it is a multilinear
form field $\T(\M)\times\cdots\times\T(\M)\longrightarrow 
\mathcal{C}^\infty(\M,\mathbb{R})$
which is fully antisymmetric.

We follow the convention of MTW \cite{MisneTW73}, Wald \cite{Wald84},
and Straumann \cite{Strau84} textbooks
for the {\em exterior product} (wedge product) between
p-forms: if $\w{\omega}$ and $\w{\sigma}$ are two 1-forms
[i.e. two elements of $\T^*(\M)$], $\w{\omega}\wedge\w{\sigma}$ is
the 2-form defined by 
\be \label{e:NH:ext_prod}
	\w{\omega}\wedge\w{\sigma} := \w{\omega}\otimes \w{\sigma}
		- \w{\sigma}\otimes \w{\omega} .
\ee 
Note that this definition disagrees with that of 
Hawking \& Ellis \cite{HawkiE73}, 
which would require a factor $1/2$ in front
of the r.h.s. of (\ref{e:NH:ext_prod}) [cf. the equation on page
21 of Ref.~\cite{HawkiE73}, and Ref.~\cite{Carte79} for a discussion].

The {\em exterior derivative} of a differential form is defined by induction
starting from $\dd f$ being the 1-form gradient of $f$ for any scalar
field (0-form) $f$. For any $(p+q)$-form that can be written as the exterior
product of a $p$-form $\w{\alpha}$ by a $q$-form $\w{\beta}$, the 
exterior derivative is the $(p+q+1)$-form defined by
\be \label{e:NH:ext_deriv}
	\dd (\w{\alpha}\wedge\w{\beta}) = 
	\dd \w{\alpha} \wedge\w{\beta} + (-1)^p 
	\w{\alpha} \wedge \dd \w{\beta} .
\ee
This equation agrees with that in Box 4.1 of MTW \cite{MisneTW73}. 
It constitutes a version of Leibnitz rule altered by the factor
$(-1)^p$; for this reason the exterior derivative is sometimes
called an {\em antiderivation} (e.g. Definition 4.2 of 
Ref.~\cite{Strau84}).

The components of the exterior derivative of a 1-form $\w{\omega}$ with
respect to some coordinate system $(x^\alpha)$ on $\M$ are 
\be \label{e:NH:d1form_comp}
	(\mathrm{d}\omega)_{\alpha\beta} = \partial_\alpha \omega_\beta
		- \partial_\beta \omega_\alpha, 
\ee
where the partial derivative $\partial_\alpha$ can be replaced
by any covariant derivative operator without torsion on $\M$.
(for instance the spacetime derivative $\nabla_\alpha$).
Taking into account Eqs.~(\ref{e:IN:grad_1form}) and 
(\ref{e:IN:direc_deriv_1form}), we can then write
\bea 
    \forall (\w{u},\w{v}) \in \T(\M)\times\T(\M), \quad 
    \dd\w{\omega}(\w{u},\w{v}) & = & \w{\nabla}\w{\omega}(\w{v},\w{u}) 
        - \w{\nabla}\w{\omega}(\w{u},\w{v}) \label{e:IN:d1form_der_cov} \\
        & = & \langle \w{\nabla}_{\w{u}} \, \w{\omega},\w{v} \rangle
        - \langle \w{\nabla}_{\w{v}} \, \w{\omega},\w{u} \rangle .
                                \label{e:IN:d1form_cov_der}
\eea

A very useful relation that we shall employ throughout the article is
{\em Cartan identity}, which relates the 
Lie derivative of a p-form $\w{\omega}$ along a vector field $\w{v}$
to the exterior derivative of $\w{\omega}$:
\be \label{e:IN:Cartan_id}
   \Lie{v} \, \w{\omega}
   	= \w{v} \cdot \dd \w{\omega} + \dd(\w{v} \cdot \w{\omega}) .
\ee

Given a 1-form $\w{\omega}\in\T^*(\M)$ and a connection operator
$\w{\tilde\nabla}$ on $\M$ (not necessarily the spacetime connection 
$\w{\nabla}$ associated with the metric $\w{g}$), the exterior derivative
$\dd\w{\omega}$ can be viewed as (minus two times) the antisymmetric part of the 
gradient $\w{\tilde\nabla}\w{\omega}$. The symmetric part is given by
(half of) the {\em Killing operator} $\Kil{\w{\tilde\nabla}}{.}$ such that
$\Kil{\w{\tilde\nabla}}{\w{\omega}}$ is the symmetric bilinear form
$\T(\M)\times\T(\M)\rightarrow \mathcal{C}^\infty(\M,\mathbb{R})$ defined by
\be \label{e:IN:def_Killing}
    \Kil{\w{\tilde\nabla}}{\w{\omega}}(\w{u},\w{v}) = 
        \w{\tilde\nabla}\w{\omega}(\w{u},\w{v})
        + \w{\tilde\nabla}\w{\omega}(\w{v},\w{u}) , 
\ee
for any $(\w{u},\w{v}) \in \T(\M)\times\T(\M)$.
Combining Eqs.~(\ref{e:IN:def_Killing}) and (\ref{e:IN:d1form_der_cov}),
we have the decomposition 
\be
    \forall \w{\omega}\in\T^*(\M),\quad
    \w{\tilde\nabla}\w{\omega} = 
    \frac{1}{2} \left[ \Kil{\w{\tilde\nabla}}{\w{\omega}} 
        - \dd \w{\omega} \right]. 
\ee
As stated before, the antisymmetric part, $\dd \w{\omega}$, is independent
of the choice of the connection $\w{\tilde\nabla}$.

%% file: geom_null.tex
%
%
\section{Basic properties of null hypersurfaces} \label{s:NH:geom_null}

There is no doubt about the  central role of null hypersurfaces 
in general relativity, and they have been extensively studied in the literature. 
We review here some of their elementary properties,
referring the reader to Refs.~\cite{Gallow00,Gallow04,Kupel87}
and \cite{BarraH98,Jezie04,JezieKC00,JezieKC02,NR00}
for further details or alternative approaches.
Let us mention that the properties described here,
as well as in the subsequent 
sections~\ref{s:FO}, \ref{s:IN}, \ref{s:KI} and \ref{s:DY},
are valid for any kind of null hypersurface
and do not require any link with a
black hole horizon. For instance they are perfectly valid
for a light cone in Minkowski spacetime. 

\begin{figure}
\centerline{\includegraphics[width=0.9\textwidth]{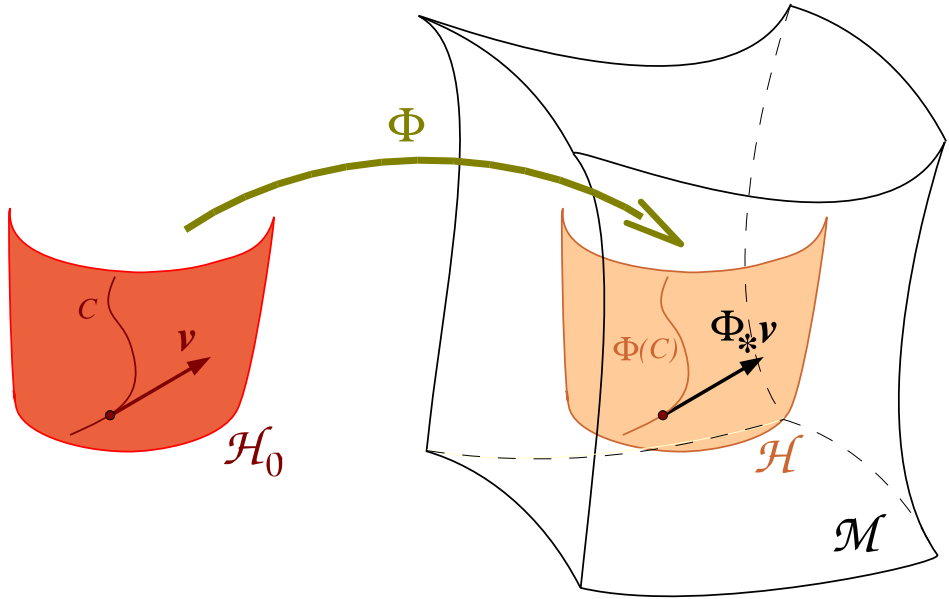}}
\caption[]{\label{f:NH:embed} 
Embedding $\Phi$ of the 3-dimensional manifold $\Hor_0$ into the 
4-dimensional manifold $\M$, 
defining the hypersurface $\Hor = \Phi(\Hor_0)$. The push-forward $\Phi_*\w{v}$
of a vector $\w{v}$ tangent to some curve $C$ in $\Hor_0$ is a vector
tangent to $\Phi(C)$ in $\M$.}
\end{figure}

\subsection{Definition of a hypersurface} \label{s:NH:def_hyp}

A {\em hypersurface} $\Hor$ of $\M$ is the image of a 3-dimensional
manifold $\Hor_0$ by an embedding $\Phi\,:\, \Hor_0 \rightarrow \M$
(Fig.~\ref{f:NH:embed}) :
\be
	\Hor = \Phi(\Hor_0) . 
\ee
Let us recall that {\em embedding} means that 
$\Phi\,:\, \Hor_0 \rightarrow \Hor$ is a homeomorphism, 
i.e. a one-to-one mapping such that both $\Phi$
and $\Phi^{-1}$ are continuous.
The one-to-one character guarantees that $\Hor$ does not ``intersect itself''.
A hypersurface can be defined locally as the set of points for
which a scalar field on $\M$, $u$ let say, is constant:
\be \label{e:NH:r_const}
	\forall p \in \M,\quad p \in \Hor \iff u(p) = 1 .
\ee
For instance, let us assume that $\Hor$ is a connected submanifold
of $\M$ with topology\footnote{This is the case we will consider in Sec.~\ref{s:NE} and 
in the subsequent
ones, whereas all results up to Sec.~\ref{s:NE} are independent of
the topology of $\Hor$.}
$\mathbb{R}\times\mathbb{S}^2$.
Then we may
introduce locally a coordinate system of $\M$, $x^\alpha=(t,u,\theta,\varphi)$,
such that $t$ spans $\mathbb{R}$ and $(\theta,\varphi)$ are spherical 
coordinates spanning $\mathbb{S}^2$. $\Hor$ is then defined by
the coordinate condition $u=1$ [Eq.~(\ref{e:NH:r_const})] and 
an explicit form of the mapping $\Phi$ can be obtained by 
considering $x^A=(t,\theta,\varphi)$ as coordinates on the 3-manifold $\Hor_0$ :
\be
	\begin{array}{cccc}
	\Phi \ : & \Hor_0 & \longrightarrow & \M \\
		& (t,\theta,\varphi) & \longmapsto & (t,1,\theta,\varphi) .
	\end{array} 
\ee
In what follows, we identify $\Hor_0$ and $\Hor=\Phi(\Hor_0)$
(consequently, $\Phi$ can be seen as the inclusion map $\Phi: \Hor
\longrightarrow \M$).

The embedding $\Phi$ ``carries along'' curves in $\Hor$ to
curves in $\M$. Consequently it also ``carries along'' vectors on $\Hor$
to vectors on $\M$ (cf. Fig.~\ref{f:NH:embed}).
In other words, it defines a {\em push-forward}
mapping $\Phi_*$ between $\T_p(\Hor)$ and $\T_p(\M)$. Thanks to the adapted 
coordinate systems $x^\alpha=(t,u,\theta,\varphi)$, the push-forward 
mapping can be  explicited as follows
\be \label{e:NH:push_forward}
	\begin{array}{cccc}
	\Phi_* \ : & \T_p(\Hor) & \longrightarrow & \T_p(\M) \\
		& \w{v} = (v^t,v^\theta,v^\varphi) & \longmapsto & 
			\Phi_*\w{v} = (v^t,0,v^\theta,v^\varphi) ,
	\end{array} 
\ee
where $v^A=(v^t,v^\theta,v^\varphi)$ denotes the components of the vector $\w{v}$ with
respect to the natural basis $\partial/\partial x^A$ of $\T_p(\Hor)$
associated  with the coordinates $(x^A)$.

Conversely, the embedding $\Phi$ induces a {\em pull-back} mapping
$\Phi^*$  between the linear forms on $\T_p(\M)$ and those on $\T_p(\Hor)$
as follows
\be \label{e:NH:pull_back}
	\begin{array}{ccccccc}
	\Phi^* \ : & \T_p^*(\M) & \longrightarrow & \T_p^*(\Hor) & & & \\
		& \w{\omega}  & \longmapsto & 
			\Phi^*\w{\omega} \ :
				& \T_p(\Hor) & \rightarrow & \mathbb{R} \\
			& & & 	 & \w{v} & \mapsto & 
				\langle \w{\omega} , \Phi_*\w{v}\rangle .
	\end{array} 
\ee
Taking into account (\ref{e:NH:push_forward}), the pull-back mapping
can be explicited:
\be \label{e:NH:pull_back_comp}
	\begin{array}{cccc}
	\Phi^* \ : & \T_p^*(\M) & \longrightarrow & \T_p^*(\Hor)  \\
		& \w{\omega} = (\omega_t,\omega_u,\omega_\theta,\omega_\varphi)  
			& \longmapsto &
		 \Phi^*\w{\omega} = (\omega_t,\omega_\theta,\omega_\varphi) ,
	\end{array} 
\ee
where $\omega_\alpha$ denotes the components of the 1-form $\w{\omega}$ with
respect to the basis $\dd x^\alpha$ associated 
with the coordinates $(x^\alpha)$.
The pull-back operation can be extended to the multi-linear forms
on $\T_p(\M)$ in an obvious way: if $\w{T}$ is a $n$-linear form
on $\T_p(\M)$, $\Phi^*\w{T}$ is the $n$-linear form on $\T_p(\Hor)$ defined
by 
\be \label{e:NH:def_pull-back_multi}
	\forall (\w{v}_1,\ldots,\w{v}_n) \in \T_p(\Hor)^n,\quad
	\Phi^*\w{T}(\w{v}_1,\ldots,\w{v}_n) = 
	\w{T} (\Phi_* \w{v}_1,\ldots,\Phi_* \w{v}_n) .
\ee

\begin{rem} \label{rem:NH:emb_map}
By itself, the embedding $\Phi$ induces a mapping
from vectors on $\Hor$ to vectors on $\M$ (push-forward mapping $\Phi_*$)
and a mapping from 1-forms on $\M$ to 1-forms on $\Hor$
(pull-back mapping $\Phi^*$),
but not in the reverse way. For instance, one may define ``naively'' 
a reverse mapping $F:\; \T_p(\M) \longrightarrow \T_p(\Hor)$ by
$\w{v} = (v^t,v^u,v^\theta,v^\varphi) \longmapsto 
F\w{v} = (v^t,v^\theta,v^\varphi)$, but it would then depend on 
the choice of coordinates $(t,u,\theta,\varphi)$, which is not the 
case of the push-forward mapping defined by Eq.~(\ref{e:NH:push_forward}). 
For spacelike or timelike hypersurfaces, the
reverse mapping is unambiguously provided by the {\em orthogonal projector}
(with respect to the ambient metric $\w{g}$) onto the hypersurface. 
In the case of a null hypersurface, there is no such a thing as an
orthogonal projector, as we shall see below (Remark~\ref{rem:NH:no_ortho_proj}). 
\end{rem}

A very important case of pull-back operation is that of the bilinear
form $\w{g}$ (i.e. the spacetime metric), which defines the 
{\em induced metric on $\Hor$} : 
\be \label{e:NH:def_q}
	\encadre{\w{q} := \Phi^* \w{g} }
\ee
$\w{q}$ is also called the {\em first fundamental form of $\Hor$}.
In terms of the coordinate system\footnote{Let us recall that by
convention capital Latin indices run in $\{0,2,3\}$.}
$x^A=(t,\theta,\varphi)$ of $\Hor$, the components of $\w{q}$ are
deduced from (\ref{e:NH:pull_back_comp}):
\be \label{e:NH:qAB}
	q_{AB} = g_{AB} .
\ee

\subsection{Definition of a null hypersurface}

The hypersurface $\Hor$ is said to be {\em null} (or {\em lightlike}, 
or {\em characteristic} or to be a {\em wavefront}) if, and only if, the 
induced metric
$\w{q}$ is degenerate. This means if, and only if, there
exists a non-vanishing vector field $\el$ in $\T(\Hor)$ which is orthogonal
(with respect to $\w{q}$) to all vector fields in $\T(\Hor)$:
\be
	\forall \w{v}\in \T(\Hor),\quad
	\w{q}(\el,\w{v}) = 0 . 
\ee
The signature of $\w{q}$ is then necessarily $(0,+,+)$.
An equivalent definition of a null hypersurface demands
any vector field $\el$ in $\T(\M)$ which is normal to $\Hor$
[i.e. orthogonal to all vectors in $\T(\Hor)$] to be a null vector
with respect to the metric $\w{g}$:
\be \label{e:NH:ell_null}
	\encadre{ \w{g}(\el,\el) = \el\cdot\el = 0 }.
\ee
We adopt the same notation $\el$ than in the previous definition,
since this $\el$ is nothing but the pushed-forward by $\Phi_*$
of the $\el$ in  $\T(\Hor)$. Indeed, 
by saying that $\el$ is orthogonal 
to itself, Eq.~(\ref{e:NH:ell_null}) states that $\el$ is tangent to $\Hor$.
A distinctive property of null hypersurfaces is that their
normal vectors are both orthogonal and tangent to them. 

Since the hypersurface $\Hor$ is defined by a constant value of
the scalar field $u$ [Eq.~(\ref{e:NH:r_const})], the gradient 
1-form $\dd u$ is normal to $\Hor$, i.e.
\be \label{e:NH:dr_normal}
	\forall \w{v}\in\T(\M),\quad
	\w{v}\in\T(\Hor) \iff \langle \dd u, \w{v} \rangle = 0 .
\ee
As a side remark notice that,
in terms of the components $v^\alpha$ of $\w{v}$
with respect to the natural basis associated with the 
coordinates $(x^\alpha)$, $\langle \dd u, \w{v} \rangle = v^u$
and the above property is equivalent to
\be
	\forall \w{v}\in\T(\M),\quad
	\w{v}\in\T(\Hor) \iff v^u = 0 ,
\ee
which agrees with (\ref{e:NH:push_forward}). 
From (\ref{e:NH:dr_normal}), it is obvious that
the 1-form $\uel$ associated with the normal vector
$\el$ by the standard metric duality [cf. notation (\ref{e:IN:underbar})]
must be collinear to $\dd u$:
\be \label{e:NH:l_grad_r}
	\encadre{ \uel = \e^\rho \dd u }, 
\ee
where $\rho$ is some scalar field on $\Hor$. We have chosen the 
coefficient relating $\uel$ and $\dd u$ to be strictly positive, 
i.e. under the form of an exponential.
This is always possible by a suitable
choice of the scalar field $u$. 

The characterization of $\T_p(\Hor)$ as a hyperplane of the vector space
$\T_p(\M)$ can then be expressed as follows:
\be \label{e:NH:l_ortho_TH}
	\encadre{ \forall \w{v}\in\T_p(\M),\quad
	\w{v}\in\T_p(\Hor) \iff \langle \uel, \w{v} \rangle = 
	\el \cdot \w{v} = 0 }.
\ee

\begin{rem} \label{rem:NH:no_norm}
Since the scalar square of $\el$ is zero
[Eq.~(\ref{e:NH:ell_null})], there is no natural normalization
of $\el$, contrary to the case of spacelike hypersurfaces, 
where one can always choose the normal to be a unit vector
(scalar square equal to $-1$). Equivalently, there is no
natural choice of the factor $\rho$ in relation (\ref{e:NH:l_grad_r}).
In Sec.~\ref{s:IN}, we will use the extra-structure introduced
in $\M$ by the spacelike foliation of the 3+1 formalism to set
unambiguously the normalization of $\el$. 
\end{rem}

\begin{rem} \label{rem:NH:no_ortho_proj}
Another distinctive feature of null hypersurfaces, with respect
to spacelike or timelike ones, is the absence of orthogonal projector
onto them. This is a direct consequence of the fact that the normal $\el$
is tangent to $\Hor$. Indeed, suppose we
define ``naively'' $\w{\Pi} := \w{1} + a \el \langle \uel, . \rangle$ (or in index notation : $\Pi^\alpha_{\ \beta} := \delta^\alpha_{\ \beta}
+ a \ell^\alpha \ell_\beta$) as the ``orthogonal projector'' with some 
coefficient $a$ to be determined ($a=1$ for a spacelike hypersurface
and $a=-1$ for a timelike hypersurface, if $\el$ is the unit normal). 
Then it is true that for any $\w{v}\in\T_p(\Hor)$, $\w{\Pi}(\w{v}) = \w{v}$,
but if $\w{v}\not\in \T_p(\Hor)$, $\el\cdot \w{\Pi}(\w{v}) = \el\cdot\w{v}
\not = 0$, which shows that $\w{\Pi}(\w{v})\not\in \T_p(\Hor)$, hence
the endomorphism $\w{\Pi}$ is not a projector on $\T_p(\Hor)$, whatever the
value of $a$. This lack of orthogonal projector implies that there 
is no canonical way, from the null structure alone, to define a
mapping $\T_p(\M) \longrightarrow \T_p(\Hor)$ (cf. Remark~\ref{rem:NH:emb_map}). 
\end{rem}

\subsection{Auxiliary null foliation in the vicinity of $\Hor$}
\label{s:NH:extended_el}

The null normal vector field $\el$ is a priori defined on $\Hor$
only and not at points $p\not\in\Hor$. 
However within the 4-dimensional point of view adopted in this 
article, we would like to consider $\el$ as a vector field 
not confined to $\Hor$ but defined 
in some open subset of $\M$ around $\Hor$.
In particular this would permit to define the spacetime covariant
derivative $\w{\nabla}\el$, which is not possible if the 
support of $\el$ is restricted to $\Hor$. 
Following Carter \cite{Carte97}, a simple way to achieve
this is to consider not only a single null hypersurface $\Hor$, 
but a foliation of $\M$ (in the vicinity 
of $\Hor$) by a family of null hypersurfaces, such that $\Hor$ is an
element of this family. 
Without any loss of generality, 
we may select the scalar field $u$ to label these hypersurfaces and 
denote the family by $(\Hor_u)$. The null hypersurface $\Hor$
is then nothing but the element $\Hor = \Hor_{u=1}$ of this family 
[Eq.~(\ref{e:NH:r_const})]. 
The vector field $\el$ can then be viewed as defined in the part of $\M$
foliated by $(\Hor_u)$, such that at each point in this region, $\el$
is null and normal to $\Hor_u$ for some value of $u$. The identity (\ref{e:NH:l_grad_r})
is then valid for this ``extended'' $\el$, and $\rho$ is now a scalar
field defined not only on $\Hor$ but in the open region of $\M$ around $\Hor$
which is foliated by $(\Hor_u)$. 

Obviously the family $(\Hor_u)$ is non-unique but all geometrical 
quantities that we shall introduce hereafter do not depend upon the choice
of the foliation $\Hor_u$ once they are evaluated at $\Hor$.

\subsection{Frobenius identity} \label{s:NH:Frobenius}

The identity (\ref{e:NH:l_grad_r}) which expresses that the 1-form 
$\uel$ is normal to a hypersurface $u={\rm const}$, leads to a
particular form for the exterior derivative of $\uel$.
Indeed, taking the exterior derivative of (\ref{e:NH:l_grad_r})
(considering $\uel$ defined in a open neighborhood of $\Hor$ in $\M$,
cf. Sec.~\ref{s:NH:extended_el}) 
and applying rule (\ref{e:NH:ext_deriv}) (with  $\e^\rho$ = 0-form) leads to 
\be \label{e:NH:duel}
	\dd\uel = \e^\rho\dd\rho \wedge \dd u 
			+ \e^\rho \dd \dd u .
\ee 
Since $\dd \dd = 0$ is a basic property of the exterior derivative, 
the last term on the 
right-hand side of (\ref{e:NH:duel}) vanishes 
[this is also obvious by applying Eq.~(\ref{e:NH:d1form_comp})
to the 1-form $\dd u$].
Hence, after 
replacing $\dd u$ 
by $\e^{-\rho} \uel$, one is left with
\be \label{e:NH:Frobenius_l}
	\encadre{\dd\uel = \dd\rho \, \wedge \uel }.
\ee
This reflects the Frobenius theorem in its dual formulation (see e.g.
Theorem B.3.2 in Wald's textbook \cite{Wald84}): the exterior derivative of
the 1-form $\uel$ is the exterior product of $\uel$ itself with some
1-form ($\dd\rho$ in the present case) if, and only if, 
$\uel$ defines hyperplanes of $\T(\M)$ [by Eq.~(\ref{e:NH:l_ortho_TH})]
which are integrable in some hypersurface ($\Hor$ in the present case). 

\subsection{Generators of $\Hor$ and non-affinity coefficient $\kappa$}
\label{s:NH:generators}

Let us establish a fundamental property of null hypersurfaces:
they are ruled by null geodesics.
Contracting Eq.~(\ref{e:NH:Frobenius_l}) with $\el$ 
and using the fact that $\el$ is null, gives
\be \label{e:NH:l_dot_Frob}
	\el \cdot \dd \uel = \langle \dd\rho, \el \rangle \,\uel 
    - \underbrace{\langle \uel, \el \rangle}_{=0}
        \dd\rho = \langle \dd\rho, \el \rangle \,\uel .
\ee
On the other side if we express the exterior derivative $\dd \uel$ in terms of the 
covariant derivative $\w{\nabla}$ associated with the spacetime metric
$\w{g}$, the left-hand side of the above equation becomes
\be
	(\el \cdot \dd \uel)_\alpha  =  \ell^\mu \nabla_\mu \ell_\alpha
		- \ell^\mu \nabla_\alpha \ell_\mu 
		=  \ell^\mu \nabla_\mu \ell_\alpha
        = (\w{\nabla}_{\el}\, \uel)_\alpha ,
\ee
where we have used $\ell^\mu \nabla_\alpha \ell_\mu = 
1/2\; \nabla_\alpha(\ell^\mu \ell_\mu) = 0$. 
Hence Eq.~(\ref{e:NH:l_dot_Frob}) leads to
\be
	\w{\nabla}_{\el}\, \uel = \langle \dd\rho, \el \rangle 
	\, \uel,
\ee
or by the metric duality between 1-forms and vectors: 
\be \label{e:NH:der_l_kappa}
	\encadre{ \w{\nabla}_{\el} \, \el = \kappa \, \el } ,
\ee 
where $\kappa$ is the scalar field defined on $\Hor$ by
\be \label{e:NH:def_kappa}
	\encadre{ \kappa := \w{\nabla}_{\el} \, \rho  
        = \langle \dd\rho, \el \rangle } .
\ee
In the case where $\Hor$ is the horizon of a 
Kerr black hole, $\el$ can be normalized to become a
Killing vector of $(\M,\w{g})$, of the form
$\el = \w{\xi}_0 + \Omega_{\Hor} \w{\xi}_1$, where 
$\Omega_{\Hor}={\rm const}$ and $\w{\xi}_0$
and $\w{\xi}_1$ are the Killing vectors associated with respectively
the stationarity and axisymmetry of Kerr spacetime and normalized so
that the parameter length of $\w{\xi}_1$'s orbits is $2\pi$ and
$\w{\xi}_0$ asymptotically coincides with the 4-velocity of an inertial
observer. $\kappa$ is then called the
{\em surface gravity} of the black hole (see Appendix~\ref{s:KE} for
further details). 

Since Eq.~(\ref{e:NH:der_l_kappa}) involves only the derivative of $\el$
along $\el$, i.e. within $\Hor$, the definition of $\kappa$ is intrinsic
to $(\Hor,\el)$ and does not depend upon the choice of the auxiliary null 
foliation $(\Hor_u)$. 

\begin{figure}
\centerline{\includegraphics[width=0.5\textwidth]{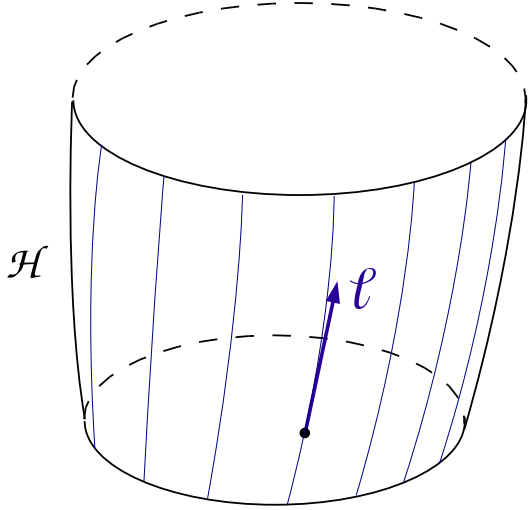}}
\caption[]{\label{f:NH:generators} 
Null hypersurface $\Hor$ with some null normal $\el$ and the
null generators (thin lines).}
\end{figure}
 
Equation~(\ref{e:NH:der_l_kappa}) means that $\el$ remains colinear to itself
when it is parallely transported along its field lines. 
This implies that these field lines are spacetime geodesics. 
Indeed, by a suitable
choice of the renormalization factor $\alpha$ such that 
$\el' = \alpha \el$,
Eq.~(\ref{e:NH:der_l_kappa}) can be brought to the classical
``equation of geodesics'' form:
\be \label{e:NH:geod_affine}
	\w{\nabla}_{\el'} \, \el' = 0 . 
\ee
This is immediate since 
\be \label{e:NH:der_lprime}
   \w{\nabla}_{\el'} \, \el' = \alpha \left[ \alpha \w{\nabla}_{\el} \, \el
   + \left( \w{\nabla}_{\el} \, \alpha \right) \el \right]
   = \alpha^2 \left( \kappa +  \w{\nabla}_{\el} \, \ln\alpha \right) \el  	
\ee
and one can choose $\alpha$ to get Eq.~(\ref{e:NH:geod_affine})
by requiring it to be a solution of the following first order differential 
equation along the field lines of $\el$
\be
	\w{\nabla}_{\el} \, \ln\alpha  = - \kappa .
\ee

If $\kappa\not = 0$, Eq.~(\ref{e:NH:der_l_kappa}) means that the parameter
$\tau$ associated with $\el$ by
$\ell^\alpha = dx^\alpha/d\tau$ 
is not an affine parameter of the geodesics. For this reason, we may
call $\kappa$ the {\em non-affinity coefficient}. 
Note that (\ref{e:NH:der_lprime}) gives the following scaling 
law for $\kappa$:
\be \label{e:NH:scale_kappa}
	\el \rightarrow \el'=\alpha\el \quad\Longrightarrow \quad
	\kappa \rightarrow \kappa' = \alpha \left( \kappa 
		+ \w{\nabla}_{\el} \, \ln\alpha \right) .
\ee

Having established that the field lines of $\el$ are geodesics, 
it is obvious that they are {\em null geodesics} (for $\el$ is null). 
They are called the {\em null generators} of $\Hor$. Note that whereas
$\el$ is not uniquely defined, being subject to the rescaling law
$\el \rightarrow \el'=\alpha\el$, the null generators, considered as 1-dimensional
curves in $\M$, are unique (see Fig.~\ref{f:NH:generators}).
In other words, they depend only upon $\Hor$. 

\begin{figure}
\centerline{\includegraphics[width=0.5\textwidth]{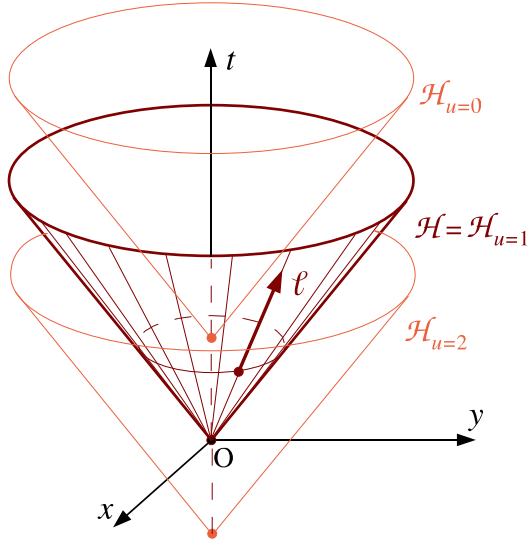}}
\caption[]{\label{f:NH:light_cone} 
Outgoing light cone in Minkowski spacetime. The null hypersurface $\Hor$ under
consideration is the member $u=1$ of the family $(\Hor_u)$ of light cones
emitted from the origin $(x,y,z)=(0,0,0)$ at successive times $t=1-u$.}
\end{figure}

\begin{exmp}        \label{ex:NH:cone}
\textbf{Outgoing light cone in Minkowski spacetime.}\\
The simplest example of a null hypersurface one may think of
is the light cone in Minkowski spacetime (Fig.~\ref{f:NH:light_cone}). 
More precisely, let us
consider for $\Hor$ the outgoing light cone from a given point $O$, 
excluding $O$ itself to keep $\Hor$ smooth. 
If $(x^\alpha)=(t,x,y,z)$ denote standard Minkowskian coordinates with
origin $O$, the scalar field $u$ defining $\Hor$ as the level set $u=1$ 
is then 
\be
   u(t,x,y,z) := r - t + 1 
   \quad \mbox{with} \quad
    r := \sqrt{x^2+y^2+z^2}.
\ee
Note that $u$ generates not only $\Hor$, but a full
null foliation $(\Hor_u)$ as the level sets of $u$
(cf. Sec.~\ref{s:NH:extended_el}). 
The member $\Hor_u$ of this foliation is then
nothing but the light cone emanating from the point $(-u+1,0,0,0)$
(cf. Fig.~\ref{f:NH:light_cone}).
In terms of components with respect to the 
coordinates $(x^\alpha)$, the gradient 1-form $\dd u$ is 
$\nabla_\alpha u = (-1, x/r, y/r, z/r)$. Hence, from Eq.~(\ref{e:NH:l_grad_r}),
the null normal to $\Hor$ is $\ell^\alpha = e^\rho(1, x/r, y/r, z/r)$.
For simplicity, let us select $\rho=0$. Then
\be \label{e:NH:ell_cone}
    \ell^\alpha = \left(1, \frac{x}{r}, \frac{y}{r}, \frac{z}{r} \right)
    \qquad \mbox{and} \qquad
    \ell_\alpha = \left(-1, \frac{x}{r}, \frac{y}{r}, \frac{z}{r} \right) .
\ee
The gradient bilinear form $\w{\nabla}\uel$ is easily computed
since, for the coordinates $(t,x,y,z)$, 
$\nabla_\beta \ell_\alpha= \partial \ell_\alpha / \partial x^\beta$:
\be \label{e:NH:nab_ell_cone}
    \nabla_\beta \ell_\alpha = \left(
    \begin{array}{cccc}
        0 & 0 & 0 & 0 \\
        0 & \frac{y^2+z^2}{r^3} & -\frac{xy}{r^3} & -\frac{xz}{r^3} \\
        0 & -\frac{xy}{r^3} & \frac{x^2+z^2}{r^3} & -\frac{yz}{r^3} \\
        0 & -\frac{xz}{r^3} & -\frac{yz}{r^3} & \frac{x^2+y^2}{r^3} 
    \end{array}
    \right) \qquad 
    \begin{array}{ll}
    ( & \mbox{\rm $\alpha$ = row index;} \\
      & \mbox{\rm $\beta$ = column index).}
      \end{array} 
\ee
We may check immediately on this formula that 
$\ell^\mu\nabla_\mu \ell_\alpha=0$, which leads to
\be \label{e:NH:kappa_cone}
    \kappa = 0 , 
\ee
in accordance with $\kappa = \w{\nabla}_{\el}\,\rho$ 
[Eq.~(\ref{e:NH:def_kappa})]
and our choice $\rho=0$. Actually it is easy to check that the coordinate
$t$ is an affine parameter
of the null geodesics generating $\Hor$ and that $\el$ is the associated 
tangent vector, hence the vanishing of the non-affinity coefficient $\kappa$.
\end{exmp}

\begin{figure}
\centerline{\includegraphics[width=0.7\textwidth]{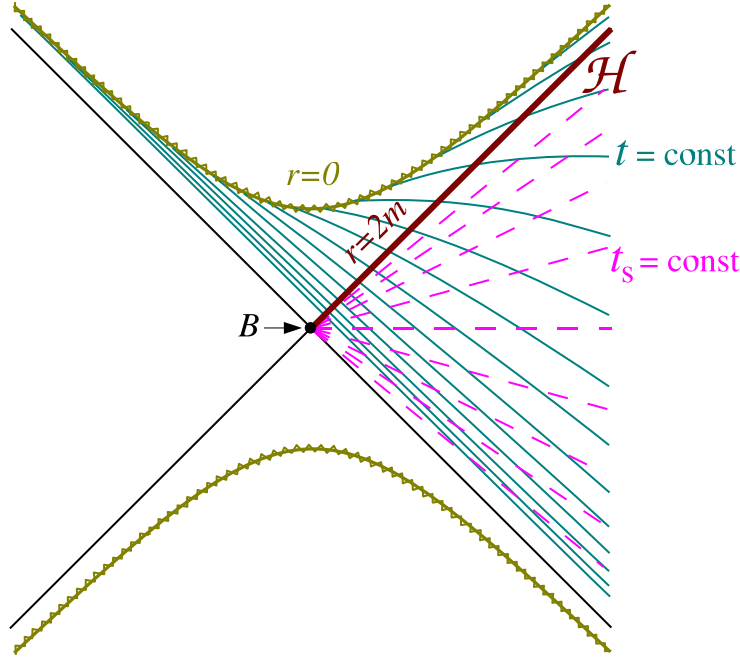}}
\caption[]{\label{f:NH:Schwarz_kruskal} 
Kruskal diagram representing the Schwarzschild spacetime; the hypersurfaces
of constant Schwarzschild time $t_{\rm S}$ (dashed lines)
do not intersect the future event
horizon $\Hor$, except at the bifurcation 2-sphere $B$ (reduced to 
a point in the figure), 
whereas the hypersurfaces of constant 
Eddington-Finkelstein time $t$ (solid lines) intersect it in such a way
that $t$ can be used as a regular coordinate on $\Hor$
(figure adapted from Fig.~3.1 of Ref.~\cite{Thornb93}).}
\end{figure}

\begin{exmp} \label{ex:NH:EF}
\textbf{Schwarzschild horizon in Eddington-Finkelstein coordinates.}\\
The next example of null surface one might think about is the (future)
event horizon of a Schwarzschild black hole. The corresponding spacetime
is often (partially) described by two sets of coordinates:
(i) the {\em Schwarzschild coordinates} $(t_{\rm S}, r, \theta,\varphi)$,
in which the metric components are given by 
\bea
    g_{\mu\nu} dx^\mu dx^\nu  &= &  - \left( 1 - \frac{2m}{r} \right) dt_{\rm S}^2 
    + \left( 1 - \frac{2m}{r} \right) ^{-1} dr^2 
    + r^2 (d\theta^2 + \sin^2\theta d\varphi^2) , \nonumber \\
    & &  
\eea
where $m$ is the mass of the black hole, 
and (ii) the {\em isotropic coordinates} $(t_{\rm S}, \tilde r, \theta,
\varphi)$, resulting in the metric components
\bea
    g_{\mu\nu} dx^\mu dx^\nu  &= &  - \left( 
    \frac{1 - \frac{m}{2\tilde r}}{1 + \frac{m}{2\tilde r}} \right) ^2
         dt_{\rm S}^2 
    + \left( 1 + \frac{m}{2\tilde r} \right) ^4 \left[ d{\tilde r}^2 
    + {\tilde r}^2 (d\theta^2 + \sin^2\theta d\varphi^2) \right] . \nonumber \\
    & &  
\eea
The relation between the two sets of coordinates is given by 
$r = {\tilde r} \left( 1 + \frac{m}{2\tilde r} \right) ^2$.
As it is well known, the above two coordinate systems are singular
at the event horizon $\Hor$, which corresponds to 
$r=2m$, $\tilde r = m/2$ and $t_{\rm S}\rightarrow+\infty$.
In particular the hypersurfaces of constant 
time $t_{\rm S}$, which constitutes a well known example of \emph{maximal
slicing} (cf. Sec.~\ref{s:FO}), do not intersect $\Hor$, except at 
a 2-sphere (named the \emph{bifurcation sphere}), where they also
cross each other (this is illustrated by
the Kruskal diagram in Fig.~\ref{f:NH:Schwarz_kruskal}).

A coordinate system, well known for being regular at $\Hor$, is constructed with
the \emph{ingoing Eddington-Finkelstein coordinates} $(V,r,\theta,\varphi)$,
where the coordinate $V$ is constant on each ingoing radial
null geodesic and is related to the Schwarzschild coordinate time $t_{\rm S}$
by $V = t_{\rm S} + r + 2m \ln\left| \frac{r}{2m} - 1\right|$. The coordinate
$V$ is null, but if we introduce
\be \label{ex:NH:t_EF}
    t := V - r = t_{\rm S} + 2m \ln\left| \frac{r}{2m} - 1\right|,
\ee
we get a timelike coordinate. The system $(t,r,\theta,\varphi)$
is called the \emph{3+1 Eddington-Finkelstein coordinates}. 
These coordinates are well behaved in the vicinity of $\Hor$, 
as shown in Fig.~\ref{f:NH:Schwarz_kruskal}, and yields to 
the following metric components:
\bea
    g_{\mu\nu} dx^\mu dx^\nu  &= &  - \left( 1 - \frac{2m}{r} \right) dt^2 
    + \frac{4m}{r} \, dt\, dr + \left( 1 + \frac{2m}{r} \right)  dr^2 
    \nonumber \\
   & &  + r^2 (d\theta^2 + \sin^2\theta d\varphi^2) 
                                        \label{e:NH:metric_edd_fink}. 
\eea
It is clear on this expression that the 3+1 Eddington-Finkelstein coordinates
are regular at the event horizon $\Hor$, which is located at $r=2m$
(cf. Fig.~\ref{f:NH:Schwarz_eddfink}). However, we cannot use
$u=r-2m+1$ for the scalar field defining $\Hor$, because the hypersurfaces
$r={\rm const}$ are not null, except for $r=2m$, whereas we have 
required in Sec.~\ref{s:NH:extended_el} all the hypersurfaces $u={\rm const}$
to be null. Actually,
a family of null hypersurfaces encompassing $\Hor$
is given by the constant values
of the outgoing Eddington-Finkelstein coordinate 
\be
    U = t_{\rm S} - r - 2m \ln\left| \frac{r}{2m} - 1\right|
     = t -r  - 4m \ln\left| \frac{r}{2m} - 1\right| .
\ee
The event horizon corresponds to $U= +\infty$; to get 
finite values, let us replace $U$ by the null Kruskal-Szekeres coordinate
(shifted by 1)
\be
    u := \pm \exp\left( -\frac{U}{4m} \right) + 1 ,
\ee
where the $+$ sign (resp. $-$ sign) is for $r\geq 2m$ (resp. $r<2m$). 
Then
\be \label{e:NH:u_EF}
    u = \left(\frac{r}{2m}-1\right) \exp\left( \frac{r-t}{4m} \right) + 1   
\ee
and all the hypersurfaces $\Hor_u$ defined by $u={\rm const}$ are 
null (there are drawn in Fig.~\ref{f:NH:Schwarz_eddfink}),
the event horizon $\Hor$ corresponding to $u=1$.

\begin{figure}
\centerline{\includegraphics[width=0.5\textwidth]{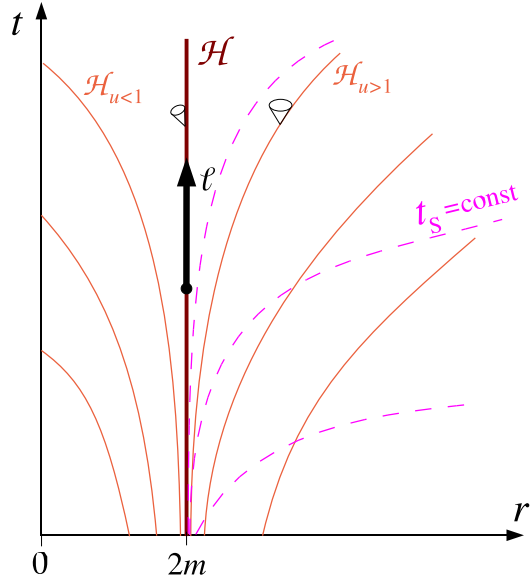}}
\caption[]{\label{f:NH:Schwarz_eddfink} 
Event horizon $\Hor$ of a Schwarzschild black hole in 3+1
Eddington-Finkelstein coordinates. 
The dashed lines represents the hypersurfaces
of constant Schwarzschild time $t_{\rm S}$ shown in 
Fig.~\ref{f:NH:Schwarz_kruskal}.}
\end{figure}

The null normals to $\Hor_u$ are deduced from the gradient of $u$
by $\ell_\alpha = \e^\rho \nabla_\alpha u$ [Eq.~(\ref{e:NH:l_grad_r})].
We get the following components with respect to the 3+1 Eddington-Finkelstein
coordinates $(x^\alpha)=(t,r,\theta,\varphi)$:
\be
    \ell^\alpha = \frac{1}{4m} \exp\left( \rho+\frac{r-t}{4m} \right)
    \, \left( 1+\frac{r}{2m},\ \frac{r}{2m}-1,\ 0,\ 0 \right) .
\ee
Let us choose $\rho$ such that $\ell^t=1$. Then
\be \label{e:NH:rho_EF}
    \rho = \frac{t-r}{4m} - \ln\left( 1+\frac{r}{2m} \right) 
    + \ln(4m) ,
\ee
\be \label{e:NH:comp_ell_EF}
    \ell^\alpha = \left( 1, \frac{r-2m}{r+2m}, 0, 0 \right) 
    \qquad\mbox{and}\qquad
    \ell_\alpha = \left( \frac{2m-r}{r+2m}, 1, 0, 0 \right) . 
\ee
Note that on the horizon, 
$\ell^\alpha \equalH (1,0,0,0)$, i.e. 
\be \label{e:NH:el_t_EF}
    \el \equalH \w{t} , 
\ee
where $\w{t}=\partial/\partial t$ is a Killing vector associated with
the stationarity of Schwarz\-schild solution.
The gradient of $\el$ is obtained by a straightforward computation,
after having evaluated the connection coefficients from the metric
components given by Eq.~(\ref{e:NH:metric_edd_fink}):
\bea
    \nabla_\beta \ell_\alpha  & = & \left(
    \begin{array}{cccc}
        \frac{m}{r^2}\frac{2m-r}{r+2m} & 
            \frac{m(3m^2+4mr-3r^2)}{r^2(r+2m)^2} & 0 & 0 \\
        \frac{m}{r^2} & \frac{m(3r+2m)}{r^2(r+2m)} & 0 & 0 \\
        0 & 0 & \frac{r(r-2m)}{r+2m} & 0 \\
        0 & 0 & 0 & \frac{(r-2m)r\sin^2\theta}{r+2m}
    \end{array}
    \right)             \label{e:NH:gradel_EF} \\
    & &  
    \mbox{\rm ($\alpha$ = row index;} \qquad 
     \mbox{\rm  $\beta$ = column index)} .  \nonumber
\eea
We deduce from these values that
\be
    \ell^\mu \nabla_\mu \ell_\alpha = 
    \left( \frac{4m}{(r+2m)^2}, \frac{4m(r-2m)}{(r+2m)^3}, 0, 0 \right) .
\ee
Comparing with the expression (\ref{e:NH:comp_ell_EF}) for $\ell_\alpha$,
we deduce the value of the non-affinity coefficient 
[cf. Eq.~(\ref{e:NH:der_l_kappa})]:
\be \label{e:NH:kappa_EF}
    \kappa = \frac{4m}{(r+2m)^2} .
\ee
As a check, we can recover $\kappa$ by means of formula
(\ref{e:NH:def_kappa}), evaluating $\w{\nabla}_{\el}\rho$
from expression (\ref{e:NH:rho_EF}) for $\rho$. Note that on the horizon,
\be \label{e:NH:kappaH_EF}
    \kappa \equalH \frac{1}{4m},
\ee
which is the standard value for the surface gravity
of a Schwarzschild black hole. 
\end{exmp}

\subsection{Weingarten map} \label{s:NH:Weingarten}

As for any hypersurface, the ``bending'' of $\Hor$ in $\M$ 
(also called {\em extrinsic curvature of $\Hor$}) is 
described by the {\em Weingarten map} (sometimes called
the {\em shape operator}), which is the endomorphism
of $\T_p(\Hor)$ which associates with each vector tangent to $\Hor$
the variation of the normal along that vector, with respect to
the spacetime connection $\w{\nabla}$:
\be \label{e:NH:Weingarten_def} \encadre{
	\begin{array}{cccc}
	\w{\chi}: & \T_p(\Hor) & \longrightarrow & \T_p(\Hor) \\
		& \w{v} & \longmapsto & \w{\nabla}_{\w{v}} \, \el 
	\end{array} }
\ee
This application is well defined (i.e. its image is in $\T_p(\Hor)$) since
\be
	\el\cdot \w{\chi}(\w{v}) = \el\cdot \w{\nabla}_{\w{v}} \, \el 
		= {1\over 2} \w{\nabla}_{\w{v}} (\el\cdot\el) = 0 ,
\ee
which shows that $\w{\chi}(\w{v}) \in \T_p(\Hor)$ 
[cf. Eq.~(\ref{e:NH:l_ortho_TH})]. 
Moreover, since it involves only the derivative of $\el$ along 
vectors tangent to $\Hor$, the definition of $\w{\chi}$ is clearly independent
of the choice of the auxiliary null foliation $(\Hor_u)$ introduced in 
Sec.~\ref{s:NH:extended_el}. 

\begin{rem}
The Weingarten map depends on the specific choice of the normal $\el$, 
in contrast with the timelike or
spacelike case, where the unit length of the normal
fixes it unambiguously. 
Indeed a rescaling of $\el$ acts as follows on $\w{\chi}$:
\be \label{e:NH:scale_chi}
	\el \rightarrow \el'=\alpha\el \quad\Longrightarrow \quad
	\w{\chi} \rightarrow \w{\chi}' = \alpha \w{\chi}
		+ \langle \dd \alpha , \cdot \rangle \, \el , 
\ee
where the notation $\langle \dd \alpha , \cdot \rangle \, \el$
stands for the endomorphism $\T_p(\Hor) \longrightarrow \T_p(\Hor)$, 
$\w{v} \longmapsto \langle \dd \alpha , \w{v} \rangle \, \el$.
\end{rem}

The fundamental property of the Weingarten map is to be {\em self-adjoint}
with respect to the metric $\w{q}$ [i.e. the pull-back of $\w{g}$ on $\T(\Hor)$,
cf. Eq.~(\ref{e:NH:def_q})]:
\be
	\encadre{ \forall (\w{u},\w{v}) \in \T(\Hor)\times\T(\Hor),\quad
	\w{u} \cdot \w{\chi}(\w{v}) = \w{\chi}(\w{u})\cdot \w{v} },
\ee
where the dot means the scalar product with respect to $\w{q}$ [considering
$\w{u}$ and $\w{v}$ as vectors of $\T(\Hor)$] or
$\w{g}$ [considering
$\w{u}$ and $\w{v}$ as vectors of $\T(\M)$].
Indeed, one obtains from the definition of $\w{\chi}$
\bea
	\w{u}\cdot\w{\chi}(\w{v}) & = & 
		\w{u}\cdot \w{\nabla}_{\w{v}} \, \el = 
                 \w{\nabla}_{\w{v}}\, (\w{u}\cdot\el) -
			\el\cdot \w{\nabla}_{\w{v}} \, \w{u} =
            0 - \el\cdot \left(   \w{\nabla}_{\w{u}} \, \w{v}
			- [\w{u},\w{v}] \right) \nonumber \\
		& = & - \w{\nabla}_{\w{u}} \, (\el\cdot \w{v})
			+ \w{v} \cdot \w{\nabla}_{\w{u}} \, \el
			+ \el \cdot [\w{u},\w{v}] \nonumber \\
            &= & 0 + \w{v}\cdot \w{\chi}(\w{u}) 
		+ \el \cdot [\w{u},\w{v}] , \label{e:NH:u_dot_chi_v}
\eea
where use has been made of $\el\cdot\w{u}=0$ and $\el\cdot\w{v}=0$.
Now the Frobenius theorem states that the commutator $[\w{u},\w{v}]$
of two vectors of the hyperplane $\T(\Hor)$ belongs to $\T(\Hor)$
since $\T(\Hor)$ is surface-forming (see e.g. Theorem B.3.1 in Wald's textbook
\cite{Wald84}). It is straightforward to establish it: 
\bea
	\el \cdot [\w{u},\w{v}] & = & \langle \uel , [\w{u},\w{v}] \rangle
			= \ell_\mu u^\nu \nabla_\nu v^\mu
			- \ell_\mu v^\nu \nabla_\nu u^\mu 
                       =  - u^\nu v^\mu \nabla_\nu \ell_\mu
			+ v^\nu u^\mu \nabla_\nu \ell_\mu \nonumber \\
                       & = & (\nabla_\mu \ell_\nu - \nabla_\nu \ell_\mu)
			 u^\nu v^\mu = 
                 ({\rm d}\underline{\ell})_{\mu\nu} \, u^\nu v^\mu  
                 =  (\nabla_\mu \rho \; \ell_\nu
			-  \nabla_\nu \rho \; \ell_\mu)
			u^\nu v^\mu \nonumber \\
	\el \cdot [\w{u},\w{v}]	& = & 0 , \label{e:NH:Frobenius_commut}
\eea
where use has been made of expression (\ref{e:NH:Frobenius_l})
for the exterior derivative of $\uel$ and the last equality
results from $\ell_\nu u^\nu = 0$ and $\ell_\mu v^\mu = 0$.
Inserting (\ref{e:NH:Frobenius_commut}) into (\ref{e:NH:u_dot_chi_v})
establishes the self-adjointness of the Weingarten map. 

Let us note that the non-affinity coefficient $\kappa$ is an eigenvalue of the
Weingarten map, corresponding to the eigenvector $\w{\el}$,
since Eq.~(\ref{e:NH:der_l_kappa}) can be written
\be \label{e:NH:l_eigen_chi}
	\encadre{ \w{\chi}(\el) = \kappa \el }. 
\ee

\subsection{Second fundamental form of $\Hor$} \label{s:NH:2ndfund}

The self-adjointness of $\w{\chi}$ implies that the 
bilinear form defined on $\Hor$'s tangent space by
\be \label{e:NH:2ndform_def}
	\encadre{
	\begin{array}{cccc}
	\w{\Theta}: & \T_p(\Hor)\times\T_p(\Hor) & \longrightarrow & \mathbb{R} \\
		& (\w{u},\w{v}) & \longmapsto & \w{u} \cdot \w{\chi}(\w{v})
	\end{array} }
\ee
is symmetric. 
It is called the {\em second fundamental form of
$\Hor$ with respect to $\el$}.
Note that $\w{\Theta}$ could have been defined for any vector field $\el$,
but it is symmetric only because $\el$ is normal to some
hypersurface (since the self-adjointness of $\w{\chi}$
originates from this last property). 
If we make explicit the value of $\w{\chi}$ in the 
definition (\ref{e:NH:2ndform_def}), we get 
[see Eq.~(\ref{e:IN:direc_deriv_1form})]
\bea
  \forall (\w{u},\w{v}) \in \T_p(\Hor)\times\T_p(\Hor),\quad
  \w{\Theta}(\w{u},\w{v}) & = & \w{u} \cdot\w{\chi}(\w{v})
   =  \w{u} \cdot\w{\nabla}_{\w{v}}\el = \langle   
    \w{\nabla}_{\w{v}}\uel, \w{u} \rangle \nonumber \\
    & = & \w{\nabla}\uel (\w{u},\w{v}) , 
\eea 
from which we conclude that $\w{\Theta}$ is nothing but the
pull-back of the bilinear form $\w{\nabla}\uel$ onto $\T_p(\Hor)$,
pull-back induced by the embedding $\Phi$ of $\Hor$ in $\M$ 
[cf. Eq.~(\ref{e:NH:def_pull-back_multi})]:
\be \label{e:NH:Theta_pb_grad_uel}
    \encadre{ \w{\Theta} = \Phi^* \w{\nabla}\uel } . 
\ee
It is worth to note that although the bilinear form $\w{\nabla}\uel$
is a priori not symmetric on $\T_p(\M)$, its pull-back $\w{\Theta}$
on $\T_p(\Hor)$ is symmetric, as a consequence of the hypersurface-orthogonality
of $\el$ (which yields the self-adjointness of $\w{\chi}$). 

The bilinear form $\w{\Theta}$ is degenerate, with a degeneracy
direction along $\el$ (as the
first fundamental form $\w{q}$), since
\be \label{e:NH:Theta_degen_l}
	\forall \w{v} \in \T_p(\Hor),\quad
	\w{\Theta}(\el,\w{v}) = \w{v}\cdot \w{\chi}(\el)
			= \kappa \w{v}\cdot\el = 0 . 
\ee 

\begin{rem}
As for $\w{\chi}$,  $\w{\Theta}$ depends on the choice of 
the normal $\el$. However its transformation under
a rescaling of $\el$ is simpler than that of $\w{\chi}$:
from Eq.~(\ref{e:NH:scale_chi}) and the orthogonality of
$\el$ with respect to $\T_p(\Hor)$, we get
\be \label{e:NH:scale_Theta}
	\el \rightarrow \el'=\alpha\el \quad\Longrightarrow \quad
	\w{\Theta} \rightarrow \w{\Theta}' = \alpha \w{\Theta} .
\ee
\end{rem}

\begin{rem} \label{r:NH:quotient}
To get rid of the dependence upon the normalization of $\el$
in the definitions of $\w{\chi}$ and $\w{\Theta}$, some 
authors \cite{Kupel87,Gallow00,Gallow04,Jezie04} introduce the
following equivalence class $\mathcal R$ on $\T_p(\Hor)$: $\w{u} \sim \w{v}$
iff $\w{u}$ and $\w{v}$ differ only by a vector collinear to $\el$.
Then the Weingarten map and 
the second fundamental form can be defined as unique geometric objects
in the quotient space $\T_p(\Hor)/{\mathcal R}$. However we do not 
adopt such an approach here because we plan to use some spacetime slicing by
spacelike hypersurfaces (the so-called \emph{3+1 formalism}) to fix in 
a natural way the normalization of $\el$, as we shall see in Sec.~\ref{s:IN}. 
\end{rem}

%% file: foliat.tex
%
%
\section{3+1 formalism}
\label{s:FO}

\subsection{Introduction}

The 3+1 formalism of general relativity
is aimed at reducing the resolution of Einstein equation
to a Cauchy problem, namely (coordinate) time evolution from initial
data specified on a given spacelike hypersurface. 
This formalism originates in the works of Lichnerowicz (1944) \cite{Lichn44},
Choquet-Bruhat (1952) \cite{Foure52}, Arnowitt, Deser \& Misner (1962)
\cite{ArnowDM62} and has many applications, in particular in 
numerical relativity.  
We refer the reader to York's seminal
article \cite{York79} for an introduction to the 3+1 formalism
and to Baumgarte \& Shapiro \cite{BaumgS03} for 
a recent review of applications in numerical relativity.  
Here we simply recall the most relevant features of the 3+1 formalism
which are necessary for our purpose.

\begin{figure}
\centerline{\includegraphics[width=0.5\textwidth]{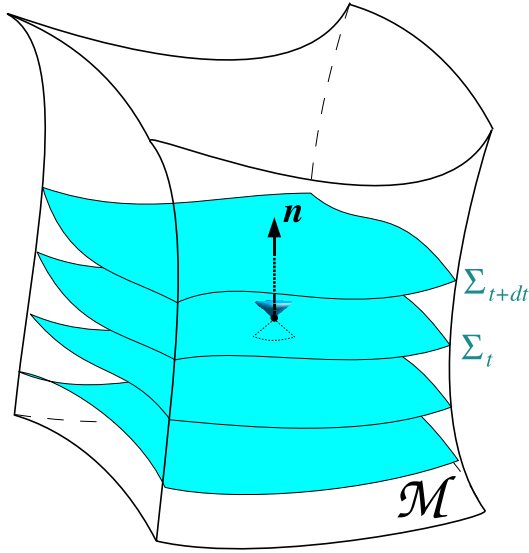}}
\caption[]{\label{f:NH:3p1slicing} 
Spacetime foliation by a family of spacelike hypersurfaces
$\Sigma_t$. The $\Sigma_t$'s can be considered as the level sets of 
some smooth scalar field $t$, such that the gradient $\dd t$ is
timelike.}
\end{figure}


\subsection{Spacetime foliation $\Sigma_t$} \label{s:FO:foliation}

The spacetime (or at least the part of it under study, in the vicinity
of the null hypersurface $\Hor$)
is supposed to be foliated by a continuous family of spacelike 
hypersurfaces $(\Sigma_t)$,
labeled by the time coordinate $t$
(Fig.~\ref{f:NH:3p1slicing}). The $\Sigma_t$'s can be considered as the level 
sets of some smooth scalar field $t$, such that the gradient $\dd t$ is
timelike.
We denote by $\w{n}$ the future directed timelike unit vector normal to 
$\Sigma_t$. It can be identified with the 4-velocity of the class of 
observers whose worldlines are orthogonal to $\Sigma_t$ {\em (Eulerian
observers)}. 
By definition the 1-form $\underline{\w{n}}$ dual to $\w{n}$ 
[cf. notation (\ref{e:IN:underbar})]
is parallel to the gradient of the scalar field $t$:
\be \label{e:FO:def_n}
	\encadre{ \underline{\w{n}} = - N \, \dd t} . 
\ee
The proportionality factor $N$ is called the {\em lapse function}.
It ensures that $\w{n}$ satisfies the normalization relation
\be
	\w{n} \cdot \w{n} = \langle \underline{\w{n}}, \w{n} \rangle = - 1  . 
\ee

The metric $\w{\gamma}$ induced by $\w{g}$ on each hypersurface $\Sigma_t$
({\em first fundamental form} of $\Sigma_t$)
is given by 
\be \label{e:FO:def_gamma}
	\encadre{
	\w{\gamma} = \w{g} + \underline{\w{n}} \otimes \underline{\w{n}} }.
\ee 
Since $\Sigma_t$ is assumed to be spacelike, $\w{\gamma}$ is
a positive definite (i.e. Riemannian) metric.  
Let us stress that the writing (\ref{e:FO:def_gamma})
is fully 4-dimensional and does not restrict the definition of $\w{\gamma}$
to $\T_p(\Sigma_t)$: it is a bilinear form on $\T_p(\M)$. 
The endomorphism $\T_p(\M)\rightarrow\T_p(\M)$ canonically associated 
with the bilinear form $\w{\gamma}$ by the metric $\w{g}$ [cf. notation
(\ref{e:IN:arrow_endo})] is the {\em orthogonal projector onto $\Sigma_t$}:
\be \label{e:NH:gamma_ortho_proj}
    \vec{\w{\gamma}} = \w{1} + \langle \underline{\w{n}} , . \rangle
    \w{n}
\ee
(in index notation: 
$\gamma^\alpha_{\ \, \beta} = \delta^\alpha_{\ \, \beta}
+ n^\alpha n_\beta$, whereas (\ref{e:FO:def_gamma}) writes
$\gamma_{\alpha\beta} = g_{\alpha\beta}
+ n_\alpha n_\beta$).

The existence of the orthogonal projector $\vec{\w{\gamma}}$ makes
a great difference with the case of null hypersurfaces, for which 
such an object does not exist (cf. Remark~\ref{rem:NH:no_ortho_proj}).
In particular we can use it to map any multilinear form on 
$\T_p(\Sigma_t)$ into a multilinear form on $\T_p(\M)$, 
which is in the direction inverse of that of the pull-back 
mapping induced by the embedding of $\Sigma_t$
in $\M$. We denote this mapping 
$\T_p^*(\Sigma_t)\rightarrow \T_p^*(\M)$
by $\vec{\w{\gamma}}^*$
and make it explicit as follows: given a $n$-linear form $\w{A}$
on $\T_p(\Sigma_t)$, $\vec{\w{\gamma}}^*\w{A}$ is the $n$-linear form
acting on $\T_p(\M)^n$ defined by
\be \label{e:FO:def_gamma_star}
	\encadre{
	\begin{array}{cccc}
	\vec{\w{\gamma}}^*\w{A} : & \T_p(\M)^n & \longrightarrow & \mathbb{R} \\
		& (\w{v}_1,\ldots,\w{v}_n) & \longmapsto & 
		\w{A} \left( \vec{\w{\gamma}}(\w{v}_1),\ldots,
                \vec{\w{\gamma}}(\w{v}_n) \right) .
	\end{array} } 
\ee
Actually we extend the above definition to all multilinear forms $\w{A}$
on $\T_p(\M)$ and not only those restricted to $\T_p(\Sigma_t)$. 
The index version of this definition is
\be \label{e:FO:def_gamma_star_index}
    (\vec{\gamma}^* A)_{\alpha_1\ldots\alpha_n}
    	= A_{\mu_1\ldots \mu_n}
	\gamma^{\mu_1}_{\ \ \ \alpha_1} \cdots \gamma^{\mu_n}_{\ \ \ \alpha_n}. 	
\ee
In particular, we have
\be
    \vec{\w{\gamma}}^* \w{g} = \w{\gamma} \qquad \mbox{and}
    \qquad \vec{\w{\gamma}}^* \underline{\w{n}} = 0 . 
\ee

There exists a unique (torsion-free) connection on $\Sigma_t$ associated 
with the metric $\w{\gamma}$, which we denote by $\w{D}$:
$\w{D}\w{\gamma}=0$. If we consider a generic tensor field $\w{T}$ 
of type $\left({p\atop q}\right)$ lying on 
$\Sigma_t$ (i.e. such that its contraction with the normal $\w{n}$
on any of its indices vanishes), then, from 
a 4-dimensional point of view, the covariant derivative
$\w{D}\w{T}$ can be expressed as the full orthogonal projection 
of the spacetime covariant derivative $\w{\nabla}\w{T}$ on
$\Sigma_t$ [see Eq.~(\ref{e:IN:cov_der_comp})]:
\be \label{e:FO:3der}
\encadre{ D_\gamma T^{\alpha_1\ldots\alpha_p}_{\ \qquad\beta_1\ldots\beta_q}
		= \gamma_{\ \ \, \mu_1}^{\alpha_1} \, \cdots 
		 \gamma_{\ \ \, \mu_p}^{\alpha_p} \,
		  \gamma_{\ \ \, \beta_1}^{\nu_1} \, \cdots
		  \gamma_{\ \ \, \beta_q}^{\nu_q} \,
		  \gamma_{\ \ \, \gamma}^{\sigma} \, \nabla_\sigma
		  T^{\mu_1\ldots\mu_p}_{\ \qquad\nu_1\ldots\nu_q} } .		  	
\ee  
In the following, we shall make extensive use of this formula,
without making explicit mention. 
In the special case of a tensor of type $\left({0\atop q}\right)$,
i.e. a multilinear form, the definition of $\w{D}\w{T}$ amounts
to, thanks to Eq.~(\ref{e:FO:def_gamma_star_index}), 
\be \label{e:FO:DT_gamma_star_nabT}
    \w{D}\w{T} =  \vec{\w{\gamma}}^* \w{\nabla} \w{T} . 
\ee  


\subsection{Weingarten map and extrinsic curvature}

As for the hypersurface $\Hor$, the ``bending'' of each hypersurface
$\Sigma_t$ in $\M$ is described by the {\em Weingarten map} which 
associates with each vector tangent to $\Sigma_t$ the covariant 
derivative (with respect to the ambient connection $\w{\nabla}$)
of the unit normal $\w{n}$ along this vector  
[compare with Eq.~(\ref{e:NH:Weingarten_def})]:
\be
	\begin{array}{cccc}
	\w{\mathcal{K}}: & \T_p(\Sigma_t) & \longrightarrow & \T_p(\Sigma_t) \\
		& \w{v} & \longmapsto & \w{\nabla}_{\w{v}} \, \w{n} .
	\end{array} 
\ee
The computations presented in Sec.~\ref{s:NH:Weingarten} for the
Weingarten map $\w{\chi}$ of $\Hor$ can be repeated 
here\footnote{Indeed the
computations in Sec.~\ref{s:NH:Weingarten} did not make use of
the fact that $\Hor$ is null, i.e. that 
$\el$ is tangent to $\Hor$.}, by simply
replacing the normal $\el$ by the normal $\w{n}$, the field $u$
by the field $t$ and the coefficient 
$\e^\rho$ by $-N$ [compare Eqs.~(\ref{e:NH:l_grad_r})
and (\ref{e:FO:def_n})]. They then show that $\w{\mathcal{K}}$
is well defined (i.e. its image is in $\T_p(\Sigma_t)$) and that it is 
self-adjoint with respect to the metric $\w{\gamma}$. 
A difference with the Weingarten map $\w{\chi}$ of $\Hor$
is that the Weingarten map $\w{\mathcal{K}}$ can be naturally
extended to $\T_p(\M)$ thanks to the orthogonal projector 
$\vec{\w{\gamma}}$ [Eq.~(\ref{e:NH:gamma_ortho_proj})],
which did not exist for $\Hor$ 
(cf. Remark~\ref{rem:NH:no_ortho_proj}), by setting
\be
	\begin{array}{cccc}
	\w{\mathcal{K}}: & \T_p(\M) & \longrightarrow & \T_p(\Sigma_t) \\
		& \w{v} & \longmapsto & 
                \w{\nabla}_{\vec{\w{\gamma}}(\w{v})} \, \w{n} ,
	\end{array} 
\ee
or in index notation:
\be \label{e:FO:calK_index}
    \mathcal{K}^\alpha_{\ \, \beta} = \nabla_\mu n^\alpha 
        \; \gamma^\mu_{\ \, \beta} .
\ee
We then define the {\em extrinsic curvature tensor} $\w{K}$ of the 
hypersurface $\Sigma_t$ as minus the second fundamental form
[compare with Eq.~(\ref{e:NH:2ndform_def})]: 
\be \label{e:FO:def_K}
	\begin{array}{cccc}
	\w{K}: & \T_p(\M)\times\T_p(\M) & \longrightarrow & \mathbb{R} \\
		& (\w{u},\w{v}) & \longmapsto & - \w{u} 
                \cdot \w{\mathcal{K}}(\w{v}) ,
	\end{array} 
\ee
or in index notation
\be \label{e:NH:Kab_proj}
    K_{\alpha\beta} = - \nabla_\mu n_\alpha 
        \; \gamma^\mu_{\ \, \beta} .
\ee
Since the image of $\w{\mathcal{K}}$ is in $\T_p(\Sigma_t)$, we can write
$\w{K}(\w{u},\w{v}) =  - \, \vec{\w{\gamma}}(\w{u}) \cdot 
\w{\mathcal{K}}(\vec{\w{\gamma}}(\w{v}))$. It follows then immediately
from the self-adjointness of $\w{\mathcal{K}}$ that $\w{K}$
is symmetric and that the following relation holds:
\be \label{e:FO:Kab_n}
    K_{\alpha\beta} = - \nabla_\mu n_\nu 
        \; \gamma^\mu_{\ \, \alpha} \gamma^\nu_{\ \, \beta} ,
\ee
which we can write, thanks to Eq.~(\ref{e:FO:def_gamma_star_index})
and the symmetry of $\w{K}$, 
\be
    \encadre{ \w{K} = - \, \vec{\w{\gamma}}^* \w{\nabla} \underline{\w{n}} } . 
\ee
Replacing in Eq.~(\ref{e:NH:Kab_proj}) 
$n_\alpha$ by its expression (\ref{e:FO:def_n})
in terms of the gradient of $t$ leads to
\begin{eqnarray}
    K_{\alpha\beta} & = & \nabla_\mu( N \nabla_\alpha t)
        \; \gamma^\mu_{\ \, \beta}  
     =  \left( \nabla_\mu N \nabla_\alpha t  
    + N \nabla_\mu \nabla_\alpha t \right) \gamma^\mu_{\ \, \beta} \nonumber \\
    & = & \left( \nabla_\mu N \nabla_\alpha t  
    + N \nabla_\alpha \nabla_\mu t \right) \gamma^\mu_{\ \, \beta} 
     =   D_\beta N \nabla_\alpha t  
    + N \nabla_\alpha \left( - N^{-1} n_\mu \right) \gamma^\mu_{\ \, \beta} 
            \nonumber \\
    & = & - n_\alpha N^{-1} D_\beta N 
        - N \nabla_\alpha (N^{-1}) \; 
        \underbrace{n_\mu \gamma^\mu_{\ \, \beta}}_{=0}
        - \nabla_\alpha n_\mu \gamma^\mu_{\ \, \beta}
            \nonumber \\
    & = & - n_\alpha D_\beta \ln N - \nabla_\alpha n_\mu \, \delta^\mu_{\ \, \beta}
        - \underbrace{\nabla_\alpha n_\mu \, n^\mu}_{=0} n_\beta .
\end{eqnarray}
Hence
\be \label{e:FO:K_grad_n_index}
    \encadre{ K_{\alpha\beta} = - \nabla_\alpha n_\beta
        - n_\alpha \, D_\beta \ln N } .
\ee
or, 
taking into account the symmetry of $\w{K}$ and Eq.~(\ref{e:IN:grad_1form})
\be \label{e:FO:K_grad_n}
    \encadre{ \w{K} = - \w{\nabla}\underline{\w{n}}
        - \w{D} \ln N \otimes \underline{\w{n}}  } .
\ee
In the following, we will make extensive use of this formula,
without explicitly mentioning it. Inserting $\w{n}$ as the second
argument in the bilinear form (\ref{e:FO:K_grad_n}) and using 
$\w{K}(.,\w{n}) = 0$ as well as 
$\langle \underline{\w{n}},\w{n}\rangle=-1$ results in the important 
formula giving the 4-acceleration of the Eulerian observers:
\be \label{e:FO:acceler_n}
    \encadre{\w{\nabla}_{\w{n}} \, \underline{\w{n}} = \w{D} \ln N }.
\ee
Another useful formula relates $\w{K}$ to the 
Lie derivative of 
the spatial metric $\w{\gamma}$ along the normal $\w{n}$:
\be
\label{e:FO:K_Lie_n}
    \encadre{\w{K} = - {1\over 2} \Lie{\w{n}} \w{\gamma} }.
\ee
This formula follows from Eq.~(\ref{e:FO:K_grad_n}) and the 
symmetry of $\w{K}$ by a direct computation, provided that the Lie derivative
along $\w{n}$ is expressed in terms of the connection $\w{\nabla}$
via Eq.~(\ref{e:IN:Lie_derivative}):
$(\Liec{\w{n}} \gamma)_{\alpha\beta} = n^\mu \nabla_\mu \gamma_{\alpha\beta}
    + \gamma_{\mu\beta} \nabla_\alpha n^\mu 
    + \gamma_{\alpha\mu} \nabla_\beta n^\mu$.

\begin{figure}
\centerline{\includegraphics[width=0.7\textwidth]{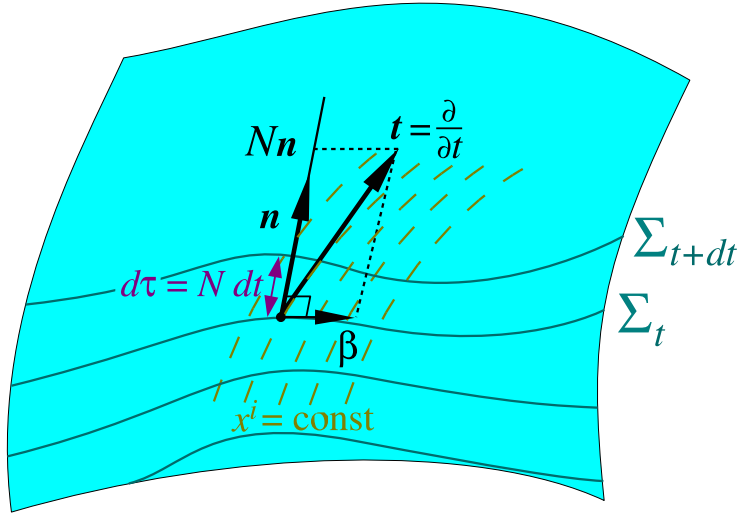}}
\caption[]{\label{f:FO:lapse_shift} 
Constant spatial coordinate lines $x^i={\rm const}$ cutting across
the foliation  $(\Sigma_t)$ and defining the coordinate time vector
$\tv$ and the shift vector $\w{\beta}$. Also represented are the 
the unit normal to each hypersurface $\Sigma_t$, $\w{n}$,
and the lapse function $N$
as giving the metric distance $d\tau$ between two neighboring hypersurfaces
$\Sigma_t$ and $\Sigma_{t+dt}$ via $d\tau = N \, dt$.}
\end{figure}


\subsection{3+1 coordinates and shift vector} \label{s:FO:coord}

We may introduce on $\M$ a coordinate system adapted to the $(\Sigma_t)$
foliation by considering on each hypersurface
$\Sigma_t$ a coordinate system $(x^i)$, such that $(x^i)$ varies smoothly 
from one hypersurface to the next one. Then,
$(x^\alpha) = (x^0 = t,x^i)$ constitutes a well behaved coordinate
of $\M$. 
The {\em coordinate time vector} of this system is 
\be
    \tv := \der{}{t}   
\ee
and is such that each spatial coordinate $x^i$ is constant along its
field lines. 
$\tv$ can be seen as a vector ``dual'' to the gradient 1-form $\dd t$,
in the sense that 
\be
    \langle \dd t, \tv \rangle = 1 . 
\ee
Then, from Eq.~(\ref{e:FO:def_n}), $\w{n}\cdot\tv = -N$ and we have the
orthogonal 3+1 decomposition 
\be \label{e:FO:def_shift}
    \encadre{\tv = N \w{n} + \w{\beta}} \qquad \mbox{with} \quad 
        \w{n}\cdot\w{\beta} = 0 . 
\ee
The vector $\w{\beta} := \vec{\w{\gamma}}(\tv)$ is called the
{\em shift vector} of the coordinate system $(x^\alpha)$.
The vectors $\tv$ and $\w{\beta}$ are represented in 
Fig.~\ref{f:FO:lapse_shift}. Given a choice of the coordinates
$(x^i)$ in an initial slice $\Sigma_0$, fixing the lapse
function $N$ and shift vector $\w{\beta}$ on every $\Sigma_t$ fully 
determines the coordinates $(x^\alpha)$ in the portion
of $\M$ covered by these coordinates.
We refer the reader to Ref.~\cite{SmarrY78b} for an extended discussion
of the choice of coordinates based on the lapse and the shift. 

The components $g_{\alpha\beta}$ of the metric tensor $\w{g}$
with respect to the coordinates $(t,x^i)$ are expressible in terms of
the lapse $N$, the components $\beta^i$ of the shift vector and the components
$\gamma_{ij}$ of the spatial metric, according to
\be \label{e:FO:g_N_beta_gam}
	g_{\mu\nu} \, dx^\mu\, dx^\nu
	= - N^2 dt^2 + \gamma_{ij} (dx^i + \beta^i dt)
		(dx^j + \beta^j dt) . 
\ee

\begin{exmp} \label{ex:FO:EF}
\textbf{Lapse and shift of Eddington-Finkelstein coordinates.}\\
Returning to Example~\ref{ex:NH:EF} (Schwarzschild spacetime in 
Eddington-Finkelstein 
coordinates), the lapse function and shift vector of the 3+1 
Eddington-Finkel\-stein coordinates are obtained by comparing 
Eqs.~(\ref{e:FO:g_N_beta_gam}) and (\ref{e:NH:metric_edd_fink}):
\be \label{e:FO:lapse_EF}
    N = \frac{1}{\sqrt{1+2m/r}} ,
\ee
\be \label{e:FO:shift_EF}
    \beta^\alpha = \left( 0, \frac{1}{1+r/(2m)}, 0, 0 \right)
    \quad \mbox{and} \quad
    \beta_\alpha = \left( \frac{4m^2}{r(r+2m)}, \frac{2m}{r},0,0\right) . 
\ee
Note that, on $\Hor$ ($r=2m$), $N\equalH 1/\sqrt{2}$ and $\beta^r\equalH 1/2$.
The expression for the unit timelike normal to the hypersurfaces $\Sigma_t$
is deduced from $N$ and $\w{\beta}$:
\be \label{e:FO:n_EF}
    n^\alpha  = \left( \sqrt{1+\frac{2m}{r}},
    -\frac{2m}{r\sqrt{1+\frac{2m}{r}}}, 0, 0\right) , \quad
    n_\alpha = \left( - \frac{1}{\sqrt{1+\frac{2m}{r}}},0,0,0 \right).
\ee
\end{exmp}

\subsection{3+1 decomposition of the Riemann tensor} \label{s:FO:Riem_3p1}

We present here the expression of the spacetime Riemann tensor 
$\mathrm{\bf Riem}$ (cf. Sec.~\ref{s:IN:curvat}) in terms of 
3+1 objects, in particular the Riemann tensor ${}^3\mathrm{\bf Riem}$
of the connection $\w{D}$ associated with the spatial metric $\w{\gamma}$.
This is a step required to get a 3+1 decomposition
of the Einstein equation in next section. Moreover, this allows 
to gain intuition on the analogous (but null) decomposition that will be 
introduced in Sec.~\ref{s:DY}, when studying the dynamics of a null 
hypersurface.
 
As a general strategy, calculations start from the 3-dimensional
objects and then use is made of Eqs. (\ref{e:FO:3der}) and
(\ref{e:FO:def_gamma}), together with the Ricci identity
(\ref{e:IN:Ricci_ident}). 
Since these techniques will be explicitly exposed in 
Sec.~\ref{s:DY}, we present the following results without
proof (see, for instance, Ref.~\cite{York79}).
The 3+1 writing of the spacetime Riemann tensor thus obtained
can be viewed as various orthogonal projections 
of $\mathrm{\bf Riem}$ onto
the hypersurface $\Sigma_t$ and along the normal 
$\w{n}$:
\bea
\gamma^\alpha_{\ \, \mu} \gamma^\nu_{\ \, \beta}
\gamma^\rho_{\ \, \gamma} \gamma^\sigma_{\ \, \delta} \, 
 R^\mu_{\ \, \nu\rho\sigma} &=&
{}^{3}\!R^\alpha_{\ \, \beta\gamma\delta}
+ K^\alpha_{\ \, \gamma} K_{\beta\delta} 
- K^\alpha_{\ \, \delta} K_{\beta\gamma} .      \label{e:FO:Gauss} \\
\gamma^\alpha_{\ \, \mu} \gamma^\nu_{\ \, \beta}
\gamma^\rho_{\ \, \gamma} n^\sigma  \, R^\mu_{\ \, \nu\rho\sigma}  &=& 
D_\beta K^\alpha_{\ \, \gamma} - D^\alpha K_{\beta\gamma} 
                                        \label{e:FO:Codazzi} \\
\gamma_{\alpha\mu} n^\nu \gamma^\rho_{\ \, \beta} n^\sigma
\, R^\mu_{\ \, \nu\rho\sigma} & = &
\frac{1}{N} \mathcal{L}_{(N \w{n})} K_{\alpha\beta} + 
 K_{\alpha\mu} K^\mu_{\ \, \beta} 
+ \frac{1}{N} D_\alpha D_\beta N . \label{e:FO:Rienorm}
\eea
From the symmetries of the Riemann tensor, 
all the other contractions involving $\w{n}$
either are equivalent to one of the 
Eqs.~(\ref{e:FO:Codazzi})-(\ref{e:FO:Rienorm}),
or vanish. For instance, a contraction with three times $\w{n}$
would be zero.  
Equation (\ref{e:FO:Gauss}) is known as the {\em Gauss equation}, 
and Eq.~(\ref{e:FO:Codazzi}) as the {\em Codazzi equation}. 
The third equation, (\ref{e:FO:Rienorm}), is sometimes called
the {\em Ricci equation} [not to be confused with the {\em Ricci
identity} (\ref{e:IN:Ricci_ident})].

The Gauss and Codazzi equations do not involve any second 
order derivative of the
metric tensor $\w{g}$ in a timelike direction. 
They constitute the necessary and sufficient 
conditions for the hypersurface $\Sigma_t$, 
endowed with a 3-metric $\w{\gamma}$
and an extrinsic curvature $\w{K}$,
to be a submanifold
of $(\mathcal{M},\w{g})$. 
Contracted versions of the Gauss and Codazzi equations turn out to be
very useful, especially in the 3+1 writing of the Einstein equation.
Contracting the Gauss equation (\ref{e:FO:Gauss}) on the indices
$\alpha$ and $\gamma$ leads to an expression that makes appear the 
Ricci tensors $\w{R}$ and ${}^3\!\w{R}$ associated 
with $\w{g}$ and $\w{\gamma}$, respectively 
[cf. Eq.~(\ref{e:IN:def_Ricci})]
\be \label{e:FO:Gauss_contracted_1}
  \gamma^\mu_{\ \, \alpha} \gamma^\nu_{\ \, \beta}  R_{\mu\nu} 
  + \gamma_{\alpha\mu} n^\nu \gamma^\rho_{\ \, \beta} n^\sigma
   \, R^\mu_{\ \, \nu\rho\sigma}  = {}^3\! R_{\alpha\beta}
   + K K_{\alpha\beta} - K_{\alpha\mu} K^\mu_{\ \, \beta} ,  
\ee
where $K$ is the trace of $\w{K}$, $K^\mu_\mu$.
Taking the trace of this equation with respect to
$\w{\gamma}$, leads to
an expression that involves the Ricci scalars $R:=g^{\mu\nu} R_{\mu\nu}$ and 
${}^3\! R  := \gamma^{\mu\nu} {}^3\! R_{\mu\nu}$, again   respectively
associated with $\w{g}$ and $\w{\gamma}$:
\be \label{e:FO:Gauss_contracted_2}
    R + 2 R_{\mu\nu} n^\mu n^\nu = {}^3\!R + K^2 - K_{\mu\nu} K^{\mu\nu} .
\ee
This formula, which relates the intrinsic curvature ${}^3\!R$
and the extrinsic curvature $\w{K}$ of $\Sigma_t$, can be seen 
as a generalization to the 4-dimensional case of Gauss' famous
{\em Theorema egregium} (see e.g. Ref.~\cite{BergeG87}).
On the other side, contracting the Codazzi equation on the indices 
$\alpha$ and $\gamma$ leads to 
\be \label{e:FO:Codazzi_contracted}
    \gamma^\mu_{\ \, \alpha} n^\nu R_{\mu\nu} = D_\alpha K
     - D^\mu K_{\alpha\mu} . 
\ee


\subsection{3+1 Einstein equation} \label{s:FO:3p1Einstein}

We are now in position of presenting the 3+1 splitting of
Einstein equation:
\be \label{e:FO:Einstein}
   \encadre{ \w{R} - \frac{1}{2} R \w{g} = 8 \pi \w{T} }, 
\ee
where $\w{T}$ is 
the total (matter + electromagnetic field) energy-momentum tensor. 
The 3+1 decomposition of the latter is
\be
	\w{T} = E\, \w{n}\otimes \w{n} + \w{n} \otimes \w{J}
		+ \w{J}\otimes \w{n} + \w{S} , 
\ee
where the energy density $E$, the momentum density $\w{J}$ and
the strain tensor $\w{S}$, all of them as measured by the Eulerian
observer of 4-velocity $\w{n}$, are given by the following projections
$E := T_{\mu\nu} n^\mu n^\nu$, 
$J_\alpha  :=  - \gamma_\alpha^{\ \,\mu} T_{\mu\nu} n^\nu$,
$S_{\alpha\beta}  :=  \gamma_\alpha^{\ \,\mu} 
\gamma_\beta^{\ \,\nu} T_{\mu\nu}$.

Einstein equation (\ref{e:FO:Einstein}) splits into three equations
by using, respectively, (i) the twice contracted Gauss equation 
(\ref{e:FO:Gauss_contracted_2}), (ii)
the contracted Codazzi equation (\ref{e:FO:Codazzi_contracted}), 
(iii) the combination of the Ricci equation 
(\ref{e:FO:Rienorm}) with the contracted Gauss equation 
(\ref{e:FO:Gauss_contracted_1}):
\be
   \encadre{ {}^3\!R + K^2 - K_{\mu\nu} K^{\mu\nu} =  16\pi E } ,
                                        \label{e:FO:Ham_constraint}
\ee
\be
   \encadre{	D^\mu K_{\alpha\mu} - D_\alpha K =  8\pi J_\alpha} ,
                                        \label{e:FO:mom_constraint} 
\ee 
\be
    \encadre{
    \begin{array}{rcl}
    \mathcal{L}_{(N \w{n})} K_{\alpha\beta} & = &\displaystyle
    - D_\alpha D_\beta N + N \Big\{ {}^3\!R_{\alpha\beta} 
    - 2 K_{\alpha\mu} K^\mu_{\ \, \beta}  + K K_{\alpha\beta} \\
    & & \displaystyle + 4\pi \left[ (S-E)\gamma_{\alpha\beta}
    - 2 S_{\alpha\beta} \right] \Big\}    
     \end{array}                \label{e:FO:evol_K_n}
     } \ .
\ee
These equations are known as the {\em Hamiltonian constraint}, 
the {\em momentum constraint} and the {\em dynamical 3+1 equations},
respectively.

The Hamiltonian and momentum constraints do not contain any second
order derivative of the metric in a timelike direction,
contrary to Eq.~(\ref{e:FO:evol_K_n}) [remember that $\w{K}$ is already
a first order derivative of the metric in the timelike direction $\w{n}$, 
according to Eq.~(\ref{e:FO:K_Lie_n})].
Therefore they are not associated with the dynamical
evolution of the gravitational field and represent
{\em constraints} to be satisfied by $\w{\gamma}$ and $\w{K}$
on each hypersurface $\Sigma_t$.

The dynamical equation (\ref{e:FO:evol_K_n}) can be written explicitly
as a time evolution equation, once a 3+1 coordinate system $(t,x^i)$
is introduced, as in Sec.~\ref{s:FO:coord}. Then 
$N\w{n}$ is expressible in terms of the coordinate time vector $\tv$ and
the shift vector $\w{\beta}$ associated with these coordinates:
$N\w{n} = \tv - \w{\beta}$ [cf. Eq.~(\ref{e:FO:def_shift})], so that
the Lie derivative in the left-hand side of Eq.~(\ref{e:FO:evol_K_n})
can be written as
\be
    \w{\mathcal{L}}_{(N\w{n})} = \Lie{\tv} - \Lie{\w{\beta}} .
\ee
Now, if one uses tensor components with respect to the coordinates $(x^i)$,
$\Liec{\tv} K_{ij} = \dert{K_{ij}}{t}$, Eq.~(\ref{e:FO:evol_K_n})
becomes
\be
    \encadre{
    \begin{array}{rcl}
    \displaystyle \der{}{t} K_{ij} - \Liec{\w{\beta}} K_{ij} & = &\displaystyle
    - D_i D_j N + N \Big\{ {}^3\!R_{ij} 
    - 2 K_{ik} K^k_{\ \, j}  + K K_{ij} \\
    & & \displaystyle + 4\pi \left[ (S-E)\gamma_{ij}
    - 2 S_{ij} \right] \Big\}    
     \end{array}                
     } .                            \label{e:FO:evol_K_t}
\ee
Similarly, the relation (\ref{e:FO:K_Lie_n}) between $\w{K}$ and
$\Lie{\w{n}}\w{\gamma}$ becomes 
\be
\label{e:FO:gam_evol}
    \encadre{
	\der{}{t}\gamma_{ij} - \Liec{\w{\beta}} \gamma_{ij} 
	= - 2N K_{ij} 
    } ,	
\ee
where one may use the following identity [cf. Eq. (\ref{e:IN:Lie_derivative})]:
$ \Liec{\w{\beta}} \gamma_{ij} = D_i \beta_j + D_j \beta_i$. 


\subsection{Initial data problem}
\label{s:FO:ini_data}

In view of the above equations, the standard procedure of numerical 
relativity consists in 
firstly specifying the values of $\w{\gamma}$ and $\w{K}$ on some
initial spatial hypersurface $\Sigma_0$ 
(Cauchy surface), and then evolving them according to 
Eqs. (\ref{e:FO:evol_K_t}) and
(\ref{e:FO:gam_evol}). 
For this scheme to be valid, the initial data  
must satisfy the constraint Eqs.
(\ref{e:FO:Ham_constraint})-(\ref{e:FO:mom_constraint}).
The problem of finding pairs 
$(\w{\gamma}, \w{K})$ on $\Sigma_0$
satisfying these constraints constitutes the {\em initial data problem}
of 3+1 general relativity.

The existence of a well-posed initial value formulation for Einstein equation,
first established by Y. Choquet-Bruhat more than fifty years ago 
\cite{Foure52},
provides fundamental insight for a number of issues in general relativity
(see e.g. Refs.~\cite{FM79,BartnI04} for a mathematical account).
In this article we aim at underlining those aspects related with the
numerical construction of astrophysically relevant spacetimes containing
black holes. In this sense, the
3+1 formalism constitutes a particularly convenient and widely extended approach
to the problem (for other numerical approaches, see for instance 
\cite{Inver97a,Inver97b,Winic01}).
Consequently, the  first step in this numerical approach
consists in generating appropriate initial data which correspond to astrophysically
realistic situations. For a review on the numerical aspects of 
this initial data 
problem see \cite{Cook00,Pfeif04}.

If one chooses to excise a sphere in the spatial surface $\Sigma_t$ for it
to represent the horizon of a black hole, 
appropriate boundary conditions in this inner boundary
must be imposed when solving the
constraint Eqs. (\ref{e:FO:Ham_constraint})-(\ref{e:FO:mom_constraint}).
This particular aspect of the initial data problem constitutes one of the main
applications of the subject studied here, and will be developed in Sec. \ref{s:BC}. 
In order to carry out such a discussion, the conformal decomposition of the initial
data introduced by Lichnerowicz \cite{Lichn44}, particularly successful
in the generation of initial data, will be presented in 
Sec.~\ref{s:TP}.

%% file: indfoliat.tex
%
%
\section{3+1-induced foliation of null hypersurfaces}
\label{s:IN}

\subsection{Introduction} \label{s:IN:intro}

In Secs.~\ref{s:NH:Weingarten} and \ref{s:NH:2ndfund}, we have introduced
two geometrical objects on the 3-manifold $\Hor$: the Weingarten
map $\w{\chi}$ and the second fundamental form $\w{\Theta}$. These
objects are unique up to some rescaling of the null normal $\el$ to $\Hor$.
Following Carter \cite{Carte92a,Carte92b,Carte97}, we would like 
to consider these objects as 4-dimensional quantities, i.e. to 
extend their definitions from the 3-manifold $\Hor$ to the 4-manifold $\M$
(or at least to the vicinity of $\Hor$, as discussed in 
Sec.~\ref{s:NH:extended_el}). 
The benefit of such an extension is an easier manipulation of these
objects, as ordinary tensors on $\M$, which will facilitate the
connection with the geometrical objects of the 3+1 slicing.
In particular this avoids the introduction of special coordinate systems and complicated notations. 
For instance, one would like to define easily something like the
type $\left( 0\atop 3 \right)$ tensor
$\w{\nabla} \w{\Theta}$, where $\w{\nabla}$ is the spacetime 
covariant derivative. At the present stage, this
not possible even when restricting the definition of 
$\w{\nabla} \w{\Theta}$ to $\Hor$, because there is no unique 
covariant derivation associated
with the induced metric $\w{q}$, since the latter is degenerate. 

We have already noticed that, from the null structure of $\Hor$
alone, there is no canonical mapping from vectors of $\M$ to
vectors of $\Hor$, and in particular no
orthogonal projector (Remarks~\ref{rem:NH:emb_map} and 
\ref{rem:NH:no_ortho_proj}). 
Such a mapping would have provided
natural four-dimensional extensions of the forms defined on $\Hor$.
Actually in order to define a projector onto $\T_p(\Hor)$,
we need some direction {\em transverse} to $\Hor$, i.e. some
vector of $\T_p(\M)$ not belonging to $\T_p(\Hor)$. We may then define a
projector {\em along} this transverse direction. The problem with null
hypersurfaces is that there is no canonical transverse direction since
the normal direction is not transverse but tangent. 

However if we take into account the foliation provided by some family of spacelike
hypersurfaces $(\Sigma_t)$ in the standard 3+1 formalism introduced in 
Sec.~\ref{s:FO}, we have
some extra-structure on $\M$. We may then use it to define 
unambiguously a transverse direction to $\Hor$
and an associated projector $\w{\Pi}$. 
Moreover this transverse direction will be, by construction,
well suited to the 3+1 
decomposition.

\subsection{3+1-induced foliation of $\Hor$ and normalization of $\el$}
\label{s:IN:normal_l}

\begin{figure}
\centerline{\includegraphics[width=0.6\textwidth]{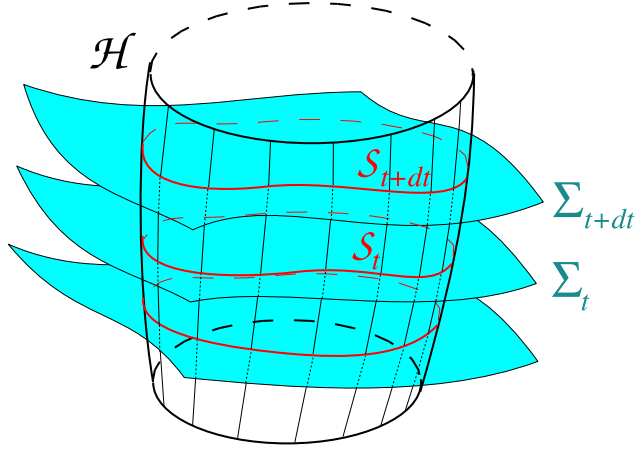}}
\caption[]{\label{f:IN:indfol} 
Foliation $(\Sp_t)$ of the null hypersurface $\Hor$ induced by
the spacetime foliation $(\Sigma_t)$ of the 3+1 formalism.}
\end{figure}

In the general case, each spacelike hypersurface $\Sigma_t$ of the 3+1 slicing 
discussed in Sec.~\ref{s:FO} intersects\footnote{Note however the existence 
of slicings of the ``exterior'' region of $\Hor$ which actually 
do not intersect $\Hor$, such as the standard 
maximal slicing of Schwarzschild spacetime defined 
by the Schwarzschild time $t_{\rm S}$ 
and illustrated in Fig.~\ref{f:NH:Schwarz_kruskal}.}  
the null hypersurface
$\Hor$ on some 2-dimensional surface $\Sp_t$ (cf. Fig.~\ref{f:IN:indfol}):
\be
	\encadre{ \Sp_t := \Hor \cap \Sigma_t }. 
\ee
More generally, considering some null foliation $(\Hor_u)$ in the vicinity of 
$\Hor$ (cf. Sec.~\ref{s:NH:extended_el}), 
we define the 2-surface family
$(\Sp_{t,u})$ by
\be 
    \Sp_{t,u} :=   \Hor_u \cap \Sigma_t  .       \label{e:IN:def_Str}
\ee
$\Sp_t$ is then nothing but the element $\Sp_{t,u=1}$ of this family. 
$(\Sp_{t,u})$ constitutes a foliation of $\M$ (in the vicinity of $\Hor$)
by 2-surfaces. This foliation is of type {\em null-timelike} in the 
terminology of the 2+2 formalism \cite{Inver97a,InverS80}. 

A local characterization of $\Sp_t$ follows from Eq.~(\ref{e:NH:r_const})
and the definition of $\Sigma_t$ as the level set of some scalar field $t$:
\be \label{e:IN:charac_St}
	\forall p \in \M,\quad p \in \Sp_t \iff  u(p) = 1 
        \ \ \mbox{and}\ \ t(p) = t  .
\ee
The subspace $\T_p(\Sp_t)$ of vectors tangent to $\Sp_t$ 
at some point $p\in\Sp_t$ is then 
characterized in terms of the gradients of the scalar fields $u$ and $t$
by
\be \label{e:IN:charac_TSt}
\forall \w{v}\in\T_p(\M),\quad
	\w{v}\in\T_p(\Sp_t) \iff \langle \dd u, \w{v} \rangle = 0 
        \ \ \mbox{and}\ \ \langle \dd t, \w{v} \rangle = 0 .
\ee

As a submanifold of $\Sigma_t$, each $\Sp_t$ is necessarily a 
spacelike surface. Until Sec.~\ref{s:NE}, we make no assumption on
the topology of $\Sp_t$, although we picture it as a closed 
(i.e. compact without boundary) manifold 
(Fig.~\ref{f:IN:indfol}). In the absence of global assumptions
on $\Sp_t$ or $\Hor$, we define the {\em exterior} 
(resp. {\em interior}) of $\Sp_t$, as the region of $\Sigma_t$
for which $u>1$ (resp. $u<1$). In the case another criterion
is available 
to define the exterior of $\Sp_t$ (e.g. if $\Sp_t$ has the topology
of $\mathbb{S}^2$, $\Sigma_t$ is asymptotically
flat and the exterior of $\Sp_t$ is defined as the connected 
component of $\Sigma_t \setminus \Sp_t$ which contains the asymptotically flat
region), we can always change the definition  of $u$ to make coincide
the two definitions of {\em exterior}.

\begin{figure}
\centerline{\includegraphics[width=0.7\textwidth]{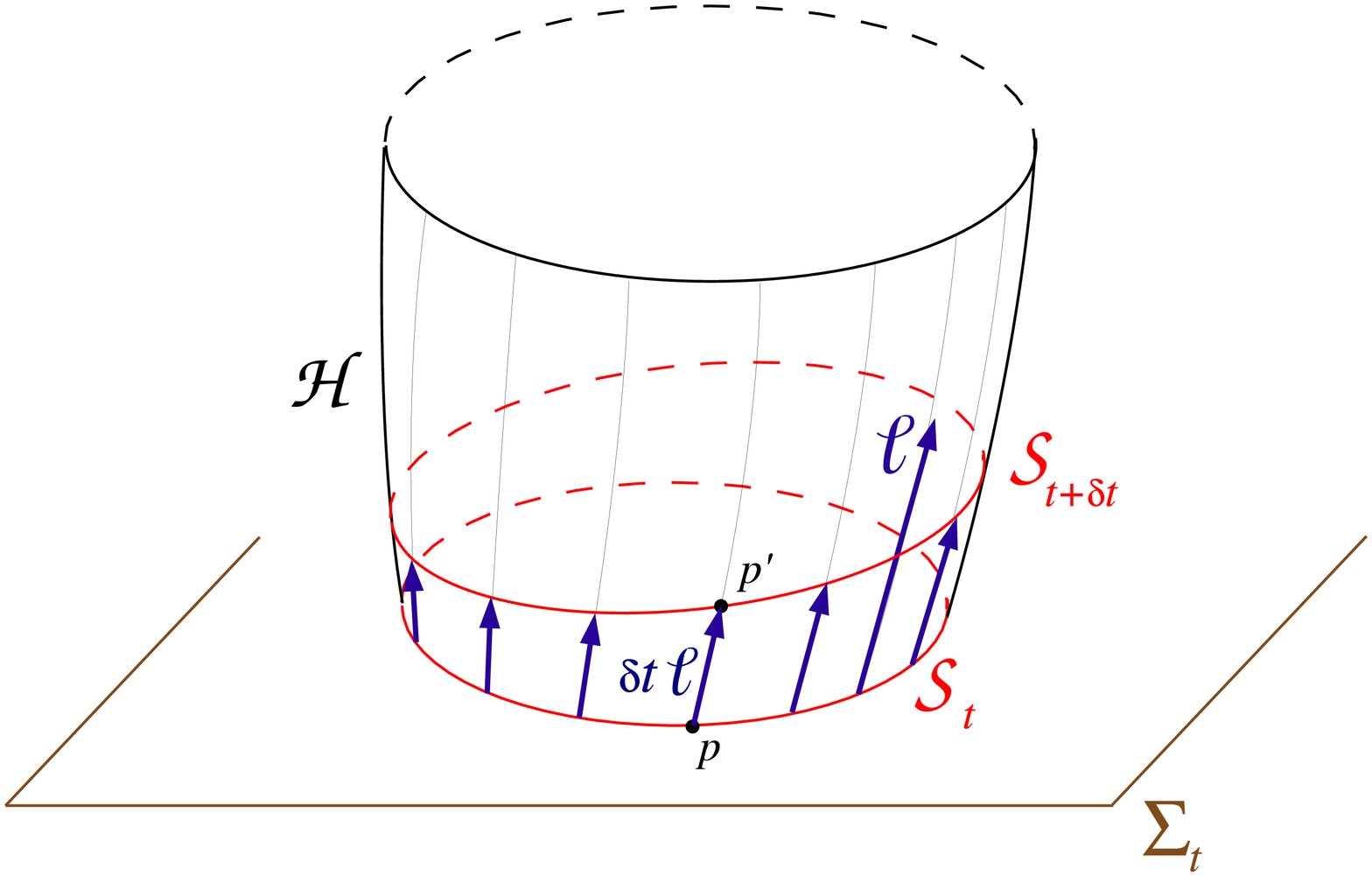}}
\caption[]{\label{f:IN:Lie_St} 
Lie transport of the surfaces $\Sp_t$ by the vector $\el$.}
\end{figure}

The $\Sp_t$'s constitute a foliation of $\Hor$.
The coordinate $t$ can then be used as a parameter, in general non-affine,
along each null geodesic
generating $\Hor$ (cf. Sec.~\ref{s:NH:generators}). 
Thanks to it, we can normalize the null normal $\el$ of $\Hor$
by demanding that $\el$ is the tangent vector associated with this
parametrization of the null generators:
\be \label{e:NH:l_norm1}
	\ell^\alpha = {dx^\alpha\over dt} . 
\ee
An equivalent phrasing of this is demanding that $\el$ is
a vector field ``dual'' to the 1-form $\dd t$ (equivalently, the function
$t$ can be regarded as a coordinate compatible with $\el$):
\be \label{e:NH:l_norm2}
	\encadre{ \langle \dd t , \el \rangle = \w{\nabla}_{\el}\, t = 1 }. 
\ee
A geometrical consequence of this choice is that the 
2-surface $\Sp_{t+\delta t}$ is obtained
from the 2-surface $\Sp_t$ by a displacement 
$\delta t\el$
at each point of $\Sp_t$, as depicted in Fig.~\ref{f:IN:Lie_St}. 
Indeed consider a point $p$ in $\Sp_t$
and displace it by a infinitesimal quantity $\delta t \el$ to the
point $p' = p + \delta t \el$ (cf. Fig.~\ref{f:IN:Lie_St}). 
From the very definition of the gradient
1-form $\dd t$, the value of the scalar field $t$ at $p'$
is
\bea
    t(p') & = & t(p + \delta t \el) 
          = t(p) + \langle \dd t, \delta t \el \rangle
          = t(p) + \delta t 
          \underbrace{\langle \dd t,  \el \rangle}_{=1} \nonumber \\
          & = & t(p) + \delta t . 
\eea
This last equality shows that $p' \in \Sp_{t+\delta t}$.
Hence the vector $\delta t \el$ carries the surface $\Sp_t$
into the neighboring one $\Sp_{t+\delta t}$. One says equivalently that
the 2-surfaces $(\Sp_t)$ are {\em Lie dragged} by the null normal $\el$.

\begin{figure}
\centerline{\includegraphics[width=0.7\textwidth]{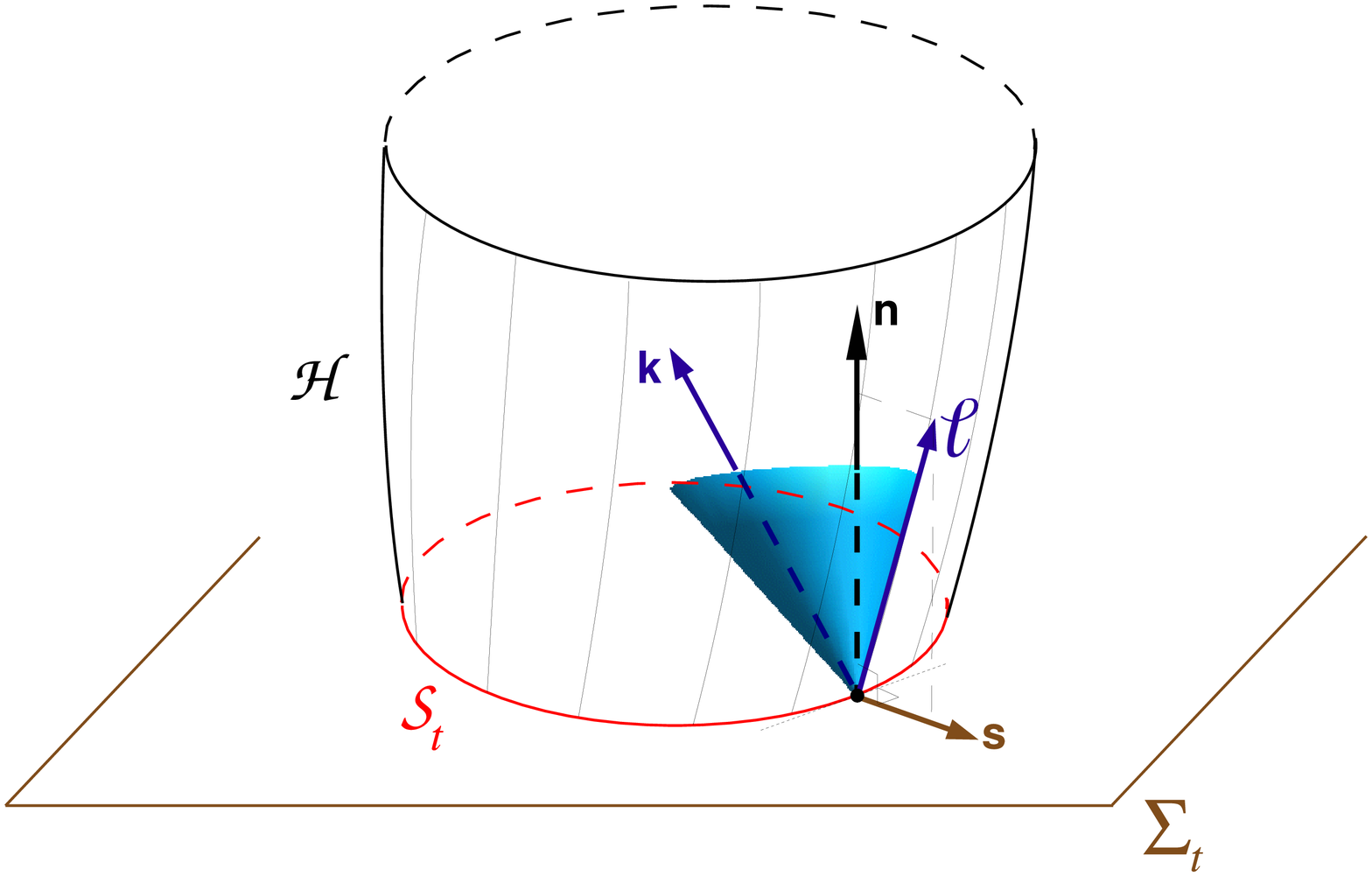}}
\caption[]{\label{f:def_s} 
Null vector $\el$ normal to $\Hor$, unit timelike vector $\w{n}$ normal to
$\Sigma_t$, unit spacelike vector $\w{s}$ normal to $\Sp_t$ and ingoing
null vector $\w{k}$ normal to $\Sp_t$.}
\end{figure}

Let $\w{s}$ be the unit vector of $\Sigma_t$, normal to $\Sp_t$ and  
directed toward the exterior of $\Sp_t$ (cf. Fig.~\ref{f:def_s});  
$\w{s}$ obeys to the following properties:
\bea
    & & \w{s}\cdot\w{s} = 1, \\
    & & \w{n}\cdot\w{s} = 0, \\
    & & \langle \dd u, \w{s} \rangle > 0 , \label{e:IN:dr_s_gt0} \\
    & & \forall \w{v}\in\T_p(\Sigma_t),\quad
	\w{v}\in\T_p(\Sp_t) \iff \w{s}\cdot\w{v} = 0 .
\eea
Let us establish a simple expression of the null normal $\el$ 
in terms of the unit vectors $\w{n}$ and $\w{s}$. 
Let $\w{b} \in\T_p(\Sigma_t)$ be the orthogonal projection of 
$\el$ onto $\Sigma_t$: $\w{b} := \vec{\w{\gamma}}(\el)$
[cf. Eq.~(\ref{e:NH:gamma_ortho_proj})]. Then $\el = a \w{n} + \w{b}$, with
a coefficient $a$ to be determined. 
By means of Eq.~(\ref{e:FO:def_n}), 
$\langle \dd t, \el \rangle = a / N$, so that the normalization condition
(\ref{e:NH:l_norm2}) leads to $a = N$, hence
\be
    \el = N \w{n} + \w{b} . 
\ee
For any vector $\w{v}\in\T_p(\Sp_t)$, $\el\cdot\w{v}=0$. Replacing $\el$
by the above expression and using
$\w{n}\cdot\w{v} = 0$ results in $\w{b}\cdot\w{v}=0$. Since this equality
is valid for any $\w{v}\in\T_p(\Sp_t)$, we deduce that $\w{b}$ is a vector
of $\Sigma_t$ which is normal to $\Sp_t$. It is then necessarily collinear
to $\w{s}$: $\w{b} = \alpha \w{s}$,
with $\alpha = \w{s}\cdot\w{b} = \w{s}\cdot\el
= \langle\uel, \w{s}\rangle = e^\rho \langle \dd u, \w{s} \rangle
> 0$, thanks to Eq.~(\ref{e:IN:dr_s_gt0}). 
The condition 
$\w{\el}\cdot\w{\el}=0$ then leads to $\alpha = N$, so that
finally
\be \label{e:IN:el_nps}
    \encadre{\el = N(\w{n} + \w{s}) } .
\ee
In particular, the three vectors $\el$, $\w{n}$ and $\w{s}$ are
coplanar (see Fig.~\ref{f:def_s}).
Moreover, since $\vec{\w{\gamma}}(\el) = N \w{s}$ with $N>0$, 
$\vec{\w{\gamma}}(\el)$ is directed toward the exterior of $\Sp_t$.
We say that $\el$ is an {\em outgoing} null vector with respect to
$\Sp_t$.

\begin{rem} \label{r:IN:norm_el}
Since $\w{n}$ is a unit timelike vector, $\w{s}$ a unit spacelike vector
and they are orthogonal, it is immediate that 
the vector $\w{n}+\w{s}$ is null. The relation (\ref{e:IN:el_nps}) implies
that this vector is tangent to $\Hor$. Therefore, another 
natural normalization of the null normal to $\Hor$  would have been to
consider $\el = \w{n} + \w{s}$, instead of 
$\langle \dd t , \el \rangle = 1$ [Eq.~(\ref{e:NH:l_norm2})]. 
Both normalizations are induced by the foliation 
$(\Sigma_t)$. Only the normalization (\ref{e:NH:l_norm2}) has the property
of Lie dragging the surfaces $(\Sp_t)$. 
On the other side, at a given point $p\in\Hor$, the normalization
$\el = \w{n} + \w{s}$ can be defined by a single spacelike hypersurface
$\Sigma$ intersecting $\Hor$, whereas the normalization (\ref{e:NH:l_norm2})
requires the existence of a family $(\Sigma_t)$ in the neighborhood
of $p$. In what follows, we will denote by $\w{\hat\ell}$ the null vector
\be \label{e:IN:def_hat_ell}
    \w{\hat\ell} := \frac{1}{\sqrt{2}} (\w{n} + \w{s}) ,
\ee
where the factor $1/\sqrt{2}$ is introduced for later convenience. 
\end{rem}

\subsection{Unit spatial normal to $\Sp_t$}

Equation (\ref{e:IN:el_nps}) can be inverted to express the unit
spatial normal to the surface $\Sp_t$\footnote{Note that the definition
of the vector $\w{s}$ can be extended to the 2-surfaces $\Sp_{t,u}$
in the vicinity of $\Hor$. This permits to extend the objects constructed
by using $\w{s}$ to a neighborhood of $\Hor$, in the spirit of 
Sec. \ref{s:NH:extended_el}. We will refer in the following to $\Sp_t$, 
keeping in mind that
the results can be extended to the whole foliation $(\Sp_ {t,u})$.}, 
$\w{s}$,
in terms of $\el$ and $\w{n}$:
\be
    \w{s} = \frac{1}{N}\, \el - \w{n} . 
\ee
When combined with 
$\uel = \e^\rho \dd u$ [Eq.~(\ref{e:NH:l_grad_r})] and 
$\underline{\w{n}} = - N \, \dd t$ [Eq.~(\ref{e:FO:def_n})] this
leads to the following expression of the 1-form $\underline{\w{s}}$
associated with the normal $\w{s}$:
\be \label{e:us_dt_dr}
    \encadre{ \underline{\w{s}} = N \dd t + M \dd u } , 
\ee
where we have introduced the factor 
\be \label{e:IN:def_M}
 M := \frac{\e^\rho}{N} , \qquad \mbox{so that} \quad
    \encadre{\rho = \ln (MN)} . 
\ee
Equation (\ref{e:us_dt_dr}) implies [cf. the definition 
(\ref{e:FO:def_gamma_star}) of the operator $\vec{\w{\gamma}}^*$]
\be
    \vec{\w{\gamma}}^* \underline{\w{s}} = M \vec{\w{\gamma}}^* \dd u , 
\ee
because 
$\vec{\w{\gamma}}^* \dd t = 
- N^{-1} \; \vec{\w{\gamma}}^* \underline{\w{n}} = 0$. Now, since
$\w{s}\in\T_p(\Sigma_t)$, $\vec{\w{\gamma}}^* \underline{\w{s}} = 
\underline{\w{s}}$. Moreover, from Eq.~(\ref{e:FO:DT_gamma_star_nabT}),
$\vec{\w{\gamma}}^* \dd u = \w{D} u$, so that we get 
\be \label{e:IN:us_MDr}
    \encadre{ \underline{\w{s}} = M \vec{\w{\gamma}}^* \dd u = M \w{D} u }. 
\ee

\subsection{Induced metric on $\Sp_t$} \label{s:IN:metric_q}

The metric $\w{h}$ induced by $\Sigma_t$'s metric $\w{\gamma}$ on the 2-surfaces 
$\Sp_t$ is given by a formula analogous to Eq.~(\ref{e:FO:def_gamma}),
except for the change of the $+$ sign into a $-$ one, to take into 
account the spacelike character of the normal $\w{s}$ (whereas the
normal $\w{n}$ was timelike):
\be
    \w{h} := \w{\gamma} - \underline{\w{s}} \otimes \underline{\w{s}} 
         = \w{g} + \underline{\w{n}} \otimes \underline{\w{n}}
                - \underline{\w{s}} \otimes \underline{\w{s}} .  
                                                    \label{e:y_ind_metric}           
\ee
Let us consider a pair of vectors $(\w{u},\w{v})$ in $\T_p(\Hor)$.
Denoting by $\w{u}_0$ and $\w{v}_0$ their respective projections along $\el$
on the vector plane $\T_p(\Sp_t)$, we have the unique decompositions
\be
    \w{u} = \w{u}_0 + \lambda \el \quad \mbox{and} \quad
    \w{v} = \w{v}_0 + \mu \el ,
\ee 
where $\lambda$ and $\mu$ are two real numbers. 
Since $\w{n}\cdot\w{u}_0 = \w{n}\cdot\w{v}_0 
= \w{s}\cdot\w{u}_0 = \w{s}\cdot\w{v}_0 = 0$, one has
\bea
    \w{h}(\w{u},\w{v}) & = &
      \w{g}(\w{u},\w{v}) + \langle \underline{\w{n}}, \w{u} \rangle
                    \langle \underline{\w{n}},\w{v} \rangle
                    -\langle \underline{\w{s}}, \w{u} \rangle
                     \langle \underline{\w{s}},\w{v} \rangle \nonumber \\
   & = &  \w{g}(\w{u},\w{v}) + [\w{n}\cdot(\w{u}_0 + \lambda \el)]
        \times [\w{n}\cdot(\w{v}_0 + \mu \el)] 
        - [\w{s}\cdot(\w{u}_0 + \lambda \el)]
        \times \nonumber \\
       & & [\w{s}\cdot(\w{v}_0 + \mu \el)] \nonumber \\
  & = & \w{g}(\w{u},\w{v}) + \lambda \mu 
    (\underbrace{\w{n}\cdot\el}_{=-N})^2
     - \lambda\mu (\underbrace{\w{s}\cdot\el}_{=N})^2 \nonumber \\
  \w{h}(\w{u},\w{v}) & = &  \w{g}(\w{u},\w{v}) .
\eea
This last equality shows that $\w{h}$ and $\w{g}$ coincide on $\T_p(\Hor)$.  
In other words, the pull-back of $\w{h}$ on $\Hor$
equals the pull-back of $\w{g}$, 
that we have denoted $\w{q}$ in Sec.~\ref{s:NH:geom_null}
[see Eq.~(\ref{e:NH:def_q})]:
$\Phi^*\w{h} = \Phi^*\w{g} = \w{q}$.
We may then take $\w{h}$ as the
4-dimensional extension of $\w{q}$ and write Eq.~(\ref{e:y_ind_metric})
as
\bea 
 & & \encadre{ \w{q} = \w{\gamma} 
                - \underline{\w{s}} \otimes \underline{\w{s}} }
                        \label{e:q_gam_ss} \\
   & &  \encadre{ \w{q} = \w{g} + \underline{\w{n}} \otimes \underline{\w{n}}
                - \underline{\w{s}} \otimes \underline{\w{s}} } .
                    \label{e:q_g_nn_ss}
\eea 
Consequently we abandon from now on the notation $\w{h}$ in profit of $\w{q}$.
To summarize, on $\T_p(\M)$, $\w{q}$ is the symmetric bilinear form
given by Eq.~(\ref{e:q_g_nn_ss}), on $\T_p(\Sigma_t)$ it is the symmetric bilinear 
form given by Eq.~(\ref{e:q_gam_ss}), on $\T_p(\Hor)$ it is the degenerate
metric induced by $\w{g}$, and on $\T_p(\Sp_t)$ it is the positive definite
(i.e. Riemannian) metric induced by $\w{g}$.

The endomorphism $\T_p(\M)\rightarrow\T_p(\M)$ canonically associated with
the bilinear form $\w{q}$ by the metric $\w{g}$ [cf. notation
(\ref{e:IN:arrow_endo})] is the {\em orthogonal projector onto
the 2-surface $\Sp_t$}:
\be \label{e:IN:vecq_ns}
    \encadre{ \vec{\w{q}} = \w{1} + \w{n} \, \langle \underline{\w{n}} , . \rangle
        -  \w{s} \, \langle \underline{\w{s}} , . \rangle  } ,
\ee
in the very same manner in which $\vec{\w{\gamma}}$ defined by 
Eq.~(\ref{e:NH:gamma_ortho_proj}) was the orthogonal projector 
onto $\Sigma_t$. 


\subsection{Ingoing null vector}

As mentioned in Sec.~\ref{s:IN:intro}, we need some direction transverse
to $\Hor$ to define a projector $\T_p(\M)\rightarrow \T_p(\Hor)$. 
The $(\Sigma_t)$ slicing
has already provided us with two different transverse directions: 
the timelike direction $\w{n}$ and 
the spacelike direction $\w{s}$, both  normal to the 2-surfaces $\Sp_t$ 
(cf. Fig.~\ref{f:def_s}). These two directions are indeed transverse to $\Hor$ 
since $\w{n}\not\in\T_p(\Hor)$ (otherwise $\Hor$ would be
a timelike hypersurface) and $\w{s}\not\in\T_p(\Hor)$ (otherwise $\Hor$ would be
a spacelike hypersurface, coinciding locally with $\Sigma_t$).
However $\w{n}$ and $\w{s}$ are not the only natural choices 
linked with the $(\Sigma_t)$ foliation: we may also think about the 
{\em null} directions normal to $\Sp_t$, i.e. the trajectories of the light 
rays emitted in the radial 
directions from points on $\Sp_t$. The light rays emitted in the 
{\em outgoing
radial} direction (as defined in Sec.~\ref{s:IN:normal_l})
define the null vector $\el$ tangent to $\Hor$ already introduced. 
But those emitted in the {\em ingoing radial} direction define 
(up to some normalization factor) another null vector:
\be
\label{e:IN:hat_k}
    \w{\hat k} := \frac{1}{\sqrt{2}} (\w{n} - \w{s}) 
\ee
[compare with Eq.~(\ref{e:IN:def_hat_ell})].
$\w{\hat k}$ is transverse to $\Hor$,
since $\el\cdot\w{\hat k} = -\sqrt{2} N \not = 0$. 
In fact we will favor this transverse
direction, rather than those arising from $\w{n}$ or $\w{s}$, because
its null character leads to simpler formul\ae\ for the description
of the null hypersurface $\Hor$. 

\begin{figure}
\centerline{\includegraphics[width=0.6\textwidth]{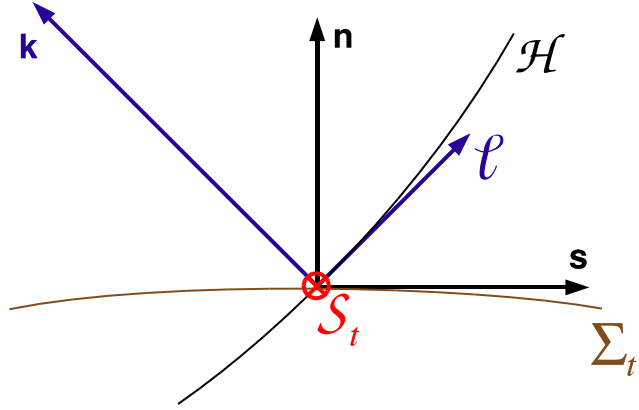}}
\caption[]{\label{f:coupe_ortho_S} 
View of the plane orthogonal to the 2-surface $\Sp_t$: the timelike
vector $\w{n}$, the spacelike vector $\w{s}$ and the null vectors
$\el$ and $\w{k}$ all belong to this plane. The dyad
$(\el,\w{k})$ defines the intersection of this plane with 
the light cone emerging from $\Sp_t$'s points. 
Intersections of the hypersurfaces $\Hor$ and $\Sigma_t$ with this plane
are also shown.}
\end{figure}

Let us renormalize the vector $\w{\hat k}$ by dividing it by $\sqrt{2} N$
to get the null vector
\be \label{e:NH:k_n_s}
	\encadre{ \w{k} = {1\over 2N} (\w{n} - \w{s})  } .
\ee
The normalization has been chosen so that $\w{k}$ satisfies the relation
\be \label{e:NH:l_k_m1}
	\encadre{ \el \cdot \w{k} = - 1} ,
\ee
which will simplify some of the subsequent formul\ae. 
Equations (\ref{e:IN:el_nps}) and (\ref{e:NH:k_n_s}) can be inverted
to express $\w{n}$ and $\w{s}$ in terms of the null vectors
$\el$ and $\w{k}$:
\bea
	\w{n} & = & {1\over 2N} \, \el + N \, \w{k} \label{e:IN:n_l_k} \\
	\w{s} & = & {1\over 2N} \, \el - N \, \w{k} \label{e:IN:s_l_k}.
\eea
Each pair $(\w{n},\w{s})$ or $(\el,\w{k})$ forms a basis of the 
vectorial plane orthogonal to $\Sp_t$:
\be \label{e:NH:span_perp_S}
	\T_p(\Sp_t)^\perp = {\rm Span}(\w{n},\w{s}) 
		= {\rm Span}(\el,\w{k}) .
\ee
This plane is shown in Fig.~\ref{f:coupe_ortho_S}.
\begin{rem}
\label{r:NH:k_foliation}
All the null vectors at a given point $p\in\Hor$, except those
collinear to $\el$ are transverse to $\Hor$ (see Fig.~\ref{f:def_s}
where all these vectors form the light cone emerging from $p$).
It is the slicing $(\Sp_t)$ of $\Hor$ which has enabled us to 
select a preferred transverse null direction $\w{k}$, as the unique null 
direction {\em normal to $\Sp_t$} and different from 
$\el$.
\end{rem}

Let us consider the 1-form $\underline{\w{k}}$ canonically associated with
the vector $\w{k}$ by the metric $\w{g}$. 
By combining Eqs.~(\ref{e:NH:k_n_s}), (\ref{e:FO:def_n}) and
(\ref{e:us_dt_dr}), one gets
\be \label{e:NH:uk_dt_dr}
	\encadre{ \underline{\w{k}} = - \dd t - {M\over 2N} \dd u}.
\ee
\begin{rem}
Equations~(\ref{e:NH:l_grad_r}), (\ref{e:FO:def_n}), (\ref{e:us_dt_dr})
and (\ref{e:NH:uk_dt_dr}) show that the 1-forms $\uel$, 
$\underline{\w{n}}$, $\underline{\w{s}}$ and $\underline{\w{k}}$
are all linear combinations of the exact 1-forms $\dd t$ and $\dd u$. 
This simply reflects the fact that the vectors $\el$, 
$\w{n}$, $\w{s}$ and $\w{k}$ are all orthogonal to $\Sp_t$
[Eq.~(\ref{e:NH:span_perp_S}) above] and that $(\dd t, \dd u)$ form
a basis of the 2-dimensional space of 1-forms normal to $\Sp_t$
[see Eq.~(\ref{e:IN:charac_TSt})]. 
\end{rem}
An immediate consequence of (\ref{e:NH:uk_dt_dr}) is that 
the action of $\underline{\w{k}}$ on vectors tangent to $\Hor$
is identical (up to some sign) to the action of the gradient 1-form
$\dd t$:
\be \label{e:NH:uk_dt_H}
	 \forall \w{v}\in\T_p(\Hor),\quad \langle \underline{\w{k}}, \w{v} \rangle
		= - \langle \dd t, \w{v} \rangle . 
\ee
An equivalent phrasing of this is: the pull-back of the 1-form
$\uk$ on $\Hor$ and that of $-\dd t$ coincide:
\be \label{e:IN:k_dt_H}
    \encadre{  \Phi^* \uk = - \Phi^* \dd t } .
\ee
\begin{rem} \label{r:dual_k_l}
The null vector $\w{k}$ can be seen as ``dual'' to the null
vector $\el$ in the following sense: (i) $\el$ belongs to $\T_p(\Hor)$,
while $\w{k}$ does not, and (ii) 
$\Phi^* \uk$ is a 
non-trivial exact 1-form in 
$\T^*(\Hor)$, while $\Phi^*\uel$ is zero.
\end{rem}

\begin{exmp} \label{ex:IN:cone}
\textbf{Slicing of Minkowski light cone.} \\
In continuation with Example~\ref{ex:NH:cone} ($\Hor$ = light cone
in Minkowski spacetime), the simplest 3+1 slicing
we may imagine is that constituted by hypersurfaces $t={\rm const}$, 
where $t$ is a standard Minkowskian time coordinate.
The lapse function $N$ is then identically one and the unit normal
to $\Sigma_t$ has trivial components with respect to the 
Minkowskian coordinates $(t,x,y,z)$:
$n^\alpha = (1,0,0,0)$ and $n_\alpha = (-1,0,0,0)$.
In Example~\ref{ex:NH:cone}, we have already normalized $\el$
so that $\ell^t = \langle \dd t , \el \rangle = 1$ [Eq.~(\ref{e:NH:l_norm2})],
[cf. Eq.~(\ref{e:NH:ell_cone})]. The 2-surface $\Sp_t$ is the
sphere $r:=\sqrt{x^2+y^2+z^2}=t$ in the hyperplane $\Sigma_t$
and its outward unit normal has the following components with respect to
$(t,x,y,z)$:
\be
    s^\alpha = \left(0, \frac{x}{r}, \frac{y}{r}, \frac{z}{r} \right),
       \quad
    s_\alpha = \left(0, \frac{x}{r}, \frac{y}{r}, \frac{z}{r} \right) .
\ee
We then deduce the components of the ingoing null vector $\w{k}$ from 
Eq.~(\ref{e:NH:k_n_s}):
\be \label{e:IN:k_comp_cone}
    k^\alpha = \left(\frac{1}{2}, -\frac{x}{2r}, -\frac{y}{2r}, 
        -\frac{z}{2r} \right),
       \quad
    k_\alpha = \left(-\frac{1}{2}, -\frac{x}{2r}, -\frac{y}{2r}, 
        -\frac{z}{2r} \right),    
\ee
and the components of $\w{q}$ from Eq.~(\ref{e:q_g_nn_ss}):
\be \label{e:IN:q_comp_cone}
    q_{\alpha\beta} = \left( \begin{array}{cccc}
        0 & 0 & 0 & 0 \\
        0 & \frac{y^2+z^2}{r^2} & - \frac{xy}{r^2} & - \frac{xz}{r^2} \\
        0 & - \frac{xy}{r^2} & \frac{x^2+z^2}{r^2} & - \frac{yz}{r^2} \\
        0 & - \frac{xz}{r^2} & - \frac{yz}{r^2} & \frac{x^2+y^2}{r^2} 
        \end{array} \right) . 
\ee
\end{exmp} 

\begin{figure}
\centerline{\includegraphics[width=0.5\textwidth]{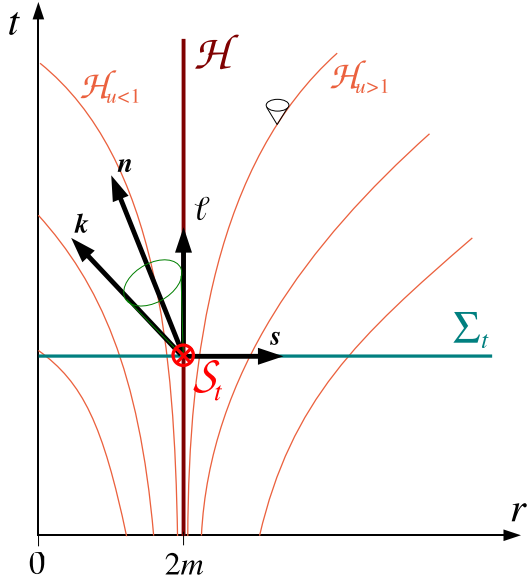}}
\caption[]{\label{f:IN:EF_slic} 
Null vector $\el$ normal to $\Hor$, unit timelike vector $\w{n}$ normal to
$\Sigma_t$, unit spacelike vector $\w{s}$ normal to $\Sp_t$ and ingoing
null vector $\w{k}$ normal to $\Sp_t$ for the 3+1 Eddington-Finkelstein
slicing of Schwarzschild horizon.}
\end{figure}

\begin{exmp} \label{ex:IN:EF}
\textbf{Eddington-Finkelstein slicing of Schwarzschild horizon.} \\
As a next example, let us consider the 3+1 slicing of Schwarzschild spacetime 
by the hypersurfaces $t={\rm const}$, where $t$ is the 
Eddington-Finkelstein time coordinate considered in 
Example~\ref{ex:NH:EF}. This slicing has been already represented in 
Fig.~\ref{f:NH:Schwarz_kruskal}. 
The corresponding lapse function has been exhibited in 
Example~\ref{ex:FO:EF}. In 
Example~\ref{ex:NH:EF}, we have already normalized the null vector $\el$
to ensure $\ell^t =  \langle \dd t , \el \rangle = 1$, so 
Eq.~(\ref{e:NH:comp_ell_EF}) provides the correct expression for
the null normal induced by the 3+1 slicing.
From the metric components given by Eq.~(\ref{e:NH:metric_edd_fink}), 
we obtain immediately the expression of the unit normal 
to $\Sp_t$ lying in $\Sigma_t$: 
\be \label{e:IN:s_comp_EF}
   s^\alpha = \left(0, \frac{1}{\sqrt{1+\frac{2m}{r}}},0,0 \right),
   \quad
   s_\alpha = \left(\frac{2m}{r\sqrt{1+\frac{2m}{r}}}, 
   \sqrt{1+\frac{2m}{r}}, 0,0\right). 
\ee
Inserting this value into formula (\ref{e:NH:k_n_s}) and making use
of expression (\ref{e:FO:lapse_EF}) for $N$ and
(\ref{e:FO:n_EF}) for $\w{n}$, we get the ingoing null vector $\w{k}$:
\be \label{e:IN:k_comp_EF}
    k^\alpha = \left( \frac{1}{2}+\frac{m}{r}, -\frac{1}{2}-\frac{m}{r},
    0, 0 \right) , \quad
    k_\alpha = \left( -\frac{1}{2}-\frac{m}{r}, -\frac{1}{2}-\frac{m}{r},
    0, 0 \right) . 
\ee
Note that on $\Hor$, $k_\alpha \equalH (-1,-1,0,0)$, so that we
verify property (\ref{e:IN:k_dt_H}), which is equivalent
to $(k_t,k_\theta,k_\varphi)\equalH (-1,0,0)$. 
The vectors $\w{n}$, $\w{s}$, $\el$ and $\w{k}$ are represented in 
Fig.~\ref{f:IN:EF_slic}. The 2-surface $\Sp_t$ is spanned by the coordinates
$(\theta,\varphi)$ and the expression of the induced metric on $\Sp_t$ 
is obtained readily from the line element (\ref{e:NH:metric_edd_fink}):
\be \label{e:IN:q_comp_EF}
    q_{\alpha\beta} = {\rm diag}(0,0,r^2, r^2\sin^2\theta) . 
\ee
\end{exmp}


\subsection{Newman-Penrose null tetrad} \label{s:IN:NP}

\subsubsection{Definition}

The two null vectors $\el$ and $\w{k}$ are the first two pieces
of the so-called {\em Newman-Penrose null tetrad}, which we briefly present
here. 
We complete the null pair $(\el,\w{k})$ by two orthonormal vectors
in $\T_p(\Sp_t)$, $(\we_a) = (\w{e}_2,\w{e}_3)$ let's say, to
get a basis of $\T_p(\M)$, such that
\be
	\begin{array}{lll}
   \el \cdot \el = 0 \quad  & \el \cdot \w{k} = - 1 \quad & \el \cdot \w{e}_{a} = 0 \\
    &	\w{k} \cdot \w{k} = 0 & \w{k} \cdot \w{e}_{a} = 0 \\
    & & \w{e}_{a} \cdot \w{e}_{b} = \delta_{ab} .
    	\end{array}
\ee
The basis $(\el,\w{k},\w{e}_{2},\w{e}_{3})$ is formed by 
two null vectors and two spacelike vectors. At the price of introducing 
complex vectors, we can modify it into a basis of four null vectors.
Indeed let us introduce the following combination of $\w{e}_{2}$ 
and $\w{e}_{3}$ with complex coefficients:
\be
\label{e:NH:m}
	\w{m} := {1\over\sqrt{2}} 
	\left( \w{e}_{2} + i\w{e}_{3} \right) . 
\ee
Then the complex conjugate defines another vector, which is linearly
independent from $\w{m}$:
\be
\label{e:NH:m_bar}
	\w{\bar m} = {1\over\sqrt{2}} 
	\left( \w{e}_{2} - i\w{e}_{3} \right) . 
\ee
Both $\w{m}$ and $\w{\bar m}$ are null vectors (with respect to the
metric $\w{g}$). 

The tetrad $(\el,\w{k},\w{m},\w{\bar m})$ constitutes a basis of $\T_p(\M)$
made of null vectors only: any vector of $\T_p(\M)$ admits a unique 
expression as a linear combination (possibly
with complex coefficients) of these four vectors. 
$(\el,\w{k},\w{m},\w{\bar m})$ is called a {\em Newman-Penrose null tetrad}
\cite{NewmaP62} (see also p.~343 of Ref.~\cite{HawkiE73} or 
p.~72 of \cite{Stewa90}). 
This tetrad obeys to 
\be
	\begin{array}{llll}
  \el \cdot \el = 0 \quad  & \el \cdot \w{k} = - 1 \quad & 
  	\el\cdot\w{m}=0 \quad & \el\cdot\w{\bar m}=0 \\ 
  & \w{k} \cdot \w{k} = 0 \quad & 
  	\w{k}\cdot\w{m}=0 \quad & \w{k}\cdot\w{\bar m}=0 \\ 
  &  & \w{m}\cdot\w{m}=0 \quad & \w{m}\cdot\w{\bar m}=1 \\ 
  & & & \w{\bar m}\cdot\w{\bar m}=0 .
    	\end{array}
\ee

Since $(\w{e}_{2},\w{e}_{3})$ is an orthonormal basis
of $\T_p(\Sp_t)$, the metric induced in $\Sp_t$ can be written
\be \label{e:NH:q_m_m}
	\w{q} = \underline{\w{e}}_{2} \otimes \underline{\w{e}}_{2}
		+ \underline{\w{e}}_{3} \otimes \underline{\w{e}}_{3}
		= \underline{\w{m}}\otimes\underline{\w{\bar m}}
		   + \underline{\w{\bar m}}\otimes\underline{\w{m}} .
\ee 

Moreover $(\w{n},\w{s},\w{e}_{2},\w{e}_{3})$ is an orthonormal
tetrad of $\T_p(\M)$. The spacetime metric can then be written
\be
	\w{g} = - \underline{\w{n}}\otimes\underline{\w{n}}
		+  \underline{\w{s}}\otimes\underline{\w{s}}
		+ \underline{\w{e}}_{2} \otimes \underline{\w{e}}_{2}
		+ \underline{\w{e}}_{3} \otimes \underline{\w{e}}_{3}.
\ee
It can also be expressed in terms of the Newman-Penrose null tetrad:
\be \label{e:NH:g_l_k_m_m}
	\w{g} = - \uel \otimes \uk - \uk \otimes \uel
		+ \underline{\w{m}}\otimes\underline{\w{\bar m}}
		   + \underline{\w{\bar m}}\otimes\underline{\w{m}} .
\ee
Comparing Eqs.~(\ref{e:NH:g_l_k_m_m}) and (\ref{e:NH:q_m_m}), we get an
expression of $\w{q}$ in terms of $\w{g}$ and the null dyad $(\el,\w{k})$:
\be \label{e:NH:q_g_l_k}
	\encadre{\w{q} = \w{g} + \uel \otimes \uk + \uk \otimes \uel} .
\ee
This expression is alternative to Eq.~(\ref{e:q_g_nn_ss}). It can 
be obtained directly by inserting Eqs.~(\ref{e:IN:n_l_k}) and 
(\ref{e:IN:s_l_k}) in Eq.~(\ref{e:q_g_nn_ss}).
The related expression for the orthogonal projector 
$\vec{\w{q}}$ onto the 2-surface $\Sp_t$ is
\be \label{e:IN:vec_q_k_l}
    \encadre{\vec{\w{q}} = \w{1} +  \el \,\langle \uk, .\rangle
        +  \w{k}\,\langle \uel, .\rangle },
\ee
which constitutes an alternative to Eq.~(\ref{e:IN:vecq_ns}).

\subsubsection{Weyl scalars} \label{s:IN:Weyl_scal}

In Section \ref{s:IN:curvat} we have introduced the Weyl tensor
$\w{C}$ and have 
indicated that it encodes ten of the twenty independent components of the
Riemann tensor. The null tetrad previously introduced permits
to write these free components as five independent complex scalars 
$\Psi_n$ ($n\in\{0,1,2,3,4\}$),
known as {\it Weyl scalars}. They are defined as
\bea
\label{e:NH:Weyl_scalars}
\Psi_0 &:=& \w{C}(\uel,\w{m},\el,\w{m})
 = C^\mu_{\ \, \nu\rho\sigma}\; \ell_\mu m^\nu \ell^\rho m^\sigma \nn \\
\Psi_1 &:=& \w{C}(\uel,\w{m},\el,\w{k})
 = C^\mu_{\ \, \nu\rho\sigma}\; \ell_\mu m^\nu \ell^\rho k^\sigma \nn \\
\Psi_2 &:=& \w{C}(\uel,\w{m},\w{\bar m},\w{k})
 = C^\mu_{\ \, \nu\rho\sigma}\; \ell_\mu m^\nu \bar{m}^\rho k^\sigma \\
\Psi_3 &:=& \w{C}(\uel,\w{k},\w{\bar m},\w{k})
 = C^\mu_{\ \, \nu\rho\sigma}\; \ell_\mu k^\nu \bar{m}^\rho k^\sigma \nn \\
\Psi_4 &:=& \w{C}(\underline{\w{\bar m}},\w{k},\w{\bar m},\w{k})
 = C^\mu_{\ \, \nu\rho\sigma}\; \bar{m}_\mu k^\nu \bar{m}^\rho k^\sigma \nn .
\eea
As we will see 
in the following sections, some relevant geometrical quantities 
are naturally expressed in terms of (some of) these scalars.
For an account of the Newman-Penrose formalism in which they are naturally
defined, see \cite{KSMH80,Steph90,Chandra92,Stewa90} and references therein.


\subsection{Projector onto $\Hor$}  \label{s:IN:proj}

Having introduced the transverse null direction $\w{k}$, we can now
define the {\em projector onto $\Hor$ along $\w{k}$} by
\be \encadre{
	\begin{array}{cccc}
	\w{\Pi}: & \T_p(\M) & \longrightarrow & \T_p(\Hor) \\
		& \w{v} & \longmapsto & \w{v} + (\el\cdot\w{v})\, \w{k}
	\end{array} } \label{e:IN:def_Pi}
\ee
This application is well defined, i.e. its image is in $\T_p(\Hor)$, 
since
\be
	\forall \w{v}\in \T_p(\M),\quad \el\cdot \w{\Pi}(\w{v}) = 
	\el \cdot \w{v} + (\el\cdot\w{v})\, 
	\underbrace{(\el\cdot\w{k})}_{=-1} = 0
\ee
Moreover, $\w{\Pi}$ leaves invariant any vector in $\T_p(\Hor)$ :
\be \label{e:NH:P_inv_H}
	\forall \w{v}\in \T_p(\Hor),\quad \w{\Pi}(\w{v}) = \w{v} 	
\ee
and
\be \label{e:NH:P_k_0}
	\w{\Pi}(\w{k}) = 0 .
\ee
These last two properties show that the operator $\w{\Pi}$ is the
projector onto $\Hor$ along $\w{k}$. 
The projector $\w{\Pi}$ can be written as 
\be \label{e:NH:P_l_k}
	\w{\Pi} = \w{1} + \w{k} \, \langle \uel, . \rangle  . 
\ee
It can be considered as a type 
$\left( {1\atop 1} \right)$ tensor, whose components are
\be \label{e:NH:P_comp}
	\Pi^\alpha_{\ \beta} = \delta^\alpha_{\ \beta}
		+ k^\alpha \ell_\beta . 
\ee
Comparing Eqs.~(\ref{e:NH:P_l_k}) and (\ref{e:IN:vec_q_k_l}) leads
to the following relation 
\be \label{e:NH:q_proj_P}
    \vec{\w{q}} = \w{\Pi} +  \el\, \langle \uk, .\rangle .    
\ee

\begin{rem}
The definition of the projector $\w{\Pi}$ does not depend 
on the normalization of $\el$ and $\w{k}$ as long as they
satisfy the relation $\el\cdot\w{k}=-1$ [Eq.~(\ref{e:NH:l_k_m1})].
Indeed a rescaling $\el \mapsto \w{\ell'} = \alpha \el$
would imply a rescaling $\w{k} \mapsto \w{k'} = \alpha^{-1} 
\w{k}$, leaving $\w{\Pi}$ invariant.  
In other words, $\w{\Pi}$ is determined only by the foliation 
$\Sp_t$ of $\Hor$ and not by the scale of $\Hor$'s null normal. 
Note that such a foliation-induced tranverse projector onto 
a null hypersurface has been already used in the literature
(see e.g. Ref.~\cite{Booth01}). 
\end{rem}

\begin{figure}
\centerline{\includegraphics[width=0.6\textwidth]{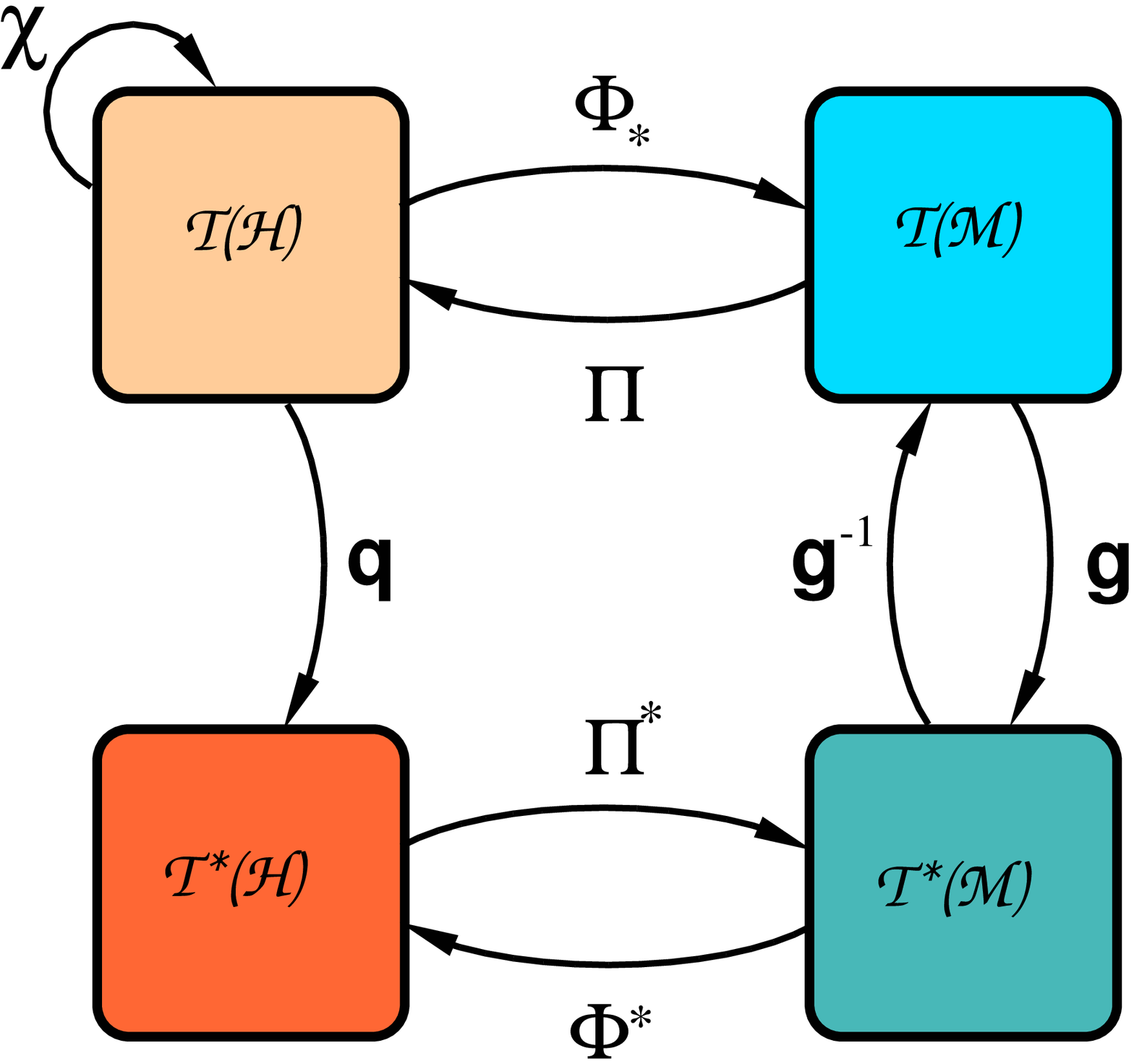}}
\caption[]{\label{f:NH:applications} \protect\footnotesize
Mappings between the space $\T_p(\Hor)$ (resp. $\T_p(\M)$) of vectors tangent
to $\Hor$ (resp. $\M$) and the space $\T_p^*(\Hor)$ (resp. $\T_p^*(\M)$) 
of 1-forms on $\Hor$ (resp. $\M$): $\Phi_*$ and $\Phi^*$ are respectively
the push-forward and the pull-back mapping canonically induced by
the embedding of $\Hor$ in $\M$; $\w{\Pi}$ is the projector onto $\Hor$ along
the null transverse direction $\w{k}$ and
$\w{\Pi}^*$ the induced mapping of 1-forms; $\w{g}$ and $\w{g}^{-1}$
denote the standard duality between vectors and 1-forms induced by the
spacetime metric $\w{g}$. Note that since the metric $\w{q}$ on $\Hor$
is degenerate, it provides a mapping $\T_p(\Hor) \rightarrow \T_p^*(\Hor)$,
but not in the reverse way. $\w{\chi}$ is the Weingarten map,
defined in Sec.~\ref{s:NH:Weingarten}, which
is an endomorphism of $\T_p(\Hor)$.}
\end{figure}

Since $\w{\Pi}$ is a well defined application $\T_p(\M)\rightarrow \T_p(\Hor)$, 
we may use it to map any linear form in $\T_p^*(\Hor)$ to a 
linear form in $\T_p^*(\M)$, in the very same way that in 
Sec.~\ref{s:NH:def_hyp} we used the application 
$\Phi_*:\ \T_p(\Hor) \rightarrow \T_p(\M)$ to map linear forms in 
the opposite way, i.e. from $\T_p^*(\M)$ to $\T_p^*(\Hor)$. 
Indeed, and more generally, if $\w{T}$ is a $n$-linear form
on $\T_p(\Hor)^n$, we define $\w{\Pi}^*\w{T}$ as the $n$-linear form 
\be \label{e:NH:def_P_star}
	\encadre{
	\begin{array}{cccc}
	\w{\Pi}^*\w{T}: & \T_p(\M)^n & \longrightarrow & \mathbb{R} \\
		& (\w{v}_1,\ldots,\w{v}_n) & \longmapsto & 
		\w{T} \left( \w{\Pi}(\w{v}_1),\ldots,\w{\Pi}(\w{v}_n) \right)
	\end{array} } .
\ee
Note that since any multilinear form on $\T_p(\M)^n$ can also be regarded
as a multilinear form on $\T_p(\Hor)^n$ thanks to the pull-back mapping
$\Phi^*$ if we identify $\Phi^*\w{T}$ with $\w{T}$ abusing of the notation
[cf. Eq.~(\ref{e:NH:def_pull-back_multi})], 
we may extend the definition (\ref{e:NH:def_P_star}) to {\it any}
multilinear form $\w{T}$ on $\T_p(\M)$. 
In index notation, we have then 
\be \label{e:NH:P_star_comp}
    (\Pi^* T)_{\alpha_1\ldots\alpha_n}
    	= T_{\mu_1\ldots \mu_n}
	\Pi^{\mu_1}_{\ \ \ \alpha_1} \cdots \Pi^{\mu_n}_{\ \ \ \alpha_n}. 	
\ee
Note that we are again abusing of the notation, since $\w{\Pi}^*$ here
should be properly denoted as $(\w{\Pi}^* \circ \Phi^*)\, \w{T}$. 
In particular, for a 1-form, the expression (\ref{e:NH:P_l_k})
for $\w{\Pi}$ yields:
\be \label{e:IN:Pi_star_form}
    \forall\w{\varpi}\in\T_p^*(\M),\quad
    \w{\Pi}^*\w{\varpi} = \w{\varpi} + \langle \w{\varpi}, \w{k} \rangle
        \, \uel . 
\ee 
For $\w{\varpi}=\uel$, we get immediately
\be \label{e:NH:P_star_l_0}
	\w{\Pi}^* \uel = 0 , 
\ee
which reflects the fact that $\uel$ restricted to $\T_p(\Hor)$ vanishes. 
On the contrary, for the 1-form $\uk$ we have
\be \label{e:NH:P_star_k}
	\w{\Pi}^* \uk = \uk . 
\ee
Collecting together Eqs.~(\ref{e:NH:P_inv_H}) (for $\w{v}=\el$), 
(\ref{e:NH:P_k_0}), (\ref{e:NH:P_star_l_0}) and 
(\ref{e:NH:P_star_k}), 
we recover the duality between $\el$ and $\w{k}$ mentioned in 
Remark~\ref{r:dual_k_l}:
\be
	\encadre{\w{\Pi}(\el) = \el} \quad \mbox{and} \quad
	\encadre{\w{\Pi}(\w{k}) = 0} ,
\ee
\be \label{e:NH:P_star_k_l}
	\encadre{\w{\Pi}^*\uel = 0} \quad \mbox{and} \quad
	\encadre{\w{\Pi}^*\uk = \uk} . 
\ee
In index notation, the above relations write respectively
\be
	\Pi^\alpha_{\ \, \mu} \ell^\mu = \ell^\alpha
		\quad \mbox{and} \quad
	\Pi^\alpha_{\ \, \mu} k^\mu = 0,
\ee
\be \label{e:NH:P_star_k_l_comp}
	\ell_\mu \Pi^\mu_{\ \, \alpha}  = 0 
		\quad \mbox{and} \quad
	k_\mu \Pi^\mu_{\ \, \alpha}  = k_\alpha .  
\ee

The various mappings introduced so far
between the vectorial spaces $\T_p(\Hor)$ and $\T_p(\M)$
and their duals are represented in Fig.~\ref{f:NH:applications}.

\begin{rem} \label{r:rigging_vector}
The vector $\w{k}$ is a special case of what is called more generally
a {\em rigging vector} \cite{MarsS93}, i.e. a vector transverse to $\Hor$
everywhere, which allows to define a projector onto $\Hor$ whatever
the character of $\Hor$
(i.e. spacelike, timelike, null or changing from point to point).
\end{rem}


\subsection{Coordinate systems stationary with respect to $\Hor$}
\label{s:IN:stacoord}

Let us consider a 3+1 coordinate system $(x^\alpha)=(t,x^i)$, 
with the associated 
coordinate time vector $\tv$ and shift vector $\w{\beta}$, as defined
in Sec.~\ref{s:FO:coord}.  
It is useful to perform an orthogonal 2+1 decomposition
of the shift vector with respect to the surface $\Sp_t$, according to
\be \label{e:IN:shift_2p1}
    \encadre{\w{\beta} = b\w{s} - \w{V}} \qquad \mbox{with} \quad 
        \w{s}\cdot\w{V} = 0 .
\ee
In other words, $b=\w{s}\cdot\w{\beta}$ and 
$\w{V} = -\vec{\w{q}}(\w{\beta}) \in \T_p(\Sp_t)$ (the minus sign is chosen
for later convenience).

Combining the two 3+1 decompositions $\el=N(\w{n}+\w{s})$ 
[Eq.~(\ref{e:IN:el_nps})] and 
$\tv = N \w{n} +\w{\beta}$ [Eq.~(\ref{e:FO:def_shift})], we get 
\be \label{e:IN:el_tau_V}
    \encadre{ \el = \tv + \w{V} + (N-b)\w{s}} . 
\ee

We say that $(x^\alpha)$ is a coordinate system {\em stationary with 
respect to the null hypersurface $\Hor$} iff the equation of $\Hor$
in this coordinate system involves only the 
spatial coordinates $(x^i)$ and does not depend upon $t$, i.e. 
iff there exist a scalar function $f(x^1,x^2,x^3)$ such that 
\be \label{e:IN:def_coord_st}
	\forall\;  p = (t,x^1,x^2,x^3) \in \M,\quad 
        p \in \Hor \iff f(x^1,x^2,x^3) = 1 .
\ee
This means that the location of the 2-surface $\Sp_t$ is fixed with 
respect to the coordinate system $(x^i)$ on $\Sigma_t$, as $t$ varies.
The gradient of $f$ is normal to $\Hor$ and thus parallel to 
$\dd u$:
\be \label{e:IN:du_df}
    \dd u \equalH \alpha \, \dd f , 
\ee
where $\equalH$ means that this identity is valid only at points on $\Hor$
and $\alpha$ is some scalar field on $\Hor$.
Equation (\ref{e:IN:du_df}) and the independence of $f$ from $t$
imply
\be
    \der{u}{t} \equalH  0 . 
\ee
This has an immediate consequence on the coordinate time vector
$\tv$:
\be
    \langle \dd u, \tv \rangle = \der{u}{x^\mu} \tvc^\mu    
        =  \der{u}{x^\mu} \delta^\mu_{\ \, t} = \der{u}{t} \equalH 0 , 
\ee
which implies that $\tv$ is tangent to $\Hor$ [cf.
Eq.~(\ref{e:NH:dr_normal})]. Consequently, for a coordinate system
stationary with respect to $\Hor$,
\be \label{e:IN:e_tau_0}
    \hat{\el} \cdot \tv = 0 . 
\ee
Replacing\footnote{Expression (\ref{e:IN:e_tau_0}) is equivalent to 
$ \el \cdot \tv = 0$ whenever $N\neq 0$ on $\Hor$.}  $\hat{\el}$
and $\tv$ by their respective 3+1 decompositions 
(\ref{e:IN:def_hat_ell}) and (\ref{e:FO:def_shift}) and using 
$b=\w{s}\cdot\w{\beta}$ leads to 
\be \label{e:IN:s_beta_N}
    b = N . 
\ee
Thus, for a coordinate system stationary with respect to $\Hor$,
the decomposition (\ref{e:IN:el_tau_V}) simplifies to 
\be
  \el = \tv + \w{V} . 
\ee
In the case where $\Hor$ is the event horizon of some black hole and
$(x^\alpha)$ is stationary with respect to $\Hor$, $\w{V}\in\T_p(\Sp_t)$ 
is called the {\em surface velocity of the black hole} 
by Damour \cite{Damou78,Damou79}. More generally, we will call $\w{V}$
the {\em surface velocity of $\Hor$ with respect to the coordinate
system $(x^\alpha)$} stationary with respect to $\Hor$. 

To summarize, we have the following :
\bea
    \mbox{( $(x^\alpha)$ stationary w.r.t. $\Hor$\ )}
        & \iff & \der{u}{t} \equalH 0 \label{e:IN:dudt_coord_st}\\
        & \iff & \tv \in \T_p(\Hor)  \label{e:IN:t_tangent_H} \\
        & \iff & \hat{\el}\cdot\tv \equalH 0   \\
        & \iff &   b \equalH N \label{e:IN:b_N_station} \\
        & \iff & \el \equalH \tv + \w{V} \label{e:IN:el_t_V_station}
\eea
Notice that for a coordinate system stationary with respect to $\Hor$,
the scalar field $u$ defining $\Hor$ is not necessarily 
such that $u=u(x^1,x^2,x^3)$ everywhere in $\M$,
but only on $\Hor$ [Eq.~(\ref{e:IN:dudt_coord_st})].

\begin{figure}
\centerline{\includegraphics[width=0.7\textwidth]{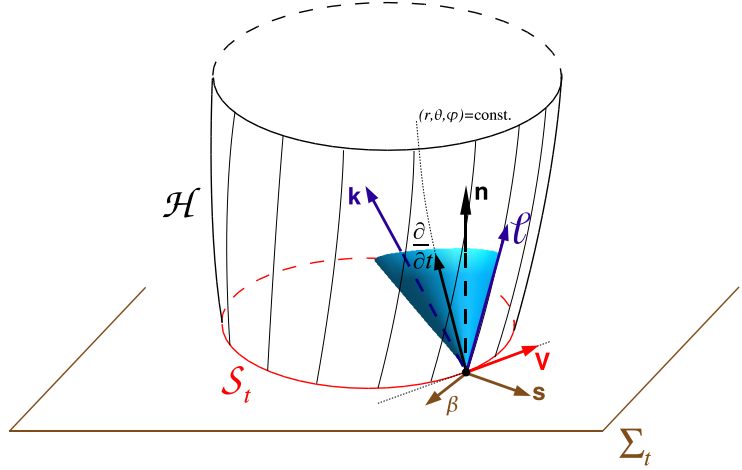}}
\caption[]{\label{f:IN:station_coord} 
Same as Fig.~\ref{f:def_s} but with the addition of 
the coordinate time vector $\tv$, the shift vector $\w{\beta}$ 
and $\Hor$'s surface velocity vector $\w{V}$ with respect to  
a coordinate system $(x^\alpha)$ stationary with respect to $\Hor$.}
\end{figure}

The vectors $\tv$, $\w{\beta}$ and $\w{V}$ of a coordinate system stationary with 
respect to $\Hor$ are shown in Fig.~\ref{f:IN:station_coord}. 

A special case of a coordinate system stationary with respect to 
$\Hor$ is a coordinate
system $(x^\alpha)$ for which the function $f$ in 
Eq.~(\ref{e:IN:def_coord_st}) is simply one of the coordinates,
$x^1$ let's say: $f(x^1,x^2,x^3)=x^1$.
We call such a system {\em a coordinate system adapted to $\Hor$}.
For instance, if the topology of $\Hor$ is $\mathbb{R}\times\mathbb{S}^2$,
an adapted coordinate system can be of spherical type
$(x^i) = (r,\vartheta,\varphi)$, 
where $r$ is such that $\Hor$ corresponds to $r=1$. 

Another special case of coordinate system stationary with respect to 
$\Hor$ is a coordinate system $(x^\alpha)$ for which $\w{V}=0$
(in addition to the stationarity condition $\w{t}\in\T_p(\Hor)$). 
We call such a system {\em a coordinate system comoving with 
$\Hor$}. From Eq.~(\ref{e:IN:el_t_V_station}) this implies
\be
    \tv \equalH \el , 
\ee
which shows that the null generators of $\Hor$ are some lines 
$x^i={\rm const}$. 

\begin{exmp} \label{ex:IN:coord}
The Minkowskian coordinates $(t,x,y,z)$ introduced in 
Examples~\ref{ex:NH:cone} and \ref{ex:IN:cone} are not stationary
with respect to the light cone $\Hor$. In particular, 
$\partial u/\partial t = -1 \not = 0$ and $N=1\ \not=\ b=0$ for these
coordinates. On the contrary, the 
Eddington-Finkelstein coordinates $(t,r,\theta,\varphi)$ introduced in 
Examples~\ref{ex:NH:EF}, \ref{ex:FO:EF} and \ref{ex:IN:EF} are
stationary with respect to the event horizon $\Hor$ of a Schwarzschild
black hole. In particular, from the expression 
(\ref{e:NH:u_EF}) for $u$, we notice that the requirement 
(\ref{e:IN:dudt_coord_st}) is fulfilled, and  
from Eqs.~(\ref{e:FO:shift_EF}) 
and (\ref{e:IN:s_comp_EF}), we get 
$b = 2m/r\, (1+2m/r)^{-1/2}$ so that $b\equalH N \equalH 1/\sqrt{2}$,
in agreement with (\ref{e:IN:b_N_station}). Moreover, 
the Eddington-Finkelstein coordinates are both adapted to $\Hor$
and comoving with $\Hor$. Indeed, the equation for $\Hor$
can be defined by $r=2m$ (instead of $u=1$), which shows the
adaptation, and we have already noticed that 
$\tv\equalH\el$ [Eq.~(\ref{e:NH:el_t_EF})], which shows the
comobility (this can also be seen from the shift vector 
which is colinear to $\w{s}$, according to Eqs.~(\ref{e:FO:shift_EF}) 
and (\ref{e:IN:s_comp_EF}), implying $\w{V}=0$). 
\end{exmp}

If $(x^\alpha)$ is a coordinate system adapted to $\Hor$, 
then\footnote{Remember the index convention given in
Sec.~\ref{s:notations}.} 
$(x^A) = (t,x^a) = (t,x^2,x^3)$ is a coordinate system
for $\Hor$. 
In terms of it, the induced
metric element on $\Hor$ is
\be
	\left. ds^2 \right| _{\Hor} = q_{AB} \, dx^A \, dx^B 
		= g_{tt} \, dt^2 + 2 g_{ta} \, dt\, dx^a +
		g_{ab} \, dx^a \, dx^b .	
\ee
Now, from Eqs.~(\ref{e:IN:el_tau_V}) and (\ref{e:IN:el_t_V_station}),
\be
	g_{tt} = \tv \cdot \tv = (\el-\w{V})\cdot(\el-\w{V}) 
        = \w{V}\cdot\w{V} = V_a V^a  
\ee
and, from Eqs.~(\ref{e:IN:shift_2p1}) and (\ref{e:us_dt_dr}),
\be
	g_{ta} = \beta_a = b s_a - V_a = - V_a . 
\ee
Besides, $g_{ab} = q_{ab}$ [cf. Eq.~(\ref{e:NH:qAB})]. 
Thus the above line element can be written
\be \label{e:IN:dsH}
	\encadre{ \left. ds^2 \right| _{\Hor} = q_{AB} \, dx^A \, dx^B 
	= q_{ab} (dx^a - V^a dt) (dx^b - V^b dt) }
\ee
This equation agrees with Eq.~(I.50c) of Damour \cite{Damou79}
(or Eq.~(6) of Appendix of Ref.~\cite{Damou82})\footnote{
Note that Damour's convention for indices $A$, $B$, ... is
the same than ours for indices $a$, $b$, ..., namely they run in
$\{2,3\}$ (whereas our convention for $A$, $B$, ... is that they
run in $\{0,2,3\}$, same as Damour's $\bar A$, $\bar B$, ...).}.

\begin{exmp}
For the Eddington-Finkelstein coordinates $(t,r,\theta,\varphi)$
considered in Examples~\ref{ex:NH:EF}, \ref{ex:FO:EF}, \ref{ex:IN:EF}
and \ref{ex:IN:coord},
$(x^A) = (t,\theta,\varphi)$ constitutes a coordinate system 
for the event horizon $\Hor$. Taking into account that
$r=2m$ on $\Hor$,  we read from the line element
(\ref{e:NH:metric_edd_fink}) that
\be
   \left. ds^2 \right| _{\Hor} = r^2 (d\theta^2 + \sin^2\theta d\varphi^2) ,
\ee 
in agreement with Eq.~(\ref{e:IN:dsH}), with, in addition $V^a=0$,
since the Eddington-Finkelstein coordinates are comoving with 
$\Hor$.
\end{exmp}

%% file: kinema.tex
%
%
\section{Null geometry in 4-dimensional version. Kinematics}
\label{s:KI}

In this section we consider the spacetime first derivatives of 
the null vectors $\el$ and
$\w{k}$ and of the associated 1-forms $\uel$ and $\uk$, as well as the
Lie derivatives of the induced metric $\w{q}$ along $\el$ and $\w{k}$.
This is what we mean by ``kinematics''. 
The first derivative of $\el$ has been represented by the Weingarten map
of $\Hor$ in Sec.~\ref{s:NH:Weingarten}.
We start by extending the definition of this map 
to the whole 4-dimensional vector space $\T_p(\M)$, whereas its original 
definition was restricted to the 3-dimensional subspace $\T_p(\Hor)$. 

\subsection{4-dimensional extensions of the Weingarten map and 
the second fundamental form of $\Hor$}

Having introduced in Sec.~\ref{s:IN:proj} the projector $\w{\Pi}$ onto 
$\Hor$, we can extend the definition 
of the Weingarten map of $\Hor$ (with respect to the null normal
$\el$) to all vectors of $\T_p(\M)$ at any point of $\Hor$, by setting
\be \label{e:NH:def_ext_Weingar} 
   \encadre{
	\begin{array}{cccc}
	\w{\chi}: & \T_p(\M) & \longrightarrow & \T_p(\Hor) \\
		& \w{v} & \longmapsto & 
		\w{\chi}_{\Hor} (\w{\Pi}(\w{v})) 
	\end{array} } ,
\ee
where $\w{\chi}_{\Hor}$ denotes the Weingarten
map introduced on $\T_p(\Hor)$ in Sec.~\ref{s:NH:Weingarten}. 
The image of $\w{\chi}$ is in $\T_p(\Hor)$ because the image of 
$\w{\chi}_{\Hor}$ is. 
Explicitly, one has [cf. Eq.~(\ref{e:NH:Weingarten_def})]
\be \label{e:IN:chi_ext_Pi}
	\forall \w{v}\in\T_p(\M),\quad
		\w{\chi}(\w{v}) = \w{\nabla}_{\w{\Pi}(\w{v})}\, \el . 
\ee
In index notation
\be
	\chi^\alpha_{\ \mu} v^\mu = \Pi(v)^\nu \nabla_\nu \ell^\alpha
	= (\delta^\nu_{\ \mu} +  k^\nu \ell_\mu) v^\mu \nabla_\nu \ell^\alpha,
\ee
hence the matrix of $\w{\chi}$:
\be \label{e:NH:Weing_comp}
	\encadre{\chi^\alpha_{\ \beta} = \nabla_\beta \ell^\alpha
	+ k^\mu \nabla_\mu \ell^\alpha \, \ell_\beta }. 
\ee

We have already noticed that $\el$ is an eigenvector of the Weingarten
map, with the eigenvalue $\kappa$ (the non-affinity coefficient)
[cf. Eq.~(\ref{e:NH:l_eigen_chi})].
Since $\w{\Pi}(\w{k})=0$, $\w{k}$ constitutes another eigenvector
of the (extended) Weingarten map, with the eigenvalue zero:
\be \label{e:NH:chi_eigen_l_k}
	\encadre{\w{\chi}(\el) = \kappa \, \el }
	\qquad \mbox{and} \qquad 
	\encadre{ \w{\chi}(\w{k}) = 0 } . 
\ee

Similarly, we make use of the projector $\w{\Pi}$ to extend the
definition of the second fundamental form of $\Hor$ with respect
to the normal $\el$ by [cf. the definition (\ref{e:NH:def_P_star})
of $\w{\Pi}^*$]
\be \label{e:NH:def_ext_Theta}
	\encadre{
	\w{\Theta} := \w{\Pi}^* \w{\Theta}_{\Hor} },
\ee
where $\w{\Theta}_{\Hor}$ denotes the second
fundamental
form of $\Hor$ with respect to $\el$ introduced 
in Sec.~\ref{s:NH:2ndfund}. 
Explicitly, $\w{\Theta}$
writes
\be \label{e:IN:def_ext_Theta2}
	\begin{array}{cccc}
	\w{\Theta}: & \T_p(\M)\times\T_p(\M) & \longrightarrow & \mathbb{R} \\
		& (\w{u},\w{v}) & \longmapsto & 
		\w{\Theta}_{\Hor} 
		\left(\w{\Pi}(\w{u}),\w{\Pi}(\w{v}) \right) .
	\end{array}  
\ee
Since $\w{\Theta}_{\Hor}$ is symmetric,
the bilinear form $\w{\Theta}$ defined above is symmetric. 
Moreover, from the relation (\ref{e:NH:q_proj_P}), we have
$\w{\Pi}(\w{u}) = \vec{\w{q}}(\w{u}) - \langle \uk, \w{u} \rangle \, \el$.
Since $\el$ is a degeneracy direction of 
$\w{\Theta}_{\Hor}$ [cf. Eq.~(\ref{e:NH:Theta_degen_l})], 
we get 
\be
	\forall (\w{u},\w{v}) \in \T_p(\M)\times\T_p(\M),\quad
	\w{\Theta}(\w{u},\w{v}) = 
        \w{\Theta}_{\Hor} (\vec{\w{q}}(\w{u}),
            \vec{\w{q}}(\w{v})) .
\ee
Replacing $\w{\Theta}_{\Hor}$ by its definition (\ref{e:NH:2ndform_def}),
we get 
\bea 
	\forall (\w{u},\w{v}) \in \T_p(\M)\times\T_p(\M),\quad
	\w{\Theta}(\w{u},\w{v}) & =  & 
		\vec{\w{q}}(\w{u}) \cdot \w{\nabla}_{\vec{\w{q}}(\w{v})}
                \, \el \nonumber \\
               & = & \w{\nabla}\uel (\vec{\w{q}}(\w{u}), \vec{\w{q}}(\w{v})) .
                \label{e:KI:Theta_grad_l_qq}
\eea
We write this relation as 
\be \label{e:KI:Theta_qstar_gradl}
    \encadre{ \w{\Theta} = \vec{\w{q}}^* \, \w{\nabla}\uel } , 
\ee
where $\vec{\w{q}}^*$ is the operator on multilinear forms
induced by the projector $\vec{\w{q}}$, in a manner similar
to $\w{\Pi}^*$ [cf. Eq.~(\ref{e:NH:def_P_star})]: for any
$n$-linear form $\w{T}$ 
on $\T_p(\M)$ or on $\T_p(\Sp_t)$, 
$\vec{\w{q}}^*\, \w{T}$ is the $n$-linear form on $\T_p(\M)$ defined by
\be \label{e:KI:def_q_star}
	\encadre{
	\begin{array}{cccc}
	\vec{\w{q}}^*\, \w{T}: & \T_p(\M)^n & \longrightarrow & \mathbb{R} \\
		& (\w{v}_1,\ldots,\w{v}_n) & \longmapsto & 
		\w{T} \left( \vec{\w{q}}(\w{v}_1),\ldots,\vec{\w{q}}(\w{v}_n) 
                \right)
	\end{array} } .
\ee
In index notation:
\be
    (\vec{q}^* T)_{\alpha_1\ldots\alpha_n}
    	= T_{\mu_1\ldots \mu_n}
	\, q^{\mu_1}_{\ \ \ \alpha_1} \cdots q^{\mu_n}_{\ \ \ \alpha_n}, 	
\ee
so that Eq.~(\ref{e:KI:Theta_qstar_gradl}) writes (taking into 
account the symmetry of $\w{\Theta}$)
\be \label{e:NH:Thetaqq}
	\encadre{ \Theta_{\alpha\beta} 
	 = \nabla_\mu \ell_\nu \; q^\mu_{\ \alpha} q^\nu_{\ \beta} } . 
\ee
The identity (\ref{e:KI:Theta_qstar_gradl})
strengthens Eq.~(\ref{e:NH:def_ext_Theta}): 
not only $\w{\Theta}$
``acts only'' in $\Hor$ (in the sense that $\w{\Theta}$ starts by a
projection onto $\Hor$), but it  ``acts only'' in the submanifold
$\Sp_t$ of $\Hor$.

The bilinear form $\w{\Theta}$ is degenerate, with at least two
degeneracy directions : $\el$ [see Eq.~(\ref{e:NH:Theta_degen_l})]
and $\w{k}$ (since $\w{\Pi}(\w{k})=0$):
\be \label{e:NH:Theta_l_k_0}
	\encadre{\w{\Theta}(\el, .) = 0 } 
	\qquad \mbox{and} \qquad
	\encadre{\w{\Theta}(\w{k}, .) = 0 } .
\ee
From Eq.~(\ref{e:NH:span_perp_S})
we conclude that any vector in the plane orthogonal 
to $\Sp_t$ is also a degeneracy direction for $\w{\Theta}$:
\be \label{e:NH:Theta_Sperp_0}
	\forall\w{v}\in\T_p(\Sp_t)^\perp,\quad 
		\w{\Theta}(\w{v}, .) = 0 . 
\ee


\subsection{Expression of $\w{\nabla}\el$: rotation 1-form and
H\'a\'\j i\v{c}ek 1-form} \label{s:KI:rot_haji}

A quantity which appears very often in our study is the spacetime 
covariant derivative of the null normal: $\w{\nabla}\uel$. 
Let us recall that, thanks to some null foliation $(\Hor_u)$,
we have extended the definition of $\el$ to an
open neighborhood of $\Hor$ (cf. Sec.~\ref{s:NH:extended_el}). 
Consequently the covariant derivatives
$\w{\nabla}\el$ and $\w{\nabla}\uel$ are well defined. 
$\w{\nabla}\uel$ is a bilinear form on $\T_p(\M)$. 
Let us express it in terms of the bilinear form $\w{\Theta}$ 
which we have just extended to the whole space $\T_p(\M)$. 
For two arbitrary vectors  $\w{u}$ and $\w{v}$ of $\T_p(\M)$,
by combining the definition (\ref{e:IN:def_ext_Theta2}) of $\w{\Theta}$
with the definition (\ref{e:NH:2ndform_def}) of 
$\w{\Theta}_{\Hor}$ and making use of the 
expression (\ref{e:IN:def_Pi}) of $\w{\Pi}$, one has
\bea
    \w{\Theta}(\w{u},\w{v}) & = & \w{\Pi}(\w{u}) \cdot 
        \w{\nabla}_{\w{\Pi}(v)} \el
        \ = \ \left( \w{u} + \langle \uel, \w{u} \rangle \w{k} \right)
            \cdot \w{\nabla}_{\w{\Pi}(v)} \el \nonumber \\
            &= &
             \w{u} \cdot \w{\nabla}_{\w{\Pi}(v)} \el
            + \langle \uel, \w{u} \rangle \w{k}  
            \cdot 
            \underbrace{\w{\nabla}_{\w{\Pi}(v)} \el}_{=\w{\chi}(\w{v})}
             =  \w{u} \cdot 
        \w{\nabla}_{\w{v}+\langle \uel, \w{v}\rangle \w{k}}
            \el + \langle \uel, \w{u} \rangle \w{k}  
            \cdot \w{\chi}(\w{v}) \nonumber \\
        & = & \w{u} \cdot \w{\nabla}_{\w{v}} \el + 
           \langle \uel, \w{v}\rangle \w{u} \cdot \w{\nabla}_{\w{k}}\el
            + \langle \uel, \w{u} \rangle \w{k}  
            \cdot \w{\chi}(\w{v}) \nonumber \\
        & = & \w{\nabla}\uel(\w{u},\w{v}) +
            \langle \uel, \w{v}\rangle 
            \langle \w{\nabla}_{\w{k}}\, \uel, \w{u} \rangle
            + \langle \uel, \w{u} \rangle \w{k}  
            \cdot \w{\chi}(\w{v}) . \label{e:IN:Theta_uv}
\eea
In this expression appears the 1-form
\be \label{e:IN:def_omega}
	\encadre{
	\begin{array}{cccc}
	\w{\omega}: & \T_p(\M) & \longrightarrow & \mathbb{R} \\
		& \w{v} & \longmapsto & 
		- \w{k} \cdot \w{\chi}(\w{v})
	\end{array} } ,
\ee
which we call the {\em rotation 1-form} (for reasons which will become clear
later; see Sec. \ref{s:IH:ang_mom}). 
The relation (\ref{e:IN:Theta_uv}) then reads 
\be
   \w{\Theta}(\w{u},\w{v}) =  \w{\nabla}\uel(\w{u},\w{v}) +
            \langle \uel, \w{v}\rangle 
            \langle \w{\nabla}_{\w{k}}\, \uel, \w{u} \rangle
            - \langle \w{\omega}, \w{v} \rangle
             \langle \uel, \w{u} \rangle  .
\ee
Since this equation is valid whatever $\w{u}$ and $\w{v}$ in $\T_p(\M)$,
we obtain the relation we were looking for:
\be \label{e:KI:grad_uel}
   \encadre{ \w{\nabla}\uel = \w{\Theta}
     + \uel \otimes  \w{\omega} - 
     \w{\nabla}_{\w{k}}\, \uel  \otimes \uel }  .
\ee
Taking into account the symmetry of $\w{\Theta}$, the 
`index' version of the above relation is
[see Eq.~(\ref{e:IN:grad_1form})]:
\be  \label{e:KI:grad_uel_index}
  \nabla_\alpha \ell_\beta = \Theta_{\alpha\beta}
		+ \omega_\alpha \ell_\beta
		- \ell_\alpha k^\mu \nabla_\mu \ell_\beta   .
\ee
An equivalent form of Eq.~(\ref{e:KI:grad_uel}), obtained via the
standard metric duality [or by raising the last index of 
Eq.~(\ref{e:KI:grad_uel_index})], gives the covariant derivative
of the vector field $\el$
\be \label{e:KI:grad_el}
   \w{\nabla}\el = \vec{\w{\Theta}}
     + \el\otimes \w{\omega}  - \w{\nabla}_{\w{k}}\, \el \otimes  \uel  , 
\ee
where $\vec{\w{\Theta}}$ is the endomorphism canonically associated with 
the bilinear form $\w{\Theta}$ by the metric $\w{g}$
[see the notation (\ref{e:IN:arrow_endo})]. Its components
are $\Theta^\alpha_{\ \, \beta} = g^{\alpha\mu} \Theta_{\mu\beta}$
and it is related to 
the Weingarten map $\w{\chi}$ by 
\be \label{e:def_vec_Theta}
    \vec{\w{\Theta}} := \vec{\w{q}} \circ \w{\chi} \circ \vec{\w{q}}=
\vec{\w{q}} \circ \w{\chi} .
\ee
By combining Eqs.~(\ref{e:IN:chi_ext_Pi}), (\ref{e:KI:grad_el})
and using the fact that $\langle \w{\omega}, \w{k}\rangle=0$,
so that $\langle \w{\omega}, \w{\Pi}(\w{v}) \rangle=
\langle \w{\omega}, \w{v} \rangle$ for any $\w{v}\in\T_p(\M)$,
we get a simple expression relating the extended Weingarten map
to the endomorphism $\vec{\w{\Theta}}$ and the rotation 1-form
$\w{\omega}$:
\be \label{e:KI:chi_Theta_omega}
    \encadre{\w{\chi} = \vec{\w{\Theta}}
        + \langle \w{\omega}, . \rangle \, \el } . 
\ee

Let us now discuss further the rotation 1-form $\w{\omega}$. 
First, from the expressions (\ref{e:IN:chi_ext_Pi}) for $\w{\chi}$
and (\ref{e:IN:def_Pi}) for $\w{\Pi}$, one has
\bea
   \forall\w{v}\in\T_p(\M),\quad  \langle \w{\omega}, \w{v} \rangle
   & = & - \w{k}\cdot \w{\nabla}_{\w{\Pi}(\w{v})} \, \el 
    =  - \w{k}\cdot \w{\nabla}_{\w{v} + \langle \uel, \w{v} \rangle \w{k}}
        \, \el \nonumber \\
  & = & - \w{k}\cdot\w{\nabla}_{\w{v}}\, \el - 
        \langle \uel, \w{v} \rangle \w{k}\cdot \w{\nabla}_{\w{k}}\, \el . 
\eea
Hence
\be \label{e:KI:omega_grad_l}
    \w{\omega} = - \w{k} \cdot \w{\nabla}\uel  
        - (\w{k}\cdot\w{\nabla}_{\w{k}}\el) \; \uel .
\ee

Next, for any vector $\w{v}\in\T_p(\M)$ we have  
$\langle \w{\omega}, \w{v} \rangle = - \langle \uk, \w{\chi}(\w{v}) \rangle$.
Since the image of the extended Weingarten map $\w{\chi}$ is in 
$\T_p(\Hor)$ and the action of the 1-forms $\uk$ and $-\dd t$ coincide on 
$\T_p(\Hor)$ [cf. Eq.~(\ref{e:IN:k_dt_H})], we get the following 
alternative expressions for $\w{\omega}$:
\be \label{e:IN:omega_dt_chi}
   \forall\w{v}\in\T_p(\M),\quad  \langle \w{\omega}, \w{v} \rangle =
   \langle \dd t, \w{\chi}(\w{v}) \rangle 
   = \langle \dd t, \w{\nabla}_{\w{\Pi}(v)} \el \rangle, 
\ee
which we can write in terms of the function composition operator 
$\circ$ as
\be \label{e:KI:omega_dt_chi}
    \encadre{ \w{\omega} = \dd t \; \circ \; \w{\chi} } .
\ee

Besides, from the very definition (\ref{e:IN:def_omega}) of $\w{\omega}$,
the eigenvector expressions
(\ref{e:NH:chi_eigen_l_k}) lead immediately to the following 
action on the null vectors $\el$ and $\w{k}$:
\be \label{e:KI:omega_l_kappa}
    \encadre{\langle \w{\omega}, \el \rangle = \kappa}
    \qquad \mbox{and} \qquad
    \encadre{\langle \w{\omega}, \w{k} \rangle = 0} .
\ee
 
The pull-back of the rotation 1-form to the 2-surfaces $\Sp_t$ 
by the inclusion mapping of $\Sp_t$ into $\M$
is called the {\em \hajicek\  1-form} and is denoted
by the capital letter $\w{\Omega}$. Following the 4-dimensional point of view
adopted in this article, we can extend the
definition of $\w{\Omega}$ to all vectors in $\T_p(\M)$ 
thanks to the orthogonal projector $\vec{\w{q}}$ and set 
[see definition (\ref{e:KI:def_q_star})]
\be \label{e:KI:def_hajicek}
    \encadre{\w{\Omega} := \w{\omega} \, \circ \, \vec{\w{q}} }
    \qquad \mbox{or} \qquad
    \encadre{\w{\Omega} := \vec{\w{q}}^* \w{\omega} } .    
\ee 
Replacing $\w{\omega}$ by its definition (\ref{e:IN:def_omega})
leads to 
\be
 \forall\w{v}\in\T_p(\M),\quad  
 \encadre{\langle \w{\Omega}, \w{v} \rangle =
 - \w{k} \cdot \w{\nabla}_{\vec{\w{q}}(\w{v})} \, \el } .
\ee
The 1-form $\w{\Omega}$ has been introduced by \hajicek\  
\cite{Hajic73,Hajic75} in the special case $\w{\Theta}=0$
(non-expanding horizons, to be discussed in Sec.~\ref{s:NE})
and in the general case by Damour \cite{Damou79,Damou82}. 
$\w{\Omega}$ is considered by \hajicek\  as a ``gravimagnetic field'',
whereas it is viewed by Damour as a surface momentum 
density, as we shall see in Sec.~\ref{s:DN:DNS}. The form
$\w{\Omega}$ has also been used in the subsequent {\em membrane paradigm}
formulation of Price \& Thorne \cite{PriceT86}. Actually, 
restricting the action of $\w{\Omega}$ to $\T_p(\Sp_t)$, on 
which $\vec{\w{q}}$ is the identity operator, and using 
Eq.~(\ref{e:IN:omega_dt_chi}), we get 
\be
	\forall \w{v}\in\T_p(\Sp_t),\quad 
	\langle	\w{\Omega}, \w{v} \rangle =  \langle
	\dd t, \,  \w{\nabla}_{\w{v}} \, \el \rangle ,  
\ee
This expression agrees with Eq.~(2.6) of Ref.~\cite{PriceT86},
used by Price \& Thorne as the definition of the H\'a\'\j i\v{c}ek 1-form.
By construction (because of the orthonormal projector $\vec{\w{q}}$
onto $\Sp_t$), the H\'a\'\j i\v{c}ek 1-form vanishes for any vector orthogonal
to the 2-surface $\Sp_t$:
\be
	\forall \w{v}\in\T_p(\Sp_t)^\perp,\quad
		\langle \w{\Omega}, \w{v} \rangle = 0 . 
\ee
In particular, it vanishes on the null dyad $(\el,\w{k})$:
\be \label{e:NH:Omega_l_k_0}
	\encadre{ \langle \w{\Omega}, \el \rangle = 0 }
\qquad \mbox{and}\qquad
	\encadre{ \langle \w{\Omega}, \w{k} \rangle = 0 }. 
\ee
Actually the 1-form $\w{\Omega}$ can be viewed as a 1-form
intrinsic to the 2-surface $\Sp_t$, independently of the fact that
$\Sp_t$ is a submanifold of $\Hor$ or $\Sigma_t$. It describes some part
of the extrinsic geometry of $\Sp_t$ as a submanifold of $(\M,\w{g})$
and is called generically a {\em normal fundamental form} of the 
2-surface $\Sp_t$ \cite{Haywa94,Epp00,Gourg05}.
The remaining part of the extrinsic geometry of $\Sp_t$ is described
by the second fundamental tensor $\w{\mathcal{K}}$ discussed in 
Remark~\ref{r:KI:2nd_fund_tensor} below. 

We have, thanks to the expression (\ref{e:NH:q_proj_P}) for 
$\vec{\w{q}}$,
\bea
 \forall\w{v}\in\T_p(\M),\quad  \langle \w{\Omega}, \w{v} \rangle 
  & = & \langle \w{\omega}, \vec{\w{q}}(\w{v}) \rangle 
    = \langle \w{\omega}, \w{\Pi}(\w{v}) + \langle \uk, \w{v}\rangle\, \el 
    \rangle \nonumber \\
    & = & 
    \underbrace{\langle \w{\omega}, \w{\Pi}(\w{v}) \rangle}_{=
    \langle \w{\omega}, \w{v} \rangle}
        + \langle \uk, \w{v}\rangle 
        \underbrace{\langle \w{\omega}, \el \rangle}_{=\kappa} .
\eea
Hence it follows the simple relation between the rotation 1-form $\w{\omega}$,
the H\'a\'\j i\v{c}ek 1-form $\w{\Omega}$ and the ``transverse'' 1-form $\uk$:
\be \label{e:KI:Omega_omega_k}
    \encadre{ \w{\omega} = \w{\Omega} - \kappa \, \uk } . 
\ee


\subsection{Frobenius identities}
\label{s:NH:Frobenius_k}

From the expression (\ref{e:KI:grad_uel}) of the spacetime
covariant derivative $\w{\nabla}\uel$,
we can compute the exterior derivative  
of $\Hor$'s normal 1-form $\uel$ 
following Eq.~(\ref{e:IN:d1form_der_cov}).
We get, taking into account the 
symmetry of $\w{\Theta}$, 
\bea
    \dd \uel & = & \w{\omega} \otimes \uel - \uel \otimes \w{\omega}
        - \uel \otimes  \w{\nabla}_{\w{k}}\, \uel
        + \w{\nabla}_{\w{k}}\, \uel \otimes \uel =
        \w{\omega} \wedge \uel
            + \w{\nabla}_{\w{k}}\, \uel  \wedge \uel \nonumber \\
     \dd \uel  & = & \left( \w{\omega} + \w{\nabla}_{\w{k}}\, \uel \right)
            \wedge \uel .   \label{e:KI:Frobenius_l}
\eea 
The fact that the exterior derivative of $\uel$ is the exterior product
of some 1-form  with $\uel$
itself reflects the fact that $\uel$ is normal to some 
hypersurface ($\Hor$): this is the dual formulation of Frobenius theorem 
already noticed in Sec.~\ref{s:NH:Frobenius}. Actually 
Eq.~(\ref{e:KI:Frobenius_l}) has the same structure as 
Eq.~(\ref{e:NH:Frobenius_l}). Let us rewrite the latter by 
expressing  $\rho$ is terms of the lapse function $N$
and the factor $M$ [Eq.~(\ref{e:IN:def_M})]:
\be \label{e:KI:dl_MN_l}
    \encadre{ \dd \uel = \dd \ln (MN) \wedge \uel } . 
\ee

Let us now evaluate the exterior derivative of the 1-form $\uk$
dual to the ingoing null vector $\w{k}$. 
First of all, the definition of $\w{k}$ 
is extended to an open  neighborhood of $\Hor$ in $\M$ as the 
vector field which satisfies (i) $\w{k}$ is an ingoing null normal 
to the 2-surface $\Sp_{t,u}$ [cf. Eq.~(\ref{e:IN:def_Str})]
and (ii) $\w{k}\cdot\el=-1$.  
The exterior derivative $\dd\uk$
is then well defined. 
Starting from expression 
(\ref{e:NH:uk_dt_dr}) for $\uk$ and using $\dd \dd = 0$, we have immediately
[cf. formula~(\ref{e:NH:ext_deriv})]
\be 
    \dd \uk  =  -\frac{1}{2} \; \dd \left( \frac{M}{N} \right) \wedge \dd u 
        = -\frac{1}{2MN} \; \dd \left( \frac{M}{N} \right) \wedge \uel , 
\ee
where we have used $\uel = MN \dd u$ [cf. Eq.~(\ref{e:NH:l_grad_r})].
Hence
\be \label{e:KI:dk}
    \encadre{ \dd \uk = \frac{1}{2N^2} \; \dd \ln \left( \frac{N}{M} \right)
     \wedge \uel } .
\ee 
\begin{rem}
\label{r:NH:k_normal_surface}
Since a priori the 1-form $\dd \ln (N/M)$ is not of the form 
$\alpha \uk + \beta \uel$ (in which case Eq.~(\ref{e:KI:dk})
would write $\dd \uk = \alpha/(2N^2)\, \uk\wedge\uel$), 
we deduce from the (dual formulation of) of Frobenius theorem 
(see e.g. Theorem B.3.2 in Wald's textbook \cite{Wald84})
and Eq.~(\ref{e:KI:dk}) that the hyperplane normal to $\uk$ is
not integrable into some hypersurface. 
On the other side, the Frobenius theorem and
relations (\ref{e:KI:dl_MN_l}) and (\ref{e:KI:dk}) imply that 
the 2-planes normal to both $\uel$ and $\uk$ are integrable into
a 2-surface: it is $\Sp_t$ [cf. the property (\ref{e:NH:span_perp_S})].
\end{rem}


\subsection{Another expression of the rotation 1-form} \label{s:KI:another}

Let us show that the Frobenius identity (\ref{e:KI:dk}) leads to the identification
of the rotation 1-form $\w{\omega}$ with 
the covariant derivative of the 1-form $\uk$ along the vector $\el$.
Starting from the definition of $\w{\omega}$
[Eq.~(\ref{e:IN:def_omega})], any vector $\w{v}\in\T_p(\M)$ satisfies
\bea
    \langle \w{\omega},\w{v}\rangle & = &
        - \w{k}\cdot\w{\nabla}_{\w{\Pi}(\w{v})} \el 
        = \el\cdot\w{\nabla}_{\w{\Pi}(\w{v})}  \w{k}
        = \langle \w{\nabla}_{\w{\Pi}(\w{v})} \uk, \el \rangle
        = \langle \w{\nabla} \uk\cdot \w{\Pi}(\w{v}),\; \el \rangle
        \nonumber \\
        & = & \langle - \dd \uk \cdot \w{\Pi}(\w{v}) 
            +  \w{\Pi}(\w{v})\cdot \w{\nabla}\uk, \; \el \rangle 
        \nonumber \\
        & = & \left\langle \frac{1}{2N^2} \left[
            \w{\nabla}_{\w{\Pi}(\w{v})} \ln \left( \frac{N}{M} \right)
            \; \uel 
  - \underbrace{\langle \uel, \w{\Pi}(\w{v}) \rangle}_{=0}
    \dd \ln \left( \frac{N}{M} \right) \right]
    ,\ \el \right\rangle \nonumber \\
   & &  + \ \langle  \w{\Pi}(\w{v})\cdot\w{\nabla}\uk  ,\; \el \rangle
        \nonumber \\
    & = & \frac{1}{2N^2} 
        \w{\nabla}_{\w{\Pi}(\w{v})} \ln \left( \frac{N}{M} \right) 
        \; \underbrace{\langle \uel, \el \rangle}_{=0} \ + \ 
        \langle \w{\nabla}_{\el} \uk, \w{\Pi}(\w{v}) \rangle  
        \nonumber \\
    & = &  \langle \w{\nabla}_{\el} \, \uk,\;  \w{v} + (\el\cdot\w{v})\w{k} \rangle
     = \langle \w{\nabla}_{\el}\,  \uk, \w{v} \rangle
        + (\el\cdot\w{v}) 
        \underbrace{\langle \w{\nabla}_{\el}\,  \uk, \w{k}\rangle}_{=0}    
        \nonumber \\
    & = &  \langle \w{\nabla}_{\el}\,  \uk, \w{v} \rangle ,         
\eea
where the relation $\w{k}\cdot\el=-1$ has been used in the first line,
Eq.~(\ref{e:IN:d1form_der_cov}) to obtain the second line, and
Eq.~(\ref{e:KI:dk}) to get the third line.
Therefore
\be \label{e:KI:omega_gradl_uk}
    \encadre{ \w{\omega} = \w{\nabla}_{\el}\,  \uk } .
\ee
From Eq.~(\ref{e:IN:k_dt_H}), $\uk$ can be written as
$\uk=-\w{\Pi}^*\dd t$ on $\Hor$. If we make use of this, together
with the expression (\ref{e:NH:q_proj_P}) for $\w{\Pi}$, 
the pull-back of the above relation on $\Hor$ results in  
\be
    \encadre{ \Phi^* \w{\omega} = - \Phi^* \w{\nabla}_{\el} \, \dd t} ,
\ee
which provides some nice perspective on $\w{\omega}$, 
alternative to Eq.~(\ref{e:KI:omega_dt_chi}).
Combining Eqs.~(\ref{e:KI:omega_gradl_uk}) and (\ref{e:KI:omega_l_kappa})
results in the simple relation
\be
    \w{\nabla}\uk(\el,\el) = \kappa . 
\ee 

Another consequence of the Frobenius identity (\ref{e:KI:dk}) is
\bea
\label{e:KI:4_dim_gradk}
    \w{k}\cdot\dd \uk & = & \frac{1}{2N^2} \left[
        \w{\nabla}_{\w{k}}  \ln \left( \frac{N}{M} \right)  \; \uel
        - \underbrace{\langle \uel, \w{k}\rangle}_{=-1}
        \dd \ln \left( \frac{N}{M} \right) \right] \nonumber \\
   \w{\nabla}\uk \cdot\w{k}-  \underbrace{\w{k}\cdot \w{\nabla}\uk }_{=0}
    & = & 
    \frac{1}{2N^2} \left[ 
    \w{\nabla}_{\w{k}}  \ln \left( \frac{N}{M} \right)  \; \uel 
    + \dd \ln \left( \frac{N}{M} \right) \right] .
\eea
Hence [cf. Eq.~(\ref{e:IN:Pi_star_form})]
\be \label{e:KI:gradk_uk}
    \encadre{
    \w{\nabla}_{\w{k}} \, \uk =  \frac{1}{2N^2}  \w{\Pi}^* 
        \dd  \ln \left( \frac{N}{M} \right) }.
\ee
Since $\w{\nabla}\uel(\w{k},\w{k}) = \w{k}\cdot\w{\nabla}_{\w{k}}\, \el
= - \el \cdot \w{\nabla}_{\w{k}} \, \w{k}
= - \langle \w{\nabla}_{\w{k}} \, \uk , \el\rangle$
and $\w{\Pi}(\el)=\el$, we deduce from the
above relation that
\be \label{e:KI:graduel_kk}
    \w{\nabla}\uel(\w{k},\w{k}) = - \frac{1}{2N^2} \w{\nabla}_{\el}
        \ln \left( \frac{N}{M} \right) .
\ee

Let us now evaluate $\w{\nabla}_{\w{k}}\, \uel$. 
Contracting the Frobenius relation (\ref{e:KI:dl_MN_l}) for $\uel$
with the vector $\w{k}$ yields
\bea
    \w{k}\cdot\dd\uel & = & \w{\nabla}_{\w{k}}  \ln (MN) \; \uel
        - \underbrace{\langle \uel, \w{k}\rangle}_{=-1} \dd  \ln (MN)
            \nonumber \\
     \w{\nabla}\uel \cdot \w{k}
    - \w{k}\cdot\w{\nabla}\uel   & = & \w{\Pi}^* \dd  \ln (MN) \nonumber \\
      \w{\nabla}_{\w{k}}\, \uel & = &  \w{k}\cdot\w{\nabla}\uel 
        + \w{\Pi}^* \dd  \ln (MN)  \ .  
\eea
Substituting Eq.~(\ref{e:KI:omega_grad_l}) for $\w{k}\cdot\w{\nabla}\uel$
and using Eq.~(\ref{e:KI:graduel_kk}) gives
\be 
   \encadre{ \w{\nabla}_{\w{k}}\, \uel = -\w{\omega} + \frac{1}{2N^2} \w{\nabla}_{\el}
        \ln \left( \frac{N}{M} \right)\; \uel 
         + \w{\Pi}^* \dd  \ln (MN) }. 
\ee
From this relation and Eqs.~(\ref{e:KI:graduel_kk}) and 
(\ref{e:IN:Pi_star_form}), we get the following expression for the 
rotation 1-form:
\be \label{e:KI:omega_Pi_something}
    \w{\omega} = \w{\Pi}^* \left(
        \dd\ln(MN) - \w{\nabla}_{\w{k}}\, \uel \right),
\ee
which clearly shows that the action of $\w{\omega}$ vanishes in the direction 
$\w{k}$.


\subsection{Deformation rate of the 2-surfaces $\Sp_t$} \label{s:KI:deform_rate}

The choice of $\el$ as the tangent vector to $\Hor$'s null generators
corresponding to the parameter $t$ (cf. Sec.~\ref{s:IN:normal_l})
makes it the natural vector field to describe the evolution of
$\Hor$'s fields with respect to $t$. 
Following Damour \cite{Damou79,Damou82}, we define the {\em tensor of
deformation rate with respect to $\el$} of the 2-surface $\Sp_t$ as 
half the Lie derivative of $\Sp_t$'s metric $\w{q}$
along the vector field $\el$:
\be \label{e:KI:def_Q}
	\w{Q} := \frac{1}{2} \, \LieS{\el} \w{q} ,
\ee
where $\w{q}$ is considered as a bilinear form field on $\Sp_t$
and $\LieS{\el}$ is the Lie derivative intrinsic to $(\Sp_t)$
which arises from the Lie-dragging of $\Sp_t$ by $\el$
(cf. Sec.~\ref{s:IN:normal_l} and Fig.~\ref{f:IN:Lie_St}). 
The precise definition of $\LieS{\el}$ is given in Appendix~\ref{s:LF}.
The relation with the Lie derivative along $\el$ within the
manifold $\M$, $\Lie{\el}$, is given in 4-dimensional form
by Eq.~(\ref{e:KI:qLqT})
\be \label{e:KI:qstar_LieS_q}
    \vec{\w{q}}^* \LieS{\el} \w{q} =  \vec{\w{q}}^* \Lie{\el} \w{q}, 
\ee
where we have used $\vec{\w{q}}^* \w{q}=\w{q}$.
Thus Eq.~(\ref{e:KI:def_Q}) becomes
\be \label{e:KI:qD_qLq}
    \vec{\w{q}}^* \w{Q} = \frac{1}{2} \vec{\w{q}}^* \Lie{\el} \w{q} , 
\ee
where $\w{q}$ is now considered as a bilinear form on $\M$
[as given by Eq.~(\ref{e:q_g_nn_ss}) or Eq.~(\ref{e:NH:q_g_l_k})].
Let us evaluate the (4-dimensional) Lie derivative in the right-hand side
of the above equation, by substituting Eq.~(\ref{e:NH:q_g_l_k})
for $\w{q}$:
\bea    
    \Lie{\el} \w{q} & = & \Lie{\el}(\w{g} + \uel \otimes \uk 
    + \uk \otimes \uel) \nonumber \\
    & = & \Lie{\el} \w{g} + \Lie{\el}\uel \otimes \uk 
    + \uel \otimes \Lie{\el}\uk 
    + \Lie{\el}\uk \otimes \uel + \uk \otimes \Lie{\el}\uel .
\eea
Since $\vec{\w{q}}^* \uel = 0$ and $\vec{\w{q}}^* \uk = 0$, 
only the term $\Lie{\el} \w{g}$ remains in the right-hand side when
applying the operator $\vec{\w{q}}^*$, so that 
Eq.~(\ref{e:KI:qD_qLq}) becomes
\be
   \vec{\w{q}}^* \w{Q} = \frac{1}{2} \vec{\w{q}}^* \Lie{\el} \w{g} . 
\ee
Now $\Lie{\el} \w{g}$ is the Killing operator applied to the 1-form
$\uel$ : 
$\Liec{\el}g_{\alpha\beta} = \nabla_\alpha \ell_\beta 
+ \nabla_\beta \ell_\alpha$. Then from 
Eq.~(\ref{e:NH:Thetaqq}) and taking into account the symmetry of $\w{\Theta}$,
we get
\be
    \vec{\w{q}}^* \w{Q} = \w{\Theta} . 
\ee
Replacing $\w{Q}$ by its definition (\ref{e:KI:def_Q}), 
we conclude that the
second fundamental form of $\Hor$ is related to the deformation rate
of the 2-surface $\Sp_t$ by 
\be \label{e:KI:Theta_deform}
    \encadre{ \w{\Theta} =  \frac{1}{2} \, \vec{\w{q}}^* \; \LieS{\el} \w{q} 
    = \frac{1}{2} \, \vec{\w{q}}^* \; \Lie{\el} \w{q} } ,  
\ee
where the second equality follows from Eq.~(\ref{e:KI:qstar_LieS_q}). 

Let us consider a coordinate system $(x^\alpha)=(t,x^i)$.
Then, according to Eq.~(\ref{e:IN:el_tau_V}),
$\el = \tv + \w{V} + (N-b)\w{s}$, where $\tv$ is the coordinate time vector
associated with $(x^\alpha)$, $\w{V} = \vec{\w{q}}(\el-\tv)$, and $b$
is the component of the shift vector of $(x^\alpha)$ along the spatial
normal $\w{s}$ to $\Sigma_t$. Equation (\ref{e:KI:Theta_deform})
can then be written
\bea
    \Theta_{\alpha\beta} & = & \frac{1}{2} 
        \left( \Liec{\tv} q_{\mu\nu}
        + \Liec{\w{V}} q_{\mu\nu} 
        + {\mathcal L}_{(N-b)\w{s}}\,  q_{\mu\nu} \right) 
        q^\mu_{\ \, \alpha} q^\nu_{\ \, \beta}
            \nonumber \\
    & = & \frac{1}{2} 
        \Big\{ \Liec{\tv} q_{\mu\nu}
            + V^\sigma \nabla_\sigma q_{\mu\nu}
                + q_{\sigma\nu}\nabla_\mu V^\sigma
                + q_{\mu\sigma}\nabla_\nu V^\sigma
                + (N-b) s^\sigma \nabla_\sigma q_{\mu\nu} \nonumber \\
            &&    + q_{\sigma\nu}\nabla_\mu [(N-b) s^\sigma]
                + q_{\mu\sigma}\nabla_\nu [(N-b) s^\sigma]
                \Big\} \, q^\mu_{\ \, \alpha} q^\nu_{\ \, \beta}  \nonumber \\
    & = & \frac{1}{2} \Big\{
        \left( \Liec{\tv} q_{\mu\nu}  \right) 
                q^\mu_{\ \, \alpha} q^\nu_{\ \, \beta} 
        + q^\mu_{\ \, \alpha} q_{\sigma\beta} \nabla_\mu V^\sigma
        + q_{\alpha\sigma} q^\nu_{\ \, \beta} \nabla_\nu V^\sigma
      \nonumber \\
      & & + (N-b)  \left(  q^\mu_{\ \, \alpha} q_{\sigma\beta} \nabla_\mu s^\sigma
      + q_{\alpha\sigma} q^\nu_{\ \, \beta} \nabla_\nu s^\sigma \right)
      \Big\} \nonumber \\
   \Theta_{\alpha\beta}  & = & \frac{1}{2} \left[
        \Liec{\tv} q_{\mu\nu} 
        + \nabla_\mu V_\nu  +  \nabla_\nu V_\mu
        + (N-b) \left( \nabla_\mu s_\nu + \nabla_\nu s_\mu \right) \right]
         q^\mu_{\ \, \alpha} q^\nu_{\ \, \beta} 
                            \label{e:KI:Theta_qqdv},
\eea 
where the last but one equality results from the identities
$q_{\mu\sigma} s^\sigma=0$
and $q^\mu_{\ \, \alpha} q^\nu_{\ \, \beta} \nabla_\sigma q_{\mu\nu} = 0$.
This last identity follows immediately from Eq.~(\ref{e:NH:q_g_l_k}). 
Now, similarly to Eq.~(\ref{e:FO:3der}) and thanks to the fact
that $\w{V}\in\T(\Sp_t)$,
\be \label{e:KI:qqnab_d2}
    q^\mu_{\ \, \alpha} q^\nu_{\ \, \beta} \nabla_\mu V_\nu = 
    \DS_\alpha V_\beta , 
\ee
where $\w{\DS}$ denotes the covariant derivative in the surface
$\Sp_t$ compatible with the induced metric $\w{q}$. 
More generally the relation between $\w{\DS}$ derivatives and
$\w{\nabla}$ derivatives is given by a formula analogous to 
Eq.~(\ref{e:FO:3der}), with the projector $\vec{\w{\gamma}}$
simply replaced by the projector $\vec{\w{q}}$:
\be \label{e:KI:2der}
\encadre{\DS_\gamma T^{\alpha_1\ldots\alpha_p}_{\ \qquad\beta_1\ldots\beta_q}
		= q_{\ \ \, \mu_1}^{\alpha_1} \, \cdots 
		 q_{\ \ \, \mu_p}^{\alpha_p} \,
		  q_{\ \ \, \beta_1}^{\nu_1} \, \cdots
		  q_{\ \ \, \beta_q}^{\nu_q} \,
		  q_{\ \ \, \gamma}^{\sigma} \, \nabla_\sigma
		  T^{\mu_1\ldots\mu_p}_{\ \qquad\nu_1\ldots\nu_q} }, 		  	
\ee  
where $\w{T}$ is any tensor of type $\left({p\atop q}\right)$ lying in 
$\Sp_t$ (i.e. such that its contraction with the normal vectors 
$\w{n}$ and $\w{s}$ (or $\el$ and $\w{k}$)
on any of its indices vanishes).
On the other side 
\be \label{e:KI:qnab_s_H}
    \left( \nabla_\mu s_\nu + \nabla_\nu s_\mu \right) 
         q^\mu_{\ \, \alpha} q^\nu_{\ \, \beta} = H_{\beta\alpha}
         + H_{\alpha\beta} = 2 H_{\alpha\beta} , 
\ee
where $\w{H}$ is the extrinsic curvature of the surface $\Sp_t$
considered as a hypersurface embedded in the Riemannian space
$(\Sigma_t,\w{\gamma})$. $\w{H}$ is a symmetric bilinear form which 
vanishes in the directions orthogonal to $\Sp_t$. It will be discussed
in a greater extent in Sec.~\ref{s:TP:H}. In particular the formula
(\ref{e:KI:qnab_s_H})
is a direct consequence of Eq.~(\ref{e:TP:H_qstar_nabla_s}) established
in that section. 
 
Thanks to Eqs.~(\ref{e:KI:qqnab_d2}) and (\ref{e:KI:qnab_s_H}), 
Eq.~(\ref{e:KI:Theta_qqdv}) becomes
\be \label{e:KI:Theta_Kil_V_index}
    \encadre{ 
      \Theta_{\alpha\beta} =  \frac{1}{2} \left[ 
      \left(
        \Liec{\tv} q_{\mu\nu} \right)  
        q^\mu_{\ \, \alpha} q^\nu_{\ \, \beta} 
        + \DS_\alpha V_\beta + \DS_\beta V_\alpha \right] 
        + (N-b) H_{\alpha\beta} } ,
\ee
or, in index-free notation (cf. the definition (\ref{e:IN:def_Killing})
of the Killing operator):
\be \label{e:KI:Theta_Kil_V}
    \encadre{ \w{\Theta} = \frac{1}{2} \left[ 
       \vec{\w{q}}^* \Lie{\tv} \w{q} + \Kil{\w{\DS}}{\underline{\w{V}}} \right] 
       + (N-b) \w{H} } .
\ee
In particular, if $(x^\alpha)$ is a coordinate system adapted to $\Hor$,
then $N-b=0$ [Eq.~(\ref{e:IN:b_N_station})] and
$q^\mu_{\ \, a} = \delta^\mu_{\ \, a}$, so that when restricting
Eq.~(\ref{e:KI:Theta_Kil_V_index}) to $\Sp_t$ (i.e. $\alpha = a \in\{2,3\}$,
$\beta = b \in\{2,3\}$) one obtains
\be \label{e:KI:Theta_ab_der_V}
   \encadre{
	\Theta_{ab} = {1\over 2} \left( \der{q_{ab}}{t}
	+  \DS_a V_b + \DS_b V_a \right) } , 
\ee
which agrees with Eq.~(I.52b) of Damour \cite{Damou79}.


\subsection{Expansion scalar and shear tensor of  
the 2-surfaces $\Sp_t$} \label{s:KI:expans_shear}

Let us split the second fundamental form 
$\w{\Theta}$ (now considered as the deformation rate of the 
2-surfaces $\Sp_t$) into a trace part and a traceless part with 
respect to $\Sp_t$'s metric $\w{q}$
\be \label{e:KI:Theta_split}
	\encadre{ \w{\Theta} = \frac{1}{2} \theta \, \w{q} + \w{\sigma}  },
\ee
where 
\be \label{e:KI:def_theta}
	\encadre{ \theta := \, {\rm tr} \, \vec{\w{\Theta}} 
	= \Theta^\mu_{\ \ \mu} = g^{\mu\nu} \Theta_{\mu\nu} 
    = q^{\mu\nu} \Theta_{\mu\nu} = q^{ab} \Theta_{ab} 
	= \Theta^a_{\ \, a} }
\ee
is the trace of the endomorphism
$\vec{\w{\Theta}}$ canonically associated with $\w{\Theta}$
by the metric $\w{g}$ [see also Eq.~(\ref{e:def_vec_Theta})]
and $\w{\sigma}$ is the traceless 
part of $\w{\Theta}$
\be \label{e:KI:def_sigma}
    \w{\sigma}  := \w{\Theta} - \frac{1}{2} \theta \, \w{q} , 
\ee
which satisfies  
\be \label{e:KI:tr_shear_0}
	\sigma^\mu_{\ \ \mu} = q^{\mu\nu} \sigma_{\mu\nu} 
	= \sigma^a_{\ \, a} = 0 . 
\ee
The trace $\theta$ is called the {\em expansion scalar of $\Sp_t$} and
$\w{\sigma}$ the {\em shear tensor of $\Sp_t$}.

The expansion $\theta$ is linked to the divergence of $\el$; indeed taking the
trace of Eq.~(\ref{e:KI:grad_el}) results in 
\be
    \w{\nabla}\cdot \el = \theta 
    + \underbrace{\langle \w{\omega},\el\rangle }_{=\kappa}
    -  \underbrace{\langle \uel,\w{\nabla}_{\w{k}}\el\rangle}_{=0} , 
\ee
hence
\be \label{e:KI:divl_k_th}
	\encadre{ \w{\nabla}\cdot \el = \kappa + \theta } . 
\ee
Another relation is obtained by combining $\theta= q^{\mu\nu} \Theta_{\mu\nu}$
with the expression (\ref{e:KI:Theta_grad_l_qq}) for $\Theta_{\mu\nu}$:
\be
	\theta = q^{\mu\nu} q^\rho_{\ \ \mu} q^\sigma_{\ \ \nu}
			\nabla_\rho \ell_\sigma
		= q^{\nu\rho} q^\sigma_{\ \ \nu} \nabla_\rho \ell_\sigma
		= q^{\rho\sigma} \nabla_\rho \ell_\sigma ,
\ee
i.e. 
\be \label{e:KI:theta_qnabl}
	\encadre{\theta = q^{\mu\nu} \nabla_\mu \ell_\nu} .
\ee
Another expression of $\theta$ is obtained as follows. 
Let $(x^\alpha)$ be a coordinate system adapted to $\Hor$; 
$(x^a)_{a=2,3}$ is then a coordinate system on $\Sp_t$. We
have $\theta= q^{ab} \Theta_{ab}$ [cf. Eq.~(\ref{e:KI:def_theta})]
and let us use Eq.~(\ref{e:KI:Theta_deform})
restricted to $\T(\Sp_t)$, i.e. under the
form $\Theta_{ab} = 1/2\; \LieS{\el} q_{ab}$. We get
\be \label{e:KI:theta_qLq}
    \theta =  \frac{1}{2}\,  q^{ab} \; \LieS{\el} q_{ab}
        =  \frac{1}{2}\,  \LieS{\el} \ln q ,  
\ee
where $q$ is the determinant of the components $q_{ab}$ of the
metric $\w{q}$ with respect to the coordinates $(x^a)$ in $\Sp_t$:
\be
	\encadre{ q := \det q_{ab}} .
\ee
The second equality in Eq.~(\ref{e:KI:theta_qLq}) follows from the
standard formula for the variation of a determinant. 
Hence we have 
\be \label{e:KI:theta_Lie_detq}
    \encadre{\theta = \LieS{\el}\ln\sqrt{q}} . 
\ee
This relation justifies the name of {\em expansion scalar} given to 
$\theta$, for $\sqrt{q}$ is related to the surface element 
${}^2 \w{\epsilon}$ of $\Sp_t$ by 
\be \label{e:KI:2-epsilon}
	{}^2 \w{\epsilon} = \sqrt{q} \, \dd x^2 \wedge \dd x^3 . 
\ee

Another expression of $\theta$ is obtained by contracting 
Eq.~(\ref{e:KI:Theta_ab_der_V}) with $q^{ab}$:
\be \label{e:KI:theta_dsdt_detq}
    \encadre{ \theta = \der{}{t} \ln \sqrt{q}
		+ \DS_a V^a } . 
\ee
More generally, contracting Eq.~(\ref{e:KI:Theta_Kil_V_index}) with
$q^{\alpha\beta}$ leads to 
\be \label{e:KI:theta_q_Lie_t_q}
    \encadre{ \theta = q^{\mu\nu} \Liec{\tv} q_{\mu\nu}
     + \DS_a V^a + (N-b) H }, 
\ee
where $H$ is twice $\Sp_t$'s mean curvature 
within $(\Sigma_t,\w{\gamma})$ [Eq.~(\ref{e:TP:def_trH}) below]. 

\begin{rem} \label{rem:KI:theta_indep_St}
Equations~(\ref{e:KI:theta_Lie_detq}) and (\ref{e:KI:theta_dsdt_detq}), by
relating $\theta$ to the rate of expansion of the 2-surfaces $\Sp_t$,
might suggest that the scalar $\theta$ depends quite sensitively 
upon the foliation of spacetime by the spacelike hypersurfaces $\Sigma_t$, 
since the surfaces $\Sp_t$ are defined by this foliation. 
Actually the dependence is pretty weak: as shown by 
Eq.~(\ref{e:KI:divl_k_th}), $\theta$ depends
only upon the null normal $\el$ to $\Hor$ (since $\kappa$ depends only 
upon $\el$). Hence the dependence of $\theta$ with
respect to the foliation $(\Sp_t)$ is only through the normalization
of $\el$ induced by the $(\Sp_t)$ slicing and not on the precise shape
of this slicing. 
\end{rem}


\subsection{Transversal deformation rate}

By analogy with the expression (\ref{e:KI:Theta_deform}) of $\w{\Theta}$, 
we define the 
{\em transversal deformation rate} of the 2-surface $\Sp_t$ 
as the projection onto $\T(\Sp_t)$ of 
the Lie derivative of $\Sp_t$'s metric $\w{q}$ along the null transverse 
vector $\w{k}$:
\be \label{e:KI:def_Xi}
    \encadre{ \w{\Xi} := \frac{1}{2} \, \vec{\w{q}}^* \; \Lie{\w{k}} \w{q} } .  
\ee

\begin{rem}
In the above definition $\w{q}$ is considered as the 4-dimensional 
bilinear form given by Eq.~(\ref{e:q_g_nn_ss}) or Eq.~(\ref{e:NH:q_g_l_k}),
rather than as the 2-dimensional metric of $\Sp_t$,
and $\Lie{\w{k}} \w{q}$ is its Lie derivative within the 4-manifold $\M$. 
Indeed since the vector field $\w{k}$ does not Lie drag the surfaces
$(\Sp_t)$, we do not have an object such as the 2-dimensional
Lie derivative ``$\; \LieS{\w{k}}$'' (the analog of $\LieS{\el}$)
which could have been applied 
to the 2-metric of $\Sp_t$ in the strict sense.  
\end{rem}

From its definition, it is obvious that $\w{\Xi}$ is a symmetric bilinear form. 
Replacing $\w{q}$ by its expression (\ref{e:NH:q_g_l_k}), we 
get
\bea    
    \w{\Xi}  & = & \frac{1}{2} \, \vec{\w{q}}^* \; 
        \Lie{\w{k}}(\w{g} + \uel \otimes \uk 
    + \uk \otimes \uel) \nonumber \\
    & = & \frac{1}{2} \, \vec{\w{q}}^* \; \left(
        \Lie{\w{k}} \w{g} + \Lie{\w{k}}\uel \otimes \uk 
    + \uel \otimes \Lie{\w{k}}\uk 
    + \Lie{\w{k}}\uk \otimes \uel + \uk \otimes \Lie{\w{k}}\uel \right).
\eea
Since $\vec{\w{q}}^* \uel = 0$ and $\vec{\w{q}}^* \uk = 0$, 
only the term $\Lie{\w{k}} \w{g}$ remains in the right-hand side after
the operator $\vec{\w{q}}^*$ has been applied. 
Now $\Lie{\w{k}} \w{g}$
is nothing but the Killing operator applied to the 1-form 
$\uk$, so that the above equation becomes
\bea
    \w{\Xi}  & = & \frac{1}{2} \, \vec{\w{q}}^* \; 
        \Lie{\w{k}} \w{g} = \frac{1}{2} \, \vec{\w{q}}^* \; 
        \Kil{\w{\nabla}}{\uk}
        = \vec{\w{q}}^*   \left( 
            \w{\nabla}\uk + \frac{1}{2} \dd\uk \right) \nonumber \\
        & = & \vec{\w{q}}^*   \left[ 
            \w{\nabla}\uk + 
            \frac{1}{4N^2} \; \dd \ln \left( \frac{N}{M} \right)
     \wedge \uel \right] = \vec{\w{q}}^* \; \w{\nabla} \uk
      +   \frac{1}{4N^2} \; \vec{\w{q}}^* \, \dd \ln \left( \frac{N}{M} \right)
     \wedge \underbrace{\vec{\w{q}}^* \,\uel}_{=0} , \nonumber
\eea
where use has been made of the Frobenius identity (\ref{e:KI:dk})
to get the second line. 
We conclude that 
\be \label{e:KI:Xi_qstar_gradk}
    \encadre{ \w{\Xi} = \vec{\w{q}}^* \; \w{\nabla} \uk },  
\ee 
which is an expression completely analogous to the expression 
(\ref{e:KI:Theta_qstar_gradl}) for $\w{\Theta}$ in terms of $\w{\nabla} \uel$.

\begin{rem} \label{r:KI:2nd_fund_tensor}
The metric $\w{q}$ induced by $\w{g}$ on the 2-surface $\Sp_t$ is 
a Riemannian metric (i.e. positive definite) (cf. Sec.~\ref{s:IN:metric_q}). 
It is called the {\em first fundamental form of $\Sp_t$} and describes
fully the {\em intrinsic geometry} of $\Sp_t$. The way $\Sp_t$ is embedded in
the spacetime $(\M,\w{g})$ constitutes the {\em extrinsic geometry}
of $\Sp_t$. For a non-null hypersurface of $\M$, this extrinsic
geometry is fully described by a single bilinear form, the so-called
{\em second fundamental form} (for instance $\w{K}$ for the
hypersurface $\Sigma_t$). For the 2-dimensional surface $\Sp_t$,
a part of the extrinsic geometry is described by a normal fundamental
form, like the \hajicek\ 1-form $\w{\Omega}$ as discussed in 
Sec.~\ref{s:KI:rot_haji}. The remaining part is described by 
a type (1,2) tensor:
the {\em second fundamental tensor} $\w{\mathcal{K}}$ 
\cite{Carte92a,Carte97} (also called {\em shape tensor} \cite{Senov04}),
which relates the covariant derivative of 
a vector tangent to $\Sp_t$ taken with the spacetime connection $\w{\nabla}$
to that taken with the connection $\w{\DS}$
in $\Sp_t$ compatible with the induced metric $\w{q}$:
\be
    \forall (\w{u},\w{v}) \in \T(\Sp_t)^2,\quad
    \w{\nabla}_{\w{u}} \w{v} = \w{\DS}_{\w{u}} \w{v} 
        + \w{\mathcal{K}}(\w{u},\w{v}). 
\ee
It is easy to see that $\w{\mathcal{K}}$ is related to the spacetime
derivative of $\vec{\w{q}}$ by
\be \label{e:KI:CarterK_grad_vq}
	 {\mathcal K}^\gamma_{\ \, \alpha\beta} =
		q^\mu_{\ \, \alpha} q^\nu_{\ \, \beta}
		\nabla_\mu q^\gamma_{\ \, \nu} . 
\ee
${\mathcal K}^\gamma_{\ \, \alpha\beta}$ 
is tangent to $\Sp_t$ with respect to the indices
$\alpha$ and $\beta$ and orthogonal to $\Sp_t$ with respect to 
the index $\gamma$. Moreover, it is symmetric in $\alpha$ and $\beta$
[although this is not obvious on Eq.~(\ref{e:KI:CarterK_grad_vq})].
From Eqs.~(\ref{e:IN:vec_q_k_l}),  (\ref{e:NH:Thetaqq}) and 
(\ref{e:KI:Xi_qstar_gradk}), we have
\be \label{e:KI:CarterK_Theta_Xi}
	 {\mathcal K}^\gamma_{\ \, \alpha\beta} =
		\Theta_{\alpha\beta} \, k^\gamma
		+ \Xi_{\alpha\beta}  \, \ell^\gamma  .
\ee
Accordingly the bilinear forms $\w{\Theta}$ and $\w{\Xi}$ can be viewed
as two facets of the same object: the second
fundamental tensor $\w{\mathcal K}$:  
\be
	\Theta_{\alpha\beta} = - \ell_\mu {\mathcal K}^\mu_{\ \, \alpha\beta} 
        \qquad \mbox{and} \qquad 
	\Xi_{\alpha\beta} = - k_\mu {\mathcal K}^\mu_{\ \, \alpha\beta}   .
\ee
\end{rem}

By substituting Eq.~(\ref{e:IN:vec_q_k_l}) for the projector $\vec{\w{q}}$ 
in Eq.~(\ref{e:KI:Xi_qstar_gradk}), using the identity
$\ell^\mu \nabla_\alpha k_\mu = - k_\mu \nabla_\alpha \ell^\mu$
(which follows from $\el\cdot\w{k}= -1$), expressing 
$\nabla_\alpha \ell^\mu$ via Eq.~(\ref{e:KI:grad_el}) 
and using Eqs.~(\ref{e:KI:Omega_omega_k}) and (\ref{e:KI:omega_gradl_uk}), 
we get the following 
expression for the spacetime derivative of the 1-form $\uk$, in terms
of $\w{\Xi}$ and the H\'a\'\j i\v{c}ek 1-form $\w{\Omega}$:
\be \label{e:KI:grad_uk_index}
    \nabla_\alpha k_\beta = \Xi_{\alpha\beta}
        - \Omega_\alpha k_\beta - \ell_\alpha k^\mu \nabla_\mu k_\beta
            - k_\alpha \omega_\beta ,  
\ee
or, taking into account the symmetry of $\w{\Xi}$,
\be \label{e:KI:grad_uk}
    \encadre{ \w{\nabla} \uk = \w{\Xi} - \uk \otimes \w{\Omega}
        - \w{\nabla}_{\w{k}} \uk \otimes \uel
        -  \w{\omega} \otimes \uk} . 
\ee

Similarly to the definition of the expansion scalar $\theta$ 
as the trace of the deformation rate $\w{\Theta}$
[cf. Eqs.~(\ref{e:KI:def_theta}) and (\ref{e:KI:theta_qnabl})], 
we define the {\em transversal
expansion scalar} $\theta_{(\w{k})}$ as the trace of $\w{\Xi}$:
\be \label{e:KI:def_theta_k}
	\encadre{ \theta_{(\w{k})} := \, {\rm tr} \, \vec{\w{\Xi}} 
	= \Xi^\mu_{\ \, \mu} = g^{\mu\nu} \Xi_{\mu\nu} 
    = q^{\mu\nu} \Xi_{\mu\nu} = q^{ab} \Xi_{ab} 
	= q^{\mu\nu} \nabla_\mu k_\nu } . 
\ee

\begin{rem}
The reader will have noticed a certain dissymmetry in our notations, since we use
$\theta_{(\w{k})}$ for the expansion of the null vector $\w{k}$ and
merely $\theta$ the expansion of the null vector $\el$. 
From the point of view of the 2-dimensional spacelike surface
$\Sp_t$, $\el$ and $\w{k}$ play perfectly symmetric roles, $\el$ (resp. $\w{k}$)
being the unique -- up to some rescaling -- outgoing (resp. ingoing)
null normal to $\Sp_t$. However, $\el$ is in addition normal to the
null hypersurface $\Hor$, whereas $\w{k}$ has not any specific relation 
to $\Hor$. In particular, there is not a unique transverse null direction
to $\Hor$, so that $\w{k}$ is defined only thanks to the extra-structure
$(\Sp_t)$. The dissymmetry in our notations accounts therefore for the 
privileged status of $\el$ with respect to $\w{k}$.
\end{rem}

\begin{exmp} \label{ex:KI:cone}
\textbf{$\w{\Theta}$, $\w{\omega}$, $\w{\Omega}$ and $\w{\Xi}$ 
for a Minkowski light cone.}\\
Let us proceed with Example~\ref{ex:IN:cone}, namely a light cone
in Minkowski spacetime, sliced according to the standard Minkowskian time
coordinate $t$. Comparing the components of $\w{\nabla}\uel$ given by
Eq.~(\ref{e:NH:nab_ell_cone}) with those of $\w{q}$ given by 
Eq.~(\ref{e:IN:q_comp_cone}), we realize that 
\be \label{e:KI:grad_uel_cone}
    \w{\nabla}\uel = \frac{1}{r} \, \w{q} . 
\ee
The second fundamental form $\w{\Theta}=\vec{\w{q}}^*\, \w{\nabla}\uel$
[Eq.~(\ref{e:KI:Theta_qstar_gradl})] follows then immediately:
\be \label{e:KI:Theta_cone}
    \w{\Theta} = \frac{1}{r} \, \w{q} . 
\ee
We deduce from this relation and Eq.~(\ref{e:KI:Theta_split})
that the expansion scalar is 
\be \label{e:KI:theta_cone}
    \theta = \frac{2}{r}
\ee
and the shear tensor vanishes identically:
\be \label{e:KI:sigma_cone}
    \w{\sigma} = 0 . 
\ee
Note that $\theta >0$, in accordance with the fact that the light cone
is expanding. 
From expression (\ref{e:KI:grad_uel_cone}) and the orthogonality
of $\w{q}$ with $\w{k}$, we deduce by means of Eq.~(\ref{e:KI:omega_grad_l})
that the rotation 1-form vanishes identically:
\be \label{e:KI:omega_cone}
    \w{\omega} = 0 . 
\ee
Consequently, its pull-back on $\Sp_t$, the \hajicek\ 1-form, 
vanishes as well:
\be \label{e:KI:Omega_cone}
    \w{\Omega}=0 . 
\ee  
Similarly, from the expression (\ref{e:IN:k_comp_cone}) for $\w{k}$,
we get $\w{\nabla}\uk = -1/(2r) \, \w{q}$. From this relation 
and Eq.~(\ref{e:KI:Xi_qstar_gradk}), we deduce that
the transversal deformation rate has the simple expression
\be \label{e:KI:Xi_cone}
    \w{\Xi} = - \frac{1}{2 r} \, \w{q},  
\ee
on which we read immediately  the transversal expansion scalar:
\be 
    \theta_{(\w{k})} = - \frac{1}{r} . 
\ee
\end{exmp}

\begin{exmp} \label{ex:KI:kin_EF}
\textbf{$\w{\Theta}$, $\w{\omega}$, $\w{\Omega}$ and $\w{\Xi}$ 
associated with the Eddington-Finkels\-tein slicing of Schwarzschild 
horizon.} \\
Let us continue the Example~\ref{ex:IN:EF} about the event horizon of 
Schwarzschild spacetime, with the 3+1 slicing provided by Eddington-Finkelstein
coordinates. The second fundamental form $\w{\Theta}$ is obtained from
Eqs.~(\ref{e:NH:Thetaqq}), (\ref{e:NH:gradel_EF}) and (\ref{e:IN:q_comp_EF}):
\be
 \Theta_{\alpha\beta} = {\rm diag}\left(0,\ 0,\   
    \frac{r-2m}{r+2m}\, r , 
    \frac{r-2m}{r+2m} \, r \sin^2\theta \right) .
\ee
Accordingly, the expansion scalar is
\be
    \theta = \frac{2}{r}\,  \frac{r-2m}{r+ 2m}
\ee
and the shear tensor vanishes identically
\be
    \sigma_{\alpha\beta} = 0 .
\ee
The transversal deformation rate is deduced from 
Eq.~(\ref{e:KI:Xi_qstar_gradk}) and expression (\ref{e:IN:k_comp_EF})
for $\w{k}$:
\be
 \Xi_{\alpha\beta} = {\rm diag}\left(0,\ 0,\
    -\frac{r+2m}{2},\ -\frac{r+2m}{2}\,  \sin^2\theta \right) ,
\ee
so that the transversal expansion scalar is
\be
    \theta_{(\w{k})} = - \frac{1}{r} - \frac{2m}{r^2} . 
\ee
The rotation 1-form is obtained from Eq.~(\ref{e:KI:omega_grad_l})
combined with expression (\ref{e:NH:gradel_EF}) for $\w{\nabla}\uel$
and expression (\ref{e:IN:k_comp_EF}) for $\w{k}$:
\be
    \omega_\alpha = \left(\frac{2m}{r(r+2m)},\ \frac{2m}{r(r+2m)},\
    0,\ 0 \right) .  
\ee
We deduce immediately from this expression and Eq.~(\ref{e:KI:def_hajicek})
that the \hajicek\ 1-form vanishes identically:
\be
    \w{\Omega} = 0. 
\ee
As a check, we verify that, from the obtained values for $\w{\omega}$,
$\w{\Omega}$, $\kappa$ and $\uk$, Eq.~(\ref{e:KI:Omega_omega_k})
is satisfied. It is of course instructive to specify the above results
on the event horizon $\Hor$ ($r=2m$):
\bea
    &\w{\Theta} \equalH 0, \qquad& \theta \equalH 0, \qquad
    \w{\sigma}\equalH 0 ,       \label{e:KI:thetaH_EF} \\
     &\w{\Xi} \equalH - \frac{1}{2m} \, \w{q}, \qquad &
    \theta_{(\w{k})} \equalH - \frac{1}{m},  \label{e:KI:XiH_EF} \\
     &\w{\omega} \equalH - \frac{1}{4m} \, \uk, \qquad&
    \w{\Omega} \equalH 0 .  \label{e:KI:OmegaH_EF}
\eea
Note that for a rotating black hole, described by the Kerr metric,
$\w{\Omega}$ is no longer zero, as shown in Appendix~\ref{s:KE}.
\end{exmp}


\subsection{Behavior under rescaling of the null normal}
\label{s:KI:rescaling}

As stressed in Sec.~\ref{s:NH:geom_null}, from the null structure only, 
the normal $\el$ to the hypersurface $\Hor$
is defined up to some normalization factor
(cf. Remark~\ref{rem:NH:no_norm}), i.e.
one can change $\el$ to 
\be \label{e:KI:rescale_l}
	\el' = \alpha \el,
\ee
where $\alpha$ is any strictly positive scalar field on $\Hor$
($\alpha>0$ ensures that $\el'$ is future oriented). 
In the present framework, the extra-structure on $\Hor$
induced by the spacelike foliation $\Sigma_t$ of the 3+1 formalism
provides a way to normalize $\el$: we have demanded $\el$ to be
the tangent vector corresponding to the parametrization by $t$
of the null geodesics generating $\Hor$ [cf. Eq.~(\ref{e:NH:l_norm1})],
or equivalently that $\el$ be 
a dual vector to the gradient $\dd t$ of the $t$ field
[cf. Eq.~(\ref{e:NH:l_norm2})]. It is however instructive
to examine how the various quantities introduced so far change
under a rescaling of the type (\ref{e:KI:rescale_l}). 
In Sec.~\ref{s:NH:geom_null}, we have already exhibited the behavior
of the non-affinity parameter $\kappa$ [cf. Eq.~(\ref{e:NH:scale_kappa})], 
as well as of the Weingarten map and the second fundamental form, 
both restricted to $\Hor$
[cf. Eqs.~(\ref{e:NH:scale_chi}) and (\ref{e:NH:scale_Theta})].

\begin{table}
\begin{center}
\begin{tabular}{cc}
\hline 
\hline 
\multicolumn{2}{c}{$\el'=\alpha\el$} \\
\hline 
$ \kappa' = \alpha \left( \kappa 
		+ \w{\nabla}_{\el} \, \ln\alpha \right) $ 
        & $\w{\Theta}' = \alpha \Theta$ \\
$\displaystyle \w{k}' = {1\over \alpha} \w{k} $ 
   & $\theta' = \alpha \theta$ \\
$\w{\Pi}' = \w{\Pi}$ 
  & $\w{\sigma}' = \alpha \w{\sigma}$ \\
$\vec{\w{q}}\,' = \vec{\w{q}}$ 
    & $\displaystyle \w{\Xi}' = {1\over \alpha} \w{\Xi}$ \\
$\w{\chi}' = \alpha \w{\chi}
		+ \langle \dd \alpha , \w{\Pi}(\cdot) \rangle \, \el$ & \\
    $\w{\omega}' = \w{\omega} + \w{\Pi}^* \dd\ln\alpha$ & \\
    $\w{\Omega}' = \w{\Omega} + \w{\DS}\ln\alpha$  & \\
\hline
\hline
\end{tabular}
\end{center}
\caption[]{\label{t:KI:scaling} 
Behavior under a rescaling $\el\rightarrow \el'=\alpha\el$
of $\Hor$'s null normal.}
\end{table}

In view of Eq.~(\ref{e:NH:l_k_m1}) the scaling properties of
the transverse null vector $\w{k}$ are simply 
$\w{k}' = \alpha^{-1} \w{k}$. 
From the expression (\ref{e:NH:P_l_k}) of the projector $\w{\Pi}$ onto $\Hor$
along $\w{k}$, in conjunction with $\el' = \alpha \el$
and $\w{k}' = \alpha^{-1} \w{k}$, 
we get that $\w{\Pi}$ is invariant under the
rescaling (\ref{e:KI:rescale_l}):
\be \label{e:KI:rescale_P}
	\w{\Pi}' = \w{\Pi} . 
\ee
This is not surprising since $\left. \w{\Pi} \right| _{\T_p(\Hor)}$ is the
identity and therefore does not depend upon $\el$. 
Similarly the orthogonal projector $\vec{\w{q}}$ onto $\Sp_t$
does not depend upon $\el$ [this is obvious from its definition and
is clear in expression (\ref{e:IN:vec_q_k_l})] so that 
\be \label{e:KI:rescale_q}
	\vec{\w{q}}\,' = \vec{\w{q}} .
\ee 

From its definition (\ref{e:NH:def_ext_Weingar}) and
the scaling properties (\ref{e:NH:scale_chi}) and (\ref{e:KI:rescale_P}), 
we get the following scaling behavior of the extended Weingarten map
\be \label{e:KI:scaling_chi}
	\w{\chi}' = \alpha \w{\chi}
		+ \langle \dd \alpha , \w{\Pi}(\cdot) \rangle \, \el ,
\ee
where the notation $\langle \dd \alpha , \w{\Pi}(\cdot) \rangle \, \el$
stands for the endomorphism $\T_p(\M) \longrightarrow \T_p(\M)$, 
$\w{v} \longmapsto \langle \dd \alpha , \w{\Pi}(\w{v}) \rangle \, \el$.
From its definition (\ref{e:NH:def_ext_Theta}) and the
scaling properties (\ref{e:NH:scale_Theta}) and (\ref{e:KI:rescale_P}), 
we get the following scaling behavior of the extended second fundamental 
form of $\Hor$ with respect to $\el$:
\be
	\w{\Theta}' = \alpha \w{\Theta}.
\ee
The scaling property of the rotation 1-form $\w{\omega}$ is
deduced from its definition (\ref{e:IN:def_omega}) and the scaling
law (\ref{e:KI:scaling_chi}) for $\w{\chi}$: 
\bea
	\forall \w{v}\in \T_p(\M),\quad
	\langle \w{\omega}', \w{v} \rangle & = &
	- \w{k}' \cdot \w{\chi}'(\w{v}) =
    - \alpha^{-1} \w{k} \cdot \left[ \alpha \w{\chi}(\w{v})
		+ \langle \dd \alpha , \w{\Pi}(\w{v}) \rangle \, \el \right]
            \nonumber\\
    & = & -\w{k}\cdot\w{\chi}(\w{v}) - \alpha^{-1} 
           \langle \dd \alpha , \w{\Pi}(\w{v}) \rangle \, 
           \underbrace{\w{k}\cdot\el}_{=-1} \nonumber\\
           & = & \langle \w{\omega}, \w{v} \rangle
            + \langle \dd \ln \alpha , \w{\Pi}(\w{v}) \rangle . 
\eea
Hence:
\be \label{e:KI:rescale_omega}
    \w{\omega}' = \w{\omega} + \w{\Pi}^* \dd\ln\alpha . 
\ee   
Since $\w{\Omega} = \vec{\w{q}}^* \w{\omega}$ 
[Eq.~(\ref{e:KI:def_hajicek})], the scaling law for the \hajicek 1-form
is immediate:
\be
	\w{\Omega}' = \w{\Omega} + \w{\DS}\ln\alpha .
\ee
The scaling properties of the expansion scalar $\theta$ and the shear
tensor $\w{\sigma}$ are deduced from that of $\w{\Theta}$ via
their definitions (\ref{e:KI:def_theta}) and (\ref{e:KI:def_sigma}):
\be \label{e:KI:rescale_theta_sigma}
	\theta' = \alpha \theta \qquad \mbox{and} \qquad 
	\w{\sigma}' = \alpha \w{\sigma} .
\ee
Finally the scaling law for the transversal expansion rate $\w{\Xi}$
is easily deduced from the Eq.~(\ref{e:KI:Xi_qstar_gradk}) and 
the scaling laws $\w{k}' = \alpha^{-1} \w{k}$
and (\ref{e:KI:rescale_q}):
\be
    \w{\Xi}' = \alpha^{-1} \w{\Xi} . 
\ee
For further reference, the various scaling laws are
summarized in Table~\ref{t:KI:scaling}.

%% file: dynam.tex
%
%
\section{Dynamics of null hypersurfaces}
\label{s:DY}

In the previous section, we have considered only first order
derivatives of the null vector fields $\el$ and $\w{k}$,
as well as of the metric $\w{q}$. In the present section
we consider second order derivatives 
of these fields. 
Some of these second order derivatives are 
written as Lie derivatives along $\el$ of the first order
quantities, like the second fundamental form $\w{\Theta}$, 
the \hajicek\ 1-form $\w{\Omega}$ and the transversal deformation 
rate $\w{\Xi}$. The obtained equations can be then qualified
as {\em evolution equations} along the future directed null normal $\el$.
(cf. the discussion at the beginning of Sec.~\ref{s:KI:deform_rate}).
Some other second order derivatives of $\el$ and $\w{k}$ are
rearranged to let appear the spacetime Riemann tensor, 
via the Ricci identity (\ref{e:IN:Ricci_ident}). 
The totality of the components of the Riemann tensor with respect
to a tetrad adapted to our problem, i.e. involving $\el$, $\w{k}$,
and two vectors tangent to $\Sp_t$, are derived in Appendix~\ref{s:CA}. 
Here we will focus only on those components related to 
the evolution of $\w{\Omega}$, $\w{\Theta}$ and $\w{\Xi}$. 

Some of the obtained evolution equations involve the Ricci part 
of the Riemann tensor. At this point, the Einstein equation enters into scene
in contrast with all results from previous
sections (except Sec.~\ref{s:FO:3p1Einstein}), 
which are independent of whether the spacetime 
metric $\w{g}$ is a solution or not 
of Einstein equation. 
This concerns the evolution equations for the expansion scalar $\theta$, 
the \hajicek\ 1-form $\w{\Omega}$ and the transversal expansion 
rate $\w{\Xi}$. On the contrary the evolution equation for the shear
tensor $\w{\sigma}$ involves only the traceless part of the Riemann tensor, 
i.e. the Weyl tensor, and consequently is independent from 
the Einstein equation.

\subsection{Null Codazzi equation}

Let us start by deriving the null analog of the contracted Codazzi
equation of the spacelike 3+1 formalism, i.e. 
Eq.~(\ref{e:FO:Codazzi_contracted}) presented in Sec.~\ref{s:FO:Riem_3p1}.
The idea is to obtain an equation involving 
the quantity $R_{\mu\nu}\ell^\mu\Pi^\nu_{\ \,\alpha}$, which is  
similar to the
left-hand side of Eq.~(\ref{e:FO:Codazzi_contracted}) with the
normal $\w{n}$ replaced by the normal $\el$, and the projector
$\vec{\w{\gamma}}$ replaced by the projector $\w{\Pi}$.

\begin{rem} 
\label{rem:DN:cod_gau}
We must point out that, from the very fact that  $\el$ 
is simultaneously normal and tangent to $\Hor$, 
the standard classification in terms 
{\em Codazzi} and {\em Gauss} equations employed
in Sec.~\ref{s:FO:Riem_3p1}, is not completely adapted to the present case. 
In particular, the trace of  
$R_{\mu\nu}\ell^\mu\Pi^\nu_{\ \,\alpha}$, 
associated with the contracted Codazzi equation, 
can also be interpreted as a component of the 
null analog of the contracted Gauss equation, as we shall see in 
Sec.~\ref{s:DN:tidal}.
\end{rem}

The starting point for the null contracted Codazzi equation is the
Ricci identity (\ref{e:IN:Ricci_ident}) applied to the null normal $\el$.
Contracting this identity on the indices $\gamma$ and $\alpha$,
we get:
\be
	\nabla_\mu \nabla_\alpha \ell^\mu 
	- \nabla_\alpha \nabla_\mu \ell^\mu
	= R_{\mu\alpha} \, \ell^\mu , 
\ee
where $R_{\mu\alpha}$ is the Ricci 
tensor of the connection $\w{\nabla}$ [cf. Eq.~(\ref{e:IN:def_Ricci_comp})].
Substituting Eq.~(\ref{e:KI:grad_el}) for $\nabla_\alpha \ell^\mu$
and Eq.~(\ref{e:KI:divl_k_th}) for $\nabla_\mu \ell^\mu$, yields
\be
	\nabla_\mu \left[ \Theta^\mu_{\ \, \alpha}
		+ \omega_\alpha \ell^\mu
		- \ell_\alpha k^\nu \nabla_\nu \ell^\mu
	 \right] 
	- \nabla_\alpha (\kappa + \theta) = R_{\mu\alpha} \, \ell^\mu .
\ee
Expanding the left-hand side and using again Eqs.~(\ref{e:KI:grad_el})
and (\ref{e:KI:divl_k_th}) leads to 
\bea
  R_{\mu\alpha} \, \ell^\mu & = & \nabla_\mu \Theta^\mu_{\ \, \alpha}
  + \ell^\mu \nabla_\mu \omega_\alpha 
	+ (\kappa + \theta) \omega_\alpha  
    - \nabla_\alpha (\kappa + \theta)
  - \Theta_{\alpha\mu} k^\nu \nabla_\nu \ell^\mu \nonumber \\
  & & - \left(\omega_\mu k^\nu \nabla_\nu \ell^\mu
	   + \nabla_\mu k^\nu \, \nabla_\nu \ell^\mu
	   + k^\nu \nabla_\mu \nabla_\nu \ell^\mu \right) \ell_\alpha
                                                   . \label{e:DN:Ricci_l}
\eea
The null contracted Codazzi equation is the contraction of this equation 
with the projector $\Pi^\alpha_{\ \, \beta}$ onto $\Hor$. 
A difference with the spacelike case is 
that this projection can be divided in 
two pieces: a projection along $\el$ itself, since the normal $\el$ is also
tangent to $\Hor$, and a projection onto the 2-surfaces $\Sp_t$.
This is clear if one expresses the projector $\w{\Pi}$ in terms of the
orthogonal projector $\vec{\w{q}}$ via Eq.~(\ref{e:NH:q_proj_P}):
\be \label{e:DN:Codazzi_split}
    R_{\mu\nu}\ell^\mu\Pi^\nu_{\ \,\alpha} = 
    - R_{\mu\nu}\ell^\mu \ell^\nu \, k_{\alpha}
      +   R_{\mu\nu}\ell^\mu q^\nu_{\ \,\alpha} .  
\ee
We will examine the two parts successively: 
the first one, $R_{\mu\nu}\ell^\mu \ell^\nu$,
will provide the null Raychaudhuri equation (Sec.~\ref{s:DN:Raychaud}), 
whereas the second
one, $R_{\mu\nu}\ell^\mu q^\nu_{\ \,\alpha}$ will lead to 
an evolution equation for the 
\hajicek\ 1-form which is analogous to a 2-dimensional Navier-Stokes 
equation (Sec.~\ref{s:DN:DNS}).

\subsection{Null Raychaudhuri equation} \label{s:DN:Raychaud}

The first part of the null contracted Codazzi equation is the one along
$\el$.
It is obtained by contracting Eq.~(\ref{e:DN:Ricci_l}) with $\ell^\alpha$:
\be
  R_{\mu\nu} \, \ell^\mu\ell^\nu = \ell^\nu \nabla_\mu 
  \Theta^\mu_{\ \, \nu} + \ell^\mu \ell^\nu \nabla_\mu \omega_\nu
    + (\kappa+\theta)\ell^\mu \omega_\mu 
    - \ell^\mu \nabla_\mu (\kappa+\theta) . 
\ee
Taking into account the identities $\Theta^\mu_{\ \, \nu} \ell^\nu = 0$
[Eq.~(\ref{e:NH:Theta_l_k_0})], 
$\ell^\mu \omega_\mu = \kappa$
[Eq.~(\ref{e:KI:omega_l_kappa})], 
$\ell^\mu \nabla_\mu\ell^\nu = \kappa \ell^\nu$ [Eq.~(\ref{e:NH:der_l_kappa})]
and expression (\ref{e:KI:grad_el}) for $\nabla_\mu \ell^\nu$,
we get
\be \label{e:DN:Raychaud_prov}
  R_{\mu\nu} \, \ell^\mu \ell^\nu  = - \Theta_{\mu\nu} \Theta^{\mu\nu}
  + \kappa\theta - \ell^\mu \nabla_\mu \theta .
\ee
As shown in Appendix~\ref{s:CA}, this relation can also been obtained
by computing the components of the Ricci tensor from the curvature
2-forms and Cartan's structure equations [cf. Eq.~(\ref{e:CA:Raychaud})].
We may express $\Theta_{\mu\nu} \Theta^{\mu\nu}$ in terms of
the shear $\w{\sigma}$ and the expansion scalar
$\theta$, thanks to Eqs.~(\ref{e:KI:Theta_split})
and (\ref{e:KI:tr_shear_0}):
\be
\label{e:DN:Theta_Theta}
     \Theta_{\mu\nu} \Theta^{\mu\nu} = \sigma_{\mu\nu} \sigma^{\mu\nu}
     	+ {1\over 2} \theta^2 = \sigma_{ab} \sigma^{ab} + {1\over 2} \theta^2 , 
\ee
to get finally
\be \label{e:DN:Raychaud}
	\encadre{ \w{\nabla}_{\el}\, \theta - \kappa\theta
	+  {1\over 2} \theta^2 + \sigma_{ab} \sigma^{ab}  
	+ \w{R}(\el,\el) = 0 } .
\ee
This is the well-known Raychaudhuri equation for a null congruence
with {\em vanishing vorticity} or {\em twist},
i.e. a congruence which is orthogonal to some hypersurface
(see e.g. Eq.~(4.35) in Ref.~\cite{HawkiE73} \footnote{Equation (4.35) in
Ref.~\cite{HawkiE73} assumes $\kappa=0$.} or Eq.~(2.21) in Ref.~\cite{Carte87}). 

If one takes into account the Einstein equation (\ref{e:FO:Einstein}),
the Ricci tensor $\w{R}$ can be replaced by the 
stress-energy tensor $\w{T}$ (owing to the fact that $\el$ is null):
\be \label{e:DN:Raychaud_T}
	\encadre{ \w{\nabla}_{\el}\, \theta - \kappa\theta
	+  {1\over 2} \theta^2 + \sigma_{ab} \sigma^{ab}  
	+ 8\pi \w{T}(\el,\el) = 0 } .
\ee


\subsection{Damour-Navier-Stokes equation}\label{s:DN:DNS}

Let us now consider the second part of the Codazzi equation
$R_{\mu\nu}\ell^\mu\Pi^\nu_{\ \,\alpha} = \cdots$, i.e. the part 
lying in the 2-surface  $\Sp_t$ [second term in the right-hand side of
Eq.~(\ref{e:DN:Codazzi_split})]. It is obtained by contracting 
Eq.~(\ref{e:DN:Ricci_l}) with $q^\alpha_{\ \, \beta}$:
\bea
  R_{\mu\nu} \, \ell^\mu q^\nu_{\ \, \alpha}  & = & 
  q^\nu_{\ \, \alpha} \nabla_\mu \Theta^\mu_{\ \, \nu}
  + q^\nu_{\ \, \alpha} \ell^\mu \nabla_\mu \omega_\nu
	+ (\kappa + \theta) \Omega_\alpha  
     - \DS_\alpha (\kappa + \theta) \nonumber \\
   & &  - \Theta_{\alpha\mu} k^\nu \nabla_\nu \ell^\mu  . \label{e:DN:startDNS}	  
\eea
The first term on the right-hand side is related to the divergence of
$\w{\Theta}$ with respect to the connection $\w{\DS}$ in $\Sp_t$ by
\bea             
    q^\nu_{\ \, \alpha} \nabla_\mu \Theta^\mu_{\ \, \nu} & = &
        \DS _\mu \Theta^\mu_{\ \, \alpha}
        + \Theta^\mu_{\ \, \alpha} 
        ( k^\nu \nabla_\nu \ell_\mu + \ell^\nu \nabla_\nu k_\mu) \nonumber \\
        & = & \DS _\mu \Theta^\mu_{\ \, \alpha}
        + \Theta^\mu_{\ \, \alpha} 
        ( k^\nu \nabla_\nu \ell_\mu + \Omega_\mu ) . 
                                                \label{e:DN:qnabThe}
\eea
The first line results from the relation (\ref{e:KI:2der})
between the derivatives $\w{\DS}$ and $\w{\nabla}$ 
for objects living on $\Sp_t$, 
whereas the second line follows from Eqs.~(\ref{e:KI:omega_gradl_uk}) and
(\ref{e:KI:def_hajicek}).

Besides, the second term on the right-hand side of
Eq.~(\ref{e:DN:startDNS}) can be expressed as 
[cf. Eq.~(\ref{e:KI:Omega_omega_k})]
\bea
    q^\nu_{\ \, \alpha} \ell^\mu \nabla_\mu \omega_\nu 
        & = & q^\nu_{\ \, \alpha} \ell^\mu \nabla_\mu (\Omega_\nu
            - \kappa k_\nu) = 
           q^\nu_{\ \, \alpha} ( \ell^\mu  \nabla_\mu \Omega_\nu
            - \kappa \ell^\mu  \nabla_\mu k_\nu) \nonumber \\
      &= &  q^\nu_{\ \, \alpha} \left( \Liec{\el} \Omega_\nu
            - \Omega_\mu \nabla_\nu \ell^\mu 
            - \kappa \omega_\nu \right) \nonumber \\
      &= &  q^\nu_{\ \, \alpha}  \Liec{\el} \Omega_\nu
            - \Theta_\alpha^{\ \, \mu} \Omega_\mu 
            - \kappa \Omega_\alpha , 
            \label{e:DN:qlnabomeg}
\eea
where, to get the last line, use has been made of 
Eq.~(\ref{e:NH:Thetaqq}) to let appear $\Theta_\alpha^{\ \, \mu}$
and of Eqs.~(\ref{e:KI:omega_gradl_uk}) and
(\ref{e:KI:def_hajicek}) to let appear $\Omega_\alpha$.
Inserting expressions (\ref{e:DN:qnabThe}) and (\ref{e:DN:qlnabomeg})
in Eq.~(\ref{e:DN:startDNS}) results immediately in
\be
  R_{\mu\nu} \, \ell^\mu q^\nu_{\ \, \alpha}  =
    q^\mu_{\ \, \alpha}  \Liec{\el} \Omega_\mu
        + \theta \, \Omega_\alpha - \DS_\alpha (\kappa + \theta)
        + \DS _\mu \Theta^\mu_{\ \, \alpha} .   \label{e:DN:Rlq_DNS}
\ee
An alternative derivation of this relation, based on 
Cartan's structure equations, is given in Appendix~\ref{s:CA} 
[cf. Eq.~(\ref{e:CA:DNS})].
Expressing $\w{\Theta}$ in terms of the expansion scalar $\theta$
and the shear tensor $\w{\sigma}$ [Eq.~(\ref{e:KI:Theta_split})],
we get
\be \label{e:DN:Ricci_lq3}
  R_{\mu\nu} \, \ell^\mu q^\nu_{\ \, \alpha}  = 
    q^\mu_{\ \, \alpha}  \Liec{\el} \Omega_\mu
        + \theta \, \Omega_\alpha 
        - \DS_\alpha \left( \kappa + \frac{\theta}{2} \right)
        + \DS _\mu \sigma^\mu_{\ \, \alpha} . 
\ee

Taking into account the Einstein equation (\ref{e:FO:Einstein}),
the Ricci tensor can be replaced by the 
stress-energy tensor (owing to the fact that 
$g_{\mu\nu}\, \ell^\mu q^\nu_{\ \, \alpha} =0$) to write
Eq.~(\ref{e:DN:Ricci_lq3}) as an evolution equation 
for the H\'a\'\j i\v{c}ek 1-form:
\be \label{e:DN:DNS}
    \encadre{
    q^\mu_{\ \, \alpha}  \Liec{\el} \Omega_\mu + \theta \, \Omega_\alpha =
       8\pi T_{\mu\nu} \, \ell^\mu q^\nu_{\ \, \alpha}
       +  \DS_\alpha \left( \kappa + \frac{\theta}{2} \right)
       -  \DS _\mu \sigma^\mu_{\ \, \alpha} } .
\ee
\be \label{e:DN:DNS_if}
    \encadre{
    \vec{\w{q}}^*  \Lie{\el} \w{\Omega} + \theta \, \w{\Omega} =
       8\pi \vec{\w{q}}^*  \w{T}\cdot\el 
       + \w{\DS}  \left( \kappa + \frac{\theta}{2} \right)
       - \w{\DS} \cdot \vec{\w{\sigma}} } .
\ee
The components $\alpha=a\in\{2,3\}$ of this equation agree with
Eq.~(I.30b) of Damour \cite{Damou79}.
Equation~(\ref{e:DN:DNS}) can also be compared with Eq.~(2.14) of
Price \& Thorne \cite{PriceT86},
after one has noticed that their operator $D_{\bar t}$ 
acting on the H\'a\'\j i\v{c}ek 1-form is 
$D_{\bar t} \Omega_\alpha = q^\nu_{\ \, \alpha} \ell^\mu \nabla_\mu \Omega_\nu$
and therefore is related to our Lie derivative along $\el$ by the 
relation $D_{\bar t} \Omega_\alpha = q^\mu_{\ \, \alpha}  \Liec{\el} \Omega_\mu
- \Theta_\alpha^{\ \, \mu} \Omega_\mu$, which can be deduced from
(our) Eq.~(\ref{e:DN:qlnabomeg}). Then 
 the components $\alpha=a\in\{2,3\}$ of
(our) Eq.~(\ref{e:DN:DNS}) coincide with
their Eq.~(2.14), called by them the ``H\'a\'\j i\v{c}ek equation''.

Let $(x^\alpha)$ be a coordinate system adapted to $\Hor$; 
$(x^a)_{a=2,3}$ is then a coordinate system on $\Sp_t$.
We can write
$\Lie{\el} \w{\Omega} = \Lie{\tv} \w{\Omega} + \Lie{\w{V}} \w{\Omega}$,
where $\tv$ is the coordinate time vector associated with $(x^\alpha)$
and $\w{V}\in\T(\Sp_t)$ is the surface velocity of $\Hor$ with respect
to $(x^\alpha)$ [cf. Eq.~(\ref{e:IN:el_t_V_station})]. The projection of this relation onto $\Sp_t$ gives
\be 
    q^\mu_{\ \, a}  \Liec{\el} \Omega_\mu = \der{\Omega_a}{t}
    + V^b \, \DS_b \Omega_a + \Omega_b \, \DS_a V^b .
\ee
Inserting this relation into
Eq.~(\ref{e:DN:DNS}) yields
\be \label{e:DN:DNS2}
 \encadre{ 
 \begin{array}{ll}
 \displaystyle \der{\Omega_a}{t} + V^b \, \DS_b \Omega_a 
	+\Omega_b \, \DS_a V^b 
 + \theta \Omega_a = & 8\pi q^\mu_{\ \, a} T_{\mu\nu} \, \ell^\nu
 + \DS_a \kappa \\
 & \displaystyle  - \DS_b \sigma^b_{\ \, a}
  + \frac{1}{2} \DS_a \theta 
  \end{array} } ,  
\ee
with, of course, $T_{\mu\nu}=0$ in the vacuum case. 
Noticing that 
\be \label{e:DN:force_surface}
	f_a := - q^\mu_{\ \, a} T_{\mu\nu} \, \ell^\nu 
\ee
is a force surface density (momentum per unit surface of $\Sp_t$
and per unit coordinate time $t$), Damour \cite{Damou79,Damou82} 
has interpreted
Eq.~(\ref{e:DN:DNS2}) as a 2-dimensional Navier-Stokes equation
for a viscous ``fluid''. 
The H\'a\'\j i\v{c}ek 1-form $\w{\Omega}$ is then interpreted as a 
momentum surface density $\w{\pi}$ (up to a factor $-8\pi$):
\be
	\pi_a := - {1\over 8\pi} \Omega_a . 
\ee
$V^a$ represents then the (2-dimensional) velocity of the ``fluid'', 
$\kappa/(8\pi)$ the ``fluid'' pressure, $1/(16\pi)$ the shear
viscosity ($\sigma_{ab}$ is then the shear tensor) 
and $-1/(16\pi)$ the bulk viscosity. This last fact holds for
$\theta$ is the divergence of the velocity field in the stationary
case: consider Eq.~(\ref{e:KI:theta_dsdt_detq}) with 
$\partial/\partial t=0$. 
We refer the reader to Chap.~VI of the {\em Membrane Paradigm} book
\cite{ThornPM86} for an extended discussion of this ``viscous fluid''
viewpoint.


\subsection{Tidal-force equation} \label{s:DN:tidal}

The null Raychaudhuri equation (\ref{e:DN:Raychaud}) has provided
an evolution equation for the trace $\theta$ of the second fundamental
form $\w{\Theta}$. Let us now derive an evolution equation for
the traceless part of $\w{\Theta}$, i.e. the 
shear tensor $\w{\sigma}$. For this purpose we evaluate $\ell^\mu \nabla_\mu
\left(\nabla _\alpha \ell_\beta\right)$ in two ways.
Firstly, we express it in terms of the Riemann tensor
by means of the Ricci identity (\ref{e:IN:Ricci_ident}): 
\bea
\ell^\rho \nabla_\rho
\left(\nabla _\alpha \ell_\beta\right) = \ell^\rho \left(R_{\beta\gamma\rho\alpha} \ell^\gamma
+  \nabla_\alpha\nabla_\rho \ell_\beta\right) \ .
\eea
Making repeated use of Eq.~(\ref{e:KI:grad_el}) to expand 
$\nabla_\rho \ell_\beta$ and employing $\ell^\mu \Theta_{\mu\nu}=0$
we find
\bea
\ell^\rho \nabla_\rho
\left(\nabla _\alpha \ell_\beta\right) &=& \ell^\rho\ell^\gamma R_{\beta\gamma\rho\alpha} 
- {\Theta^\rho}_\beta\Theta_{\alpha\rho} + \kappa \Theta_{\alpha\beta} \nn \\
&-& \ell_\alpha\left[\kappa k^\rho \nabla_\rho \ell_\beta 
   - k^\mu(\nabla_\mu \ell_\rho)
  {\Theta^\rho}_\beta\right] 
+ \ell_\beta\left[\ell^\rho \nabla_\alpha\omega_\rho + \kappa \omega_\alpha\right] \ .
\eea
On the other hand, using directly  Eq.~(\ref{e:KI:grad_el}) to expand 
$\nabla_\alpha \ell_\beta$ we find
\bea
\ell^\rho \nabla_\rho
\left(\nabla _\alpha \ell_\beta\right) =\ell^\rho\nabla_\rho \Theta_{\alpha\beta}
&-& \ell_\alpha\left[\kappa(k^\mu\nabla_\mu \ell_\beta) \nn
+ \ell^\rho\nabla_\rho(k^\mu\nabla_\mu \ell_\beta)\right] \\
&+&\ell_\beta\left[\kappa \omega_\alpha + \ell^\rho \nabla_\rho \omega_\alpha\right]
\eea
From both expressions for 
$\ell^\rho \nabla_\rho\left(\nabla _\alpha \ell_\beta\right)$, 
and projecting on $\Sp_t$ we obtain
\bea
q^\mu_{\ \, \alpha} q^\nu_{\ \, \beta} \, \ell^\rho\nabla_\rho \Theta_{\mu\nu}=
\kappa \Theta_{\alpha\beta}
- \Theta_{\alpha\rho}  {\Theta^\rho}_\beta
-q^\mu_{\ \, \alpha} q^\nu_{\ \, \beta} \, 
   R^\gamma_{\ \, \mu\rho\nu} \ell_\gamma \ell^\rho\ ,
\eea
i.e. 
\be
    \vec{\w{q}}^* \w{\nabla}_{\el} \w{\Theta} = \kappa \w{\Theta}
     - \w{\Theta}\cdot\vec{\w{\Theta}}
     -  \vec{\w{q}}^* \mathrm{\bf Riem}(\uel, ., \el, .) . 
                                \label{e:DN:qLieTheta_prov}
\ee
Now, expressing the Lie derivative $\Lie{\el} \w{\Theta}$ in terms
of the covariant derivative $\w{\nabla}$ and using 
$\w{\Theta} = \vec{\w{q}}^* \, \w{\nabla}\uel$
[Eq.~(\ref{e:KI:Theta_qstar_gradl})], 
we find the relation 
\be
\label{eq:DN:lie-cov}
\vec{\w{q}}^* \Lie{\el} \w{\Theta} = \vec{\w{q}}^*\w{\nabla}_{\el} \w{\Theta}
+ 2 \w{\Theta}\cdot\vec{\w{\Theta}} ,
\ee
so that Eq.~(\ref{e:DN:qLieTheta_prov}) becomes
\be
  \encadre{\vec{\w{q}}^* \Lie{\el}\w{\Theta} = \kappa \w{\Theta}
    + \w{\Theta}\cdot\vec{\w{\Theta}} 
    - \vec{\w{q}}^*  \mathrm{\bf Riem}(\uel, . , \el , .) }.    
                                    \label{e:DN:qLieTheta}
\ee
An alternative derivation of this relation, based on Cartan's structure equations,
is given in Appendix~\ref{s:CA} [cf. Eq.~(\ref{e:CA:qLieTheta})].
The equation equivalent to (\ref{e:DN:qLieTheta}) in the framework 
of the quotient formalism (cf. Remark~\ref{r:NH:quotient}) 
is called a 'Ricatti equation' by
Galloway \cite{Gallow04}, by analogy
with the classical Ricatti ODE: $y' = a(x) y^2 + b(x) y + c(x)$. 
See also Refs. \cite{Jezie04,LewanP04}.

Expressing the 4-dimensional Riemann tensor in terms of the Weyl tensor
$\w{C}$ (its traceless part) and the Ricci tensor $\w{R}$
via Eq.~(\ref{e:IN:Weyl}),
Eq.~(\ref{e:DN:qLieTheta}) becomes
\be \label{e:DN:evol_Theta}
  \encadre{\vec{\w{q}}^* \Lie{\el}\w{\Theta} = \kappa \w{\Theta}
    + \w{\Theta}\cdot\vec{\w{\Theta}} 
    - \vec{\w{q}}^*  \w{C}(\uel, . , \el , .) 
    - \frac{1}{2} \w{R}(\el,\el) \, \w{q} }, 
\ee
or 
\be
\label{e:DN:evol_Theta_index}
\encadre{
q^\mu_{\ \, \alpha} q^\nu_{\ \, \beta}\left(\Liec{\el} \Theta_{\mu\nu}\right)
=\kappa \, \Theta_{\alpha\beta}
+ \Theta_{\alpha\mu} \Theta^\mu_{\ \, \beta}
 -q^\mu_{\ \, \alpha} q^\nu_{\ \, \beta}
    C_{\rho\mu\sigma\nu} \ell^\rho\ell^\sigma 
-\frac{1}{2} (R_{\mu\nu} \ell^\mu \ell^\nu)  q_{\alpha\beta} 
} \ ,
\ee
where we have made use of $\w{q}\cdot\el=0$ and 
$\el\cdot\el=0$.

Taking the trace of Eq.~(\ref{e:DN:evol_Theta}) and making use of 
Eq.~(\ref{e:DN:Theta_Theta}), results immediately in an evolution for 
the expansion scalar $\theta$, which is nothing but the
Raychaudhuri Eq. (\ref{e:DN:Raychaud}). 
From the components of the Riemann tensor appearing in 
Eq.~(\ref{e:DN:qLieTheta}), Eq.~(\ref{e:DN:evol_Theta}) could have been 
considered as the projection on $\Sp_t$ of the null analog of the
3+1 Ricci equation (\ref{e:FO:Rienorm}). Thus the Raychaudhuri equation
(\ref{e:DN:Raychaud}) can be derived either from the null Codazzi equation
as in Sec.~\ref{s:DN:Raychaud},  or from the null Ricci equation. 
This reflects the fact that the Gauss-Codazzi-Ricci terminology is
not well adapted to the null case, 
as anticipated in Remark~\ref{rem:DN:cod_gau}.

On the other hand, the  traceless part of Eq.~(\ref{e:DN:evol_Theta})
results in the evolution equation for the shear tensor:
\be \label{e:DN:evol_sigma}
  \encadre{\vec{\w{q}}^* \Lie{\el}\w{\sigma} = \kappa \, \w{\sigma}
    + \sigma_{ab}\sigma^{ab} \, \w{q}
    - \vec{\w{q}}^*  \w{C}(\uel, . , \el , .) }, 
\ee
or
\be
\label{e:DN:evol_sigma_index}
\encadre{
q^\mu_{\ \, \alpha} q^\nu_{\ \, \beta}\left(\Liec{\el} \sigma_{\mu\nu}\right)
=\kappa \, \sigma_{\alpha\beta}
+ \sigma_{\mu\nu}\sigma^{\mu\nu} \, q_{\alpha\beta}
 -q^\mu_{\ \, \alpha} q^\nu_{\ \, \beta}
    C_{\rho\mu\sigma\nu} \ell^\rho\ell^\sigma } ,
\ee
where we have used the fact that for a 2-dimensional symmetric tensor we have 
$\sigma_{\alpha\mu}{\sigma^\mu}_\beta=
\frac{1}{2}\sigma_{\mu\nu}\sigma^{\mu\nu}q_{\alpha\beta}$.
In particular, from
$\vec{\w{q}}^* \w{\mathcal L}_{\w{\ell}} \w{\sigma} = 
\vec{\w{q}}^*\w{\nabla}_{\ell} \w{\sigma}
+ 2 \w{\sigma}\cdot\vec{\w{\Theta}}$,
we find the equivalent expression
\be
q^\mu_{\ \, \alpha} q^\nu_{\ \, \beta} \ell^\rho
 \nabla_\rho \sigma_{\mu\nu}
-(\kappa - \theta) \, \sigma_{\alpha\beta}=
-q^\mu_{\ \, \alpha} q^\nu_{\ \, \beta}
    C_{\rho\mu\sigma\nu} \ell^\rho\ell^\sigma ,
\ee
which coincides with Eq. (2.13) of  Price \& Thorne \cite{PriceT86}, 
once we identify $q^\mu_{\ \, \alpha} q^\nu_{\ \, \beta} \ell^\rho
 \nabla_\rho \sigma_{\mu\nu}$
with $D_{\bar t}\sigma_{\alpha\beta}$ in that reference.

This equation is denominated {\em tidal equation} in Ref.~\cite{PriceT86},
since the term in the right-hand side is directly 
related to the driving force responsible for the relative acceleration 
between two null geodesics via the {\em geodesic deviation equation}
(see e.g. Ref. \cite{Wald84}).
In  other words, this force is responsible for the tidal forces
on the 2-surface $\Sp_t$.
The tidal equation (\ref{e:DN:evol_sigma}) and the null Raychaudhuri
equation (\ref{e:DN:Raychaud}) are part of the so-called 
{\em optical scalar equations} derived by Sachs within the Newman-Penrose
formalism \cite{Sachs62b}.

\subsection{Evolution of the transversal deformation rate}

Let us consider now an equation that can be
seen as the null analog of the contracted Gauss equation 
(\ref{e:FO:Gauss_contracted_1}) combined with the Ricci equation
(\ref{e:FO:Rienorm}). It is obtained 
by projecting the spacetime Ricci tensor 
$\w{R}$ onto the hypersurface $\Hor$. The difference with the spacelike case 
of Sec.~\ref{s:FO:Riem_3p1} is 
that this projection is not an orthogonal one, but instead is performed
via the projector $\w{\Pi}$ along the transverse direction $\w{k}$. 
\begin{rem}
The projection $\w{\Pi}^* \w{R}$ of the spacetime Ricci
tensor onto $\Hor$ [as defined by Eq.~(\ref{e:NH:def_P_star})]
can be decomposed in the following way, thanks to 
Eq.~(\ref{e:NH:q_proj_P}):
\bea
    R_{\mu\nu} \Pi^\mu_{\ \, \alpha} \Pi^\mu_{\ \, \beta} & = & 
        R_{\mu\nu} q^\mu_{\ \, \alpha} q^\mu_{\ \, \beta}
        - \underbrace{R_{\mu\nu} \ell^\mu 
            q^\nu_{\ \, \alpha}}_{\mbox{Codazzi 2}} k_\beta
            - \underbrace{R_{\mu\nu} \ell^\mu 
                q^\nu_{\ \, \beta}}_{\mbox{Codazzi 2}} k_\alpha \nonumber \\
           &&  + \underbrace{R_{\mu\nu} \ell^\mu 
                    \ell^\nu }_{\mbox{Codazzi 1}} k_\alpha k_\beta  . 
\eea
We notice that among the four parts of this decomposition, three of them  
are parts of the Codazzi equation and have been already considered as
the Raychaudhuri equation (Codazzi 1, Sec.~\ref{s:DN:Raychaud}) or 
the Damour-Navier-Stokes equation (Codazzi 2, Sec.~\ref{s:DN:DNS}). 
Thus the only new piece of information contained in the 
null analog of the contracted Gauss equation is $\vec{\w{q}}^* \w{R}$,
namely the (orthogonal) 
projection of the Ricci tensor onto the 2-surfaces $\Sp_t$. 
This contrasts with the spacelike case of the standard 3+1 formalism, where
the contracted Gauss equation (\ref{e:FO:Gauss_contracted_1})
is totally independent of the Codazzi equation (\ref{e:FO:Codazzi}). 
\end{rem}

In order to evaluate $\vec{\w{q}}^* \w{R}$, let us start by the 
contracted Ricci
identity applied to the connection $\w{\DS}$ induced by the spacetime
connection $\w{\nabla}$ onto the 2-surfaces $\Sp_t$:
\be
    \DS_\mu \DS_\alpha v^\mu - \DS_\alpha \DS_\mu v^\mu
        = {}^2\! R_{\alpha\mu} v^\mu ,  
\ee
where $\w{v}$ is any vector field in $\T(\Sp_t)$ and ${}^2\! \w{R}$ is the
Ricci tensor associated with $\w{\DS}$. 
Expressing each $\w{\DS}$ derivative in terms of the spacetime derivative
$\w{\nabla}$ via Eq.~(\ref{e:KI:2der})
and substituting Eq.~(\ref{e:IN:vec_q_k_l}) for the projector
$\vec{\w{q}}$ leads to
\bea
    {}^2\! R_{\alpha\mu} v^\mu  & = & \left[
        q^\mu_{\ \,\alpha} (\theta k_\nu + 
        \theta_{(\w{k})} \ell_\nu) 
        - \Theta^\mu_{\ \,\alpha} k_\nu 
        -  \Xi^\mu_{\ \,\alpha} \ell_\nu \right] \nabla_\mu  v^\nu\nonumber \\
    & & + q^\mu_{\ \,\alpha}  q^\rho_{\ \, \nu} 
    ( \nabla_\rho \nabla_\mu v^\nu - \nabla_\mu \nabla_\rho v^\nu) , 
                                            \label{e:DN:2Rv1}
\eea
where use has been made of Eqs.~(\ref{e:NH:Thetaqq}), 
(\ref{e:KI:grad_uel_index}), (\ref{e:KI:def_theta}), 
(\ref{e:KI:Xi_qstar_gradk}), (\ref{e:KI:grad_uk_index}) and
(\ref{e:KI:def_theta_k}) to let appear $\w{\Theta}$, $\theta$,
$\w{\Xi}$ and $\theta_{(\w{k})}$. 
Now the 4-dimensional Ricci identity (\ref{e:IN:Ricci_ident})
applied to the vector field $\w{v}$
yields
\be
 q^\mu_{\ \,\alpha}  q^\rho_{\ \, \nu} (
    \nabla_\rho \nabla_\mu v^\nu - \nabla_\mu \nabla_\rho v^\nu)  
    = q^\mu_{\ \,\alpha}  q^\rho_{\ \, \nu} 
        R^\nu_{\ \, \sigma\rho\mu} v^\sigma
    = q^\mu_{\ \,\alpha}  q^\rho_{\ \, \lambda} 
        R^\lambda_{\ \, \sigma\rho\mu} q^\sigma_{\ \, \nu} v^\nu , 
\ee
where 
$R^\lambda_{\ \, \sigma\rho\mu}$ denotes the spacetime Riemann curvature tensor.
Moreover, $\el\cdot\w{v}=0$ and $\w{k}\cdot\w{v}=0$ [since 
$\w{v}\in\T(\Sp_t)$], so that we can transform Eq.~(\ref{e:DN:2Rv1}) into
\bea
    {}^2\! R_{\alpha\mu} v^\mu  & = & \left[ 
    - \theta \, \Xi_{\alpha\mu} - \theta_{(\w{k})} \, \Theta_{\alpha\mu} 
    + \Xi_{\mu\nu} \, \Theta^\nu_{\ \, \alpha}
    + \Theta_{\mu\nu} \, \Xi^\nu_{\ \, \alpha} \right] v^\mu \nonumber \\
    & & 
    + q^\mu_{\ \,\alpha}  q^\rho_{\ \, \lambda} q^\sigma_{\ \, \nu}
        R^\lambda_{\ \, \sigma\rho\mu} v^\nu . 
\eea
Since this identity is valid for any vector $\w{v}\in\T(\Sp_t)$, we
deduce the following expression of the Ricci tensor of the 2-dimensional
Riemannian spaces $(\Sp_t,\w{q})$ in terms of the Riemann tensor
of $(\M,\w{g})$, the second fundamental form $\w{\Theta}$ of $\Hor$
and the transversal deformation rate $\w{\Xi}$:  
\be
   {}^2\! R_{\alpha\beta} = q^\mu_{\ \,\alpha}  q^\nu_{\ \, \beta} 
        q^\rho_{\ \, \sigma} R^\sigma_{\ \, \nu\rho\mu}
    - \theta \, \Xi_{\alpha\beta} - \theta_{(\w{k})} \, \Theta_{\alpha\beta} 
    + \Theta_{\alpha\mu} \, \Xi^\mu_{\ \, \beta}  
    + \Xi_{\alpha\mu} \, \Theta^\mu_{\ \, \beta} . \label{e:DN:2Ricci}
\ee
Let us now express the term $q^\mu_{\ \,\alpha}  q^\nu_{\ \, \beta} 
q^\rho_{\ \, \sigma} R^\sigma_{\ \, \nu\rho\mu}$ in terms of the 
spacetime Ricci tensor $R_{\alpha\beta}$. 
We have, using the symmetries of the Riemann tensor
and the Ricci identity (\ref{e:IN:Ricci_ident}) 
for the vector field $\w{k}$
\bea
  q^\mu_{\ \,\alpha}  q^\nu_{\ \, \beta} q^\rho_{\ \, \sigma} 
  R^\sigma_{\ \, \nu\rho\mu} & = & q^\mu_{\ \,\alpha}  q^\nu_{\ \, \beta} 
  \left( \delta^\rho_{\ \, \sigma} + k^\rho \ell_\sigma + \ell^\rho k_\sigma
  \right) R^\sigma_{\ \, \nu\rho\mu} \nonumber \\
  & = & q^\mu_{\ \,\alpha}  q^\nu_{\ \, \beta} 
  \left( R_{\mu\nu} - R_{\mu\sigma\rho\nu} k^\sigma \ell^\rho
  - R_{\nu\sigma\rho\mu} k^\sigma \ell^\rho \right) \nonumber \\
  & = & q^\mu_{\ \,\alpha}  q^\nu_{\ \, \beta} \bigg[ R_{\mu\nu}
        - \ell^\rho \left( 
        \nabla_\rho\nabla_\nu k_\mu - \nabla_\nu\nabla_\rho k_\mu \right)
        \nonumber \\
    & &    - \ell^\rho \left( 
        \nabla_\rho\nabla_\mu k_\nu - \nabla_\mu\nabla_\rho k_\nu \right)
        \bigg]  \nonumber \\
  & = & q^\mu_{\ \,\alpha}  q^\nu_{\ \, \beta} \bigg[ R_{\mu\nu}
        - \ell^\rho \nabla_\rho\nabla_\nu k_\mu+ \nabla_\nu \omega_\mu 
        - \nabla_\nu \ell^\rho \nabla_\rho k_\mu \nonumber \\
    & & - \ell^\rho \nabla_\rho\nabla_\mu k_\nu
    + \nabla_\mu \omega_\nu - \nabla_\mu \ell^\rho \nabla_\rho k_\nu
    \bigg] ,        \label{e:DN:qqqR_prov}
\eea
where use has been made of the relation 
$\ell^\rho\nabla_\rho k_\mu = \omega_\mu$
[Eq.~(\ref{e:KI:omega_gradl_uk})]. After expanding the 
gradient of $\uk$ by means of Eq.~(\ref{e:KI:grad_uk}) and the
gradient of $\el$ by means of Eq.~(\ref{e:KI:grad_el}), 
we arrive at 
\bea
    q^\mu_{\ \,\alpha}  q^\nu_{\ \, \beta} 
q^\rho_{\ \, \sigma} R^\sigma_{\ \, \nu\rho\mu} & = &
    q^\mu_{\ \,\alpha}  q^\nu_{\ \, \beta}  \left( R_{\mu\nu}
        - 2 \ell^\sigma \nabla_\sigma \Xi_{\mu\nu}
            + \nabla_\mu \omega_\nu + \nabla_\nu \omega_\mu \right) \nonumber \\
    & & + 2 \Omega_\alpha \Omega_\beta   
    - \Theta_{\alpha\mu} \, \Xi^\mu_{\ \, \beta}  
    -  \Xi_{\alpha\mu} \, \Theta^\mu_{\ \, \beta}. \label{e:DN:qqqR}
\eea
Now, by means of Eqs.~(\ref{e:KI:Omega_omega_k}) and (\ref{e:KI:2der}),
\be
    q^\mu_{\ \,\alpha}  q^\nu_{\ \, \beta}  \left( 
    \nabla_\mu \omega_\nu + \nabla_\nu \omega_\mu \right) =
    \DS_\alpha \Omega_\beta + \DS_\beta \Omega_\alpha
     - 2\kappa \Xi_{\alpha\beta}. 
\ee
Inserting this relation along with 
$\ell^\sigma \nabla_\sigma \Xi_{\mu\nu} = \Liec{\el} \Xi_{\mu\nu}
        - \Xi_{\sigma\nu} \nabla_\mu \ell^\sigma
        - \Xi_{\mu\sigma} \nabla_\nu \ell^\sigma$
into Eq.~(\ref{e:DN:qqqR}) results in 
\bea
    q^\mu_{\ \,\alpha}  q^\nu_{\ \, \beta} 
q^\rho_{\ \, \sigma} R^\sigma_{\ \, \nu\rho\mu} & = &
    q^\mu_{\ \,\alpha}  q^\nu_{\ \, \beta}  \left( R_{\mu\nu}
    - 2 \Liec{\el} \Xi_{\mu\nu} \right) 
    + \DS_\alpha \Omega_\beta + \DS_\beta \Omega_\alpha \nonumber \\
      & & + 2 \Omega_\alpha \Omega_\beta 
    - 2\kappa \Xi_{\alpha\beta} 
       +   \Theta_{\alpha\mu} \, \Xi^\mu_{\ \, \beta}  
    +   \Xi_{\alpha\mu} \, \Theta^\mu_{\ \, \beta} .
\eea
Replacing into Eq.~(\ref{e:DN:2Ricci}) leads to the following 
evolution equation for $\w{\Xi}$
\bea
   q^\mu_{\ \,\alpha}  q^\nu_{\ \, \beta} \,  \Liec{\el} \Xi_{\mu\nu} &=&
   \frac{1}{2} \left( \DS_\alpha \Omega_\beta + \DS_\beta \Omega_\alpha \right)
   + \Omega_\alpha \Omega_\beta 
   - \frac{1}{2}  {}^2\! R_{\alpha\beta} 
    + \frac{1}{2} q^\mu_{\ \,\alpha}  q^\nu_{\ \, \beta} R_{\mu\nu}
   \nonumber \\
   & & - \left(\kappa + \frac{\theta}{2}\right) \Xi_{\alpha\beta}
   - \frac{\theta_{(\w{k})}}{2}  \Theta_{\alpha\beta}
   + \Theta_{\alpha\mu} \, \Xi^\mu_{\ \, \beta}  
    +  \Xi_{\alpha\mu} \, \Theta^\mu_{\ \, \beta}  . 
\eea
or
\bea
   \vec{\w{q}}^*  \Lie{\el} \w{\Xi} &=&
   \frac{1}{2} \Kil{\w{\DS}}{\w{\Omega}}
   + \w{\Omega}\otimes \w{\Omega}
   - \frac{1}{2}  {}^2\!\w{R}
    + \frac{1}{2} \vec{\w{q}}^* \w{R}
   - \left(\kappa + \frac{\theta}{2}\right) \w{\Xi}
   - \frac{\theta_{(\w{k})}}{2} \w{\Theta}
    \nonumber \\
   & & + \w{\Theta}\cdot\vec{\w{\Xi}} + \w{\Xi}\cdot\vec{\w{\Theta}}  . 
                                            \label{e:DN:qLieXi}
\eea
\begin{rem}
It is legitimate to compare Eq.~(\ref{e:DN:qLieXi}) with 
Eq.~(\ref{e:CA:Ricci_ab_final})
derived in Appendix~\ref{s:CA} by means of Cartan's structure equations, 
since both equations involve $\vec{\w{q}}^*  \Lie{\el} \w{\Xi}$,
$\vec{\w{q}}^* \w{R}$ and ${}^2\!\w{R}$. The major difference
is that Eq.~(\ref{e:CA:Ricci_ab_final}) involves in addition the
Lie derivative $\Lie{\w{k}} \w{\Theta}$. Actually 
Eq.~(\ref{e:CA:Ricci_ab_final}) is completely symmetric between $\el$
and $\w{k}$ (and hence between $\w{\Theta}$ and $\w{\Xi}$). This reflects
the fact that $\vec{\w{q}}^* \w{R}$ and ${}^2\!\w{R}$ depend only upon
the 2-surface $\Sp_t$ and, from the point of view of $\Sp_t$ alone,
$\el$ and $\w{k}$ are on the same footing, being respectively the
outgoing and ingoing null normals to $\Sp_t$. However, in the derivation
of Eq.~(\ref{e:DN:qLieXi}), we have broken this symmetry,
which is apparent in Eq.~(\ref{e:DN:2Ricci}), at the step
(\ref{e:DN:qqqR_prov}) by rearranging terms
in order to consider the Ricci identity for the vector $\w{k}$ only.
Actually by a direct computation (substituting Eq.~(\ref{e:KI:grad_uel})
for $\w{\Theta}$ and permuting the derivatives of $\uel$ via Ricci
identity), one gets the following relation between the two Lie derivatives:
\bea
    \vec{\w{q}}^* \Lie{\w{k}} \w{\Theta} & = & 
    \vec{\w{q}}^* \Lie{\el} \w{\Xi} + \w{\DS}\w{\DS}\rho + 
    \w{\DS}\rho\otimes\w{\DS}\rho - \w{\Omega}\otimes\w{\DS}\rho
    - \w{\DS}\rho\otimes\w{\Omega} \nonumber \\
    & & - \Kil{\w{\DS}}{\w{\Omega}} + N^{-2} \w{\nabla}_{\el}\sigma \; 
    \w{\Theta} + \kappa\, \w{\Xi}. 
\eea 
Substituting this expression for $\vec{\w{q}}^* \Lie{\w{k}} \w{\Theta}$
into Eq.~(\ref{e:CA:Ricci_ab_final}), we recover Eq.~(\ref{e:DN:qLieXi}).
\end{rem}
If we take into account Einstein equation (\ref{e:FO:Einstein}), the 
4-dimensional Ricci term can be written
$\vec{\w{q}}^* \w{R} = 8\pi \left(\vec{\w{q}}^* \w{T}
- 1/2\; T \w{q} \right)$, where $T = {\rm tr}\, \vec{\w{T}}$ is
the trace of the energy-momentum tensor $\w{T}$. The evolution equation
for $\w{\Xi}$ becomes then
\be \label{e:DN:evol_xi}
 \encadre{ 
 \begin{array}{ll}
 \displaystyle \vec{\w{q}}^*  \Lie{\el} \w{\Xi} = &
 \displaystyle  \frac{1}{2} \Kil{\w{\DS}}{\w{\Omega}}
   + \w{\Omega}\otimes \w{\Omega}
   - \frac{1}{2}  {}^2\!\w{R}
   + 4\pi \left( \vec{\w{q}}^* \w{T} - \frac{T}{2} \w{q} \right)
   \nonumber \\
  & \displaystyle - \left(\kappa + \frac{\theta}{2}\right) \w{\Xi} 
   - \frac{\theta_{(\w{k})}}{2} \w{\Theta}
   + \w{\Theta}\cdot\vec{\w{\Xi}} + \w{\Xi}\cdot\vec{\w{\Theta}} . 
 \end{array}
 } 
\ee

\begin{exmp} \label{ex:DN:cone}
\textbf{``Dynamics'' of Minkowski light cone.}\\
As a check of the above dynamical equations, let us specify them 
to the case where $\Hor$ is a light cone in Minkowski spacetime,
as considered in Examples~\ref{ex:NH:cone}, \ref{ex:IN:cone},
and \ref{ex:KI:cone}. Since $\kappa=0$, $\w{\sigma}=0$ and $\w{T}=0$
for this case 
[Eqs.~(\ref{e:NH:kappa_cone}) and (\ref{e:KI:sigma_cone})], the 
null Raychaudhuri equation (\ref{e:DN:Raychaud_T}) reduces to
\be
    \w{\nabla}_{\el}\, \theta + \frac{1}{2} \theta^2 = 0 . 
\ee
Using the values $\ell^\alpha=(1,x/r,y/r,z/r)$ and $\theta=2/r$ 
given respectively by Eqs.~(\ref{e:NH:ell_cone}) and (\ref{e:KI:theta_cone}), 
we check that the above equation is satisfied. Besides, since 
$\w{\Omega}=0$ in this case [Eq.~(\ref{e:KI:Omega_cone})], 
the Damour-Navier-Stokes equation 
(\ref{e:DN:DNS_if}) reduces to 
\be
    \w{\DS} \theta = 0. 
\ee
Since $\theta=2/r$
is a function of $r$ only and $r$ is constant on $\Sp_t$ (being equal
to $t$), we have $\w{\DS} \theta = 0$, i.e. the Damour-Navier-Stokes
equation is fulfilled. 
The tidal force equation (\ref{e:DN:evol_sigma}) is trivially satisfied 
in the present case since both the shear tensor $\w{\sigma}$ and
the Weyl tensor $\w{C}$ vanish. On the other hand, the evolution equation 
for $\w{\Xi}$, Eq.~(\ref{e:DN:evol_xi}), reduces somewhat, but still
contains many non-vanishing terms:
\be \label{e:DN:evol_xi_cone}
 \vec{\w{q}}^*  \Lie{\el} \w{\Xi} = 
   - \frac{1}{2}  {}^2\!\w{R}
   - \frac{\theta}{2} \w{\Xi} 
   - \frac{\theta_{(\w{k})}}{2} \w{\Theta}
   + \w{\Theta}\cdot\vec{\w{\Xi}} + \w{\Xi}\cdot\vec{\w{\Theta}} . 
\ee
Using the value of $\w{\Xi}$ given by  Eq.~(\ref{e:KI:Xi_cone}) 
allows to write the left-hand side as
\bea
   \vec{\w{q}}^*  \Lie{\el} \w{\Xi} = \vec{\w{q}}^*  \Lie{\el} \left(
   - \frac{1}{2 r} \, \w{q} \right) 
   & = & - \frac{1}{2} \Bigg[ \Lie{\el} \left(\frac{1}{r}\right) \, \w{q}
     + \frac{1}{r} 
     \underbrace{\vec{\w{q}}^*  \Lie{\el} \w{q}}_{=2\w{\Theta}
     =\frac{2}{r}\w{q}} \Bigg] \nonumber \\
    & = & - \frac{1}{2} 
    \left[ \ell^\mu \der{}{x^\mu} \left(\frac{1}{r}\right)
    + \frac{2}{r^2} \right] \w{q} = - \frac{1}{2r^2}\, \w{q} , 
            \label{e:DN:qLxi_cone}
\eea
where we have used expressions (\ref{e:KI:Theta_cone}) for
$\w{\Theta}$ and (\ref{e:NH:ell_cone}) for $\ell^\mu$. 
Besides, since $\Sp_t$ is a 2-dimensional manifold, 
the Ricci tensor ${}^2\!\w{R}$ which appears on the right-hand side
of Eq.~(\ref{e:DN:evol_xi_cone}) is expressible in terms of the Ricci 
scalar ${}^2\!R$ as ${}^2\!\w{R} = \frac{1}{2} \, {}^2\!R \, \w{q}$. 
Moreover, $\Sp_t$ being a metric 2-sphere of radius $r$,
${}^2\!R = 2/r^2$. Thus
\be \label{e:DN:2R_cone}
    {}^2\!\w{R} = \frac{1}{r^2} \, \w{q} . 
\ee
Inserting Eqs.~(\ref{e:DN:qLxi_cone}) and  (\ref{e:DN:2R_cone}),
as well as the values of $\w{\Theta}$, $\theta$, $\w{\Xi}$ and
$\theta_{(\w{k})}$ obtained in Example~\ref{ex:KI:cone} into 
Eq.~(\ref{e:DN:evol_xi_cone}), and using $\w{q}\cdot\vec{\w{q}}=\w{q}$,
leads to ``$0=0$'', as it should be. 
\end{exmp}

\begin{exmp} \label{ex:DN:EF}
\textbf{Dynamics of Schwarzschild horizon.} \\
In view of the values obtained in Example~\ref{ex:KI:kin_EF} for 
$\w{\Theta}$, $\w{\omega}$, $\w{\Omega}$ and $\w{\Xi}$ corresponding 
to the Eddington-Finkelstein slicing of the event horizon of Schwarzschild 
spacetime, let us specify the dynamical equations obtained above to 
that case.
First of all, the Ricci tensor and the stress-energy tensor
vanish identically, since we are dealing with a vacuum solution 
of Einstein equation: $\w{R}=0$ and $\w{T}=0$. Taking into account 
$\theta\equalH 0$ and $\w{\sigma}\equalH 0$ [Eq.~(\ref{e:KI:thetaH_EF})],
the null Raychaudhuri equation (\ref{e:DN:Raychaud_T}) is then trivially
satisfied on $\Hor$. Similarly, since $\w{\Omega}\equalH 0$
[Eq.~(\ref{e:KI:OmegaH_EF})] and $\kappa\equalH 1/(4m)$ 
is a constant [Eq.~(\ref{e:NH:kappaH_EF})],
the Damour-Navier-Stokes equation (\ref{e:DN:DNS_if})
is trivially satisfied on $\Hor$.
On the other side, the tidal force equation (\ref{e:DN:evol_sigma}) reduces
to 
\be
    \vec{\w{q}}^* \w{C}(\uel,.,\el,.) = 0 .
\ee 
This constraint on the Weyl tensor is satisfied by 
the Schwarz\-schild solution, as a consequence of being
of Petrov type D and $\el$ a principal null direction 
(see e.g. Proposition 5.5.5 in Ref.~\cite{Oneil95}). 
Finally Eq.~(\ref{e:DN:evol_xi}) giving the 
evolution of the transversal deformation rate reduces to
(since $\w{T}=0$, $\w{\Omega}\equalH 0$ and $\w{\Theta}\equalH 0$)
\be \label{e:DN:evol_xi_EF}
    \vec{\w{q}}^*  \Lie{\el} \w{\Xi} \equalH 
    - \frac{1}{2}  {}^2\!\w{R} - \kappa \, \w{\Xi} .
\ee  
Now from Eq.~(\ref{e:KI:XiH_EF}), we have
$\vec{\w{q}}^* \Lie{\el} \w{\Xi} \equalH -(2m)^{-1} \, \vec{\w{q}}^*
\Lie{\el} \w{q} \equalH - m^{-1} \, \w{\Theta} \equalH 0$, hence
\be \label{e:DN:LieXi_EF}
    \vec{\w{q}}^* \Lie{\el} \w{\Xi} \equalH 0 . 
\ee 
On the other side, $\kappa\equalH 1/(4m)$ [Eq.~(\ref{e:NH:kappaH_EF})] and
expression (\ref{e:KI:XiH_EF}) for $\w{\Xi}$ leads to 
\be \label{e:DN:kappa_Xi_EF}
    \kappa \, \w{\Xi}  \equalH - \frac{1}{8m^2}\, \w{q} .
\ee 
Besides, since $\Sp_t$ is a metric 2-sphere, as in Example~\ref{ex:DN:cone}
above, Eq.~(\ref{e:DN:2R_cone}) holds. Since $r\equalH 2m$, it yields
\be  \label{e:DN:2Ricci_EF}
    {}^2\!\w{R} \equalH \frac{1}{4m^2} \, \w{q} . 
\ee
Gathering Eqs.~(\ref{e:DN:LieXi_EF}), (\ref{e:DN:kappa_Xi_EF}) and
(\ref{e:DN:2Ricci_EF}), we check that Eq.~(\ref{e:DN:evol_xi_EF})
is satisfied. 
\end{exmp}

%% file: neh.tex
%
\section{Non-expanding horizons}
\label{s:NE}

All results presented in the previous sections apply to 
any null hypersurface and are not specific to the event horizon 
of a black hole. For instance, 
they are perfectly valid for a light cone in Minkowski spacetime,
as illustrated by Examples~\ref{ex:NH:cone}, \ref{ex:IN:cone}, \ref{ex:KI:cone}
and \ref{ex:DN:cone}. 
In this section, we move on the way to (quasi-equilibrium) black holes by requiring  
the null hypersurface $\Hor$ to be {\em non-expanding}, in the sense that the
expansion scalar $\theta$ defined in Sec.~\ref{s:KI:expans_shear} is 
vanishing. Indeed, one should remind
that $\theta$ measures the rate of variation of the
surface element of the spatial 2-surfaces $\Sp_t$ foliating $\Hor$
[cf. Eqs.~(\ref{e:KI:theta_Lie_detq}) or (\ref{e:KI:theta_dsdt_detq})]. 
We have seen in Example~\ref{ex:KI:kin_EF} 
that $\theta=0$ for the event horizon of a Schwarzschild black hole
[cf. Eq.~(\ref{e:KI:thetaH_EF})]. On the contrary, 
in any weak gravitational field, the null hypersurfaces with compact spacelike 
sections are always expanding or contracting
(cf. Example~\ref{ex:KI:cone} for the light cone in flat spacetime). 
Therefore the existence of a non-expanding null hypersurface 
$\Hor$ with compact sections $\Sp_t$ 
is a signature of a very strong
gravitational field. 

As we shall detail below, non-expanding horizons are closely related 
to the concept of
{\em trapped surfaces} introduced by Penrose in 1965 \cite{Penro65}
and the associated notion of {\em apparent horizon} \cite{HawkiE73}. 
They constitute the first structure in the hierarchy
recently introduced by Ashtekar et al. 
\cite{AshteBF99,AshteBDFKLW00,AshteBL01,AshteBL02,AshteFK00,AshteK05} which 
leads to {\em isolated horizons}. Contrary to event horizons, isolated
horizons constitute a {\em local} concept. Moreover, contrary to 
Killing horizons
--- which are also local ---, isolated horizons are well defined even in the
absence of any spacetime symmetry.  


\subsection{Definition and basic properties}

\subsubsection{Definition} \label{s:NE:def_NEH}

Following \hajicek\ \cite{Hajic73,Hajic74} and
Ashtekar et al. \cite{AshteFK00,AshteBL02}, we say that the
null hypersurface $\Hor$ 
is a {\em non-expanding horizon (NEH)}\footnote{\hajicek\ \cite{Hajic74} used
the term ``perfect horizon'' instead of ``non-expanding horizon''.} 
if, and only if, the following properties hold\footnote{In this
review we are working with metrics satisfying Einstein equation on the 
whole spacetime $\M$, and in particular on $\Hor$ (this has been fully
employed in Sec. \ref{s:DY}). In more general contexts, 
the NEH definition must also include the enforcing of the Einstein equation 
on $\Hor$. }
\begin{enumerate}
\item $\Hor$ has the topology of $\mathbb{R}\times\mathbb{S}^2$;
\item the expansion scalar $\theta$ introduced in 
Sec.~\ref{s:KI:expans_shear} vanishes on $\Hor$:
\be \label{e:NE:theta_0}
	\encadre{\theta \equalH 0} ; 
\ee
\item the matter stress-energy tensor $\w{T}$ 
obeys the {\em null dominant energy
condition} on $\Hor$, namely the ``energy-momentum current density''
vector
\be \label{e:NE:def_W}
    \w{W} := - \vec{\w{T}}\cdot \el 
\ee 
is future directed timelike or null on $\Hor$. 
\end{enumerate}
Let us recall that, although $\theta$ can be viewed as the
rate of variation of the surface element of the 
spatial 2-surfaces $\Sp_t$ foliating $\Hor$
[cf. Eqs.~(\ref{e:KI:theta_Lie_detq}) or (\ref{e:KI:theta_dsdt_detq})],
it does not depend upon $\Sp_t$ but only on $\el$
(cf. Remark~\ref{rem:KI:theta_indep_St}). 
Moreover, thanks to the behavior $\theta\rightarrow \theta' = \alpha \theta$
[Eq.~(\ref{e:KI:rescale_theta_sigma})] under the rescaling
$\el\rightarrow \el'= \alpha\el$, the vanishing of $\theta$
does not depend upon the choice of a specific null normal $\el$.
Similarly the property (3) does not depend upon the choice of the null
normal $\el$, provided that it is future directed.  
In other words, the property of being a NEH is an
intrinsic property of the null hypersurface $\Hor$. In particular
it does not
depend upon the spacetime foliation by the hypersurfaces $(\Sigma_t)$.

\begin{rem}
The topology requirement on $\Hor$ is very important in the definition 
of an NEH, in order to capture the notion of black hole. 
Without it, we could for instance consider for $\Hor$ a 
null hyperplane of Minkowski
spacetime. Indeed let $t-x=0$ be the equation of this hyperplane in 
usual Minkowski coordinates $(x^\alpha)=(t,x,y,z)$. The components
of the null normal $\el$ with respect to these coordinates
are then $\ell^\alpha = (1,1,0,0)$, so that $\w{\nabla}\el =0$. 
Consequently, $\Hor$ fulfills condition (2) in the above definition: 
$\theta=0$, although $\Hor$ has nothing
to do with a black hole. 
\end{rem}

\begin{rem}
The null dominant energy condition (3) is trivially fulfilled in 
vacuum spacetimes. Moreover, in the non-vacuum case, this is a very 
weak condition, which is satisfied by any 
electromagnetic field or reasonable matter model (e.g. perfect fluid).
In particular, it is implied by the much stronger {\em dominant energy
condition}, which says that energy cannot travel faster than light
(see e.g. the textbook \cite{HawkiE73}, p.~91 or \cite{Wald84}, p.~219).
\end{rem}

\subsubsection{Link with trapped surfaces and apparent horizons}
\label{s:NE:trapped_app}
 
Let us first recall that a {\em trapped surface} has been 
defined by Penrose (1965) \cite{Penro65} as a closed (i.e. compact
without boundary) spacelike 2-surface $\mathcal{S}$
such that the two systems of null geodesics emerging orthogonally
from $\mathcal{S}$ converge locally at $\mathcal{S}$, i.e. they have
non-positive scalar expansions (see also the definition p.~262 of 
Ref.~\cite{HawkiE73} and Ref.~\cite{DemiaL73} for an early characterization
of black holes by trapped surfaces). In the present context, demanding 
that the spacelike 2-surface $\Sp_t=\Hor\cap \Sigma_t$ is a trapped surface 
is equivalent to 
\be
    \theta \leq 0 \qquad \mbox{and}\qquad  \theta_{(\w{k})}  \leq 0 ,
\ee
where $\theta_{(\w{k})}$ is defined by Eq.~(\ref{e:KI:def_theta_k}).
The sub-case $\theta=0$ or $\theta_{(\w{k})}=0$ is referred
to as a {\em marginally trapped surface} (or simply {\em marginal
surface} by Hayward \cite{Haywa94}). 

Penrose's definition is purely local since it involves only quantities
defined on the surface $\mathcal{S}$. 
On the contrary, Hawking \cite{Hawki73} (see also Ref.~\cite{HawkiE73}, p.~319)
has introduced the concept of {\em outer trapped surface}, the definition
of which relies on a global property of spacetime, 
namely asymptotic flatness:
an {\em outer trapped surface} is 
an orientable compact spacelike 2-surface $\mathcal{S}$ contained in
the future development of a partial Cauchy hypersurface $\Sigma_0$
and which is such that the {\em outgoing} null geodesics emerging 
orthogonally from $\mathcal{S}$ converge locally at $\mathcal{S}$. 
This requires the definition of {\em outgoing} null geodesics,
which is based on the assumption of asymptotic flatness. 
In present context, demanding 
that the spacelike 2-surface $\Sp_t=\Hor\cap \Sigma_t$ is an 
outer trapped surface is equivalent to 
\begin{enumerate}
\item the spacelike hypersurface $\Sigma_t$ is asymptotically flat 
(more precisely {\em strongly future asymptotically predictable}
and simply connected, cf. Ref.~\cite{Hawki73}, p.~26 or
Ref.~\cite{HawkiE73}, p.~319) and the scalar field $u$ defining $\Hor$ 
has been chosen so that the exterior of $\Sp_t$ (defined by $u>1$, 
cf. Sec.~\ref{s:IN:normal_l}) contains the 
asymptotically flat region, so that $\el$ is an outgoing null normal
in the sense of Hawking;         
\item the expansion scalar of $\el$ is negative or null:
\be
    \theta \leq 0 . 
\ee
\end{enumerate}
The sub-case $\theta=0$ is referred
to as a {\em marginally outer trapped surface}. 
Note that the above definition does not assume anything on $\theta_{(\w{k})}$,
contrary to Penrose's one. 

A related concept, also introduced by Hawking \cite{Hawki73}
and widely used in numerical relativity (see e.g. 
\cite{NakamOK87,York89,CookY90,Thorn96,Gundl98,Schne03,SchneHP05}
and Sec.~6.1 of Ref.~\cite{BaumgS03} for a review),
is that of {\em apparent horizon}: it is defined as a 2-surface 
$\mathcal{A}$ inside a Cauchy spacelike hypersurface $\Sigma$
such that $\mathcal{A}$ is
a connected component of the outer boundary of the trapped region of $\Sigma$.
By {\em trapped region}, it is meant the set of points $p\in\Sigma$
through which there is an outer trapped surface lying in $\Sigma$
 \footnote{Note that this definition
of apparent horizon, which is Hawking's original one \cite{Hawki73,HawkiE73}
and  which is commonly used in the numerical relativity community,
is different from that given in the recent study \cite{DreyeKSS03}
devoted to the use of isolated horizons in numerical relativity, 
which requires in addition $\theta_{(\w{k})} < 0$.}. 
From Proposition 9.2.9 of Hawking \& Ellis \cite{HawkiE73},
an apparent horizon is a marginally outer trapped surface
(but see Sec.~1.6 of Ref.~\cite{Chrus02} for an update and
refinements).  

In view of the above definitions, let us make explicit the relations with
a NEH:
if $\Hor$ is an NEH in an asymptotically 
flat spacetime, then each slice $\Sp_t$
is a marginally outer trapped surface. 
If, in addition, $\theta_{(\w{k})}  \leq 0$, then $\Sp_t$
is a marginally trapped surface. In general, 
$\w{k}$ being the inward null normal to $\Sp_t$, 
$\theta_{(\w{k})}$ is always
negative. However there exist some pathological situations for 
which $\theta_{(\w{k})} > 0$ at some points of $\Sp_t$
\cite{GerocH82}. 

Hence a NEH can be constructed by stacking
marginally outer trapped surfaces. In particular, it can be
obtained by stacking apparent horizons. However, it must be pointed out
that contrary to the black hole event horizon, nothing guarantees that the world
tube formed by a sequence of apparent horizons is
smooth. It can even be discontinuous (cf. Fig.~60 in Ref.~\cite{HawkiE73}
picturing the merger of two black holes) !
Moreover, even when it is smooth, the world tube of apparent 
horizons is generally spacelike and not null (Ref.~\cite{HawkiE73}, p.~323).
It is only in some equilibrium state that it can be null.
Note that this notion of equilibrium needs only to be local: non-expanding
horizons can exist in non-stationary spacetimes \cite{Chrus92,AshteK05}. 

It is also worth to relate NEH's to the concept
of {\em trapping horizon} introduced in 1994 by Hayward \cite{Haywa94}
(see also \cite{Haywa04b}) and aimed at providing a local description of 
a black hole. A {\em trapping horizon} is defined as a hypersurface
of $\M$ foliated by spacelike 2-surfaces $\mathcal{S}$
such that the expansion scalar $\theta_{(\el)}$ of one of the two families 
of null geodesics orthogonal to $\mathcal{S}$ vanishes. 
A trapping horizon can be either spacelike or null. 
It follows immediately from the above definition 
that NEHs are null trapping horizons.

\subsubsection{Vanishing of the second fundamental form}

Let us show that on a NEH, not only the trace $\theta$
of the second fundamental form $\w{\Theta}$ vanishes, but also 
$\w{\Theta}$ as a whole.
Setting $\theta=0$ in the null Raychaudhuri equation (\ref{e:DN:Raychaud_T})
leads to
\be \label{e:NE:Raychaud_0}
	 \sigma_{ab} \sigma^{ab}  
	+ 8\pi \w{T}(\el,\el) = 0 .
\ee
Besides,  $\w{q}$ being a positive definite metric on $\Sp_t$, one has
\be \label{e:NE:sigsig_geq_0}
\sigma_{ab} \sigma^{ab} \geq 0.
\ee 
Moreover, the null dominant energy condition (condition (3) in the
definition of a NEH) implies
\be \label{e:NE:Tll_geq_0}
	\w{T}(\el,\el) \geq 0 . 
\ee
The three relations (\ref{e:NE:Raychaud_0}), (\ref{e:NE:sigsig_geq_0}) 
and  (\ref{e:NE:Tll_geq_0}) imply
\be \label{e:NE:ss_0}	
	\sigma_{ab} \sigma^{ab} = 0  
\ee
and
\be \label{e:NE:T_ll_0}
	\w{T}(\el,\el) = 0 . 
\ee
Note that this last constraint is trivially satisfied in the vacuum case
($\w{T}=0$). 
Invoking again the positive definite character of $\w{q}$ and the
symmetry of $\sigma_{ab}$, 
Eq.~(\ref{e:NE:ss_0}) implies that $\sigma_{ab}=0$, i.e. the
vanishing of the shear tensor:
\be
	\w{\sigma} = 0 . 
\ee
Since we had already $\theta=0$,
we conclude that for a NEH, not only the scalar expansion
vanishes, but also the full tensor of deformation rate
[cf. the decomposition (\ref{e:KI:Theta_split})]:
\be \label{e:NE:Theta_null}
	\encadre{\w{\Theta} = 0 }. 
\ee
From Eq.~(\ref{e:KI:Theta_deform}), this implies
\be \label{e:NE:LieS_q_0}
	\encadre{ \LieS{\el} \w{q} = 0 }, 
\ee
which means that the Riemannian 
metric of the 2-surfaces $\Sp_t$ is invariant as $t$ evolves. 

\begin{rem} \label{r:NE:chi_not_zero}
The vanishing of the second fundamental form $\w{\Theta}$ does not imply 
the vanishing of $\Hor$'s
Weingarten map $\w{\chi}$, 
as it would do if the hypersurface $\Hor$ was not null:
Eq.~(\ref{e:KI:chi_Theta_omega}) shows clearly that the vanishing
of $\w{\chi}$ would require $\w{\omega}=0$ in addition to $\w{\Theta}=0$.
On the contrary, for the spatial hypersurface $\Sigma_t$, 
the Weingarten map $\w{\mathcal K}$ and the second
fundamental form $-\w{K}$ 
are related by 
$\mathcal{K}^\alpha_{\ \, \beta} = - g^{\alpha\mu} K_{\mu\beta}$
[cf. Eqs.~(\ref{e:FO:calK_index}) and (\ref{e:NH:Kab_proj})], 
so that $\w{K}=0\Longrightarrow\w{\mathcal K}=0$.
\end{rem}


\subsection{Induced affine connection on $\Hor$} \label{s:NE:ind_connect}

Since $\w{\Theta}$ is the pull-back of the bilinear form
$\w{\nabla}\uel$ onto $\Hor$ [Eq.~(\ref{e:NH:Theta_pb_grad_uel})], 
its vanishing is equivalent to 
\be \label{e:NE:pull-back_grad_l_0}
   \Phi^* \w{\nabla}\uel = 0 . 
\ee
An important consequence of this is that
\be
\forall (\w{u},\w{v}) \in \T(\Hor)\times\T(\Hor),\quad
	\el \cdot \w{\nabla}_{\w{u}} \, \w{v} = 
	\w{\nabla}_{\w{u}}(\underbrace{\el\cdot\w{v}}_{=0})
	- \underbrace{\w{v} \cdot  \w{\nabla}_{\w{u}} 
    \, \el}_{\Phi^* \w{\nabla}\uel(\w{v},\w{u})=0}
	= 0 .
\ee
This means that for any vectors $\w{u}$ and $\w{v}$ tangent to 
$\Hor$, $\w{\nabla}_{\w{u}} \, \w{v}$ is also a vector tangent to
$\Hor$. Therefore $\w{\nabla}$ gives birth to an affine 
connection intrinsic to $\Hor$, which we will denote $\w{\hat\nabla}$
to distinguish it from the connection on $\M$:
\be \label{e:NE:def_connect_H}
	\encadre{
	\forall (\w{u},\w{v}) \in \T(\Hor)\times\T(\Hor),\quad
	\w{\hat\nabla}_{\w{u}} \, \w{v} := \w{\nabla}_{\w{u}} \, \w{v} }.
\ee
We naturally call $\w{\hat\nabla}$ the {\em connection induced on $\Hor$} by
the spacetime connection $\w{\nabla}$. 

\begin{rem}
More generally, i.e. when $\Hor$ is not necessarily a NEH, 
the vector $\w{k}$, 
considered as a rigging
vector (cf. Remark~\ref{r:rigging_vector}), provides a torsion-free connection
on $\Hor$ via the projector $\w{\Pi}$ along $\w{k}$:
\be
   	\forall (\w{u},\w{v}) \in \T(\Hor)\times\T(\Hor),\quad
	\w{\hat\nabla}_{\w{u}} \, \w{v} := 
        \w{\Pi}(\w{\nabla}_{\w{u}} \, \w{v} )  .  
\ee
This connection is called the \emph{rigged connection} \cite{MarsS93}. 
By expressing $\w{\Pi}$ via Eq.~(\ref{e:NH:P_l_k}) it  is easy to see that 
\be
\forall (\w{u},\w{v}) \in \T(\Hor)\times\T(\Hor),\quad
	\w{\hat\nabla}_{\w{u}} \, \w{v} = \w{\nabla}_{\w{u}} \, \w{v} 
        - \w{\Theta}(\w{u},\w{v}) \, \w{k} . 
\ee
We see then clearly that in the NEH case ($\w{\Theta}=0$), $\w{\hat\nabla}$
is independent of $\w{k}$, i.e. of the choice of the slicing $(\Sp_t)$:
it becomes a connection intrinsic to $\Hor$.  
\end{rem}

As a consequence of  $\w{\nabla}_{\w{u}} \, \w{u}\in\T(\Hor)$, whatever $\w{u}\in\T(\Hor)$, 
and the fact that a geodesic passing through a given point is completely
determined by its derivative at that point, it follows that any geodesic curve of 
$\M$ which starts a some point $p\in\Hor$ and is tangent to $\Hor$ at $p$
remains within $\Hor$ for all points. For this reason,  
$\Hor$ is called a {\em totally geodesic hypersurface} of $\M$.
This explains why in \hajicek's study \cite{Hajic73}, non-expanding
horizons are called ``TGNH'' for ``totally geodesic null hypersurfaces''. 

The definition of $\w{\hat\nabla}$ can be extended to 1-forms
on $\T(\Hor)$ by means of the Leibnitz rule: given 
a 1-form field $\w{\varpi}\in\T^*(\Hor)$, the bilinear form
$\w{\hat\nabla}\w{\varpi}$ is defined by
\bea
    \forall (\w{u},\w{v}) \in \T(\Hor)\times\T(\Hor),\quad
    \w{\hat\nabla}\w{\varpi}(\w{u},\w{v})  &:= &
    \langle \w{\hat\nabla}_{\w{v}} \w{\varpi} , \w{u} \rangle 
    \nonumber \\
  & := & \w{\hat\nabla}_{\w{v}}\langle
        \w{\varpi},\w{u}\rangle
        - \langle \w{\varpi},\w{\hat\nabla}_{\w{v}} \, \w{u} \rangle . 
\eea
Now, thanks to Eq.~(\ref{e:NE:def_connect_H}), 
\bea
  \w{\hat\nabla}\w{\varpi}(\w{u},\w{v})& = &
  \w{\nabla}_{\w{v}}\langle
        \w{\varpi},\w{u}\rangle
        - \langle \w{\varpi},\w{\nabla}_{\w{v}} \, \w{u} \rangle 
  = \w{\nabla}_{\w{v}}\langle
        \w{\varpi},\w{\Pi}(\w{u})\rangle
        - \langle \w{\varpi},\w{\Pi}(\w{\nabla}_{\w{v}} \, \w{u}) \rangle \nn \\
   & = & \w{\nabla}_{\w{v}}\langle
        \w{\Pi}^* \w{\varpi},\w{u}\rangle
        - \langle \w{\Pi}^* \w{\varpi}, \w{\nabla}_{\w{v}} \, \w{u} \rangle
        \nn \\
   & = & \w{\nabla}( \w{\Pi}^* \w{\varpi}) (\w{u},\w{v}) , 
\eea
where $\w{\Pi}^* \w{\varpi}\in\T^*(\M)$ is the extension of $\w{\varpi}$ to
$\T(\M)$ provided by the projector $\w{\Pi}$ onto $\Hor$
[cf. Eq.~(\ref{e:NH:def_P_star})]. 
Since the above equation is valid for any pair of vectors
$(\w{u},\w{v})$ in $\T(\Hor)$, we conclude that
the $\w{\hat\nabla}$-derivative of the 1-form $\w{\varpi}$ is
the pull-back onto $\Hor$ of the spacetime covariant derivative
of $\w{\Pi}^* \w{\varpi}$:
\be
    \w{\hat\nabla}\w{\varpi} = \Phi^* \w{\nabla} ( \w{\Pi}^* \w{\varpi} ) . 
\ee
The above relation is extended to any multilinear form
$\w{A}$ on $\T(\Hor)$, in order to define 
$\w{\hat\nabla}\w{A}$:
\be \encadre{
   \w{\hat\nabla}\w{A} = \Phi^* \w{\nabla}(\w{\Pi}^*\w{A}) } .
\ee
In words: the intrinsic covariant derivative $\w{\hat\nabla}\w{A}$
of a multilinear form $\w{A}$ on $\T(\Hor)$ is the pull-back via the
embedding of $\Hor$ in $\M$ of the ambient spacetime covariant derivative
of the extension of $\w{A}$ to $\T(\M)$, the extension 
being provided by the projector
$\w{\Pi}$ onto $\T(\Hor)$. 
In particular for the bilinear form $\w{q}$
constituting the (degenerate)
metric on $\Hor$:
\be
  \w{\hat\nabla}\w{q} = \Phi^* \w{\nabla} \w{q} ,	
\ee
where we have used $\w{\Pi}^*\w{q}=\w{q}$.
Then
\bea
\forall (\w{u},\w{v},\w{w}) \in \T(\Hor)^3,\quad
  \w{\hat\nabla}\w{q} (\w{u},\w{v},\w{w}) & = &
    \w{\nabla} \w{q} (\w{u},\w{v},\w{w}) 
    = \w{\nabla}_{\w{w}} \, \w{q}(\w{u},\w{v}) \nonumber \\
 & = & \langle \w{\nabla}_{\w{w}}\, \uel, \w{u} \rangle 
    \langle \uk, \w{v}\rangle 
    + \langle \uk, \w{u}\rangle  \langle \w{\nabla}_{\w{w}}\, \uel, \w{v} \rangle
    \nonumber \\
 & = & \w{\Theta}(\w{u},\w{w})   \langle \uk, \w{v}\rangle 
    + \w{\Theta}(\w{v},\w{w})   \langle \uk, \w{u}\rangle \nonumber \\
    &=& 0 , 
\eea
where we have used $\w{\Theta}=\Phi^* \w{\nabla}\uel$
[Eq.~(\ref{e:NH:Theta_pb_grad_uel})] to let
appear $\w{\Theta}$ and the property $\w{\Theta}=0$ 
[Eq.~(\ref{e:NE:Theta_null})] characterizing NEHs. 
Hence 
\be\label{e:NE:hatnab_compat_q}
    \encadre{ \w{\hat\nabla}\w{q} = 0 } . 
\ee
We thus conclude that the induced connection $\w{\hat\nabla}$ is 
compatible with the metric $\w{q}$ on $\Hor$.  

\begin{rem} \label{r:NE:NEH_geometry}
Since the metric $\w{q}$ on $\Hor$ is degenerate, there is a priori no unique
affine connection compatible with it 
(i.e. a torsion-free connection $\bar{\w{\nabla}}$ such that 
$\bar{\w{\nabla}}\w{q}=0$).
The non-expanding character of $\Hor$ allows then 
a canonical choice for such a connection, namely the connection 
$\w{\hat\nabla}$ which coincides with the ambient
spacetime connection. The couple $(\w{q},\w{\hat\nabla})$ defines
an intrinsic geometry of $\Hor$. This geometrical structure, 
which was first exhibited by  \hajicek\ \cite{Hajic73,Hajic74}, is
however different from that for a spacelike or timelike hypersurface,
in so far as $\w{q}$ and $\w{\hat\nabla}$ are largely two independent
entities [apart from the relation (\ref{e:NE:hatnab_compat_q})]:
for instance the components $\hat\nabla_A$ with respect to a given coordinate
system $(x^A)$ on $\Hor$ are not deduced 
from the components $q_{AB}$ by means of some Christoffel symbols.
\end{rem}

The $\w{\hat\nabla}$-derivative of the null normal $\el$ (considered 
as a vector field in $\T(\Hor)$) takes a simple form, 
obtained by setting $\vec{\w{\Theta}}=0$ in 
Eq.~(\ref{e:KI:grad_el}) and using $\Phi^*\uel=0$:
\be \label{e:NE:hatnab_el}
    \w{\hat\nabla}\el = \el\otimes\w{\omega} . 
\ee


\subsection{Damour-Navier-Stokes equation in NEHs}

The vanishing of $\theta$ and $\w{\sigma}$ means that
for a NEH, all the ``viscous''
terms of Damour-Navier-Stokes equation [Eq.~(\ref{e:DN:DNS_if})]
disappear, so that one is left with
\be \label{e:NE:Inviscid_T}
    \vec{\w{q}}^*  \Lie{\el} \w{\Omega}  =
       8\pi \vec{\w{q}}^*  \w{T}\cdot\el 
       + \w{\DS}  \kappa  .
\ee
In this equation it appears the orthogonal projection onto the 
spatial 2-surfaces $\Sp_t$ of the ``energy-momentum current density'' vector 
$\w{W}$ defined by Eq.~(\ref{e:NE:def_W}). 
The orthogonal projection $\vec{\w{q}}(\w{W})$ on the 2-surfaces
$\Sp_t$ is the force surface density denoted by $\w{f}$ in 
Eq.~(\ref{e:DN:force_surface}).
For a NEH, Eq.~(\ref{e:NE:T_ll_0}) holds and
yields
\be
	\el \cdot \w{W} = 0 . 
\ee
This means that $\w{W}$ is tangent to $\Hor$. Then $\w{W}$ cannot be
timelike (for $\Hor$ is a null hypersurface). 
From the null dominant energy
condition (hypothesis (3) in Sec.~\ref{s:NE:def_NEH}), it cannot 
be spacelike. It is then necessarily null.
Moreover,  being tangent to $\Hor$, it must be collinear to $\el$:
\be \label{e:NE:W_w_l}
	\w{W} = w \el ,
\ee
where $w$ is some positive scalar field on $\Hor$. 
Note that in the vacuum case, this relation is trivially 
fulfilled with $w=0$.
An immediate consequence of (\ref{e:NE:W_w_l}) is the vanishing
of the force surface density, since $\vec{\w{q}}(\el) = 0$:
\be
	\vec{\w{q}}(\w{W}) = 0 . 
\ee
Accordingly, the Damour-Navier-Stokes Eq.~(\ref{e:NE:Inviscid_T})
simplifies to 
\be \label{e:NE:Inviscid}
  \encadre{
    \vec{\w{q}}^*  \Lie{\el} \w{\Omega}  = \w{\DS}  \kappa } ,  
\ee
i.e. the only term left in the right-hand side is the ``pressure''
gradient $\w{\DS}  \kappa$.

Finally, we note that Eq. (\ref{e:NE:W_w_l}) can be recast by
using Einstein equation (\ref{e:FO:Einstein}) into  
\be
    \w{R}(\el,.) = \left( \frac{1}{2} R - 8\pi w \right) \uel , 
\ee
which implies 
\be
\label{e:NE:Ricci_l=0}
    \w{\Pi}^* \w{R}(\el,\cdot) = 0 \ .
\ee


\subsection{Evolution of the transversal deformation rate in NEHs}

After having considered the non-expanding limit of the 
Raychaudhuri and Damour-Navier-Stokes equations, let us now
turn to the evolution equation for the transversal deformation
rate $\w{\Xi}$, namely Eq.~(\ref{e:DN:evol_xi}). Setting 
$\w{\Theta}=0$ in it, we get
\be \label{e:NE:Lie_Xi}
 \encadre{ 
 \vec{\w{q}}^*  \Lie{\el} \w{\Xi} = 
   \frac{1}{2} \Kil{\w{\DS}}{\w{\Omega}}
   + \w{\Omega}\otimes \w{\Omega}
   - \frac{1}{2}  {}^2\!\w{R}
   + 4\pi \left( \vec{\w{q}}^* \w{T} - \frac{T}{2} \w{q} \right)
    - \kappa \w{\Xi} }  . 
\ee
As a check, we verify that this equation agrees with Eq.~(3.9)
of Ashtekar et al. \cite{AshteBL02}, after the proper changes of
notation have been performed: Ashtekar's
$\tilde S_{ab}$ corresponds to our $\Xi_{ab}$, $\tilde{\mathcal D}_a$
to $\DS_a$, $\tilde\omega_a$ to $\Omega_a$, $\tilde{\mathcal R}_{ab}$
to ${}^2\!R_{ab}$ and ${\tilde q}_a^{\ \, c}$ to $q_a^{\ \, \mu}$. Note 
also that objects in Ashtekar et al. are generally defined only on $\Hor$
(or in an appropriate quotient space of it), whereas we are
considering 4-dimensional objects.


\subsection{Weingarten map and rotation 1-form on a NEH}
\label{s:NE:Wein_rotation_1form}
 
We have already noticed (Remark~\ref{r:NE:chi_not_zero}) that the
vanishing of the second fundamental form $\w{\Theta}$ on a non-expanding
horizon does not imply the vanishing of the Weingarten map $\w{\chi}$, 
because $\Hor$ is a null hypersurface. The expression of $\w{\chi}$ when
$\w{\Theta}=0$ however simplifies [cf. Eq.~(\ref{e:KI:chi_Theta_omega})] :
\be \label{e:NE:chi_om_el}
    \encadre{ \w{\chi} = \langle \w{\omega}, . \rangle \, \el } .
\ee

\begin{rem} \label{r:NE:omega_intrinsic_NEH}
Equation (\ref{e:NE:chi_om_el}) shows that, on a NEH, all the information 
about the Weingarten map is actually encoded in the rotation 
1-form $\w{\omega}$.
Restricted to $\T_p(\Hor)$, Eq.~(\ref{e:NE:chi_om_el}) implies
\be \label{e:NE:interp_omega}
    \forall\w{v}\in\T_p(\Hor),\quad \w{\hat\nabla}_{\w{v}}\, \el =
    \langle \w{\omega}, \w{v} \rangle \, \el .
\ee
Actually this last relation is that used by Ashtekar et al. 
\cite{AshteFK00,AshteBL02,AshteK05} 
to define $\wo$ for a NEH as a 1-form in $\T^*(\Hor)$.
It is clear from Eq.~(\ref{e:NE:interp_omega}) that, on a NEH, $\w{\omega}$
depends only upon $\el$ (more precisely upon the normalization of $\el$)
and not directely upon the 2-surfaces $\Sp_t$
induced by the 3+1 slicing. On the contrary, the \hajicek\ 1-form 
$\w{\Omega}$ depends directly upon $\Sp_t$, since its definition 
(\ref{e:KI:def_hajicek}) involves the orthogonal projector $\vec{\w{q}}$
onto $\Sp_t$: $\w{\Omega}= \vec{\w{q}}^* \w{\omega}$.
\end{rem}

We have seen that, for a NEH,
the degenerate metric $\w{q}$ does not vary along $\el$
[Eq.~(\ref{e:NE:LieS_q_0})]. It is then interesting to investigate
the evolution of $\w{\omega}$ along $\el$, i.e. to evaluate
$\Lie{\el}\w{\omega}$. Since $\el\in\T(\Hor)$, we may a priori consider
two Lie derivatives: the Lie derivative of $\w{\omega}$ along $\el$
within the manifold $\M$, denoted by $\Lie{\el}\w{\omega}$,
and the Lie derivative of $\w{\omega}$ (or more precisely of the 
pull-back $\Phi^*\w{\omega}$ of $\w{\omega}$ onto $\Hor$)
along $\el$
within the manifold $\Hor$, that we will denote by $\LieH{\el}\w{\omega}$. 
The relation between these two Lie derivatives is given in 
Appendix~\ref{s:LF}. In particular Eq.~(\ref{e:NE:LieH_Lie_form})
gives
\be \label{e:NE:LieH_om_Pi}
    \w{\Pi}^* \, \LieH{\el}\w{\omega} = \w{\Pi}^*\Lie{\el}
       (\w{\Pi}^*\w{\w{\omega}}) . 
\ee
Since $\langle \w{\omega}, \w{k}\rangle = 0$
[Eq.~(\ref{e:KI:omega_l_kappa})], we have 
$ \w{\Pi}^* \w{\omega} = \w{\omega}$, so that Eq.~(\ref{e:NE:LieH_om_Pi}) 
results in 
\bea
    \w{\Pi}^* \, \LieH{\el}\w{\omega} & = & \w{\Pi}^*\Lie{\el}\w{\omega} 
    = \w{\Pi}^*\Lie{\el} (\w{\Omega} -\kappa \uk) \nonumber \\
   & = & \w{\Pi}^* \left( \Lie{\el} \w{\Omega} - \w{\nabla}_{\el}\kappa
        \, \uk - \kappa \Lie{\el}\uk \right) , \label{e:NE:PiLieHomega}
\eea
where use has been made of Eq.~(\ref{e:KI:Omega_omega_k}).
Now, by Cartan identity (\ref{e:IN:Cartan_id}), 
$\Lie{\el}\uk = \el \cdot \dd \uk + \dd\langle \uk, \el \rangle
= \el \cdot \dd \uk$ (since $\langle \uk, \el \rangle=-1$).
Using the Frobenius relation (\ref{e:KI:dk}) to express $\dd \uk$, 
we get 
\be
    \Lie{\el}\uk = \frac{1}{2N^2} \w{\nabla}_{\el} 
        \ln \left( \frac{N}{M} \right) \; \uel . 
\ee
It is then obvious that 
\be \label{e:NE:PLie_el_uk}
   \w{\Pi}^* \Lie{\el}\uk = 0
\ee
since $\w{\Pi}^* \uel = 0$ [Eq.~(\ref{e:NH:P_star_k_l})], so that
Eq.~(\ref{e:NE:PiLieHomega}) reduces to 
\be
    \w{\Pi}^* \, \LieH{\el}\w{\omega} = \w{\Pi}^*  \Lie{\el} \w{\Omega}
    - \w{\nabla}_{\el}\kappa \, \uk , \label{e:NE:P_LieH_om_Om}
\ee
where use has been made of the property
$\w{\Pi}^* \uk = \uk$ [Eq.~(\ref{e:NH:P_star_k_l})].
Using relation (\ref{e:NH:q_proj_P}),  we have
$\w{\Pi}^*  \Lie{\el} \w{\Omega} = \vec{\w{q}}^* \Lie{\el} \w{\Omega}$,
hence
\be
    \w{\Pi}^* \, \LieH{\el}\w{\omega} = \vec{\w{q}}^* \Lie{\el} \w{\Omega}
    - \w{\nabla}_{\el}\kappa \, \uk .    
\ee
Substituting Eq.~(\ref{e:NE:Inviscid})
for $\vec{\w{q}}^* \Lie{\el} \w{\Omega}$, we obtain
\be
    \w{\Pi}^* \, \LieH{\el}\w{\omega} = \w{\DS}  \kappa
        - \w{\nabla}_{\el}\kappa \, \uk .
\ee
Expanding the relation $\w{\DS}  \kappa = \vec{\w{q}}^* \w{\nabla} \kappa$
[cf. Eq.~(\ref{e:KI:2der})] by substituting 
$\w{\Pi} + \langle \uk, .\rangle\, \el$ for $\vec{\w{q}}$
[Eq.~(\ref{e:NH:q_proj_P})], we realize that the right-hand of the above
equation is nothing but the projection on $\Hor$ of the spacetime gradient
of $\kappa$:
\be
    \w{\Pi}^* \, \LieH{\el}\w{\omega} = \w{\Pi}^* \w{\nabla}\kappa.
\ee
We can rewrite this 4-dimensional equation as a 3-dimensional 
equation entirely within $\Hor$, by means of the induced connection
$\w{\hat\nabla}$ introduced in Sec.~\ref{s:NE:ind_connect}:
\be \label{e:NE:evol_omega}
   \encadre{ \LieH{\el}\w{\omega} = \w{\hat\nabla} \kappa }. 
\ee
This simple relation, which is of course valid only for a non-expanding
horizon, has been obtained by Ashtekar, Beetle \& Lewandowski 
\cite{AshteBL02} [cf. their Eq.~(2.11)].
Its orthogonal projection onto the 2-surfaces $\Sp_t$ foliating $\Hor$
is the reduced Damour-Navier-Stokes equation (\ref{e:NE:Inviscid}).


\subsection{Rotation 2-form and Weyl tensor}

\subsubsection{The rotation 2-form as an invariant on $\Hor$}

In Remark~\ref{r:NE:omega_intrinsic_NEH}, we have noticed that
for a NEH, the rotation 1-form $\wo$ is ``almost'' 
intrinsic to $\Hor$, in the sense that it does not depend upon
the specific spacelike slicing $\Sp_t$ of $\Hor$ but only on
the normalization of $\el$. On the other side,
considering $\wo$ as a 1-form in $\T^*(\Hor)$ (more precisely
considering the pull-back 1-form $\Phi^*\wo$), its exterior derivative
within the manifold $\Hor$, which we denote by ${}^\Hor\dd\wo$,
is fully intrinsic to $\Hor$. It is indeed 
invariant under a rescaling
of the null normal $\el$, as we are going to show. 
Consider a rescaling $\el'=\alpha\el$ of the null normal, 
as in Sec.~\ref{s:KI:rescaling}. Then $\wo$ varies according to
Eq.~(\ref{e:KI:rescale_omega}), which we can write [via
Eq.~(\ref{e:IN:Pi_star_form})],
\be
    \w{\omega}' = \w{\omega} + \dd\ln\alpha + 
    (\w{\nabla}_{\w{k}} \ln\alpha) \, \uel . 
\ee
Taking the exterior derivative (within $\M$) of this relation 
and using $\dd\dd = 0$, as well as 
$\dd\uel = \dd\rho \, \wedge \uel$ [Eq.~(\ref{e:NH:Frobenius_l})],
yields
\be
    \dd\w{\omega}' = \dd \w{\omega} + \left[
        \dd\left( \w{\nabla}_{\w{k}} \ln\alpha \right) 
        + \w{\nabla}_{\w{k}} \ln\alpha \, \dd \rho \right] \wedge \uel . 
            \label{e:NE:domega'}
\ee
Let us consider the pull-back of this relation onto $\Hor$
[cf. Eq.~(\ref{e:NH:def_pull-back_multi})]. First of all,
we have that the external differential is {\it natural} with respect to the 
pull-back\footnote{More precisely, we should write 
$\Phi^* \dd\wo = {}^\Hor\dd(\Phi^* \wo)$,
but the above remark about $\wo$ ``leaving essentially'' in $\Hor$ allows us not 
to distinguish between $\Phi^*\wo$ and $\wo$.}
\be \label{e:NE:Phi_star_domega}
    \Phi^* \dd\wo = {}^\Hor\dd\wo .
\ee
It is straightforward to 
establish it by means of a coordinate
system $(x^\alpha)$ adapted to $\Hor$ (cf. Sec.~\ref{s:IN:stacoord}):
\bea
\forall (\w{u},\w{v}) \in \T_p(\Hor)^2,\quad
   \Phi^* \dd\wo (\w{u},\w{v}) &=&  \dd\wo (\Phi_*\w{u},\Phi_*\w{v})
    \nonumber \\
    & = & (\partial_\mu \omega_\nu - \partial_\nu \omega_\mu)
        (\Phi_* u)^\mu (\Phi_* v)^\nu \nonumber \\
    & = & (\partial_A \omega_B - \partial_B \omega_A)
        u^A v^B  \nonumber \\
    & = & {}^\Hor\dd\wo(\w{u},\w{v}) . 
\eea
Taking into account Eq.~(\ref{e:NE:Phi_star_domega}) and the similar
relation for $\dd\wo'$, as well as $\Phi^*\uel=0$, 
the pull-back of Eq.~(\ref{e:NE:domega'}) onto $\Hor$
results in 
\be
    \encadre{ {}^\Hor\dd\w{\omega}' = {}^\Hor\dd \w{\omega} } , 
\ee
which shows the independence of the 2-form ${}^\Hor\dd\wo$ with respect to
the choice of the null normal $\el$. We call ${}^\Hor\dd\wo$ the
{\em rotation 2-form} of the null hypersurface $\Hor$. 

\subsubsection{Expression of the rotation 2-form}

Let us compute the rotation 2-form  ${}^\Hor\dd\wo$
from Eq.~(\ref{e:NE:Phi_star_domega}), i.e. by performing the pull-back 
of the 4-dimensional exterior derivative $\dd\wo$.
The latter is given by the Cartan structure equation (\ref{e:CA:2nd_struct_00})
derived in Appendix~\ref{s:CA}. Indeed, $\wo$ is closely related to the
connection 1-form $\wo^0_{\ \, 0}$ associated with the tetrad
$(\el,\w{k},\we_2,\we_3)$, where $(\we_2,\we_3)$ is any orthonormal
basis of $\T_p(\Sp_t)$ [cf. Eq.~(\ref{e:CA:omega00})]. 
When $\w{\Theta}$ vanishes on $\Hor$ (NEH), 
Eq.~(\ref{e:CA:2nd_struct_00}) simplifies to
\be \label{e:NE:domega_4d}
    \dd \wo = \mathrm{\bf Riem}(\uel,\w{k}, . , .) + \w{A}\wedge\uel , 
\ee
where $\w{A}$ is a 1-form, the precise expression of which is given
by Eq.~(\ref{e:CA:2nd_struct_00}) and is not required here. 

The pull-back of Eq.~(\ref{e:NE:domega_4d}) on $\Hor$ yields  
[taking into account Eq.~(\ref{e:NE:Phi_star_domega}) and $\Phi^*\uel=0$]
\be \label{e:NE:Hdomega_prov}
    {}^\Hor\dd\wo = \Phi^* \mathrm{\bf Riem}(\uel,\w{k}, . , .) .
\ee
Let us consider two vectors $\w{u}$ and $\w{v}$ tangent to $\Hor$.
In the tetrad $(\w{e}_\alpha)=(\el,\w{k},\we_2,\we_3)$ of 
Appendix~\ref{s:CA}, they do not have any component along
$\w{k}$ and expand as 
\be
    \w{u} = u^0 \el + u^a\we_a \qquad\mbox{and}\qquad
    \w{v} = v^0 \el + v^a\we_a . 
\ee
Then Eq.~(\ref{e:NE:Hdomega_prov}) leads to  
\bea
     {}^\Hor\dd\wo(\w{u},\w{v}) & = & 
        \mathrm{\bf Riem}(\uel,\w{k}, u^a\we_a + u^0 \el,
         v^b\we_b + v^0 \el) \nonumber \\
         & = & \mathrm{\bf Riem}(\uel,\w{k},u^a\we_a,v^b\we_b)
         \nonumber \\
         & & + (u^0 v^a - v^0 u^a) 
            \mathrm{\bf Riem}(\uel,\w{k},\el,\we_a) ,  
                \label{e:NE:Hdomega_uv}
\eea
where we have taken into account the antisymmetry of the Riemann tensor
with respect to its last two arguments. 
Let us evaluate the last term in the above equation.   
By virtue of the symmetry property of the Riemann tensor
with respect to the permutation of the first pair of indices with the
second one, we can write
$   \mathrm{\bf Riem}(\uel,\w{k},\el,\we_a) = 
    \mathrm{\bf Riem}(\uel,\we_a,\el,\w{k}) $.
Then we may express $\mathrm{\bf Riem}(\uel,\we_a,\el,\w{k})$ by plugging
the vector pair $(\el,\w{k})$ in Cartan's structure
equation (\ref{e:CA:2nd_struct_1a}) derived in Appendix~\ref{s:CA}. 
Notice that, since we are dealing with a NEH, we can 
set to zero all the terms involving $\Theta_{ab}$ in the right-hand side
of Eq.~(\ref{e:CA:2nd_struct_1a}), but not the term in the left-hand side,
since the derivative of $\w{\Theta}$ in directions transverse to
$\Hor$ is a priori not zero. However,
$\dd(\Theta_{ab}\we^b) = \dd\Theta_{ab}\wedge\we^b + \Theta_{ab} \dd\we^b
 = \dd\Theta_{ab}\wedge\we^b$ and $\langle \we^b, \el\rangle = 0$
and $\langle \we^b, \w{k}\rangle = 0$, so that the left-hand side
of Eq.~(\ref{e:CA:2nd_struct_1a}) vanishes when applied to 
$(\el,\w{k})$. Consequently, one is left with
\bea
  \mathrm{\bf Riem}(\uel,\w{k},\el,\we_a) & = &
    - \w{\nabla}_{\el} (\w{\nabla}_{\we_a}\rho - \Omega_a)
    - \Gamma^b_{\ \, a0} (\w{\nabla}_{\we_b}\rho - \Omega_b) \nonumber \\
    & = & \left\langle \w{\nabla}_{\el}(\w{\Omega} - \w{\DS}\rho) ,\
        \we_a \right\rangle , \label{e:NE:Riem_lklea}
\eea
where we have used $\langle \wo-\dd\rho, \el \rangle = \kappa - 
\w{\nabla}_{\el} \rho = 0$ [Eq.~(\ref{e:NH:def_kappa})] 
to get the first line and 
$\w{\Omega} = \Omega_a \we^a$ and 
$\w{\DS}\rho = (\w{\nabla}_{\we_a}\rho) \, \we^a$ to get the second one. 
Now $\langle \w{\nabla}_{\el} \w{\DS}\rho, \we_a\rangle$ 
is the component along $\we^a$ of the 1-form 
$\vec{\w{q}}^* \w{\nabla}_{\el} \w{\DS}\rho$. Let us
evaluate the latter (using index notation):
\bea
    q^\mu_{\ \, \alpha} \ell^\nu \nabla_\nu\left( \DS_\mu \rho \right) 
    & = & q^\mu_{\ \, \alpha} \, \ell^\nu \nabla_\nu \left( 
    q^\sigma_{\ \, \mu} \nabla_\sigma \rho \right) \nonumber \\
    & = & q^\mu_{\ \, \alpha} \, \ell^\nu \left[
        \nabla_\nu \left( k^\sigma \ell_\mu +\ell^\sigma k_\mu \right)
            \nabla_\sigma \rho +  q^\sigma_{\ \, \mu} \nabla_\sigma 
            \nabla_\nu \rho \right] \nonumber \\
    & = & q^\mu_{\ \, \alpha} \left\{ \ell^\sigma  \ell^\nu \nabla_\nu k_\mu 
        \, \nabla_\sigma\rho + q^\sigma_{\ \, \mu}
        \left[ \nabla_\sigma (\ell^\nu \nabla_\nu \rho) 
        - \nabla_\sigma \ell^\nu \nabla_\nu \rho
        \right] \right\} \nonumber \\
    & = & q^\mu_{\ \, \alpha} \left\{ \kappa \omega_\mu
        + q^\sigma_{\ \, \nu} \left[ \nabla_\sigma \kappa
            - (\Theta^\nu_{\ \, \sigma} + \omega_\sigma \ell^\nu)
            \nabla_\nu \rho \right] \right\} \nonumber \\
    & = & \kappa \Omega_\alpha + \DS_\alpha \kappa 
        - \Theta^\nu_{\ \, \alpha} \nabla_\nu \rho
        - \kappa \Omega_\alpha \nonumber \\
    & = & \DS_\alpha \kappa - \Theta^\nu_{\ \, \alpha} \nabla_\nu \rho . 
\eea
For a NEH, the term involving 
$\Theta^\nu_{\ \, \alpha}$ vanishes, so that one is left with 
\be
    \vec{\w{q}}^* \w{\nabla}_{\el} \w{\DS}\rho = \w{\DS} \kappa . 
\ee
Now, let us use the Damour-Navier-Stokes Eq.~(\ref{e:NE:Inviscid})
to replace $\w{\DS} \kappa$ and get 
\be
    \vec{\w{q}}^* \w{\nabla}_{\el} \w{\DS}\rho = 
        \vec{\w{q}}^*  \Lie{\el} \w{\Omega} . 
\ee
Substituting this last relation for 
$\w{\nabla}_{\el} \w{\DS}\rho$ into Eq.~(\ref{e:NE:Riem_lklea}) 
yields
\be
  \mathrm{\bf Riem}(\uel,\w{k},\el,\we_a) =
   \left\langle \w{\nabla}_{\el} \w{\Omega} - \Lie{\el} \w{\Omega}  ,\
        \we_a \right\rangle .
\ee
Now, by expressing the Lie derivative in terms of the connection
$\w{\nabla}$, one has immediately the relation (using $\vec{\w{\Theta}}=0$)
\be
    \vec{\w{q}}^* \left( 
    \w{\nabla}_{\el} \w{\Omega} - \Lie{\el} \w{\Omega} \right)
    = - \w{\Omega} \cdot \vec{\w{\Theta}} = 0 .
\ee
Thus we conclude that, for a NEH, 
\be
\label{e:NE:almost_Psi_1=0}
    \mathrm{\bf Riem}(\uel,\w{k},\el,\we_a) = 0 . 
\ee
Consequently, there remains only one term in the right-hand side of 
Eq.~(\ref{e:NE:Hdomega_uv}), which we can write, taking into
account that the orthogonal projections of $\w{u}$ and $\w{v}$ 
onto $\T_p(\Sp_t)$ are expressible as $\vec{\w{q}}(\w{u})=u^a \we_a$ 
and $\vec{\w{q}}(\w{v})=v^a \we_a$,
\be
    {}^\Hor\dd\wo(\w{u},\w{v}) =  
        \mathrm{\bf Riem}(\uel,\w{k}, \vec{\w{q}}(\w{u}),
        \vec{\w{q}}(\w{v}) ) ,
\ee
Since $\w{u}$ and $\w{v}$ are any vectors in $\T_p(\Hor)$, we conclude that
the following identity between 2-forms holds:
\be
    \encadre{ {}^\Hor\dd\wo = \vec{\w{q}}^* 
        \mathrm{\bf Riem}(\uel,\w{k},.,.) }  .  \label{e:NE:Hdomega_prov2}    
\ee
This relation considerably strengthens Eq.~(\ref{e:NE:Hdomega_prov}):
the presence of the operator $\vec{\w{q}}^*$ means that 
the 2-form ${}^\Hor\dd\wo$ acts only within the subspace $\T_p(\Sp_t)$
of $\T_p(\Hor)$. Now since the vector space $\T_p(\Sp_t)$ is of dimension two,
the space of 2-forms on it is of dimension only one and is
generated by $\we^2\wedge\we^3$.
Thus, because of the antisymmetry in the last two indices of the
Riemann tensor, Eq.~(\ref{e:NE:Hdomega_prov2}) implies 
${}^\Hor\dd\wo = a \, \we^2\wedge\we^3$,
with the coefficient $a$ being simply 
$\mathrm{\bf Riem}(\uel,\w{k},\we_2,\we_3)$.
Moreover, since $(\we_2,\we_3)$
is an orthonormal basis of $\T_p(\Sp_t)$ ($\we_2$ and $\we_3$ can be 
permuted if necessary to match the volume orientation) we have 
\be
    \we^2\wedge\we^3 = {}^2 \w{\epsilon} , 
\ee
where ${}^2 \w{\epsilon}$ is the surface element of $\Sp_t$ induced by
the spacetime metric [cf. Eq.~(\ref{e:KI:2-epsilon})]. 
Consequently, we have
\be \label{e:NE:Hdomega_Riem}
    \encadre{ {}^\Hor\dd\wo = a \; {}^2 \w{\epsilon} } ,
    \qquad \mbox{with} \quad
    a := \mathrm{\bf Riem}(\uel,\w{k},\we_2,\we_3) .      
\ee
Actually, the coefficient $a$ can be completely expressed in terms of the Weyl 
tensor $\w{C}$:  replacing the
Riemann tensor in Eq.~(\ref{e:NE:Hdomega_Riem})
by its decomposition (\ref{e:IN:Weyl}) in terms of the 
Ricci and Weyl tensors yields, thanks to Eq.~(\ref{e:NE:Ricci_l=0})  
\be
 a = \w{C}(\uel,\w{k},\we_2,\we_3) .
\ee
In order to make the link with previous results in the literature, 
let us express $a$ in terms of the complex Weyl scalars of the
Newman-Penrose formalism introduced in Sec.~\ref{s:IN:Weyl_scal}.
Expanding $\we_2$ and $\we_3$ in terms of $\w{m}$ and $\w{\bar{m}}$
by inverting Eqs.~(\ref{e:NH:m}) and (\ref{e:NH:m_bar}), 
and using the cyclic property in the last three slots of the Weyl
tensor (inherited from property (\ref{e:IN:Riemann_cyclic}) of the 
Riemann tensor), we get
\be
a=\frac{1}{i}\left[\w{C}(\uel, \w{m}, \w{\bar{m}}, \w{k})
-  \w{C}(\uel, \w{\bar{m}},\w{m},\w{k})\right] .
\ee
From the definition~(\ref{e:NH:Weyl_scalars}) of the Weyl scalars $\Psi_n$,
the coefficient $a$ is written in terms of the imaginary part
of $\Psi_2$, so that 
\be  
  \label{e:NE:Hdomega}
    \encadre{ {}^\Hor\dd\wo = 2 \, \mathrm{Im} \Psi_2 \; {}^2 \w{\epsilon} } .
\ee
This relation has been firstly derived in the seventies 
by \hajicek\ (cf. Eq.~(23) in Ref.~\cite{Hajic73}) and
special emphasis has been put on it by Ashtekar et al. 
\cite{AshteFK00,AshteBL02,AshteK05}. More precisely, \hajicek\ 
has derived Eq.~(\ref{e:NE:Hdomega}) in the case $\kappa=0$,
for which $\wo$ coincides with $\w{\Omega}$. However, since (i)
${}^\Hor\dd\wo$ does not depend on the normalization of $\el$ and (ii) it is
always possible to rescale $\el$ to ensure $\kappa=0$
[cf. Eq.~(\ref{e:NH:scale_kappa})], the demonstration of
\hajicek\ is fully general. 

\begin{rem}
Let us compute the Lie derivative of $\wo$ along $\el$ within the 
manifold $\Hor$ from Eq.~(\ref{e:NE:Hdomega_Riem}), 
by means of Cartan identity (\ref{e:IN:Cartan_id}):
\be
   \LieH{\el}\wo = \el\cdot {}^\Hor\dd\wo 
   + {}^\Hor\dd \langle \wo,\el \rangle
   = a \; \underbrace{{}^2 \w{\epsilon}(\el, .)}_{=0} 
    + {}^\Hor\dd \kappa = \w{\hat\nabla} \kappa . 
\ee
Thus we recover the evolution equation (\ref{e:NE:evol_omega}) as a 
consequence of Eq.~(\ref{e:NE:Hdomega_Riem}). 
\end{rem}

\subsubsection{Other components of the Weyl tensor}
\label{s:NE:Psi_0-Psi_1}

We have just shown that the component 
$\w{C}(\uel,\w{k},\we_2,\we_3) = 2 \, \mathrm{Im} \Psi_2$ of the 
Weyl tensor with respect to the tetrad $(\el,\w{k},\we_2,\we_3)$
provides the proportionality between the 2-forms 
${}^\Hor\dd\wo$ and ${}^2 \w{\epsilon}$.
Let us now investigate some other components of the Weyl tensor. 

Setting $\w{\Theta}=0$ in the evolution equation (\ref{e:DN:qLieTheta})
for $\w{\Theta}$ yields
\be
    \vec{\w{q}}^* \mathrm{\bf Riem}(\uel, ., \el, .) = 0 .
\ee
Making use of the property (\ref{e:NE:Ricci_l=0}), we rewrite this
expression as 
\be
 \forall a,b\in\{2,3\}, \quad \w{C}(\uel, \we_a  , \el,  \we_b) = 0 .
\ee
Moreover, from 
Eq. (\ref{e:NE:almost_Psi_1=0}) and again Eq.~(\ref{e:NE:Ricci_l=0}),
we get
\be
 \forall a\in\{2,3\}, \quad \w{C}(\uel, \we_a  , \el,  \w{k}) = 0 \ ,
\ee
where we have made use of the symmetries of  the Weyl tensor.
From  Eqs.~(\ref{e:NH:m}) and (\ref{e:NH:m_bar}) together with
(\ref{e:NH:Weyl_scalars}), the above relations imply the vanishing 
of two of the complex Weyl scalars: 
\be
\label{e:NE:Psi_0-Psi_1}
\encadre{\Psi_0=\Psi_1=0} .
\ee
This means that the NEH structure sets strong constraints on the Weyl tensor 
evaluated at $\Hor$. These constraints are physically relevant. 
On the one hand, 
the Weyl components $\Psi_0$ and $\Psi_1$ are associated with the 
{\it ingoing} transversal and longitudinal parts of 
the gravitational field \cite{Sze65}.
Their vanishing is consistent with the quasi-equilibrium situation
modelled by NEHs, since no dynamical gravitational degrees of freedom
{\it fall} into the black hole by crossing the horizon. 
On the other hand, the
change $\delta \Psi_2$ of the Weyl scalar $\Psi_2$
under a Lorentz transformation
(i.e. either a boost or a rotation) of the 
tetrad $(\el, \w{k}, \we_2, \we_3)$,
turns out to be 
a linear combination of scalars $\Psi_0$ and $\Psi_1$ \cite{Chandra92}.
Consequently, as long as we choose the first vector in the null tetrad
to be the $\el$ normal to $\Hor$, $\Psi_2$ is an invariant.
In particular this means that the value of  $\Psi_2$ does
not depend on the chosen null normal, therefore guaranteeing
its invariance under the choice of the spacelike slicing. Finally, 
we point out that the vanishing of $\Psi_0$ and $\Psi_1$ 
could have been obtained directly as an application of the Goldberg-Sachs
theorem, which establishes the equivalence between $\Psi_0=\Psi_1=0$ and the 
existence of a geodesic ($\kappa=0$), shear-free null vector $\el$  
(see \cite{Chandra92}; as we mentioned above and will discuss in more
detail in Sec. \ref{s:IH:WIHdefinition}, a NEH always admits a null normal
with vanishing non-affinity coefficient $\kappa$).
In particular,  this means that the Weyl tensor is of Petrov type II 
on $\Hor$ \cite{AshteBF99,AshteBDFKLW00}.


\subsection{NEH-constraints and free data on a NEH}
\label{s:NE:NEHinitdata}

\subsubsection{Constraints of the NEH structure}
\label{s:NE:NEHconstraints}

Let us determine which part of the geometry of a NEH
can be freely specified. As we shall see, such a part is essentially given
by fields {\it living} on the spatial slices $\Sp_t$ of the horizon, that
can be considered as initial data. This is 
specially important in the present setting, since it perfectly matches 
with the 3+1 point of view we have adopted. 

As discussed in Remark~\ref{r:NE:NEH_geometry}, 
the geometry of a NEH is characterized by the pair 
$(\w{q}, \w{\hat\nabla})$. 
However, in order to build a NEH one cannot make completely arbitrary choices 
for $\w{q}$ and $\w{\hat\nabla}$ if $\Hor$ is a null hypersurface within a 
spacetime satisfying Einstein equations. The reason is that   
$\w{q}$, $\w{\hat\nabla}$
and the Ricci tensor $\w{R}$ must satisfy certain relations, as
established in the previous sections.
Following \cite{AshteBL02,LewanP04} 
any geometrical identity involving $\w{q}, \w{\hat\nabla}$ and $\w{R}$
on $\Hor$ will be referred to as a {\em constraint of the NEH structure} 
(or {\em NEH-constraint}).
Such geometrical identities can be obtained by evaluating the change of $\w{q}$ and
$\w{\hat\nabla}$ along the integral lines of a null normal $\el$.

Regarding $\w{q}$, the NEH condition (\ref{e:NE:LieS_q_0}) directly provides
the constraint $\LieS{\el}\w{q}=0$. In order to cope with
the constraints associated with 
the evolution of $\w{\hat\nabla}$, we follow an 
analysis which dwells directly on the spatial slicing of $\Hor$.
As explained in Sec.~\ref{s:IN:normal_l},
the foliation $(\Sp_t)$ of $\Hor$ is preserved by the flow
of $\el$ due to the normalization (\ref{e:NH:l_norm2}). 
The pull-back of the 1-form $\uk$ on $\Hor$, $\Phi^*\uk$,
is also preserved by the flow of $\el$:
\be
\label{e:NE:L_lk=0}
\LieH{\el} \Phi^*\uk =0 .
\ee
This follows directly from Eq.~(\ref{e:NE:PLie_el_uk}) and the 
relation (\ref{e:LF:LieH_Lie}) between $\LieH{\el}$ and $\Lie{\el}$. 
Eq.~(\ref{e:NE:L_lk=0}) can also be obtained by 
simply noticing that, 
according to Eq.~(\ref{e:IN:k_dt_H}), $\Phi^*\uk$ is 
(minus) the pull-back to $\Hor$ of the differential of $t$.

Following Ref. \cite{LewanP04}, let us determine the different objects composing
$\Hor$'s connection $\w{\hat\nabla}$ 
[this is also performed in Appendix~\ref{s:CA};
see Eqs.~(\ref{e:CA:omega00}-\ref{e:CA:omegaba})].
Considering an arbitrary vector field $\w{v}\in\T(\Hor)$, 
$\w{\hat\nabla} \w{v}$ can be decomposed in the following parts:
\be
\label{e:NE:elem_delta}
\begin{array}{ll}
q^\mu_{\ \, a} \, q^b_{\ \, \nu} \hat{\nabla}_\mu v^\nu = 
    \DS_a (q^b_{\ \, \mu} v^\mu) \ \ \ \ \ \ \ \ \ &\hbox{(spatial-spatial)} \nn\\
q^\mu_{\ \, a} \, k_\nu  \hat{\nabla}_\mu v^\nu = 
    \DS_a (v^\mu k_\mu) - q^\mu_{\ \, a} v^\nu \nabla_\mu k_\nu &
\hbox{(spatial-null)}  \\
\ell^\mu  \hat{\nabla}_\mu v^\alpha = \Liec{\el}v^\alpha + v^\mu \omega_\mu 
\ell^\alpha & \hbox{(null-arbitrary)}
\end{array}
\ee
From this decomposition and taking into account the expression 
(\ref{e:KI:grad_uk}) for the gradient of $\w{k}$,
we note that  $\w{\hat\nabla} \w{v}$ on each slice $\Sp_t$
can be reconstructed if 
$(\w{\DS}, \w{\omega}, \w{\Xi})$ are known on $\Sp_t$.
Likewise, the invariance of the foliation $(\Sp_t)$
under $\el$ permits to express
the evolution of $\w{\hat{\nabla}}$ in terms of the evolution of 
$(\w{\DS}, \w{\omega}, \w{\Xi})$ or, equivalently, of 
$(\w{q}, \w{\Omega}, \kappa$, $\w{\Xi})$. Since we want to emphasize the 3+1 
point of view, we adopt the second set of variables, which are
fields intrinsic to $\Sp_t$. 
The NEH-constraints are therefore given by the previously derived equations
(\ref{e:NE:LieS_q_0}), (\ref{e:NE:Inviscid}) [or (\ref{e:NE:evol_omega})] 
and (\ref{e:NE:Lie_Xi}):
\be
\label{e:NE:constraints}
\encadre{
\begin{array}{rcl}
\LieS{\el} \w{q} &=& \displaystyle 0   \nn \\
\vec{\w{q}}^*  \Lie{\el} \w{\Omega}  &=& \w{\DS}  \kappa 
\ \ \ \ \ \ \ \ \ \ \ \ (\LieH{\el}\w{\omega} = 
\w{\hat\nabla} \kappa) \nn \\
 \vec{\w{q}}^*  \Lie{\el} \w{\Xi} &=& 
   \frac{1}{2} \Kil{\w{\DS}}{\w{\Omega}}
   + \w{\Omega}\otimes \w{\Omega}
   - \frac{1}{2}  {}^2\!\w{R}
   + 4\pi \left( \vec{\w{q}}^* \w{T} - \frac{T}{2} \w{q} \right)
    - \kappa \w{\Xi}   \nn 
\end{array}
} 
\ee
Notice that the geometry on $\Hor$ does not enforce any evolution equation 
for the non-affinity parameter $\kappa$. 
In fact, the NEH geometry  constrains neither the   
value of $\kappa$ on $\Sp_t$ nor its evolution. This is a consequence of
the freedom to rescale $\el$ in the NEH structure 
(see below in relation with the gauge ambiguity in the choice of initial free data).

\begin{rem}
We have defined the constraints on the NEH structure as identities 
relating $\w{q}, \w{\hat{\nabla}}$ and
$\w{R}$. In fact, components of the Ricci tensor parallel to 
${\Hor}$ actually constrain the null geometry via Einstein equations. 
However, we have derived in previous subsections
other kind of geometric relations like
(\ref{e:NE:Hdomega}) or conditions like (\ref{e:NE:Psi_0-Psi_1}). 
They involve some of the components of the Weyl tensor, i.e. 
the part of the Riemann that is not fixed by Einstein equations.
In particular, once the geometry $(\w{q}, \w{\hat{\nabla}})$ is given, the
components $\Psi_0$, $\Psi_1$ and the imaginary part of $\Psi_2$ are fixed. 
In this sense, they could be considered rather as {\it constraints} for the 4-geometry
containing a NEH.
\end{rem}

\begin{figure}
\centerline{\includegraphics[width=1.0\textwidth]{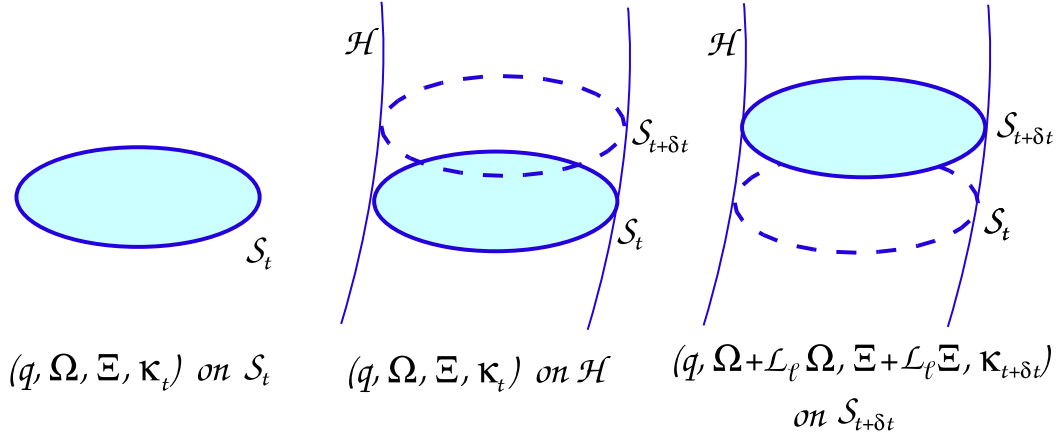}}
\caption[]{\label{f:NH:reconstfreedata} 
Reconstruction of the NEH from the free data. Left: choice of free data on 
$\Sp_t$. Center: Free data as fields on $\Hor$. Right: 
Evolution of the free data to a infinitesimally close surface
$\Sp_{t+\delta t}$.}
\end{figure}

\subsubsection{Reconstruction of $\Hor$ from data on ${\mathcal S}_t$. Free data}
\label{s:NE:NEHfreedata}

Let us describe how a NEH may be reconstructed from data on an initial
spatial slice $\Sp_t$. 
We proceed in three steps (see Fig.~\ref{f:NH:reconstfreedata}).
Firstly, a free choice for the values of 
$(\w{q}, \w{\Omega}, \kappa$, $\w{\Xi})$ 
on $\Sp_t$ is made, considering them as 
objects intrinsic to ${\mathcal S}_t$.
Secondly, these objects are regarded as fields {\it living} in $\Hor$ by imposing,
on the slice ${\mathcal S}_t$, the vanishing of their
corresponding null components:
\bea
\el\cdot \w{q} = 0 \ \ \ \ \el\cdot \w{\Omega}= 0 \ \ \ \  \el \cdot \w{\Xi} =0 \ .
\eea
Finally, the value of these fields is calculated in an infinitesimally close
slice ${\mathcal S}_{t+\delta t}$ by employing Eqs.  
(\ref{e:NE:constraints}).
The value of $\kappa$ on ${\mathcal S}_{t+\delta t}$ can be chosen freely again.
We note by passing that the expression 
$\w{\hat{\nabla}} \el = \el\otimes \wo$ [Eq.~(\ref{e:NE:hatnab_el})],
which can be seen as a constraint on the NEH geometry, is automatically
satisfied by following the above procedure, since ${\el}$ is torsion free
(being normal to $\Hor$) and we construct a vanishing $\w{\Theta}$.
It is not an independent constraint.

We conclude that the free data for the null NEH geometry are given by 
\bea
\label{e:NE:freedataNEH}
\encadre{\hbox{NEH-free data:} \ \ \ (\w{q}|_{{\mathcal S}_t}, 
\w{\Omega}|_{{\mathcal S}_t}, \w{\Xi}|_{{\mathcal S}_t}, \kappa|_{\Hor})
} \ .
\eea
That is, the initial data $\w{q}$, $\w{\Omega}$ and  $\w{\Xi}$ on  ${\mathcal S}_t$, 
together with the function $\kappa$ on $\Hor$, can be freely specified. This is
in contrast with the $3+1$ Cauchy problem, where the initial data on 
the spatial slice $\Sigma_t$
are in fact constrained (cf. Sec.~\ref{s:FO:3p1Einstein}). 
Once the free data are given, the full geometry in $\Hor$ can be 
reconstructed by evolving these quantities along $\el$, this null normal
being determined through Eq. (\ref{e:IN:el_nps})
by the additional structure provided by the slicing $(\Sigma_t)$. Therefore, 
for a given null geometry
$(\w{q},\w{\hat{\nabla}})$, their initial data  $(\w{q}|_{{\mathcal S}_t}, 
\w{\Omega}|_{{\mathcal S}_t}, \w{\Xi}|_{{\mathcal S}_t}, \kappa|_{\Hor})$
are associated with a particular $\el=N(\w{n}+\w{s})$ (see below).

\begin{figure}
\centerline{\includegraphics[width=1.0\textwidth]{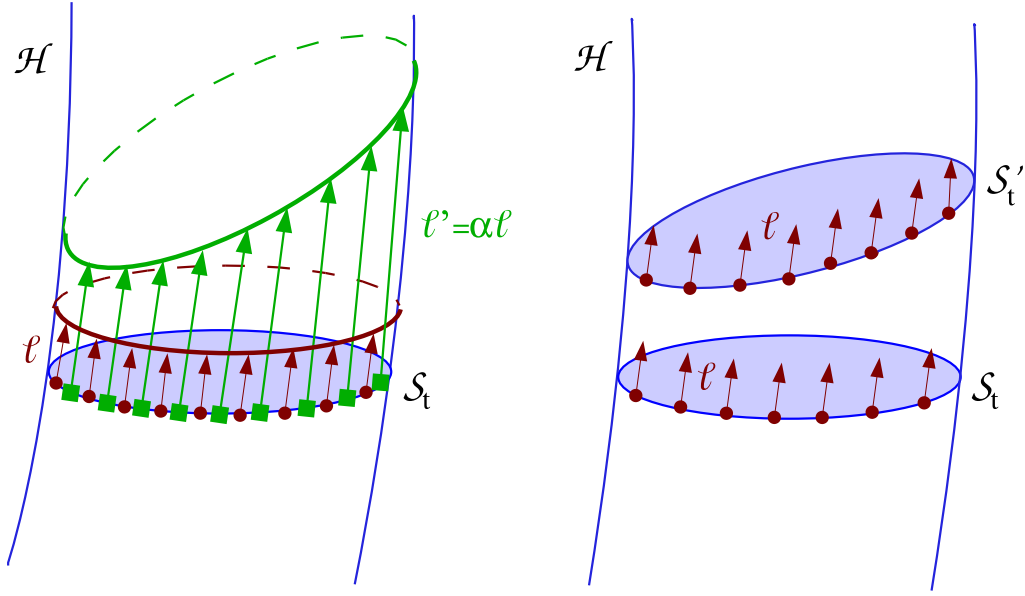}}
\caption[]{\label{f:NH:actpasgauge} 
On the left the {\it active} aspect of the gauge freedom, linked to the choice of the 
null normal, and on the right the {\it passive} aspect, associated with the 
choice of a initial slice.
}
\end{figure}

\noindent {\it Gauge freedom in the choice of the free data}

Even though each specific choice of data in (\ref{e:NE:freedataNEH}), 
together with a slicing $(\Sigma_t)$, fixes the 
NEH geometry, there are different choices that actually
define the same NEH structure: there exists a degeneracy in the 
free data that can be referred to as a {\it gauge} freedom.
This degeneration presents two aspects, the first one linked
to the choice of the null normal, and the second one to the slicing
of $\Hor$ once $\el$ is fixed (see Fig.~\ref{f:NH:actpasgauge}): 
\begin{itemize}
\item [{\it a)}]
If we start from a {\em fixed}  slice $\Sp_0$, our choice of 
$(\w{q}, \w{\Omega}, \kappa$, $\w{\Xi})$
is, as mentioned above, associated with a particular null vector $\el$. 
However, the NEH geometry 
is not changed under a rescaling $\alpha$ of the null normal, $\el'=\alpha \el$. 
Under this rescaling, the fields $(\w{q}, \w{\Omega}, \w{\Xi}, \kappa)$
change according to the transformations given in Table~\ref{t:KI:scaling}.
The resulting intrinsic null geometry $(\w{q},\w{\hat\nabla})$
is the same, even though the slicings $(\Sp_t)$ and $(\Sp_{t'})$
of $\Hor$ induced by the transport of $\Sp_0$ along $\el$ and $\el'$ will
in general differ. We will call this ambiguity the {\it active} aspect of the
gauge freedom. It is associated with the fact that different
slicings $(\Sigma_t)$ fix (in general) different null normals $\el$ via Eq. (\ref{e:IN:el_nps}),
and it is a natural gauge freedom when one actually constructs $\Hor$ 
starting from a slice $\Sp_0$ as in a Cauchy problem.
\item [{\it b)}] Even though a slicing $(\Sigma_t)$ fixes $\el$ 
the reverse is not true, since this null normal is compatible with 
different spatial slicings. Keeping $\el$ fixed, we can consider 
initial slices  $\Sp_{0}$ and $\Sp'_{0}$ in $\Hor$
belonging to different slicings $(\Sp_t)$ and 
$(\Sp_{t'})$ of $\Hor$  compatible with $\el$ via
(\ref{e:NH:l_norm2}). The corresponding level functions 
$t$ and $t'$ are related by $\w{\hat{\nabla}}t= \w{\hat{\nabla}}t' - \w{\hat{\nabla}}g$,
where $\LieH{\el}g=0$. This change in the slicing does not affect
$\w{q}$\footnote{Note that although the metric $\w{q}$  (as an object acting
on $\T\Hor$) does not change,  the projector $\vec{\w{q}}$ on the
slices $\Sp_t$ {\it does} change: $\vec{\w{q}} = \vec{\w{q}} - \el\, \w{\hat{\nabla}}g$,
as follows from  Eq. (\ref{e:NH:q_proj_P}).}, 
$\el$, $\w{\omega}$ and
consequently $\kappa$, but induces via Eq. (\ref{e:IN:k_dt_H}) 
the transformations (cf. Eq. (3.10) in 
\cite{AshteBL02})
\bea
\label{e:NE:passivegauge}
\uk\rightarrow \uk+ \w{\hat{\nabla}}g \ \ , \ \ \w{\Omega}\rightarrow  \w{\Omega}
-\kappa \w{\hat{\nabla}}g  \ \ , \ \ \w{\Xi}\rightarrow \w{\Xi} +
\w{\hat{\nabla}}\w{\hat{\nabla}}g \ .
\eea
This aspect of the gauge freedom, that we can refer to
as {\it passive} and which corresponds to a coordinate choice in $\Hor$, 
is natural when $\Hor$ is a hypersurface 
existing {\it a priori}, its slicings only entering in a second step
once $\el$ is fixed.
It is the aspect discussed in Refs. \cite{AshteBL02,Kr02}.
\end{itemize}
In brief, from a 3+1 perspective, in both cases the underlying source
of degeneracy in the free data
is related to the choice of a specific slicing of $\Hor$.
When there is no canonical manner of fixing neither the initial slice nor the
null vector $\el$, as it is the case in a purely intrinsic formulation of the 
geometry of $\Hor$,
both sources of gauge freedom are simultaneously 
present\footnote{We have discussed here the gauge freedom in the
determination of {\it intrinsic} geometrical objects on
$\Sp_0$. Of course, regarding their specific coordinate expression, there 
exists an additional underlying freedom related to the choice of coordinate 
system on $\Sp_0$.} 
(see discussion in Section VI.A.2 of Ref.  \cite{LewanP04}).

\subsubsection{Evolution of $\hat{\nabla}$ from an {\it intrinsic} 
null perspective}
\label{s:NE:intrinsic_NEH-const}
Following the 3+1 approach adopted in this article, the previous 
discussion on the NEH-constraints has made explicit use of a
spatial slicing of $\Hor$.
However, the time evolution of the connection $\w{\hat{\nabla}}$
can be described in a more intrinsic manner.
In order to prepare the notion of  (strongly) isolated horizon 
in Section \ref{s:IH:strongIH}, we briefly comment on it.

The evolution of the connection $\w{\hat{\nabla}}$ along $\el$
can be written (in components) as follows (see Refs. \cite{AshteBL02,LewanP04})
\bea
\label{e:NE:evol_deltaH}
[\LieH{l}, \hat{\nabla}_\alpha]\varpi_\beta =-N_{\alpha\beta} \ell^\mu \varpi_\mu
\eea
for any 1-form $\w{\varpi}\in \T^*(\Hor)$, where
\bea
\label{e:NE:N}
N_{\alpha\beta} = \hat{\nabla}_{(\alpha}\omega_{\beta)} + \omega_{\alpha} \omega_{\beta}
+\frac{1}{2}\left(R_{\alpha\beta} - {}^2\!R_{\alpha\beta}\right) \ .
\eea
Together with the evolution for $\w{q}$ in Eq.  (\ref{e:NE:LieS_q_0}), these evolution
equations constitute the geometrical identities involving only 
$\w{q}, \w{\hat\nabla}$ and $\w{R}$, i.e. they provide the NEH constraints.

Following the procedure in the isolated horizon literature
\cite{AshteBL02,LewanP04}, we introduce the following tensor on $\Hor$
\bea
\label{e:NE:S}
\w{S} := \Phi^* \w{\nabla} \uk \ ,
\eea
where $\uk$ is associated with a foliation compatible with $\el$.
According to Eq. (\ref{e:IN:k_dt_H}),  $\Phi^*\uk$ is exact on 
$\Hor$ implying that $\w{S}$ is symmetric.
In addition,  $\vec{\w{q}}^*\w{S}= \w{\Xi}$ 
and $\el \w{\cdot S}= \w{\omega}$ (see
Eqs. (\ref{e:KI:Xi_qstar_gradk}) and (\ref{e:KI:grad_uk}), respectively). 
From the discussion in Secs. \ref{s:NE:NEHconstraints} and
\ref{s:NE:NEHfreedata}, it follows that the
evolution of $\w{\hat{\nabla}}$ is given by that of the pair
$(\w{q}, \w{S})$. Making $\w{\varpi}$ equal to $\uk$ in  
Eq. (\ref{e:NE:evol_deltaH}), one obtains
\bea
\LieHc{\el} S_{\alpha\beta} = N_{\alpha\beta} \ ,
\eea
thus providing the evolution equation for $\w{S}$.
The NEH-constraints are now
\bea
\LieS{\el} \w{q} &=& 0 \label{e:NE:constraints_H} \\
\LieHc{\el} S_{\alpha\beta}&=&  \hat{\nabla}_{(\alpha}\omega_{\beta)} + 
\omega_{\alpha} \omega_{\beta}
+\frac{1}{2}\left(R_{\alpha\beta} - {}^2\!R_{\alpha\beta}\right)
    \label{e:NE:constraints_H2}
\eea
Contracting the second one with $\el$, Eq. (\ref{e:NE:evol_omega}) for
the evolution of $\w{\omega}$ is recovered, whereas the projection onto a 
slice ${\mathcal S}_t$ leads to the evolution equation
(\ref{e:NE:Lie_Xi}) for $\w{\Xi}$.

%% file: isolhor.tex
%
%
\section{Isolated horizons I: weakly isolated horizons}
\label{s:IH}

\subsection{Introduction}

The NEH notion introduced in the previous section,
represents a  first
step toward the quasi-local characterization of a 
black hole horizon in equilibrium.
As we have seen, this is achieved essentially 
by imposing a condition on the degenerate metric $\w{q}$,
namely to be time-independent [Eq.~(\ref{e:NE:LieS_q_0})].

This minimal geometrical condition captures some fundamental features
of a black hole in quasi-equilibrium. On the one hand, it is
sufficiently flexible so as to accommodate a variety of interesting 
physical scenarios. But on the other hand,
such a structure is not tight enough to determine
some {\it geometrical} and {\it physical} properties of a black hole horizon.
From the point of view of the geometry of $\Hor$ as
a hypersurface in $\M$, the NEH notion by itself provides
a limited set of tools to extract information about the spacetime 
containing the black hole. For instance, it does not pick up 
any particular normalization of the null normal nor suggest any 
concrete foliation of $\Hor$, something that was accomplished
in the previous sections by using the additional structure $(\Sigma_t)$.
Since in the spacetime construction as a Cauchy problem, the 3+1 foliation
is actually dynamically determined, it would be very useful to dispose of
a slicing $(\Sp_t)$ motivated from an intrinsic analysis of $\Hor$
and that one can employ as an inner boundary condition for $(\Sigma_t)$. 
Regarding the determination 
of physical black hole parameters, a point of evident astrophysical 
interest, a NEH does not determine any prescription for the mass, the angular 
momentum or, more generally, for the black hole multipole moments.

In order to address these issues, i.e. the discussion of the horizon
properties from a point of view intrinsic to $\Hor$, 
we need a finer characterization of 
the notion of quasi-equilibrium. Following Ashtekar et al. \cite{AshteBL02},
this demands the introduction of additional structures on $\Hor$.
After imposing the degenerate metric
$\w{q}$ to be time-independent, a natural way to proceed in order 
to further constrain the horizon geometry consists in extending this condition 
to the rest of the geometrical objects on $\Hor$, in particular to the
induced connection $\w{\hat\nabla}$. 
This strategy directly leads to the introduction of 
a hierarchy of quasi-equilibrium structures on the horizon, which
turns out to be very useful for keeping control of 
the physical and geometrical hypotheses actually assumed.
A particularly clear synthesis of the resulting formalism can be found in 
Refs.~\cite{LewanP04,KLP04}.

The pair of fields $(\w{q},\w{\hat{\nabla}})$, or equivalently  
$(\w{q}, \w{\Omega}, \kappa$, $\w{\Xi})$ (see Sec. \ref{s:NE:NEHinitdata}), 
characterizes the geometry of a NEH (cf. Remark~\ref{r:NE:NEH_geometry}). 
From a 3+1 perspective,
and following the strategy previously outlined,  a natural manner of 
obtaining different degrees of quasi-equilibrium for the horizon 
would consist in imposing the time independence of different combinations 
of the fields $(\w{q}, \w{\Omega}, \kappa$, $\w{\Xi})$.
However, the resulting notions of horizon quasi-equilibrium would be
slicing-dependent.
Even though in Sec. \ref{s:IH:IHhierarchy} we will revisit this 
idea, we rather proceed here by constructing the new structures on $\Hor$ on
the grounds of fields intrinsic to $\Hor$. Once the time independence
of $\w{q}$ has been used to define a NEH,  this means imposing time
independence on the full
connection $\w{\hat{\nabla}}$ or, in a 
intermediate step, on its component  $\w{\omega}$.
This defines the three levels in the (intrinsic)
isolated horizon hierarchy.

We separate the study of the isolated horizons in two sections.
In the present one, we focus on the intermediate level resulting from
the introduction, in a consistent way, of a time-independent rotation
1-form $\w{\omega}$.
This second level in the isolated horizon hierarchy leads to
the notion of {\it weakly isolated horizon}  (WIH).
After introducing in Sec. \ref{s:IH:WIHdefinition}
its definition and straightforward consequences, Secs. \ref{s:IH:WIHinitdata},
\ref{s:IH:prefWIH} and \ref{s:IH:goodcuts} are devoted to 
different implications on the null geometry of $\Hor$.
Finally Sec. \ref{s:IH:PhysParam} briefly presents how
a WIH structure permits to determine 
the physical parameters associated with the black hole horizon.
The stronger level in the isolated horizon hierarchy, which results
from a full time-independent $\w{\hat{\nabla}}$ , will be discussed in 
Sec. \ref{s:IHII}, where we will also comment on other developments
naturally related to the isolated horizon structures.

\subsection{Basic properties of weakly isolated horizons}
\label{s:IH:WIHdefinition}

\subsubsection{Definition}

As already noticed, the NEH notion is independent of the rescaling of the 
null normal $\el$. 
On the contrary, imposing the constancy of the rotation 1-form $\w{\omega}$ 
[the part of $\w{\hat\nabla}$ along $\el$, cf. Eq.~(\ref{e:NE:elem_delta})]
under the flow of the null normal, does depend on the actual choice 
for $\el$. This is a consequence of the transformation
of $\w{\omega}$ under a rescaling of $\el$ (cf. Table~\ref{t:KI:scaling}):
\be
\label{e:IH:transfomega}
\w{\omega} \stackrel{\el\to \alpha\el}
{\longrightarrow} \w{\omega} + \w{\Pi}^* \dd \ln\alpha  \ .
\ee
If we consider a null normal $\el$ such that 
$\LieH{\el}\w{\omega}=0$  then, after a rescaling of $\el$
by some non-constant $\alpha$, the new rotation 1-form will not be 
time-independent in general. 
In order to make sense of condition $\LieH{\el}\w{\omega}=0$, 
we must restrict the set of the null normals
for which it actually applies. From Eq.~(\ref{e:IH:transfomega})
we conclude that if $\w{\omega}$ is invariant under a given null
normal $\el$, then it is also invariant under any  constant rescaling
of it (this could be relaxed to {\it time-independent} functions).
This is formalized by the following definitions \cite{AshteFK00,AshteBL02}:
\begin{itemize}
\item [i)] Two null normals $\el$ and $\el'$ are said to be related to each other 
if and only if $\el' = c \,\el$ with $c$ a positive constant. This defines an 
equivalence relation whose equivalence classes are denoted by $[\el]$.
\item [ii)]  A {\em weakly isolated horizon (WIH)} $(\Hor,[\el])$ is a 
NEH $\Hor$ endowed with an equivalence class $[\el]$ of null normals such that 
\bea 
 \encadre{\LieH{\el} \w{\omega} = 0} . \label{e:IH:Lw} 
\eea
\end{itemize}

We comment on some consequences of this definition.

\noindent{\it 1. Extremal and non-extremal WIH}.
As a consequence of the transformation rule 
for $\kappa$ under a rescaling of $\el$ (cf. Table~\ref{t:KI:scaling}), 
two null normals $\el$ and $\el'$ belonging to the same WIH class
have non-affinity coefficients $\kappa_{(\el)}$ and $\kappa_{(\el')}$ 
related by\footnote{In this section, since we will deal with different 
null normals at the same time, we make explicit the dependence of the
non-affinity coefficient on $\el$.}
\be \label{e:IH:var_kappa}
    \kappa_{(\el')} = c \, \kappa_{(\el)} , 
\ee
where $c$ is the constant 
linking the two null vectors: $\el'=c\, \el$. This implies that there is no canonical value
of the non-affinity coefficient $\kappa$ on a given WIH. 
This reflects the absence of canonical 
representative in the class $[\el]$. 
We have already presented a solution to this point, relying on the
3+1 spacelike slicing in Sec.~\ref{s:IN:normal_l}. We will present another
solution, intrinsic to $\Hor$, in Sec.~\ref{s:IH:PhysParam}.
The transformation law (\ref{e:IH:var_kappa}) means 
that, on a given NEH, the WIH structures 
are naturally divided in two types: 
those with vanishing $\kappa_{(\el)}$,
that will be referred as {\it extremal} WIHs, and those with
$\kappa_{(\el)}\neq 0$, {\it non-extremal} WIHs. 
This terminology arises from the Kerr spacetime, where one can always
choose the null normal $\el$ to let it coincide with a Killing vector field. 
Then  $\kappa_{(\el)}$ is nothing but the surface gravity of the black hole
[cf. Eq.~(\ref{e:NH:kappa_EF}) in Example~\ref{ex:NH:EF} for the
non-rotating case and Eq.~(\ref{e:KE:kappa}) in Appendix~\ref{s:KE}
for the general case].
Extreme Kerr black holes are those for which the surface gravity
vanishes. As shown by Eq.~(\ref{e:KE:kappa}), 
this corresponds to the angular momentum parameter
$a/m=1$. They are not
expected to exist in the Universe, because it is not possible
to make a black hole rotate faster than
$a/m=0.998$ by standard astrophysical processes (i.e. infall 
of matter from an accretion disk \cite{Thorn74}). 
In this article, we will focus on the non-extremal case,
and refer the reader to \cite{AshteBL02,LewanP04} for 
the general case.

\noindent {\it 2. Constancy of the surface gravity on $\mathcal{H}$}.
From the NEH Eq. (\ref{e:NE:Hdomega})
\be
\el\cdot {}^\Hor\dd\w{\omega} = 0 ,
\ee
and using the Cartan identity (\ref{e:IN:Cartan_id}) on $\Hor$:
\bea 
\label{e:IH:WIHzerolaw}
0= \LieH{\el} \w{\omega}=
\el\cdot {}^\Hor\dd\w{\omega}+{}^\Hor\dd\langle \el,\w{\omega}\rangle
={}^\Hor\dd \kappa_{(\el)} \ \Leftrightarrow 
\encadre{ \w{\hat{\nabla}}\kappa_{(\el)}= 0}  . 
\eea
Therefore, the non-affinity coefficient of a given $\el$
is a constant on $\Hor$. This property, that
is referred to as the {\it zeroth law of black hole mechanics}, 
characterizes $[\el]$ as associated with a WIH structure.
It will be discussed further in Remark~\ref{r:IH:thermo}
below.

\noindent {\it 3. Any NEH admits a WIH structure}.
We have just seen that the class $[\el]$ associated with $\el$ is a
WIH structure if and only if $\kappa_{(\el)}$ is a constant on $\Hor$.
In Secs. \ref{s:NE:NEHinitdata} and \ref{s:NE:NEHfreedata}
we established that a given NEH geometry is determined by the 
set of fields $(\w{q}, \w{\Omega}, \kappa_{(\el)}$, $\w{\Xi})$,
where $\kappa_{(\el)}$ is an arbitrary function, and {\it also} by 
any other set obtained from this one by applying the transformations
in Table \ref{t:KI:scaling}. This was called the {\it active} aspect of the
gauge freedom in the NEH-free data, and it simply corresponds to a rescaling
by $\alpha$ of the null normal $\el$. Therefore if, according to 
Eq. (\ref{e:NH:scale_kappa}), we choose $\alpha$ satisfying 
\bea
\label{e:IH:eq_alpha}
\kappa'= \w{\nabla}_{\el} \alpha + \alpha \kappa_{(\el)}
\eea
with $\kappa'$ constant on $\Hor$, then $[\el']$ given by
$\el'=\alpha \el$ constitutes a WIH class 
(in particular, making $\kappa'=0$ shows that any NEH admits an extremal horizon;
Sec. III.A of Ref. \cite{AshteBL02} firstly shows the existence
of an extremal WIH for any NEH, and then constructs from it a family
of non-extremal ones).
As a consequence, the addition of a WIH {\it does not} represent
an actual constraint on the null geometry of $\Hor$. It rather 
distinguishes certain classes of null normals.

\noindent {\it 4. Infinite freedom of the WIH structure}. 
Not only it is always possible to choose a WIH structure on a NEH, but
there exists actually an infinite number of non-equivalent WIHs.
Reasoning for the non-extremal case,
if $\el$ is such that $\kappa_{(\el)}$ is a non-vanishing constant on 
$\Hor$ and  $t$ is a coordinate on $\Hor$ compatible with $\el$,
i.e. $\LieH{\el} t =1$, then the class $[\el']$ associated
with the vector $\el'$ defined by
\bea
\label{e:IH:fam_WIH}
\el'&=&
\left( 1 + B e^{-\kappa_{(\el)}t}\right)\el \\
&&\LieH{\el}B=0 \ ,
\eea
is distinct from the class $[\el]$ (for $\alpha:=1+B \e^{-\kappa_{(\el)}t}$
is not constant) and also defines a WIH.
This follows from $\kappa_{(\el')}=\kappa_{(\el)}=\mathrm{const}$.
Using the slicing $(\Sp_t)$ induced on $\Hor$ from the $3+1$ decomposition,
the functions $B$ in Eq. (\ref{e:IH:fam_WIH}) have actually support on the
sections $\Sp_t$ and parametrize the different WIH structures.
In an analogous manner for the extremal case, $\kappa_{(\el)}=0$, 
the non-equivalent rescaled null normal 
$\el'= A \el$ with $A$ a non-constant function on $\mathcal{S}_t$,
is also associated with a non-equivalent extremal WIH.

\subsubsection{Link with the 3+1 slicing}

Since a WIH structure does not further constrain the geometry of $\Hor$, 
from a quasi-equilibrium point of view a WIH is not 
{\it more isolated} than a NEH. However, this notion presents
a remarkable richness as a structural tool. In this sense its interest is
two-fold, both from a geometrical and physical point of view.
In addition, this concept provides a natural framework to discuss
the interplay between the horizon $\Hor$ and the spatial 
3+1 foliation $(\Sigma_t)$, something specially
relevant in our approach.
In this last sense, the issue of the compatibility between the structures 
intrinsically defined on $\Hor$ and the additional one provided by 
the 3+1 slicing of the spacetime, is naturally posed.

Given a WIH $({\mathcal H},[\el])$,  a 3+1 slicing $(\Sigma_t)$
is called {\em WIH-compatible} 
if there exists a representative $\el$ in $[\el]$ such that 
the associated level function $t$ evaluated on the
horizon ${\mathcal H}$ constitutes a natural coordinate for $\el$, i.e. 
${\mathcal L}_{\el}t=1$ \cite{JaramGM04}. Note that the representative
$\el$ is then nothing but the null normal associated with the slicing
$(\Sp_t)$ of $\Hor$ by the normalization (\ref{e:NH:l_norm2}). 

Given independently a NEH $\Hor$ and a 3+1 slicing of $\M$, 
there is no guarantee that there exists a WIH structure on the NEH
such that $(\Sigma_t)$ is WIH-compatible.
If such a WIH exists, $\el$ is {\it tied} to the $t$ 
function and therefore to the slicing. If not, no
relation exists between that WIH and $(\Sigma_t)$. 
Therefore, even though the choice of a specific WIH structure on our NEH 
does not affect the intrinsic geometry of the
horizon, demanding the 3+1-slicing to be WIH-compatible 
represents an actual restriction 
on the 3+1 {\it description} of spacetime (see Sec. \ref{s:BC:kappa_const}).
And this fact is crucial in our approach: some of the possible spacetimes
slicings are directly ruled out. 

We have noticed in Remark~\ref{r:NE:omega_intrinsic_NEH} that, on a NEH,
the \hajicek\ 1-form $\w{\Omega}$ (more concretely its divergence)
is an object directly depending upon
the 3+1 slicing, whereas the rotation 1-form $\w{\omega}$
depends only upon the normalization of $\el$. Let us then
investigate the consequences of the WIH condition on $\w{\Omega}$.
From Eq.~(\ref{e:NE:P_LieH_om_Om}), we have
\be
    \LieH{\el}\wo = \LieH{\el} \w{\Omega} - 
    \left( \LieH{\el} \kappa_{(\el)} \right)
    \, \Phi^* \uk , 
\ee
where we have considered both $\w{\omega}$ and $\w{\Omega}$ as a 1-forms 
in $\T^*(\Hor)$
(identifying them with their pull-back $\Phi^*\w{\omega}$
and $\Phi^*\w{\Omega}$). In view of the above relation, we deduce
from Eqs.~(\ref{e:IH:Lw}) and (\ref{e:IH:WIHzerolaw}) that
\be \label{e:IH:characWIH_Omega_kappa}
    \encadre{ (\mbox{$(\Hor,[\el])$ is a WIH})
    \iff \left\{ \begin{array}{l}
      \LieH{\el} \w{\Omega} = 0 \\
       \LieH{\el} \kappa_{(\el)} = 0
       \end{array}
       \right. }.
\ee
Note that $\LieH{\el} \kappa_{(\el)} = 0$ is listed here, along with 
$\LieH{\el} \w{\Omega} = 0$, as a sufficient condition to have a WIH,
but once the WIH structure holds, one has actually the much 
stronger property of constancy of $\kappa_{(\el)}$ on all $\Hor$
[zeroth law, Eq.~(\ref{e:IH:WIHzerolaw})].


\subsubsection{WIH-symmetries}

The discussion of the physical parameters of the black hole horizon
in Sec. \ref{s:IH:PhysParam} demands the introduction of the notion
of a symmetry related to a WIH horizon. 
We present a brief account of it, with emphasis in the 
non-extremal case and refer 
the reader to Refs.~\cite{AshteBL01,KLP04} for details and extensions.

A symmetry of a WIH is a diffeomorphism of $\Hor$
preserving the relevant structures of the WIH. Infinitesimally this is captured 
as follows:
a vector field $\w{W}$ tangent to $\Hor$ is said to be an
infinitesimal {\it WIH-symmetry} of $(\Hor,[\el])$, if it preserves
the equivalence class of null normals, the metric $\w{q}$
and the 1-form $\w{\omega}$, namely 
\bea 
\label{e:IH:WIHsymmetry} 
\LieH{W}
\el= \mathrm{const} \cdot \el , \ \ \ \LieH{W}
\w{q} = 0 \ \ \ \mbox{and} \ \ \ \LieH{W}
\w{\omega} = 0 \, . 
\eea

In the considered non-extremal case, the general form of such a WIH symmetry is given by
(see Sec. III in \cite{AshteBL01})
\bea 
\w{W} = c_{\w{W}} \el + b_{\w{W}} \w{S}
\, ,\label{e:IH:WIHsymmetry_form} 
\eea 
where $c_{\w{W}}$ and $b_{\w{W}}$ are
constant on $\Hor$ and the vector field $\w{S}$, satisfying  $\el\cdot \w{S}=0$,
is an isometry on each section $(\Sp_t,\w{q})$.
From this general form of an infinitesimal symmetry, 
and according to the number of independent
generators in the associated Lie algebra of symmetries,
one can distinguish different {\it universality classes}
of WIH-symmetries. Since 
$\el\in [\el]$ is an infinitesimal symmetry by construction, the Lie algebra 
(and therefore the Lie group) of WIH-symmetries  is always at least 
one-dimensional:
\begin{itemize}
\item[a)] Class I. The symmetry Lie algebra is generated by $\el$ together
with the infinitesimal rotations acting on the 2-sphere 
${\mathcal S}_t$. The resulting group is the direct product of 
$SO(3)$ and the translations in the $\el$ direction.
This case corresponds to the horizon of a non-rotating black hole.

\item[b)] Class II. The symmetry group is now the
direct product of the translations along  $\el$ and an axial
$SO(2)$ symmetry on ${\mathcal S}_t$. It corresponds to an axisymmetric 
horizon and represents the most interesting physical case, since it 
corresponds to a black hole with well-defined non-vanishing angular momentum 
(see Sec. \ref{s:IH:PhysParam}).

\item[c)] Class III. The symmetry group is one-dimensional (translations along $\el$).
It corresponds to the general distorted case.
\end{itemize}
Note that I is a special case of II, and the latter is a special case of III.

\subsection{Initial (free) data of a WIH}
\label{s:IH:WIHinitdata}

As commented in the point {\it 3.} after the WIH definition, the geometry of 
a WIH  as a null hypersurface is that of a NEH. 
In particular, the free data of a WIH are {\it essentially}
those presented in Sec. \ref{s:NE:NEHinitdata} for a NEH.
The only difference is the choice of a null normal $\el$ such that the 
zeroth law (\ref{e:IH:WIHzerolaw}) is satisfied and, consequently,
$\kappa_{(\el)}$ is constant on $\Hor$, $\kappa_o$:
\bea
\label{e:NE:freedataWIH}
\encadre{\hbox{WIH-free initial data:} \ \ \ (\w{q}|_{{\mathcal S}_t}, 
\w{\Omega}|_{{\mathcal S}_t}, \w{\Xi}|_{{\mathcal S}_t}, \kappa_o)
} \ .
\eea
We note that, since in a WIH the null normal $[\el]$ is fixed up 
to a constant, the gauge freedom in the WIH-free data
concerns mainly  what we called the {\it passive} aspect in Sec.
\ref{s:NE:NEHfreedata} . The {\it active} one 
reduces to constant rescalings of $\kappa_{(\el)}$ and $\w{\Xi}$.
Once these free data are fixed on ${\mathcal S}_t$, 
the reconstruction of the WIH
on $\Hor$ proceeds as in Sec. \ref{s:NE:NEHfreedata}.
The only subtlety now enters in the third step represented in Fig. 
\ref{f:NH:reconstfreedata},
when the fields on the slice $\Sp_t$ are transported to the next slice. 
In the construction of the WIH  not only the metric $\w{q}$ 
is Lie dragged by $\el$, but also the 
\hajicek\ form $\w{\Omega}$ and the non-affinity coefficient $\kappa_o$,
as shown by Eq.~(\ref{e:IH:characWIH_Omega_kappa}). 
Now, the only field evolving in time is $\w{\Xi}$.

In view of Eq.~(\ref{e:NE:Lie_Xi}) and the time-independence of $\w{q}$
(hence of $\w{\DS}$ and ${}^2\!\w{R}$), $\w{\Omega}$ and $\kappa$,
the time dependence of 
$\w{\Xi}$ can be explicitly integrated if 
we assume that the projection of the 4-dimensional Ricci tensor  
(or the matter
stress-energy tensor via the Einstein equation) is time-independent,
i.e. 
\bea
\label{e:IH:Ricci_time_ind}
\w{\Pi}^*  \Lie{\el} \w{R}=0 \ .
\eea
In fact, this condition is actually well-defined on a NEH, i.e.
it does not depend on the choice of the null normal
$\el$. This follows from property (\ref{e:NE:Ricci_l=0}).
Condition (\ref{e:IH:Ricci_time_ind}) is rather mild 
and is obviously satisfied in
a vacuum spacetime. 
If we deal with a WIH built on a NEH that fulfills (\ref{e:IH:Ricci_time_ind}),
then in the evolution of $\w{\Xi}$ dictated by
Eq. (\ref{e:NE:Lie_Xi}) the only terms which depend on time 
are $\w{\Xi}$ and $\LieH{\el}\w{\Xi}$. In this situation, and 
assuming a non-extremal WIH ($\kappa_{(\el)}\not=0$), 
this equation can be straightforwardly integrated \cite{AshteBL02},
resulting in 
\bea
\label{e:IH:Xi0}
 \w{\Xi} &= &e^{-\kappa_{(\el)} t} \, \w{\Xi}^0 +\frac{1}{\kappa_{(\el)}} \left[\frac{1}{2} 
\Kil{\w{\DS}}{\w{\Omega}}
   + \w{\Omega}\otimes \w{\Omega}
   - \frac{1}{2}  {}^2\!\w{R}
   + 4\pi \left( \vec{\w{q}}^* \w{T} - \frac{T}{2} \w{q}\right)\right] ,
    \nonumber \\
\eea
where $ \w{\Xi}^0$ is the integration constant, a 
time-independent symmetric tensor, and $t$ is a coordinate on $\Hor$
compatible  with $\el$ via Eq. (\ref{e:NH:l_norm2}), i.e. $\Lie{\el}t=1$.


\subsection{Preferred WIH class $[\el]$}
\label{s:IH:prefWIH}
We have seen that a NEH admits an infinite number of WIH structures
which, in the non-extremal case and  according to (\ref{e:IH:fam_WIH}),
are parametrized by functions $B$ defined on $\Sp_t$.
The question about the existence of a natural 
choice among them is naturally posed. 

In case we dispose of an {\it a priori} slicing $(\Sigma_t)$ of $\M$, 
a slicing $(\Sp_t)$ of the horizon is determined
independently  of the geometrical structures defined on $\Hor$,
as discussed in Sec. \ref{s:IN}. 
Such a slicing fixes the null normal $\el$ via the normalization 
(\ref{e:NH:l_norm2}) [or, equivalently, the slicing fixes the 
lapse $N$ and $N$ determines $\el$ by Eq. (\ref{e:IN:el_nps})].
If the slicing is a WIH-compatible one, a particular class $[\el]$ 
is chosen. If not, such a slicing does not help in making such a choice.

More interesting is, however, the opposite situation, which occurs
whenever the 3+1 slicing $(\Sigma_t)$ is 
determined in a dynamical way. In such a context,
an intrinsic determination of a preferred WIH class 
helps in the very construction of the 3+1 slicing. 
In fact, if such a WIH class is provided together with a definite 
initial cross section $\Sp_0$, then the slicing $(\Sp_t)$ of $\Hor$
is completely determined\footnote{We understand here the slicing
$(\Sp_t)$ of $\Hor$ as the set of slices $\Sp_t$. The global constant
ambiguity in the WIH class affects the rate at which the null
generator on $\Hor$ traverses this ensemble (thus determining the associated lapse $N$ 
on $\Hor$ up to constant), but does not change the ensemble itself.}, 
and can be used as a boundary condition to 
fix $(\Sigma_t)$.

From an intrinsic point of view, fixing the class $[\el]$
reduces to choosing
the function $B$ in Eq. (\ref{e:IH:fam_WIH}).
Such a choice can be made by imposing a condition 
on a scalar definable in terms of the fields defining the WIH 
geometry. Following \cite{AshteBL02},
an appropriate scalar in this sense is provided by the trace of $\LieH{\el}\w{\Xi}$
which, on a NEH, corresponds to the Lie derivative in the $\el$ direction
of the expansion $\theta_{(\w{k})}$ associated
with the ingoing null normal $\w{k}$
[cf. Eq.~(\ref{e:KI:def_theta_k})].
Contracting equation (\ref{e:NE:Lie_Xi}) with $\w{q}$ yields
\bea
\label{e:IH:Lie_thetak}
\Lie{\el}\theta_{(\w{k})}={}^2\!D^\mu\Omega_\mu + \Omega_\mu \Omega^\mu
-\frac{1}{2} {}^2\!R + \frac{1}{2} q^{\mu\nu}R_{\mu\nu} 
- \kappa \theta_{(\w{k})} .
\eea
We start with a WIH with null normal $\el$ and
free data $(\w{q}, \w{\Omega}, \w{\Xi}, \kappa_{(\el)})$.
Firstly we notice that, if condition (\ref{e:IH:Ricci_time_ind}) holds, 
the value of $\Lie{\el}\theta_{(\w{k})}$ at $t=0$ resulting from Eq. (\ref{e:IH:Xi0}) 
satisfies (where $\Xi^0_{\mu\nu}$ are constant)
\bea
\label{e:IH:thetak0}
\left.\Lie{\el}\theta_{(\w{k})}\right|_{t=0} = - \kappa_{(\el)} \left(q^{\mu\nu}\,\Xi^0_{\mu\nu}\right) \ .
\eea
Now we search a transformation of $(\w{q}, \w{\Omega}, \w{\Xi}, \kappa_{(\el)})$
to new free data corresponding to another WIH built on the same NEH, 
such that the new $(\w{q}', \w{\Omega}', \w{\Xi}', \kappa_{(\el')})$
make $\Lie{\el'}\theta_{(\w{k'})}$ to vanish via the 
``primed'' Eq. (\ref{e:IH:Lie_thetak}). 
In the non-extremal case, this is achieved by a rescaling 
$\el'=\alpha \el$ with
\bea
\label{e:IH:alpha}
\alpha = (1 + B e^{-\kappa_{(\el)}  t}) ,
\eea
with $t$ a coordinate compatible with $\el$,
i.e. an {\it active} transformation of the free data. 
Up to some caveats we shall mention below, this permits 
to fix the function $B$ and therefore the WIH class 
(we adapt the discussion in \cite{AshteBL02,Kr02}, where it is carried out
in terms of the  tensor $N_{ab}$ introduced in Sec. 
\ref{s:NE:intrinsic_NEH-const}). 

According to Table~\ref{t:KI:scaling} and
using the coefficient $\alpha$ given by Eq.~(\ref{e:IH:alpha}), the transformations 
of the objects in the right-hand side of Eq. (\ref{e:IH:Lie_thetak})
are parametri\-zed by the functions $B$. 
We make explicit these transformations in a first order expansion
in the parameter $B$, and evaluate the transformed fields 
on the 2-surface  $\Sp_0$  (i.e. we set  $t=0$):
\bea
\label{e:NE:inf_trans_B}
\DS_\mu &\rightarrow&\DS'_\mu = \DS_\mu \\
\Omega_\mu &\rightarrow&  \Omega'_\mu 
=\Omega_\mu + \frac{\DS_\mu B}{1+B} \approx \Omega_\mu + \DS_\mu B \\
\kappa_{(\el)}  &\rightarrow& \kappa_{(\el')}=\kappa_{(\el)} \\
\theta_{(\w{k})}  &\rightarrow& \theta_{(\w{k'})}=\frac{\theta_{(\w{k})}}{1+B}\approx 
\theta_{(\w{k})} - B \theta_{(\w{k})} .
\eea 
Introducing these transformed fields in the (transformed) 
Eq. (\ref{e:IH:Lie_thetak}) and gathering together the terms 
that expand  $\Lie{\el}\theta_{(\w{k})}$ by using again the ``non-primed'' 
Eq.  (\ref{e:IH:Lie_thetak}), we obtain
\bea
\label{e:IH:fixingB} 
&& \encadre{
\left({\DS^\mu}{\DS_\mu} + 2 \Omega_\mu{\DS^\mu} + D^\mu\Omega_\mu 
+ \Omega_\mu \Omega^\mu -\frac{1}{2} \, {}^2\!R 
+ \frac{1}{2} q^{\mu\nu}R_{\mu\nu} \right) B =-\left.\Lie{\el}\theta_{(\w{k})}\right|_{t=0}  \ . 
}\nn\\
\eea
Following Ashtekar et al. \cite{AshteBL02} we denote the operator 
acting on $B$ as
$\w{M}$. Making use of Eq. (\ref{e:IH:thetak0}), we finally find
the following condition on $B$
\be
\label{e:IH:MB}
\encadre{ 
\w{M} B = \kappa_{(\el)} \left(q^{\mu\nu}\,\Xi^0_{\mu\nu}\right)
} .
\ee
A NEH is called {\it generic} if it satisfies condition 
(\ref{e:IH:Ricci_time_ind}) and
if it admits a null normal $\el$ with constant $\kappa_{(\el)}\neq 0$
such that its associated $\w{M}$ is invertible \cite{AshteBL02}. 

If a NEH is generic, Eq.~(\ref{e:IH:MB}) admits a unique solution $B$
which is inserted in the rescaling (\ref{e:IH:alpha}) of $\el$ for 
finding the null vector $\el'$ which satisfies
$\Lie{\el'}\theta_{(\w{k'})}=0$.
The main point to retain from this discussion is the fact that the
condition $\Lie{\el}\theta_{(\w{k})}=0$, together with
$\kappa_{(\el)}=\mathrm{const}$, fixes a unique WIH structure on the
NEH (cf. \cite{AshteBL02}).

\subsection{Good slicings of a non-extremal WIH}
\label{s:IH:goodcuts}

Fixing the WIH class
determines the foliation of $\Hor$ if an initial cross-section is
provided. This is particularly interesting for the construction,
from a Cauchy slice, of 
a spacetime containing a WIH. However, in 
more general problems in which no initial slice is singled out, simply 
demanding the slicing of $\Hor$ to be compatible with the chosen WIH class,
is not enough to fix $(\Sp_t)$ (this corresponds to the {\it passive}
aspect of the gauge freedom discussed in Sec. \ref{s:NE:NEHfreedata}).
It is worthwhile to consider if a particular slicing associated with
a WIH class can be chosen in a natural way.
Even though in Sec. \ref{s:BC:BCWIH} we will provide a more intuitive 
presentation of this issue in terms of the 3+1 decomposition, 
we briefly show here the intrinsic approach followed in
Refs.~\cite{AshteBL02,Kr02,LewanP04}.

The foliation $(\Sp_t)$ is fixed by providing a function $t$
on $\Hor$, that can be seen as the restriction to $\Hor$ of a 
scalar field defined in the whole spacetime $\M$, and whose inverse
images do foliate $\Hor$. 
According to Eq. (\ref{e:IN:k_dt_H}), 
fixing such a function $t$ entails a specific choice of $\uk$
[see Remark \ref{r:NH:k_foliation} for further insight on the relation
between $\w{k}$ and the foliation $(\Sp_t)$].
If we fix a WIH class, for instance
following the procedure explained in the previous section,
the rotational 1-form $\w{\omega}$ is completely determined in this
class, according to transformation rule in Table \ref{t:KI:scaling}.
Consequently, for a non-extremal horizon
$\kappa=\mathrm{const}\neq 0$, and taking into account Eq. (\ref{e:KI:Omega_omega_k}),
fixing $\uk$ (or equivalently the slicing on $\Hor$)
translates into specifying the \hajicek\  form $\w{\Omega}$. 
\bea
\encadre{
\left.
\begin{array}{c}
\hbox{Fixing } (\Sp_t)  \\
(\Hor, [\w{\el}]) \hbox{ non-extremal WIH}
\end{array}\right\}
\Leftrightarrow
\hbox{fixing \hajicek\   1-form }\w{\Omega}
} .
\eea
This will be the starting point of the discussion in Sec. \ref{s:BC:fixslic}.
Here we briefly comment on an intrinsic procedure to fix $\w{\Omega}$.
Firstly we note that, on a WIH with $\uk$ satisfying 
$\LieH{\el}\uk=0$ (as it is the case), we have $\Lie{\el}\w{\Omega}=0$.
In addition $\langle\w{\Omega},\el\rangle=0$, so $\w{\Omega}$ projects to the 
sphere obtained as the quotient of $\Hor$ by the trajectories
of the vector field $\el$ [cf. Remark \ref{r:NH:quotient} for a brief comment on 
this construction, but in terms of vectors in $\T_p(\Hor)$].

In general, a 1-form $\w{\Omega}$ on a sphere 
$S^2$ can always be decomposed in
\bea
\label{e:IH:Omegadecomp}
\w{\Omega} = \w{\Omega}^{\mathrm{div-free}} +
\w{\Omega}^{\mathrm{exact}} \ ,
\eea
where $\w{\DS\cdot}\w{\Omega}^{\mathrm{div-free}}=0$ and 
$\w{\Omega}^{\mathrm{exact}} = \w{\DS}f$ for some function $f$
on $S^2$. This is a specific case of the general expression,
known as {\em Hodge decomposition}  
for p-forms defined on a compact manifold provided with 
a non-degenerate metric (see for instance Refs.~\cite{ChoquDD77} or 
\cite{Nakah03}).
The divergence-free part of the \hajicek\ 1-form is determined by Eq.
(\ref{e:NE:Hdomega}) that, together with $\kappa_{(\el)}=\mathrm{const}$ and
Eq. (\ref{e:KI:dk}), implies
\bea
\label{e:IH:dHacijek}
\w{d} \w{\Omega}^{\mathrm{div-free}} = 2 \; \mathrm{Im}\,\Psi_2 {}^2 \w{\epsilon}
\eea
Again in the context of the Hodge decomposition, 
the divergence-free part can always be written as
\bea
\label{e:IH:div_free}
\w{\Omega}^{\mathrm{div-free}}=\vec{\w{\DS}} h \w{\cdot} {}^2 \w{\epsilon}
\eea
for a certain function $h$. In terms of $h$, Eq. (\ref{e:IH:dHacijek}) 
results in the Laplace equation on the sphere
\bea
{}^2\!\Delta h = 2 \; \mathrm{Im}\,\Psi_2  \ ,
\eea
which completely fixes $\w{\Omega}^{\mathrm{div-free}}$.
In order to consider the exact part of $\w{\Omega}$, we take the divergence 
of Eq.~(\ref{e:IH:Omegadecomp}), resulting in
\bea
\label{e:IH:divOmega}
{}^2\!\Delta f = \w{\DS\cdot} \w{\Omega}^{\mathrm{exact}} 
= \w{\DS}\w{\cdot}\w{\Omega}\ .
\eea
Whereas the divergence-free part of $\w{\Omega}$ is fixed by the WIH geometry,
its divergence is not constrained by the null geometry.
Therefore, in order to fix  
the exact part we must make a choice for the value 
$\w{\DS}\w{\cdot}\w{\Omega}$. 
Therefore $f$ encodes the {\it passive} gauge freedom in the
determination of the foliation $(\Sp_t)$.

A {\it natural} condition \cite{AshteBL02} consists simply in choosing
$f=0$, i.e.
$\w{\DS\cdot}\w{\Omega}=0$, which implies the vanishing of $\w{\Omega}^{\mathrm{exact}}$. 
However, such a choice does not lead to the usual foliations in the case of 
a rotating Kerr metric \cite{Kr02,AshteK05}. A choice that permits to
recover the Kerr-Schild slicing of the
horizon (cf. Appendix~D), and which is motivated by the extremal 
Kerr black hole,
is given by the {\it Pawlowski gauge} \cite{Kr02,AshteK05} :
\bea
\w{\DS\cdot} \w{\Omega} = -\frac{1}{3} {}^2\!\Delta  \mathrm{ln} |\Psi_2| \ ,
\eea
where the complex Weyl scalar $\Psi_2$ has been defined by 
Eq.~(\ref{e:NH:Weyl_scalars}). 

\begin{rem}
\label{r:IH:IDplace}
As we saw in Sec. \ref{s:NE:NEHinitdata}, the discussion of the
free data associated with the null geometry involves a
slicing of $\Hor$. Since in this article we are working with the 
additional structure provided by the slicing $(\Sp_t)$, 
it was appropriate for us to carry out such analysis in the context
of a NEH. However, if no extra structure is added to that
intrinsically defined on $\Hor$, a WIH is needed in order to define a
slicing of $\Hor$ (in the non-extremal case) as shown above.
In such an approach, this section \ref{s:IH} on WIH would probably offer a more
natural setting for the general discussion on the free data, 
as done in Ref.~\cite{AshteBL02}.
\end{rem}


\subsection{Physical parameters of the horizon}
\label{s:IH:PhysParam}

Having discussed the applications of the WIH structure for analyzing  
the geometry of $\Hor$ as a hypersurface in $\M$, let us turn our attention
to the determination of the physical parameters associated with the black hole
horizon\footnote{As commented
    in the Introduction, in this review we do not discuss the
    electromagnetic properties of a black hole. In particular, in this
    section we restrain ourselves to solutions without matter at the horizon. For a
    more general study (incorporating electromagnetic and Yang-Mills fields) see
    Refs.  \cite{AshteBF99,AshteBL01}.}.

The introduction of a WIH structure on $\Hor$ permits to associate a 
quasi-local notion of mass and angular momentum with the black hole, 
independently of its environment. Such quasi-local notions are of fundamental
astrophysical relevance for the study of black holes. Regarding the mass, 
the ADM mass of an asymptotically
flat spacetime (see the textbooks~\cite{Wald84} or \cite{Poi04}  
for a brief presentation of the different notions of mass in general
relativity) accounts for the total mass included in a
spacelike slice $\Sigma_t$. 
However, in a multi-component system it does not 
allow to determine which part is properly associated with the black hole and 
which part corresponds to the binding energy or the gravitational radiation.

\begin{rem} 
\label{rem:IH:qlocal_mass}
There exist in the literature other quasi-local approaches to prescribe 
the physical parameters associated with spatially bounded regions. 
See in this sense the review \cite{Szaba04}. Let us highlight 
Brown \& York work \cite{BrownY93}, where 
a review of the existing literature can also be found, and
Refs.~\cite{HH96,LY03} for recent developments in the notion of 
quasi-local mass. Here we simply present the approach followed 
in Refs.~\cite{AshteFK00,AshteBL01}, developed in the framework 
of quasi-equilibrium black hole horizons modeled by null 
surfaces. For an extension of the discussion to the
dynamical regime, see Ref.~\cite{AshteK03}.
\end{rem}

The strategy to determine the quasi-local parameters is also 
geometrical, but relying on techniques which are rather
different from the ones introduced in the present article, where we 
have focused on the characterization
of the geometry of $\Hor$ as a hypersurface embedded in spacetime.
The setting for the discussion of the physical parameters is 
provided by the so-called {\it Hamiltonian} or {\it symplectic} techniques
(see for instance Refs.~\cite{AM78,Ar89,GS84} for general presentations). 
As in standard classical mechanics, physical parameters are characterized
as quantities conserved under certain transformations, which in the
present case are
related to {\it symmetries} of the horizon (see Sec. \ref{s:IH:WIHdefinition}).

More specifically, one considers the {\it phase space} $\Gamma$ of solutions 
to the Einstein equation containing a WIH $(\Hor,[\el])$ in its interior. That is,
each {\it point} of $\Gamma$ is a Lorentzian manifold $(\M,\w{g})$ endowed  
with a WIH $(\Hor, [\el])$. 
Diffeomorphisms of $\M$ preserving $\Hor$ and such that their restriction to
$\Hor$ implement a WIH-symmetry, 
induce {\it canonical} transformations on $\Gamma$ (when some
additional non-trivial conditions are fulfilled; see
Appendix\ref{s:SP}). The functions on $\Gamma$ generating these
canonical transformations are identified with the physical quantities.
A systematic discussion of these tools lays beyond
the scope of this article. A brief account of them, 
organized in terms of (relevant) examples rather than 
a formal presentation, can be found in Appendix~\ref{s:SP}.

\subsubsection{ Angular momentum}
\label{s:IH:ang_mom}

Following Ashtekar et al. \cite{AshteBL02}, we restrict ourselves to those
horizons $\Hor$ which admit a WIH of class II (see Sec.
\ref{s:IH:WIHdefinition}). 
Therefore, there exists an axial vector field $\w{\phi}$ on $\Hor$
which is a $SO(2)$ isometry of the induced metric $\w{q}$
and is normalized in order to have a $2 \pi$ affine length.
Noting that this vector field $\w{\phi}$ presents in fact 
the standard form (\ref{e:IH:WIHsymmetry_form}) (with $c_{\w{\phi}}=0,
b_{\w{\phi}}=1$), a conserved quantity in $\Gamma$  associated with the 
horizon $\Hor$ can be defined (see Appendix \ref{s:SP}). 
This quantity, denoted as $J_{\mathcal H}$ and identified 
with the {\em angular momentum of the horizon}, 
has the explicit form\footnote{In
the expressions for the physical parameters we reintroduced explicitly the
Newton constant $G$.}
\bea
\label{e:IH:angmom}
&& \encadre{
J_{\mathcal H} = 
-\frac{1}{8\pi G}\int_{\Sp_t} \langle\w{\omega}, \w{\phi}\rangle\  
        {}^2\w{\epsilon} =
-\frac{1}{8\pi G}\int_{\Sp_t} \langle\w{\Omega}, \w{\phi}\rangle\  
        {}^2\w{\epsilon} =
-\frac{1}{4\pi G}\int_{\Sp_t} f \mathrm{Im}\Psi_2 {}^2\w{\epsilon} 
} \ , \nonumber \\
\eea
where the second equality holds thanks to $\w{\phi}\in\T_p(\Sp_t)$, and
in the third equality we have used the fact that,
since $\w{\phi}$ is a divergence-free vector, 
there exists a function $f$ such that 
$\w{\phi}= {}^2\vec{\w{D}}f \w{\cdot} {}^2\w{\epsilon} $ [analogue to
Eq. (\ref{e:IH:div_free})].  Using then Eq. (\ref{e:NE:Hdomega}) and an 
integration by parts leads to the third form for $J_{\Hor}$.
We note that this expression justifies the {\it rotation 1-form} terminology
introduced in Sec. \ref{s:NE:Wein_rotation_1form} for $\w{\omega}$.

If the vector $\w{\phi}$ can be extended to a  stationary and axially
symmetric neighborhood of $\Hor$ in $\M$, representing 
the corresponding 
rotational Killing symmetry, then expression (\ref{e:IH:angmom}) 
can be shown to be equivalent to the {\it Komar}
angular momentum \cite{Komar59,AshteBL01} (see also expression 
(\ref{e:TP:angmom}) and, for instance, \cite{Poi04}). 
We point out that the integral (\ref{e:IH:angmom}) is well defined 
even if $\w{\phi}$ is not a WIH-symmetry
(in fact, the divergence-free property is enough to guarantee the independence
of (\ref{e:IH:angmom}) from the cross-section $\Sp_t$ of $\Hor$ 
\cite{AshteEPB04,AshteK05}). However, in the absence of a symmetry,
it is not so clear how to associate  physically this value with a physical 
parameter.

\subsubsection{Mass}
\label{s:IH:mass}

The definition of $\Hor$'s mass is related to the choice of an
evolution vector $\w{t}$. 
In order to have simultaneously a notion of angular momentum,
we restrict ourselves again to horizons of class II
and choose a fixed axial symmetry $\w{\phi}$ on ${\mathcal H}$.
Since we want the restriction of $\w{t}$ on $\Hor$ to generate
a WIH-symmetry of $(\Hor,[\el])$, we demand,
 according to expression (\ref{e:IH:WIHsymmetry_form}),
\bea
\label{e:IH:t_boundary_cond}
\encadre{
\left.\left(\w{t}+\Omega_{(\w{t})}\w{\phi}\right)\right|_{\mathcal H}\in [\el] \ \ ,
}
\eea
where $\Omega_{(\w{t})}$ is a constant on ${\mathcal H}$.
Once these boundary conditions for $\w{t}$ are set, the determination
of the expression for the mass proceeds in two steps.

\noindent 1. {\it First Law of Thermodynamics}

As a result of demanding $\w{t}$ to be associated with a conserved quantity $E^t_\Hor$
in the phase space $\Gamma$, it can be shown \cite{AshteBL01}
that the function $E^t_\Hor$ must depend {\it only} on two 
parameters defined entirely in terms of the horizon geometry:
the area,  $a_{\mathcal H}=\int_{\Sp_t} {}^2\w{\epsilon} $, 
of the 2-slice $\Sp_t$ (constant, as a consequence of the NEH geometry)
and the angular momentum $J_{\mathcal H}$ defined in Eq. (\ref{e:IH:angmom}).
In fact, the variation of $E^t_\Hor$ with respect to
these parameters must satisfy \cite{AshteBL01}
\bea
\label{e:IH:first_law}
\delta E^t_{\mathcal H} = 
\frac{\kappa_{(\w{t})}(a_{\mathcal H}, J_{\mathcal H})}{8\pi G}
\, \delta a_{\mathcal H}
+ \Omega_{(\w{t})}(a_{\mathcal H}, J_{\mathcal H}) \, 
\delta J_{\mathcal H} .
\eea
This expression can be interpreted as a {\it first law of black hole mechanics}
(see Remark~\ref{r:IH:thermo} below), 
where  $E^t_{\mathcal H}$ is an energy function
associated with the horizon\footnote{In Eq. (\ref{e:IH:first_law}),
$\kappa_{(\w{t})}=\langle \w{t}-\w{q}^*\w{t},\w{\omega}\rangle$, and a 
more precise meaning for the $\delta$ symbol in this
context can be found in Appendix \ref{s:SP}.}. 
Note that $\kappa_{(\w{t})}(a_{\mathcal H}, J_{\mathcal H})$ and
$\Omega_{(\w{t})}(a_{\mathcal H}, J_{\mathcal H})$ are constant on a given
horizon $\Hor$, where $a_{\mathcal H}$ and $J_{\mathcal H}$ have a
definite value;  in Eq. (\ref{e:IH:first_law}), 
$a_{\mathcal H}$ and $J_{\mathcal H}$  are rather
parameters in the phase space $\Gamma$ (see Appendix \ref{s:SP}).

However, this result does not suffice to prescribe a specific expression for the
mass of the black hole. In fact, since in condition (\ref{e:IH:t_boundary_cond})
we have not made an explicit choice  for the representative $\el\in [\el]$, the evolution
vector $\w{t}$ has not been completely specified.
Therefore, the functional forms of $\kappa_{(\w{t})}(a_{\mathcal H}, J_{\mathcal H})$,
$\Omega_{(\w{t})}(a_{\mathcal H}, J_{\mathcal H})$ and  $E^t_{\mathcal H}$ are not fixed.
However, once their dependences on $a_{\mathcal H}$ and $J_{\mathcal H}$ 
are specified, 
they turn out to be {\it the same} for every spacetime in $\Gamma$, 
no matter how distorted is the WIH or how dynamical is the neighboring spacetime. 
This is a non-trivial result.

\noindent 2. {\it Normalization of the Energy function}

The second step consists precisely in fixing the functional forms
of the physical parameters. In the space $\Gamma$ of solutions to 
the Einstein equation
containing a WIH,
there exists a subspace constituted by stationary spacetimes
(the Kerr family, in fact parametrized by the area and angular  momentum), 
where the existence of an exact rotational spacetime
Killing symmetry $\w{t}_{\mathrm{Kerr}}$ provides a natural choice for the
representative in $[\el]$. This fixes the evolution vector $\w{t}$
on ${\Hor}$ as well as the functional dependence of  $\kappa_{(\w{t})}$  
and $\Omega_{(\w{t})}$ for this family.
If we impose the functional forms of the physical parameters,
forms which are {\it the same for any spacetime} in $\Gamma$, to coincide
with those of the Kerr family when we restrict $\Gamma$ to its submanifold 
of stationary solutions, this completely determines their dependence 
on $a_{\mathcal H}$ and $J_{\mathcal H}$ (the
biparametric nature of the Kerr family is crucial for this). 
This is not an arbitrary choice but a consistent normalization.

Defining the areal {\it radius} of the horizon $R_{\mathcal H}$ by
\be
\label{e:IH:radius}
\encadre{
R_{\mathcal H}^2 := \frac{a_{\mathcal H}}{4 \pi}=\frac{1}{4\pi}
\int_{{\mathcal S}_t} {}^2\w{\epsilon} 
} \ ,
\ee
the horizon black hole physical parameters can be expressed as
\be 
\label{e:IH:physparam}
\encadre{
\begin{array}{rcl}
M_{\mathcal H}(R_{{\mathcal H}},J_{{\mathcal H}})
&:=&M_{\mathrm{Kerr}}(R_{{\mathcal H}},J_{{\mathcal H}}) =
\frac{\sqrt{R_{{\mathcal H}}^4 + 4G^2 J_{{\mathcal H}}^2}}{2 G R_{{\mathcal H}}}\
,\nonumber
\\ \kappa_{\mathcal H}(R_{{\mathcal H}},J_{{\mathcal H}})&:=&
\kappa_{\mathrm{Kerr}}(R_{{\mathcal H}},J_{{\mathcal H}}) =
\frac{R_{{\mathcal H}}^4 - 4 G^2 J_{{\mathcal H}}^2} {2 R_{{\mathcal H}}^3
\sqrt{R_{{\mathcal H}}^4 + 4G J_{{\mathcal H}}^2}}\ , \nonumber
\\ \Omega_{\mathcal H}(R_{{\mathcal H}},J_{{\mathcal H}})&:=&
\Omega_{\mathrm{Kerr}}(R_{{\mathcal H}},J_{{\mathcal H}}) =
\frac{2 G J_{{\mathcal H}}}{R_{{\mathcal H}} \sqrt{R_{{\mathcal H}}^4 + 4G
J_{{\mathcal H}}^2}} \ . 
\end{array}
}
\ee

\subsubsection{Final remarks}
\label{s:IH:phys_param_rem}

The following results can be retained from the previous discussion: 
\begin{itemize}
\item[{\it a)}] Explicit quasi-local expressions for the physical parameters
associated with the black hole. Their determination proceeds by firstly
calculating $R_{\mathcal H}$ and $J_{\mathcal H}$ from the geometry of 
$\Hor$\footnote{In this article we are not including the
electro-magnetic field. In the context of the Einstein-Maxwell theory,
the resulting expressions for the horizon angular momentum and mass
include an additional term corresponding to the electromagnetic field 
(see Ref.\cite{AshteBL01}). In the even more general Einstein-Yang-Mills case,
implications on the mass of the solitonic solutions in the
theory follow from the analysis of the first law
in the isolated horizon framework \cite{AshteK05}.}, via evaluation of expressions (\ref{e:IH:angmom}) and 
(\ref{e:IH:radius}). These values are then plugged into (\ref{e:IH:physparam}).

\item[{\it b)}] Even though we need a WIH structure on $\Hor$ in order to
{\it derive} the physical parameters, the final expressions
only depend on the NEH geometry, and not on the specific chosen WIH. 
This is straightforward for the radius $R_{\Hor}$, since it only depends 
on the 2-metric induced on $\Sp_t$. 
Regarding the angular momentum, the value of $J_{\Hor}$ through Eq. 
(\ref{e:IH:angmom}) does not depend on the null normal $\el$ chosen on 
the NEH. Given a null normal $\el$, a different 
one $\el'$ is related to $\el$ by the rescaling
$\el'=\alpha \el$ for some function $\alpha$ on $\Hor$.
Using the transformation rule for $\w{\omega}$ in Table \ref{t:KI:scaling},
the difference between $J'_{\Hor}$ and $J_{\Hor}$, calculated
respectively with $\el$ and $\el'$, is given by
\bea
J_{\mathcal H}'-J_{\mathcal H} & =& 
-\frac{1}{8\pi G}\int_{{\mathcal S}} 
\langle \w{\DS} \left(\ln\alpha\right), \w{\phi} \rangle\, 
{}^2\w{\epsilon}=
\frac{1}{8\pi G}\int_{{\mathcal S}} \left(\ln \alpha\right) 
\dd(\w{\phi}\cdot{}^2\w{\epsilon})= 0 , \nonumber \\
\eea
where we have firstly integrated by parts and then used
that $\w{\phi}$,  being an isometry
of $\w{q}$, is a divergence-free vector (or straightforwardly, 
$\dd(\w{\phi}\cdot{}^2\w{\epsilon})={\mathcal L}_{\w{\phi}}{}^2\w{\epsilon}-
\w{\phi}\cdot\dd({}^2\w{\epsilon})=0$, since ${\mathcal L}_{\w{\phi}}\w{q}=0$). 
Therefore, it makes sense to refer to $J_{\mathcal H}$ as
the angular momentum of a NEH and, in fact, its very notation makes sense.

\item[{\it c)}] A by-product of the Hamiltonian analysis with implications 
for the null geometry of a non-extremal WIH $(\Hor,[\el])$, is the 
singularization of a specific null normal $\el_0$ in $[\el]$.  
In any non-extremal WIH class $[\el]$ there
is a unique representative  such that its associated
{\it non-affinity coefficient} coincides with the surface gravity
of the Kerr family. 
In terms of an arbitrary null normal $\el$ in
$[\el]$, and according to transformations in Table \ref{t:KI:scaling},
\be
\label{e:IH:norma_l}
\el_0 =\frac{\kappa_{\Hor}(R_{{\mathcal H}},J_{{\mathcal H}})}{\kappa_{(\el)}}\el \ .
\ee
The choice of a {\it physical} normalization for the null normal permits,
on the one hand, to refer to its {\it non-affinity coefficient} $\kappa_0=
\kappa_{\mathcal H}(R_{{\mathcal H}},J_{\Hor})$ as the
horizon {\it surface gravity}. On the other hand, such a normalization 
can be conveniently exploited for the determination of 
the horizon slicing as discussed in Secs. \ref{s:IH:prefWIH} and \ref{s:IH:goodcuts}.
See in this sense Sec. \ref{s:BC:BCWIH}.
More generally, expression (\ref{e:IH:t_boundary_cond}) can be used to set boundary
conditions for certain fields on $\Hor$ (see Sec. \ref{s:BC}).

\end{itemize}

\begin{rem}
\label{r:IH:thermo} 
A major motivation for introducing the WIH structure on $\Hor$
is the extension of the black hole mechanics laws beyond the situation 
in which the horizon is embedded in a stationary spacetime \cite{BardeCH73}. 
The thermodynamical aspects of black hole horizons represent 
a cornerstone in understanding the physics of Gravity, both
at the classical and the quantum level \cite{Wald01,Padma05}
(for the case of black hole binaries, see Ref.~\cite{FriedUS02}). 
In this article,
we have focused on those applications of isolated horizons to  
the null geometry of a hypersurface representing
a black hole horizon in quasi-equilibrium inside a generally
dynamical spacetime, and mainly aiming at their astrophysical 
applications in numerical relativity. 
However, the implications of this formalism go far beyond
this aspect and, in particular, its results in black hole mechanics 
offer a link to the applications in quantum gravity. See the
review \cite{AshteK05} for a detailed account.
We have seen how the 
constancy of the surface gravity $\kappa_{(\el)}$ on the horizon,
the {\it zeroth law}, follows from the WIH definition
[see Eq. (\ref{e:IH:WIHzerolaw})], whereas the {\it first law}
results from the introduction of a consistent notion of energy
associated with the horizon [Eq. (\ref{e:IH:first_law})].
In order to discuss the {\it second law}, linked
to the increasing law of the area, one should go beyond the 
quasi-equilibrium regime and enter into the properly dynamical one
(see Sec. \ref{s:IH:dynam} for a brief outline of this regime). 
As a by-product, dynamical horizons
provide another version of the first law \cite{AshteK03}, 
associated with the evolution (a {\it process}) of a single system 
-Clausius-Kelvin' sense-, whereas 
Eq. (\ref{e:IH:first_law}) dwells on (horizon) {\it equilibrium} states
-Gibbs' sense- in the phase space $\Gamma$.
\end{rem}

%% file: isolhorII.tex
%
%
\section{Isolated horizons II: (strongly) isolated horizons and further
developments}
\label{s:IHII}

\subsection{(Strongly) isolated horizons}
\label{s:IH:strongIH}
After introducing the NEH's and WIH's, the third and final level in the 
isolated horizon hierarchy of intrinsic structures capturing the 
concept of black hole horizon in quasi-equilibrium, is provided by the notion 
of strongly isolated horizon, or simply isolated horizon (IH). 
We continue the strategy outlined at the beginning of 
Sec. \ref{s:IH}. Consequently, starting from a NEH, we demand
the full connection $\w{\hat{\nabla}}$ to be time-independent.

Following Ashtekar et al. \cite{AshteFK00}, a 
{\it strongly isolated horizon} (IH) is defined as a NEH, provided with a 
WIH-equivalence class $[\el]$ such that
\be
\label{e:IH:IH}
\encadre{[\LieH{\el}, \w{\hat{\nabla}}] = 0 } .
\ee
The consequences of imposing this structure on $\Hor$ can be analyzed
in terms of the constraints and free data of the null geometry.
From the discussion in Sec \ref{s:NE:intrinsic_NEH-const}, the time 
independence of $\w{\hat{\nabla}}$
implies the vanishing of $\LieH{\el}\w{S}$ in Eq. (\ref{e:NE:constraints_H2}),
that is
\bea
\hat{\nabla}_{(\alpha}\omega_{\beta)} + \omega_{\alpha} \omega_{\beta}
+\frac{1}{2}\left(R_{\alpha\beta} - {}^2\!R_{\alpha\beta}\right) = 0 \ .
\eea
In terms of the decomposition (\ref{e:NE:elem_delta}) of $\w{\hat{\nabla}}$, 
the time independence of $\w{S}$
implies the WIH condition $\LieH{\el}\w{\omega}=0$ (i.e.  $\LieH{\el}\w{\Omega}=0,
\LieH{\el}\kappa_{(\el)}=0$)
together with the vanishing
of $\vec{\w{q}}^*\Lie{\el}\w{\Xi}$. That is, an IH is characterized by
the constraints
\bea
\label{e:IH:3+1IH}
\encadre{
\LieH{\el}\w{q}=\LieH{\el}\w{\omega}=\vec{\w{q}}^*\Lie{\el}\w{\Xi}=0
} \ ,
\eea
where the only difference with respect to the WIH case discussed
in Sec. \ref{s:IH:WIHinitdata} is the time independence of $\w{\Xi}$. 
From Eq. (\ref{e:NE:Lie_Xi}), it follows
\bea
\label{e:IH:Xi_constraint}
\kappa \w{\Xi}=
   \frac{1}{2} \Kil{\w{\DS}}{\w{\Omega}}
   + \w{\Omega}\otimes \w{\Omega}
   - \frac{1}{2}  {}^2\!\w{R}
   + 4\pi \left( \vec{\w{q}}^* \w{T} - \frac{T}{2} \w{q} \right) \ .
\eea  
In contrast with the NEH and WIH cases, where the {\it initial data fields} 
$(\w{q}, \w{\Omega}, \kappa_o, \w{\Xi})$ can be freely specified on a given cross-section, 
in the IH case Eq. (\ref{e:IH:Xi_constraint}) sets a {\it constraint
on the IH initial data}. We note by passing that this is in complete analogy
with the 3+1 spacetime case where initial data $(\w{\gamma}, \w{K})$
are constraint by Eqs. (\ref{e:FO:Ham_constraint}) and (\ref{e:FO:mom_constraint}).

We comment on the non-extremal case and, for completeness at the 
level of basic definitions,
also on the extremal case. Regarding the non-extremal case $\kappa_o\neq 0$, 
the IH constraint can be straightforwardly solved.
In fact, once the fields $\w{q}$ and $\w{\Omega}$ and the constant $\kappa_o$ are 
freely chosen, the field $\w{\Xi}$ is fixed by Eq.(\ref{e:IH:Xi_constraint}). 
Therefore, the free data are given in this case by
\bea
\encadre{\hbox{non-extremal IH-free data:} \ \ \ (\w{q}|_{{\mathcal S}_t}, 
\w{\Omega}|_{{\mathcal S}_t}, \kappa_o\neq 0)
} \ ,
\eea
The $\w{\Xi}$ needed for reconstructing the full connection
$\w{\hat{\nabla}}$ is then given by Eq. (\ref{e:IH:Xi_constraint}).

In the extremal case, $\kappa_o = 0$, the situation changes.
The vanishing of the left-hand side in Eq. (\ref{e:IH:Xi_constraint})
leaves ${\w{\Xi}}$ as a free field on the initial cross-section
$\Sp_t$. The right-hand side becomes a constraint on 
$(\w{q}|_{{\mathcal S}_t}, \w{\Omega}|_{{\mathcal S}_t})$
\bea
\label{e:IH:IHconstraints_extremal}
 \frac{1}{2} \Kil{\w{\DS}}{\w{\Omega}}
   + \w{\Omega}\otimes \w{\Omega}
   - \frac{1}{2}  {}^2\!\w{R}
   + 4\pi \left( \vec{\w{q}}^* \w{T} - \frac{T}{2} \w{q} \right)=0 \ .
\eea
The initial data are given in this case by
\bea
\encadre{
\begin{array}{ll}
\hbox{extremal IH-initial data:} & \ \ \ (\w{q}|_{{\mathcal S}_t}, 
\w{\Omega}|_{{\mathcal S}_t},  \w{\Xi}|_{{\mathcal S}_t}, \kappa_o = 0) \\

& \hbox{where }  \w{q}|_{{\mathcal S}_t} 
 \hbox{ and }  \w{\Omega}|_{{\mathcal S}_t} \hbox{ satisfy Eq. (\ref{e:IH:IHconstraints_extremal})}
\end{array}
} \ ,
\eea

\subsubsection{General comments on the IH structure}
\label{s:IH:gen_IH}

In simple terms, an isolated horizon is a NEH in which all the objects defining
the null geometry $(\w{q},\w{\hat\nabla})$
are time-independent. It represents the maximum degree of
{\it stationarity} for the horizon defined in a quasi-local manner. 
However,
the notion of IH is less restrictive than that of a {\it Killing horizon},
which involves also the stationarity of the neighboring space-time. 
In fact, a Killing horizon is a particular case of an isolated horizon, 
but the reverse is not true. One can have an IH such that no spacetime 
Killing vector can be found in any neighborhood of the horizon. 
Consequently, an IH permits to model situations 
with a stationary horizon {\it inside} a truly dynamical spacetime.
Interestingly, non-trivial {\it exact} examples of this situation 
are provided in \cite{Le00}, in the context of a local analysis,
and globally by the Robinson-Trautman spacetime (see \cite{Chrus92}).
This flexibility of the IH structure is important for its 
applications in dynamical astrophysical situations.

In contrast with a WIH structure, an IH represents an actual restriction on 
the geometry of a NEH. In other words, if we start with an arbitrary NEH, it is not
guaranteed that a null normal $\el$ can be found such that condition 
(\ref{e:IH:IH}) is satisfied. 
An IH is in particular a WIH.
Reasoning in terms of initial data as in point {\it 3.} of Sec. (\ref{s:IH:WIHdefinition}), 
given an arbitrary $\el$ on a NEH with
associated data $(\w{q}|_{{\mathcal S}_t}, \w{\Omega}|_{{\mathcal S}_t},\w{\Xi}_{\Sp_t},\kappa_{(\el)})$, 
a function $\alpha$ can always be found such that the fields transformed 
under the rescaling $\el\rightarrow\el'=\alpha\el$ correspond to WIH free data
(cf. Table~\ref{t:KI:scaling}). 
That is, $\kappa_{(\el')}=\mathrm{const}$ (let us assume
$\kappa_{(\el')}\neq 0$ for definiteness). If the transformed fields 
satisfy (\ref{e:IH:Xi_constraint}), they correspond to the initial data of an
IH. If not, the remaining freedom in these data corresponds to a new transformation
$\el'\rightarrow\el''=\alpha'\el'$ with $\alpha'$
\bea
\alpha' =  1 + B e^{-\kappa_{(\el')}t},
\eea
where $B$ is a function on ${\mathcal S}_t$ and $\LieH{\el'}t=1$.
Substituting the transformed fields into (\ref{e:IH:Xi_constraint})
leads to
three independent equations for a single variable $B$.
If the system has no solution, this means that the NEH does not admit any IH.
In general the choice of $B$ only permits to cancel a scalar obtained 
from $\LieH{\el}\w{\Xi}$ (this was in fact the strategy in Sec. \ref{s:IH:prefWIH}
to fix the WIH class).
A similar argument applies in the extremal case.
An analysis of the necessary conditions for a NEH to admit an IH can be found
in Appendix A.2. of Ref.~\cite{AshteBL02}.

\noindent{\it A posteriori analysis of black hole spacetimes}

The interest of applying the geometrical tools 
discussed so far in numerical relativity is twofold.
On the one hand, they can be used to set constraints 
on the fields entering in the numerical construction of a spacetime. This 
will be discussed in some detail in Sec. \ref{s:BC}, mainly
involving the NEH and WIH structures.
On the other hand, they can be employed to extract physics in an 
invariant manner out of already constructed spacetimes.
In fact, even though the IH level could be not flexible enough in order
to accommodate astrophysically realistic {\it initial} situations, 
it is very well suited for the {\it a posteriori} 
analysis of the dynamical evolution toward stationarity after a stellar 
collapse or a black hole merger. 

In this sense, we briefly comment on the possibility of constructing  
a coordinate system in an invariant way for a neighborhood
of the horizon $\Hor$ (in fact, only WIH notions are involved). 
This can be specially relevant for comparing 
results between different numerical simulations.
Once a WIH class is fixed (using for instance results in Sec. \ref{s:IH:prefWIH};
see also Sec. \ref{s:BC:BCWIH}), we can choose a vector 
$\el$ in $[\el]$ [for instance via
Eq. (\ref{e:IH:norma_l})] and a compatible slicing of
$\Hor$ (see Sec. \ref{s:IH:goodcuts}). Then, coordinates $(t,\theta, \phi)$
can be chosen on $\Hor$, up to a re-parametrization of $(\theta, \phi)$
on the cross-sections $\Sp_t$. In order to construct the additional
coordinate {\it outside} the horizon, we choose the only vector $\w{k}$ normal to
$\Sp_t$ that satisfies $\w{k}\cdot \el=-1$. The affine parameter
$r$ of the only geodesic passing through a given (generic) point in $\Hor$
(with $r=r_0$ and derivative $-\w{k}$ on that point) 
provides a coordinate in a neighborhood of $\Hor$.
The rest of the coordinates $(t,\theta, \phi)$ are Lie-dragged along
these geodesics. An analogous procedure can be followed in order to 
construct invariantly a tetrad in a neighborhood of $\Hor$.
Details of this construction can be found in 
\cite{AshteBL02,AshteBDFKLW00,Kr02}. In particular, a near horizon
expansion of the metric in such a (null) tetrad can be found
in \cite{AshteBDFKLW00}.

Another example of {\it a posteriori} extraction of physics is provided
in \cite{DreyeKSS03,Kr02}. This reference presents an algorithm for
assessing the existence of the horizon axial symmetry  $\w{\phi}$
entering in the expression for the angular momentum
(\ref{e:IH:angmom}).
In case this symmetry actually exists,  $\w{\phi}$ is
explicitly reconstructed permitting the coordinate-independent
assignation of mass and angular momentum to the horizon (the algorithm
can also be used to look for {\it approximate} symmetries in case
of small departs from axisymmetry).

\subsubsection{Multipole moments}
\label{s:IH:multipoles}

In this section we address an important point from
the point of view of applications and whose physical interpretation
is quite straightforward. As in Sec. \ref{s:IH:PhysParam}, we only
consider the horizon vacuum case (see Ref. \cite{AshteEPB04} for the inclusion of
electromagnetic fields).
In analogy with the source multipoles of an extended object in Newtonian
gravity, which encode the distribution of matter of the source, the geometry of an
IH permits to define a set of {\it mass} and 
{\it angular momentum} multipoles which characterize the 
black hole (whose horizon is in quasi-equilibrium) 
as a source of gravitational field (see \cite{AshteEPB04}).

As we saw in Sec. \ref{s:IH:PhysParam}, a meaningful notion of angular momentum 
can be associated with the horizon if we impose its transversal sections 
${\mathcal S}_t$ to admit an axial symmetry $\w{\phi}$, i.e. if the horizon is
of class II in the terminology of Sec. \ref{s:IH:PhysParam} (the
requirement on the Killing character of $\w{\phi}$ can be relaxed 
to a divergence-free condition; see in this sense the discussion 
following Eq.(3.18) in Ref. \cite{AshteEPB04}). 
In the same spirit, the discussion here applies only to 
class II horizons (and its {\it subclass} I). 
We limit ourselves again to the non-extremal case.

As we have shown in Sec. \ref{s:IH:strongIH}, the geometry of a 
non-extremal IH is determined by the free data $(\w{q}, \w{\Omega})$ on
a cross-section ${\mathcal S}_t$ (the different constant values 
of $\kappa$ change the representative of $[\el]$, but not the IH geometry). 
We must therefore characterize these two geometrical
objects. The main idea is to identify two scalar
functions, associated respectively with  $\w{q}$ and  $\w{\Omega}$, 
such that they encode the geometrical information of these initial
data
[remember that only the divergence-free part of  $\w{\Omega}$ is a
geometrical object in the sense of being independent of the cross
section; see discussion in Sec. \ref{s:IH:goodcuts} or transformation 
rules (\ref{e:NE:passivegauge})]
Multipoles are then given by the coefficients
in the expansion of these scalars in a spherical harmonic basis.
This can actually be achieved in an invariant manner. 
In this section we simply present a brief account of the
results in \cite{AshteEPB04}, referring the reader to this reference 
for details.

\noindent {\it The metric $\w{q}$}

On a sphere $S^2$, the geometrical content of a metric $\w{q}$  can be 
encoded in a scalar function, such as its scalar curvature ${}^2\!R$.
The crucial remark is that, given a metric on $S^2$ with an
axial symmetry $\w{\phi}$ (i.e. $\w{\phi}$ is a Killing vector on $S^2$
with closed orbits and vanishing exactly
on two points), a particular coordinate system $(\theta, \phi)$  can be 
constructed in an {\it invariant} manner\footnote{This coordinate 
system plays a fundamental role in the discussion of
IH multipoles. However, the technical details of its construction
go beyond the scope of this section, mainly focused in presenting the final expressions
for the multipoles. We refer the reader to Ref. \cite{AshteEPB04} for a
complete presentation.}, 
where the $2$-metric is written as
\bea
\label{e:IH:preferredq}
\w{q} = (R_{\mathcal H})^2\left( f^{-1} \mathrm{sin}^2\theta d\theta\otimes d\theta +
f d\phi\otimes d\phi \right) ,
\eea
with $f=\frac{\w{q}(\w{\phi},\w{\phi})}{(R_{\Hor})^2}$ and $R_{\Hor}$ is given by 
Eq. (\ref{e:IH:radius}).
The function $f$ is related to the Ricci scalar ${}^2\!R$ by
\bea
\label{e:IH:f-R}
{}^2\!R= -\frac{1}{(R_\Hor)^2} \frac{d^2 f}{d(\mathrm{cos}\theta)^2} \ .
\eea
Therefore, from the knowledge of ${}^2\!R$ the metric $\w{q}$ can be reconstructed by using
Eqs. (\ref{e:IH:preferredq}) and (\ref{e:IH:f-R}).
The round metric $\w{q}_0$
\bea
\label{e:IH:roundmetric}
\w{q}_0 = (R_\Hor)^2\left(d\theta\otimes d\theta + \mathrm{sin}^2\theta d\phi\otimes d\phi\right)
\eea
is obtained by making $f=\mathrm{sin}^2\theta$ and has {\it the same}
volume element, $d^2V\equiv
{}^2\w{\epsilon}_0=(R_{\Hor})^2\mathrm{sin}^2\theta \; d\theta \wedge d\phi$,
than the physical metric $\w{q}$.

\noindent {\it The \hajicek\ 1-form $\w{\Omega}$}

As shown in Sec. \ref{s:IH:goodcuts} the divergence-free 
part $\w{\Omega}^{\mathrm{div-free}}$ of the \hajicek\ 1-form 
is completely characterized by $\mathrm{Im}\Psi_2$, 
whereas its exact part $\w{\Omega}^{\mathrm{exact}}$ 
is a gauge term related with the foliation in $\Hor$, but without affecting
the intrinsic geometry of the horizon.

Therefore, the geometry of the IH is encoded in the pair 
$({}^2\!R,\mathrm{Im}\Psi_2)$ (together with the radius $R_{\Hor}$). 
In order to characterize these fields, and taking advantage
of the {\it invariant} coordinate system previously introduced,
an expansion in spherical harmonics 
$Y_{lm}(\theta,\phi)$ can be performed. Due to the axial symmetry,
the only functions entering into the decomposition are $Y_{l0}(\theta)$, 
which do not depend on $\phi$.
Since the volume element ${}^2\w{\epsilon}$,
corresponding to $\w{q}$, coincides with the volume ${}^2\w{\epsilon}_0 = d^2V$
associated with
the round metric $\w{q}_0$, the spherical harmonics are the standard ones,
normalized according to
\bea
\label{e:IH:normsphhar}
\int_{{\mathcal S}_t}  Y_{n0}(\theta)Y_{m0}(\theta)d^2V = (R_{\mathcal
  H})^2 \ \ .
\delta_{nm}
\eea
We define the two series of numbers $I_n$, $L_n$ as
\bea
I_n &:=& \frac{1}{4} \int_{{\mathcal S}_t} {}^2\!R \; Y_{n0}(\theta) \;d^2V \label{e:IH:I_n}\\
L_n &:=& - \int_{{\mathcal S}_t} \mathrm{Im}\Psi_2 \; Y_{n0}(\theta)
  \;d^2V \label{e:IH:L_n} \ \ .
\eea
These are geometrical dimensionless quantities that measure, respectively,
the distortions and rotations of the horizon with respect to the round metric.
We see explicitly
that even the strongest notion of quasi-equilibrium
that we have introduced, i.e. the IH structure, is rich and flexible enough so as 
to model physically interesting scenarios.
If the two series $I_n$ and $L_n$ are given, the full isolated horizon geometry
can be reconstructed from
\bea
\label{e:IH:IHreconstruction}
{}^2\!R&=&\frac{4}{(R_{\mathcal H})^2}\sum_{n=0}^\infty I_n Y_{n0}(\theta) \\
\mathrm{Im}\Psi_2&=&-\frac{1}{(R_{\mathcal H})^2}\sum_{n=0}^\infty L_n
Y_{n0}(\theta) \ \ .
\eea
We note that in a NEH, the information about $\w{q}$ and
$\w{\Omega}$ is {\it also} encoded invariantly in $({}^2\!R,\mathrm{Im}\Psi_2)$ 
[together with the invariant coordinate
system where $\w{q}$ takes the form (\ref{e:IH:preferredq})]. Therefore,
it makes sense to define $I_n$ and $L_n$ for a NEH. 
In this case, however, the full geometry cannot be 
reconstructed since the information on $\w{\Xi}$ is missing.

In order to obtain physical magnitudes
that one can associate with the mass and rotation multipoles of
the horizon, one must rescale $I_n$ and $L_n$ with dimensionful parameters.
Motivated by heuristic considerations based in the analogy with magnetostatics
and electrostatics in flat spacetime (see \cite{AshteEPB04}), together
with the results for the angular momentum $J_{\mathcal H}$ and mass
$M_{\mathcal H}$ presented in Sec. \ref{s:IH:PhysParam}, rotation multipoles
are defined as \cite{AshteEPB04}
\bea
\label{e:IH:rotmultipol}
J_n := \sqrt{\frac{4\pi}{2 n + 1}} 
\frac{R_{\mathcal H}^{n+1}}{4\pi G} L_n=-\frac{R_{\mathcal H}^{n+1}}{8\pi G}
\int_{{\mathcal S}_t} \left({}^2\!\epsilon^{\mu\nu} \;\DS_\nu P_n(\mathrm{cos}\theta) \right)
\Omega_\mu \; d^2V ,
\eea 
resulting $J_0=0$ and $J_1=J_{\mathcal H}$. 
The mass multipoles are then introduced as 
\bea
\label{e:IH:massmultipol}
M_n := \sqrt{\frac{4\pi}{2 n + 1}} 
\frac{M_{\mathcal H} R_{\mathcal H}^{n}}{2\pi} I_n=
 \sqrt{\frac{4\pi}{2 n + 1}}\frac{M_{\mathcal H} R_{\mathcal H}^{n}}{2\pi}
\int_{{\mathcal S}_t} \left({}^2\!R \; Y_{n0}(\theta)\right) d^2V ,
\eea 
where $M_{\mathcal H}$ is given by  expression 
(\ref{e:IH:physparam}), resulting $M_0=M_{\mathcal H}$ and
$M_1=0$ ({\it centre of mass} frame).

These source multipoles present a vast domain of applications \cite{AshteEPB04,AshteK05}
ranging from the description of the motion of a black hole
inside a strong external gravitational field, the study of the effects on the
black hole induced by a companion or the invariant comparison 
at sufficiently late times of the numerical simulations of black hole
spacetimes having suffered a strongly dynamical process (numerical
simulations in Refs. \cite{BaiotHMLRSFS05,BeyerKS05} show that the
isolated horizon notion becomes a good approximation quite fastly
after the black hole formation).

\subsection{3+1 slicing and the hierarchy of isolated horizons}
\label{s:IH:IHhierarchy}

The stratification of the IH hierarchy in NEH, WIH and IH can be defined in 
terms of the null geometry  
because the structures on which the time independence is imposed,
i.e. $\w{q}, \w{\omega}$ and $\w{\hat{\nabla}}$, 
are intrinsic to $\Hor$. 

However, as indicated at the beginning of Sec. \ref{s:IH},
when a 3+1 perspective is adopted by introducing the additional
structure provided by the spatial slicing $(\Sigma_t)$, new objects which
are not intrinsic to $\Hor$ enter into scene. 
This is the case of the \hajicek\ 1-form $\w{\Omega}$ and $\w{\Xi}$.
The geometry of $\Hor$ can  now be defined in terms of the initial values of these 
fields on a given slice $\Sp_t$. In this context, it seems natural to 
introduce progressive levels of horizon quasi-equilibrium
by demanding the time independence of different combinations 
of the fields $(\w{q}, \w{\Omega},\kappa, \w{\Xi})$.

This is a {\it practical} manner to proceed in the actual construction of the
spacetime from a given initial slice.
In accordance with the discussion of NEH initial data in Sec. \ref{s:NE:NEHfreedata},
we introduce the non-affinity coefficient $\kappa$ as an initial data, even if 
this parameter can be gauged away by a rescaling of the null normal.
As we will see in Sec. \ref{s:BC}, such a rescaling actually contain relevant
information on the 3+1 description, in particular
on the lapse, when a WIH-compatible slicing is chosen.
In addition, we split tensor $\w{\Xi}$ in its trace and traceless (shear) parts
\bea
\label{e:IH:Xidecomp}
\w{\Xi} = \frac{1}{2}\theta_{(\w{k})} \w{q} + \w{\sigma}_{(\w{k})}  \ .
\eea
Given a NEH and a specific null normal $\el$, the fields 
capable of changing in time are
$(\w{\Omega},\kappa,\theta_{(\w{k})}, \w{\sigma}_{(\w{k})})$.

Let us denote by the letters $A,B, C, D$ the four conditions
\bea
\label{e:IH:ABCDcondtions}
\begin{array}{ll}
A: \ \LieH{\el}\w{\Omega}=\w{\DS}\kappa=0 ;
\ \ \ \ & B: \ \LieH{\el}\kappa=0 ;\\
C: \  \LieH{\el}\theta_{(\w{k})}=0; & D: \ \LieH{\el}\w{\sigma}_{(\w{k})}=0.
\end{array}
\eea
A NEH endowed with a null normal $\el$ will be called a {\it $(A,B,...)$-horizon} if 
conditions $A,B,...$ are satisfied.

As an example, a $(A,B)$-horizon is simply a WIH, whereas a $(A,B,C,D)$-horizon
is a (strongly) IH.
It is important to underline that this terminology makes no sense from 
a point of view intrinsic to $\Hor$. Even more, due to the gauge freedom
in the set of initial data, some of these $(A,B,..)$-horizons actually correspond
to the main intrinsic object (for instance, a $(B)-horizon$ is simply
a NEH where we have chosen $\el$ in such a way that $\kappa$ is 
time-independent,
although it can depend on the angular variables on $\Sp$).

The only aim of such a decomposition of the horizon quasi-equilibrium
conditions, is to classify
the different potential constraints on the null geometry of $\Hor$ that
one would straightforwardly find in the 3+1 spacetime construction.
In any case, it turns out to be useful for
keeping track of the structures that are actually imposed in
the construction of the horizon (see Sec. \ref{s:BC}).

\subsection{Departure from equilibrium: dynamical horizons}
\label{s:IH:dynam}

This article deals with the properties of a black hole
horizon in equilibrium, following  a quasi-local perspective. 
The basic idea is to consider an {\it apparent horizon} $\Sp$ in a 
spatial slice $\Sigma_t$, 
and then assume that this apparent horizon {\it evolves} 
smoothly into other apparent horizons (see discussion in Sec. \ref{s:NE:trapped_app}). 
The hypersurface $\Hor$ defined in this way
constitutes the quasi-local characterization of the black hole.
The {\it key} element associated with the quasi-equilibrium of
this apparent-horizon world-tube is the {\it null} character of $\Hor$.
In this section we briefly indicate how these quasi-local ideas have been 
extended in the literature to the regime in which the black hole horizon is 
dynamical (see Refs.~\cite{AshteK05} and \cite{Booth05} for recent reviews). 

The horizon is in quasi-equilibrium if neither matter nor radiation 
actually cross this horizon \cite{Booth05,BoothBGV05}. 
Motivated by Hawking's {\it black hole 
area theorem} \cite{Hawki71} for event horizons, the world-tube
$\Hor$ corresponds to a quasi-equilibrium situation if
the volume element of the {\it apparent horizons} remains constant
(this implies that the area is constant, but the converse is not 
true in general). On the contrary, the dynamical case corresponds 
to an increasing area along the evolution of the world-tube.
These considerations on the rate of change of the area, translate
into the metric type of $\Hor$ as follows.  
Under the physically reasonable assumption $\Lie{\w{k}}\theta_{(\el)}<0$ 
(a necessary condition for the spheres inside the 
apparent horizon to be future-trapped),
a vector $\w{z}$ tangent to $\Hor$ and normal to the apparent
horizon sections $\Sp_t$ is either null or spacelike
(see \cite{Haywa94,DreyeKSS03,Kr02}).
The volume element of the apparent horizons is constant, corresponding
to the quasi-equilibrium case, if and only if
$\w{z}$ is everywhere null on $\Hor$, i.e. if $\Hor$ is null hypersurface
\cite{Kr02,DreyeKSS03} (cf. Eq. (\ref{e:KI:theta_Lie_detq})). 
If $\w{z}$ is everywhere spacelike, and $\theta_{(\w{k})}<0$,
then the area of the horizons actually increases.
In the intermediate cases, with $\theta_{(\w{k})}<0$,
the area is never decreasing.

The properly dynamical regime can be quasi-locally characterized
by the notion of {\it dynamical horizon} 
introduced by Ashtekar et al. \cite{AshteK02,AshteK03}.
A {\it dynamical horizon} is a spacelike hypersurface in spacetime that  
is foliated by a family of spheres $(\Sp_t)$ and such that,
on each $\Sp_t$,
the expansion $\theta_{(\el)}$ associated with the 
outgoing null normal $\el$ vanishes and the expansion $\theta_{(\w{k})}$
associated with the ingoing null normal $\w{k}$ is strictly negative.
This is a particular case of the previous notion
of {\it trapping horizon} introduced by Hayward 
\cite{Haywa94} and commented at the end of Sec. \ref{s:NE:trapped_app}. 
In particular, it is directly related to 
{\it future outer trapping horizons}, whose definition 
coincides with that of dynamical horizons once the
{\it spacelike} nature of $\Hor$ has been substituted by the condition 
$\Lie{\w{k}}\theta_{(\el)}<0$. This last condition can be crucial
to ensure that trapped surfaces (i.e. having both $\theta_{(\el)}<0$
and $\theta_{(\w{k})}<0$) exist inside $\Hor$ \cite{Senov03}. 
A future outer trapping horizon can be a null or spacelike  hypersurface
(more generally, the vector $\w{z}$ can be either null or spacelike),
potentially permitting a clearer description of the transition from the 
equilibrium to the
dynamical situation \cite{BoothF04}.
Let us note that the Damour-Navier-Stokes equation discussed in 
Sec.~\ref{s:DN:DNS} has been recently extended to future outer trapping
horizons and dynamical horizons \cite{Gourg05}. 

We can think of implementing Hayward's dual null construction, since in our 
approach we have extended $\w{k}$ outside $\Hor$. For doing this,  
we demand the null field $\w{k}$ to be normal to a null hypersurface $\Hor'$,
in such a way that $\Sp=\Hor\cap\Hor'$. From Frobenius's identity 
(see Sec. \ref{s:NH:Frobenius_k}; in particular Eq. (\ref{e:KI:dk})
and Remark \ref{r:NH:k_normal_surface}), it follows 
\bea
\label{e:IH:k_normal}
\dd \ln \left( \frac{N}{M} \right) = \alpha \uk + \beta \uel \ . 
\eea
On the one hand, this can be used as a constraint between the lapse $N$
and $M$ [the latter being a function of the 3-metric 
$\w{\gamma}$, as a consequence of Eq. (\ref{e:IN:us_MDr})].
On the other hand, since $\w{k}$ is null and hypersurface-normal, it
is pre-geodesic (Sec. \ref{s:NH:generators}). This can be checked
explicitly in Eq. (\ref{e:KI:4_dim_gradk}) by noting that, due to (\ref {e:IH:k_normal}),
we have  $\w{\nabla}_{\w{k}}  \ln \left( \frac{N}{M} \right)=
\w{k}\cdot \dd \ln \left( \frac{N}{M} \right)=-\beta$, so
$\w{\nabla_k k} =\frac{\alpha}{N^2}\w{k}$. Consequently, 
$\Hor'$ provides the surface $t=\mathrm{const}$ parametrized
by $(r,\theta,\phi)$ in the invariant construction
of the coordinate system 
on a neighborhood of $\Hor$, presented  at the end of 
Sec.~\ref{s:IH:gen_IH}.


%% file: express3p1.tex

\section{Expressions in terms of the 3+1 fields}
\label{s:TP}

\subsection{Introduction}

Hitherto we have used the 3+1 foliation of spacetime by the 
spacelike hypersurfaces $(\Sigma_t)$ only (i) to set the normalization 
of the normal $\el$ to the null hypersurface $\Hor$ 
(by demanding that $\el$ is the tangent vector of the null generators
of $\Hor$ when parameterizing the latter by $t$ [Eq.~(\ref{e:NH:l_norm1})]),
and (ii) to introduce the ingoing null vector $\w{k}$ and the associated
projector onto $\Hor$, $\w{\Pi}$.  
In the present section, we move forward in our ``3+1 perspective''
by expressing all the fields intrinsic to the null hypersurface $\Hor$,
such as the second fundamental form $\w{\Theta}$ or the rotation 1-form
$\wo$, in terms of the 3+1 basics objects, like the extrinsic curvature
tensor $\w{K}$, the lapse function $N$ or the timelike unit normal $\w{n}$. 
In this process, we benefit from  the 4-dimensional point
of view adopted in defining $\w{\Theta}$, $\wo$, and other objects
relative to $\Hor$, thanks to the projector $\w{\Pi}$. 


\subsection{3+1 decompositions} \label{s:TP:3p1decomp}

\subsubsection{3+1 expression of $\Hor$'s fields}

We have already obtained the 3+1 decomposition of the null normal $\el$
[Eq.~(\ref{e:IN:el_nps})]. By inserting it into Eq.~(\ref{e:NH:Thetaqq}), 
we get
\bea
    \Theta_{\alpha\beta} & = &\nabla_\mu \ell_\nu q^\mu_{\ \, \alpha}
        q^\nu_{\ \, \beta} = \nabla_\mu [N(n_\nu+s_\nu)] 
         q^\mu_{\ \, \alpha} q^\nu_{\ \, \beta} \nonumber \\
         & = & N(\nabla_\mu n_\nu + \nabla_\mu s_\nu)
            q^\mu_{\ \, \alpha} q^\nu_{\ \, \beta}  
            = N (\nabla_\mu n_\nu + \nabla_\mu s_\nu) \gamma^\mu_{\ \, \rho}
                 \gamma^\nu_{\ \, \sigma}  
                 q^\rho_{\ \, \alpha} q^\sigma_{\ \, \beta}   \nonumber \\
        & = & N (-K_{\rho\sigma} + D_\rho s_\sigma) 
        q^\rho_{\ \, \alpha} q^\sigma_{\ \, \beta}  ,   
                                \label{e:TP:Theta_ab_DS_K_prov}
\eea
where we have used $n_\nu q^\nu_{\ \, \beta} =0$ and 
$s_\nu q^\nu_{\ \, \beta} =0$ to get the second line
and Eqs.~(\ref{e:FO:Kab_n}) and (\ref{e:FO:3der}) to get the third one. 
Hence the 3+1 expression of the second fundamental form:
\be \label{e:TP:Theta_ab_DS_K}
    \encadre{\Theta_{\alpha\beta} = N (D_\mu s_\nu - K_{\mu\nu}) 
    q^\mu_{\ \, \alpha} q^\nu_{\ \, \beta} } ,
\ee
or equivalently (taking into account the symmetry of $\w{\Theta}$
and $\w{K}$):
\be \label{e:TP:Theta_Ds_K}
    \encadre{\w{\Theta} = N \vec{\w{q}}^* (\w{D}\underline{\w{s}} - \w{K})} .
\ee
Contracting equation (\ref{e:TP:Theta_ab_DS_K}) with $q^{\alpha\beta}$
gives the expansion scalar [cf. Eq.~(\ref{e:KI:def_theta})]:
\bea
    \theta & = & N q^{\mu\nu} (D_\mu s_\nu - K_{\mu\nu})
    = N(\gamma^{\mu\nu} - s^\mu s^\nu) (D_\mu s_\nu - K_{\mu\nu}) \nonumber \\
    & = & N (D_\mu s^\mu - K - s^\mu \underbrace{s^\nu D_\mu s_\nu}_{=0}
        + K_{\mu\nu} s^\mu s^\nu) , 
\eea
hence
\be
    \encadre{ \theta = N (D_i s^i + K_{ij} s^i s^j - K) } \ ,
\ee
or equivalently
\be \label{e:TP:theta_3p1}
    \encadre{ \theta = N (\w{D}\!\cdot\!\w{s} + \w{K}(\w{s},\w{s}) - K) } .
\ee
We then deduce the 3+1 expression of the non-affinity parameter $\kappa$
via Eq.~(\ref{e:KI:divl_k_th}):
\bea
    \kappa & = & \w{\nabla}\cdot\el - \theta 
  = \nabla_\mu[N(n^\mu + s^\mu)] - \theta \nonumber \\
 & = & \ell^\mu \nabla_\mu \ln N + N(-K + \nabla_\mu s^\mu ) 
    - N (D_i s^i + K_{ij} s^i s^j - K) \nonumber \\
    & = & \ell^\mu \nabla_\mu \ln N  + N ( \nabla_\mu s^\mu - 
    D_i s^i - K_{ij} s^i s^j ) .  \label{e:TP:kappa_prov}
\eea
Now, by taking the trace of  
$D_\alpha s^\beta = \gamma^\mu_{\ \, \alpha} \gamma^\beta_{\ \, \nu}
    \nabla_\mu s^\nu$ [cf. Eq.~(\ref{e:FO:3der})], we get
the following relation between the 3-dimensional and 4-dimensional divergences
of the vector $\w{s}$
\bea
    D_i s^i & = & \gamma^\mu_{\ \, \nu} \nabla_\mu s^\nu
            = \nabla_\mu s^\mu + n^\mu n_\nu \nabla_\mu s^\nu 
            = \nabla_\mu s^\mu - n^\mu s^\nu \nabla_\mu n_\nu
            \nonumber \\
            & = & \nabla_\mu s^\mu - s^\nu D_\nu\ln N  ,
                    \label{e:TP:div3s_div4s}
\eea 
where we have used $n_\nu s^\nu = 0$ and Eq.~(\ref{e:FO:acceler_n}).
Substituting Eq.~(\ref{e:TP:div3s_div4s}) for $D_i s^i$ into 
Eq.~(\ref{e:TP:kappa_prov}) leads to
\be
   \encadre{ \kappa = \ell^\mu \nabla_\mu \ln N + s^i D_i N 
   - N K_{ij} s^i s^j} ,    \label{e:TP:kappa_3p1_index}
\ee
or equivalently,
\be
   \encadre{ \kappa = \w{\nabla}_{\el} \ln N + \w{D}_{\w{s}} N 
   - N \w{K}(\w{s},\w{s})} .   \label{e:TP:kappa_3p1}
\ee

To compute the 3+1 expression of the rotation 1-form, the easiest manner is 
to  start from expression (\ref{e:KI:omega_gradl_uk}) for $\wo$ and
to replace in it $\w{k}$ from Eq.~(\ref{e:IN:n_l_k}):
\bea
    \omega_\alpha & = & \ell^\mu \nabla_\mu k_\alpha =
        \ell^\mu \nabla_\mu \left( \frac{1}{N} n_\alpha
            - \frac{1}{2N^2} \ell_\alpha \right)
        \nonumber \\
        & = & - \frac{1}{N^2} \ell^\mu \nabla_\mu N \; n_\alpha
            + \frac{1}{N} \ell^\mu (-K_{\mu\alpha} - n_\mu D_\alpha\ln N)
             + \frac{1}{N^3} \ell^\mu \nabla_\mu N \; \ell_\alpha \nonumber \\
         &&  - \frac{1}{2N^2} \kappa \ell_\alpha \nonumber \\
         & = &  D_\alpha \ln N - K_{\alpha\mu} s^\mu
         + \frac{1}{N} \left(  \ell^\mu \nabla_\mu \ln N - \frac{\kappa}{2}
         \right) s_\alpha - \frac{\kappa}{2N} n_\alpha , 
\eea
where we have used Eqs.~(\ref{e:FO:K_grad_n_index}) and 
(\ref{e:NH:der_l_kappa}) to get the second line. 
Substituting Eq.~(\ref{e:TP:kappa_3p1_index}) for $\kappa$ in the 
$s_\alpha$ term yields
\be
    \encadre{\omega_\alpha = D_\alpha \ln N - K_{\alpha\mu} s^\mu
        + \frac{1}{2} \left( n^\mu \nabla_\mu \ln N + K_{ij}s^i s^j
        \right) s_\alpha -  \frac{\kappa}{2N} n_\alpha } , 
\ee
or equivalently,
\be
    \encadre{\wo= \w{D}\ln N - \w{K}(\w{s},.) 
        + \frac{1}{2} \left[ \w{\nabla}_{\w{n}} \ln N 
            + \w{K}(\w{s},\w{s}) \right] \underline{\w{s}}
            - \frac{\kappa}{2N}  \underline{\w{n}} } . 
\ee
The 3+1 expression of the \hajicek\ 1-form $\w{\Omega}$ is then 
immediately deduced from $\Omega_\alpha = \omega_\mu q^\mu_{\ \, \alpha}$
along with $n_\nu q^\mu_{\ \, \alpha} = 0$ and $s_\nu q^\mu_{\ \, \alpha} = 0$
\be
\label{e:TP:Omega3+1}
    \encadre{ \Omega_\alpha = \DS_\alpha \ln N 
        - K_{\mu\nu} s^\mu q^\nu_{\ \, \alpha} } ,
\ee
or equivalently, 
\be \label{e:TP:Omega_DN_qKs}
    \encadre{\w{\Omega}= \w{\DS}\ln N - \vec{\w{q}}^* \w{K}(\w{s},.) }. 
\ee
In Appendix~\ref{s:KE}, the above formul\ae\  are evaluated in the specific
case where $\Hor$ is the event horizon of a Kerr black hole, with a
3+1 slicing linked to Kerr coordinates. In particular the standard value
of $\kappa$ (called {\em surface gravity} in that case)
is recovered from Eq.~(\ref{e:TP:kappa_3p1}). Moreover the
\hajicek\ 1-form computed from Eq.~(\ref{e:TP:Omega_DN_qKs}), once plugged
into formula (\ref{e:IH:angmom}), leads to the angular momentum
$J_{\Hor}=am$ (where $a$ and $m$ are the standard parameters of the 
Kerr solution), as expected. 

Let us now derive the 3+1 expression of the transversal deformation rate
$\w{\Xi}$. Similarly to the computation leading to 
Eq.~(\ref{e:TP:Theta_ab_DS_K_prov}), we have, thanks to 
Eqs.~(\ref{e:KI:Xi_qstar_gradk}) and the 3+1 expression 
(\ref{e:NH:k_n_s}) of $\w{k}$,
\bea
    \Xi_{\alpha\beta} & = &\nabla_\mu k_\nu \, q^\mu_{\ \, \alpha}
        q^\nu_{\ \, \beta} = \nabla_\mu \left[
            \frac{1}{2N}(n_\nu-s_\nu) \right] 
         q^\mu_{\ \, \alpha} q^\nu_{\ \, \beta} \nonumber \\
         & = & \frac{1}{2N} \left( \nabla_\mu n_\nu 
            - \nabla_\mu s_\nu \right) q^\mu_{\ \, \alpha} q^\nu_{\ \, \beta} 
            + 0 .
\eea
Hence 
\be \label{e:TP:Xi_ab_DS_K}
    \encadre{\Xi_{\alpha\beta} = -\frac{1}{2N} (D_\mu s_\nu + K_{\mu\nu}) 
    q^\mu_{\ \, \alpha} q^\nu_{\ \, \beta} } ,
\ee
or equivalently (taking into account the symmetry of $\w{\Xi}$
and $\w{K}$):
\be \label{e:TP:Xi_Ds_K}
    \encadre{\w{\Xi} = 
    -\frac{1}{2N} \vec{\w{q}}^* (\w{D}\underline{\w{s}} + \w{K})} .
\ee
Contracting equation (\ref{e:TP:Xi_ab_DS_K}) with $q^{\alpha\beta}$
gives the transversal expansion scalar [cf. Eq.~(\ref{e:KI:def_theta_k})]:
\be
    \encadre{\theta_{(\w{k})} = 
    - \frac{1}{2N} ( D_i s^i - K_{ij} s^i s^j + K)} ,
\ee
or equivalently
\be \label{e:TP:theta_k_3p1}
     \encadre{\theta_{(\w{k})} = 
    - \frac{1}{2N} (\w{D}\!\cdot\!\w{s} - \w{K}(\w{s},\w{s}) + K )} .
\ee

\subsubsection{3+1 expression of physical parameters}

In Sec.~\ref{s:IH:PhysParam} physical parameters have been associated with 
$\Hor$, when the latter constitutes a NEH.
More specifically, we proceeded by firstly characterizing the radius 
$R_{\mathcal H}$ 
and the angular momentum $J_{\mathcal H}$ in terms of geometrical objects on
$\Hor$ through Eqs. (\ref{e:IH:radius}) and (\ref{e:IH:angmom}), respectively, and then
we introduced expressions for the mass, surface gravity and angular velocity through 
(\ref{e:IH:physparam}). In an analogous manner, in Section \ref{s:IH:multipoles}
mass and angular momentum multipoles $M_n$ and $J_n$ have been expressed in terms of
geometrical multipoles $I_n$ and $L_n$ through 
Eqs.~(\ref{e:IH:massmultipol})-(\ref{e:IH:rotmultipol})
and Eqs.~(\ref{e:IH:I_n})-(\ref{e:IH:L_n}).  

Regarding their expressions
in terms of 3+1 fields,  $R_{\mathcal H}$ and  $I_n$ are already
in an appropriate form, since they are completely defined in terms of the metric
$\w{q}$ living on $\Sp_t\subset \Sigma_t$. In order to obtain a 3+1 expression for the 
angular momentum we make use of Eq. (\ref{e:TP:Omega3+1}), obtaining
\bea
\label{e:TP:angmom}
\encadre{
J_\Hor = \frac{1}{8\pi G} \int_{\Sp_t} \phi^\mu s^\nu K_{\mu\nu} \, d^2V  \ .
}
\eea
Regarding $L_n$ and $J_n$, the relation 
${}^\Hor\dd \w{\Omega} = 2 \mathrm{Im}\Psi_2 {}^2\w{\epsilon}$, 
which follows from (\ref{e:NE:Hdomega}) and (\ref{e:NE:L_lk=0}), permits to express
\bea
\label{e:TP:Im_Psi-2}
\mathrm{Im}\Psi_2 = \frac{1}{2} 
{}^2\epsilon^{\mu\nu}\; \DS_\mu \Omega_\nu \ \ .
\eea
Making use again of (\ref{e:TP:Omega3+1}), $J_n$ reads
\bea
J_n = \sqrt{\frac{4\pi}{2 n + 1}} 
\frac{R_{\mathcal H}^{n+1}}{8\pi G}  \int_{\Sp_t} 
\DS_\mu({}^2\epsilon^{\mu\nu}s^\rho K_{\rho\nu})  Y_{n0}(\theta) d^2V \ .
\eea



\subsection{2+1 decomposition} \label{s:TP:2p1decomp}

As discussed in Sec.~\ref{s:IN:normal_l}, each spatial hypersurface $\Sigma_t$
can be foliated in the vicinity of $\Hor$ by a family of 2-surfaces $(\Sp_{t,u})$
defined by $u={\rm const}$ and such that the intersection $\Sp_t$ of 
$\Sigma_t$ with the null hypersurface $\Hor$ is the element $u=1$
of this family. The foliation $(\Sp_{t,u})$ induces an 
orthogonal {\em 2+1 decomposition} 
of the 3-dimensional Riemannian manifold $(\Sigma_t,\w{\gamma})$, 
in the same manner that the
foliation $(\Sigma_t)$ induces an orthogonal 3+1 decomposition of 
the 4-dimensional Lorentzian manifold $(\M,\w{g})$, as presented in
Sec.~\ref{s:FO}. The 2+1 equivalent of the unit normal vector $\w{n}$ is
then $\w{s}$ and the 2+1 equivalent of the lapse function $N$ is 
the scalar field $M$ defined by Eq.~(\ref{e:IN:def_M}). Indeed 
we have shown the relation $\underline{\w{s}} = M \w{D} u$ 
[Eq.~(\ref{e:IN:us_MDr})], which is similar to the relation 
(\ref{e:FO:def_n}) between $\underline{\w{n}}$ and $\dd t$. 
The only difference is a sign factor, owing to the fact that $\w{n}$
is timelike, whereas $\w{s}$ is spacelike. 

\subsubsection{Extrinsic curvature of the surfaces $\Sp_t$} \label{s:TP:H}

The {\em second fundamental form} (or {\em extrinsic curvature})
of $\Sp_t$, as a hypersurface of $(\Sigma_t,\w{\gamma})$, is
the bilinear form
\be 
	\begin{array}{cccc}
	\w{H}_0 : & \T_p(\Sigma_t)\times\T_p(\Sigma_t) & \longrightarrow & \mathbb{R} \\
		& (\w{u},\w{v}) & \longmapsto & 
                \w{u} \cdot \w{D}_{\vec{\w{q}}(\w{v})} \w{s} .
	\end{array} 
\ee
Notice the similarity
with Eq.~(\ref{e:FO:def_K}) defining the extrinsic curvature
$\w{K}$ of $\Sigma_t$ and with Eq.~(\ref{e:KI:Theta_grad_l_qq})
defining the second fundamental form $\w{\Theta}$ of $\Hor$.
Following our 4-dimensional point of view, we extend the definition 
of $\w{H}_0$ to $\T_p(\M)\times\T_p(\M)$, via the mapping  $\vec{\w{\gamma}}^*$
[cf. Eq.~(\ref{e:FO:def_gamma_star})]:
\be
    \w{H} := \vec{\w{\gamma}}^* \w{H}_0 . 
\ee 
Then, for any pair of vectors $(\w{u},\w{v})$ in $\T_p(\M)$, 
$\w{H}(\w{u},\w{v}) = \vec{\w{\gamma}}(\w{u}) 
\cdot \w{D}_{\vec{\w{q}}(\w{v})} \w{s}$. Actually the projector
$\vec{\w{\gamma}}$ in front of $\w{u}$ is not necessary since
$\w{D}_{\vec{\w{q}}(\w{v})} \w{s}$ is tangent to $\Sigma_t$, so that
we can write
\be \label{e:TP:def_H}
	\begin{array}{cccc}
	\w{H} : & \T_p(\M)\times\T_p(\M) & \longrightarrow & \mathbb{R} \\
		& (\w{u},\w{v}) & \longmapsto & 
                \w{u} \cdot \w{D}_{\vec{\w{q}}(\w{v})} \w{s} .
	\end{array} 
\ee
In index notation,
\be \label{e:TP:def_H_index}
    H_{\alpha\beta} = D_\mu s_\alpha \; q^\mu_{\ \, \beta} . 
\ee
We have, for any $(\w{u},\w{v})\in\T_p(\M)\times\T_p(\M)$,
\bea
     \w{H}(\vec{\w{q}}(\w{u}),\vec{\w{q}}(\w{v})) 
     & = & \vec{\w{q}}(\w{u}) \cdot \w{D}_{\vec{\w{q}}(\w{v})} \w{s}
        \label{e:TP:Hquqv1} \\
     & = & \left( \vec{\w{\gamma}}(\w{u}) - \langle \underline{\w{s}},\w{u}
     \rangle \w{s} \right) \cdot \w{D}_{\vec{\w{q}}(\w{v})} \w{s} \nonumber \\
     & = & \w{H}(\w{u},\w{v}) , \label{e:TP:Hquqv2}
\eea
for $\w{s}\cdot\w{s}=1$ implies 
$\w{s}\cdot \w{D}_{\vec{\w{q}}(\w{v})} \w{s}=0$.
Combining (\ref{e:TP:Hquqv1}) and (\ref{e:TP:Hquqv2}), we realize that
\be \label{e:TP:H_qstar_Ds}
    \w{H} = \vec{\w{q}}^* \w{D}\underline{\w{s}} ,
\ee
or equivalently
\be
    H_{\alpha\beta} = D_\nu s_\mu \; q^\mu_{\ \,\alpha} q^\nu_{\ \,\beta} ,  
\ee
which strengthens Eq.~(\ref{e:TP:def_H_index}). 
Since 
$\w{D}\underline{\w{s}} = \vec{\w{\gamma}}^* \w{\nabla}\underline{\w{s}}$
[cf. Eq.~(\ref{e:FO:DT_gamma_star_nabT})] and 
$\vec{\w{q}}^* \vec{\w{\gamma}}^* = \vec{\w{q}}^* $, we 
obtain from Eq.~(\ref{e:TP:H_qstar_Ds}) that
\be \label{e:TP:H_qstar_nabla_s}
    \w{H} = \vec{\w{q}}^* \w{\nabla}\underline{\w{s}} .
\ee
Substituting Eq.~(\ref{e:us_dt_dr}) for $\underline{\w{s}}$ in this expression
leads to
\bea
    H_{\alpha\beta} & = & \nabla_\nu\left( N \nabla_\mu t + M \nabla_\mu u
        \right) q^\mu_{\ \,\alpha} q^\nu_{\ \,\beta} \nonumber \\
        & = & \left( \nabla_\nu N \nabla_\mu t + N \nabla_\nu \nabla_\mu t
        + \nabla_\nu M \nabla_\mu u + M \nabla_\nu \nabla_\mu u \right)
        q^\mu_{\ \,\alpha} q^\nu_{\ \,\beta} \nonumber \\
        & = & \left(  N \nabla_\nu \nabla_\mu t
         + M \nabla_\nu \nabla_\mu u \right)
        q^\mu_{\ \,\alpha} q^\nu_{\ \,\beta} , \label{e:TP:H_sym}
\eea
where we have used $q^\mu_{\ \,\alpha} \nabla_\mu t = - N^{-1}
q^\mu_{\ \,\alpha} n_\mu = 0$ and 
$q^\mu_{\ \,\alpha} \nabla_\mu u = e^{-\rho}
q^\mu_{\ \,\alpha} \ell_\mu = 0$.
Since $\nabla_\nu \nabla_\mu f = \nabla_\mu \nabla_\nu f$ for any
scalar field $f$ (vanishing of $\w{\nabla}$'s torsion), 
Eq.~(\ref{e:TP:H_sym}) allows us to conclude that the bilinear 
form $\w{H}$ is symmetric. This property, which is shared by 
the other second fundamental forms $\w{K}$ and $\w{\Theta}$, 
arises from the orthogonality of $\w{s}$ with respect to some
surface ($\Sp_t$) and is a special case of what is
referred to as the {\em Weingarten theorem}. 

Expanding the $\vec{\w{q}}$ in the definition (\ref{e:TP:def_H_index})
of $\w{H}$ leads to
\be \label{e:TP:H_Ds_accs}
    H_{\alpha\beta} = D_\mu s_\alpha \; q^\mu_{\ \, \beta} 
    = D_\mu s_\alpha \; ( \gamma^\mu_{\ \, \beta} - s^\mu s_\beta)
    = D_\beta s_\alpha - s^\mu D_\mu s_\alpha \, s_\beta .  
\ee
Let us evaluate the ``acceleration'' term $s^\mu D_\mu s_\alpha$
which appears in this expression. We have
\be
   s^\mu D_\mu s_\alpha = s^\mu \gamma^\rho_{\ \, \mu} 
   \gamma^\sigma_{\ \, \alpha} \nabla_\rho s_\sigma
   = s^\rho \nabla_\rho s_\sigma \; \gamma^\sigma_{\ \, \alpha},
\ee
hence
\be \label{e:TP:Ds_us_prov}
    \w{D}_{\w{s}} \, \underline{\w{s}} = \vec{\w{\gamma}}^*
       \w{\nabla}_{\w{s}}\,  \underline{\w{s}} .
\ee
$\w{\nabla}_{\w{s}}\,  \underline{\w{s}}$ is easily expressed in terms
of the exterior derivative of $\underline{\w{s}}$:
indeed, $(\w{s}\cdot\dd\underline{\w{s}})_\alpha = 
s^\mu (\nabla_\mu s_\alpha - \nabla_\alpha s_\mu) = 
s^\mu \nabla_\mu s_\alpha$, since $s^\mu \nabla_\alpha s_\mu = 0$
for $\w{s}$ has a fixed norm. 
Thus Eq.~(\ref{e:TP:Ds_us_prov}) becomes
\bea
    \w{D}_{\w{s}}\,  \underline{\w{s}} & = & \vec{\w{\gamma}}^*
    (\w{s}\cdot \dd \underline{\w{s}} )
    = \vec{\w{\gamma}}^*
   \left[  \w{s}\cdot \dd ( N\dd t + M \dd u) \right] \nonumber \\
    & = & \vec{\w{\gamma}}^* \left[
    \w{s}\cdot (\dd N \wedge \dd t + \dd M \wedge \dd u)\right]
    \nonumber \\
 &= & \vec{\w{\gamma}}^* \Big( \langle \dd N, \w{s} \rangle \dd t
 -  \underbrace{\langle \dd t, \w{s} \rangle}_{=0} \dd N
 +  \langle \dd M, \w{s} \rangle \dd u
 - \underbrace{\langle \dd u, \w{s} \rangle}_{=M^{-1}} \dd M \Big) \nonumber \\  
 & = & \langle \dd N, \w{s} \rangle 
 \underbrace{\vec{\w{\gamma}}^* \dd t}_{=0} + 
 \langle \dd M, \w{s} \rangle 
    \underbrace{\vec{\w{\gamma}}^* \dd u}_{=M^{-1} \underline{\w{s}}}
      - M^{-1} \vec{\w{\gamma}}^* \dd M \nonumber \\
   & = & -  \vec{\w{\gamma}}^* \dd \ln M + 
   \langle \dd \ln M, \w{s} \rangle \underline{\w{s}} 
   = - \vec{\w{q}}^* \dd \ln M,
\eea
from which we conclude that 
\be \label{e:TP:Ds_s_DlnM}
    \encadre{\w{D}_{\w{s}} \, \underline{\w{s}} = - \w{\DS}\ln M }, 
\ee
which is a relation similar to Eq.~(\ref{e:FO:acceler_n}).
If we used it to replace $s^\mu D_\mu s_\alpha$ in Eq.~(\ref{e:TP:H_Ds_accs})
and use the symmetry of $\w{H}$,
we get [compare with (\ref{e:FO:K_grad_n_index})]
\be
 H_{\alpha\beta}= D_\alpha s_\beta +  s_\alpha \,\DS_\beta \ln M 
\ee
or equivalently, 
\be \label{e:TP:H_Ds}
    \encadre{\w{H} = \w{D}\underline{\w{s}} + \w{\DS}\ln M \otimes 
        \underline{\w{s}} } . 
\ee
Again note the sign differences with Eq.~(\ref{e:FO:K_grad_n}).

The {\em mean curvature} of $\Sp_t$, as
a surface embedded in $(\Sigma_t,\w{\gamma})$, is 
given by half the trace of $\vec{\w{H}}$:
\be \label{e:TP:def_trH}
    H := \mathrm{tr}\, \vec{\w{H}} = H^\mu_{\ \ \mu} = 
    g^{\mu\nu} H_{\mu\nu} 
    = q^{\mu\nu} H_{\mu\nu} = q^{ab} H_{ab} 
	= H^a_{\ \, a} . 
\ee
From Eq.~(\ref{e:TP:H_Ds}), $H$ is equal to 
the 3-dimensional divergence of the unit normal to $\Sp_t$:
\be \label{e:TP:trH_divs}
    H = \w{D}\!\cdot\!\w{s} . 
\ee
\begin{rem}
We recover immediately from this expression
that for a sphere in the Euclidean space $\mathbb{R}^3$, the mean
curvature is nothing but the inverse of the radius.
Indeed Eq.~(\ref{e:TP:trH_divs}) yields $H=2/R$, where $R$ is the
radius of the sphere.
In the Riemannian 3-manifold $(\Sigma_t,\w{\gamma})$, $H$ may 
vanish if $\Sp_t$ is a minimal surface, in the very same manner
that $K$ vanishes if $\Sigma_t$ is a maximal hypersurface of spacetime.
Note that minimal surfaces have been used as inner boundaries in 
the numerical construction of black initial data by many authors
\cite{Misne63,Lindq63,BowenY80,Cook91,CookCDKMO93,Cook94,PfeifTC00,DieneJKN00,%
GourgGB02,GrandGB02}.
\end{rem}

\subsubsection{Expressions of $\w{\Theta}$ and $\w{\Xi}$ in terms of $\w{H}$}
\label{s:TP:Th_Xi_H}

Combining Eqs.~(\ref{e:TP:H_qstar_Ds}) and
(\ref{e:TP:Theta_Ds_K}), we get a expression of the
second fundamental form of $\Hor$ (associated with $\el$) in terms of 
the second fundamental form of the 2-surface $\Sp_t$ embedded in $\Sigma_t$
(i.e. $\w{H}$)
and the second fundamental form of the 3-surface $\Sigma_t$ embedded in 
$\M$ (i.e. $\w{K}$):
\be \label{e:TP:Theta_H_K}
    \encadre{ \w{\Theta} = N (\w{H} - \vec{\w{q}}^* \w{K}) }. 
\ee
Similarly the expression (\ref{e:TP:Xi_Ds_K}) for the transversal 
deformation rate $\w{\Xi}$ becomes
\be \label{e:TP:Xi_H_K}
    \encadre{ \w{\Xi} = 
    -\frac{1}{2N} (\w{H} + \vec{\w{q}}^* \w{K}) }.
\ee
As a check of formulas (\ref{e:TP:Theta_H_K}) and (\ref{e:TP:Xi_H_K}), 
we can compare them with those in Eq.~(29) of 
Cook \& Pfeiffer \cite{CookP04}, after having noticed that the null 
normal vector $\el'$ used by these authors is $\el' = 
\w{\hat\ell} = (\sqrt{2} N)^{-1} \el$,
with $\w{\hat\ell}$ defined by Eq.~(\ref{e:IN:def_hat_ell}); this
results in a second fundamental form $\w{\Theta}' = (\sqrt{2} N)^{-1} \w{\Theta}$
and a transversal deformation rate 
$\w{\Xi}' = \sqrt{2} N \w{\Xi}$
[cf. the scaling laws in Table~\ref{t:KI:scaling}]\footnote{Also note
that Cook \& Pfeiffer \cite{CookP04} define $\w{\Theta}$ and
$\w{\Xi}$ (denoted
by them $\w{\Sigma}$ and $\w{\acute\Sigma}$ respectively) 
by $\w{\Theta} = 1/2\;  \vec{\w{q}}^* \Lie{\el} \w{g}$
and $\w{\Xi} = 1/2\;  \vec{\w{q}}^* \Lie{\w{k}} \w{g}$,
[their Eqs.~(24) and (25)], whereas we have established that 
$\w{\Theta} = 1/2 \; \vec{\w{q}}^* \Lie{\el} \w{q}$ and
$\w{\Xi} = 1/2 \; \vec{\w{q}}^* \Lie{\w{k}} \w{q}$
[our Eqs.~(\ref{e:KI:Theta_deform}) and (\ref{e:KI:def_Xi})]. 
It can be seen easily that both
expressions for $\w{\Theta}$ and $\w{\Xi}$ coincide, thanks to the  
operator $\vec{\w{q}}^*$ and the relation 
$\w{q} = \w{g} + \uel \otimes \uk + \uk \otimes \uel$.}.

\begin{rem}
Replacing $\w{\Theta}$ and $\w{\Xi}$ by the expressions 
(\ref{e:TP:Theta_H_K}) and (\ref{e:TP:Xi_H_K}), as well as 
$\el$ and $\w{k}$ by their expressions in terms of $\w{n}$ and $\w{s}$
[Eqs.~(\ref{e:IN:el_nps}) and (\ref{e:NH:k_n_s})], into the formula
(\ref{e:KI:CarterK_Theta_Xi}) for the second fundamental tensor
of the 2-surface $\Sp_t$ (cf. Remark~\ref{r:KI:2nd_fund_tensor})
results in 
\be \label{e:TP:CarterK_qK_H}
	 {\mathcal K}^\gamma_{\ \, \alpha\beta} =
		- (\vec{q}^* K)_{\alpha\beta} \; n^\gamma
		- H_{\alpha\beta}  \; s^\gamma  .
\ee
This expression has the same structure than Eq.~(\ref{e:KI:CarterK_Theta_Xi}),
describing $\w{\mathcal K}$ in terms of the timelike-spacelike pair of
normals $(\w{n},\w{s})$, whereas Eq.~(\ref{e:KI:CarterK_Theta_Xi})
describes $\w{\mathcal K}$ in terms of the null-null pair of
normals $(\el,\w{k})$.
\end{rem}

In Appendix~\ref{s:KE}, formul\ae\ (\ref{e:TP:Theta_H_K}) and
(\ref{e:TP:Xi_H_K}) are used to evaluate $\w{\Theta}$ and $\w{\Xi}$
in the specific case of the event horizon of a Kerr black hole. 
In particular Eq.~(\ref{e:TP:Theta_H_K}) leads to $\w{\Theta}=0$
as expected for a stationary horizon. 

Thanks to Eq.~(\ref{e:TP:trH_divs}), the expressions 
(\ref{e:TP:theta_3p1}) and (\ref{e:TP:theta_k_3p1}) for the expansion
scalars $\theta$ and $\theta_{(\w{k})}$ become
\be \label{e:TP:theta_H_K}
   \encadre{ \theta = N \left( H-K +\w{K}(\w{s},\w{s}) \right) }
\ee
and 
\be \label{e:TP:thetak_H_K}
 \encadre{ \theta_{(\w{k})} = 
    - \frac{1}{2N} \left( H+K - \w{K}(\w{s},\w{s})\right) } .
\ee

\begin{rem} \label{r:TP:already_2p1}
Equations (\ref{e:TP:Theta_H_K}) and (\ref{e:TP:theta_H_K}) constitute
a 2+1 writing of $\w{\Theta}$ and $\theta$. 
We had obtained a different (but equivalent !) 2+1 writing 
in Sec.~\ref{s:KI:deform_rate}, via Eqs.~(\ref{e:KI:Theta_Kil_V}) and
(\ref{e:KI:theta_q_Lie_t_q}).  
\end{rem}


\subsection{Conformal decomposition} \label{s:TP:conf_decomp}

\subsubsection{Conformal 3-metric}

In many modern applications of the 3+1 formalism in numerical relativity,
a conformal decomposition of the spatial metric $\w{\gamma}$ is performed. 
This includes the 3+1 initial data problem, following the works
of Lichnerowicz \cite{Lichn44} and York and collaborators
\cite{York73,OMurcY74,York99,PfeifY03} and the time evolution schemes
proposed by Shibata \& Nakamura \cite{ShibaN95} and Baumgarte \& 
Shapiro \cite{BaumgS99}, as well as the recent constrained scheme 
based on Dirac gauge proposed by Bonazzola et al. \cite{BonazGGN04}.
The conformal decomposition consists in writing
\be \label{e:TP:def_conf_metr}
    \w{\gamma} = \Psi^4 \, \w{\tilde \gamma} , 
\ee 
where $\Psi$ is some scalar field. Very often, $\Psi$ is chosen so 
that $\w{\tilde \gamma}$ is unimodular. As shown in Ref.~\cite{BonazGGN04},
this can be achieved without making $\Psi$ and $\w{\tilde \gamma}$
tensor densities by introducing a background flat metric $\w{f}$
on $\Sigma_t$. The conformal factor $\Psi$ is then defined by
\be
    \Psi = \left( \frac{\det \gamma_{ij}}{\det f_{ij}} \right) ^{1/12} , 
                                        \label{e:TP:Psi_det}
\ee
where $\det \gamma_{ij}$ (resp. $\det f_{ij}$) is the determinant of the
component of $\w{\gamma}$ (resp. $\w{f}$) with respect to a 
coordinate system $(x^i)$ on $\Sigma_t$. The quotient 
of the two determinants 
is independent of the coordinates $(x^i)$, so that $\Psi$ is a genuine scalar
field and $\w{\tilde\gamma}$ a genuine tensor field. The unimodular character
of $\w{\tilde \gamma}$ then translates into 
\be
    \det\tilde\gamma_{ij} = \det f_{ij} , 
\ee
with $\det f_{ij} = 1$ if $(x^i)$ are coordinates of Cartesian type. 

Let us denote by $\w{\tD}$ the connection on $\Sigma_t$ compatible
with the metric $\w{\tilde\gamma}$.
The $\w{D}$-derivative and $\w{\tD}$-derivative of any vector 
$\w{v}\in\T(\Sigma_t)$ or any 1-form $\w{\varpi}\in\T^*(\Sigma_t)$
are related by
\be \label{e:TP:connect_deriv}
    D_i v^j = \tD_i v^j + C^j_{\ \, ki} v^k 
    \qquad\mbox{and}\qquad
    D_i \varpi_j = \tD_i \varpi_j - C^k_{\ \, ji} \varpi_k ,  
\ee
with 
\bea
    C^k_{\ \, ij} & := & \frac{1}{2} \gamma^{kl} \left( \tD_i \gamma_{lj}
        + \tD_j \gamma_{il} - \tD_l \gamma_{ij} \right) \nonumber  \\
        & = & 2 \left( \tD_i \ln \Psi \, \delta^k_{\ \, j}
            + \tD_j \ln \Psi \, \delta^k_{\ \, i}
            - \tD^k\ln \Psi \tilde\gamma_{ij} \right) . 
\eea
For any tensor field $\w{T}$ on $\Sigma_t$, we define 
$\w{\tD}\w{T}$ as a tensor field on $\M$ by
\be \label{e:TP:tD-4D}
    \w{\tD}\w{T} = \vec{\w{\gamma}}^* \; {}^{\Sigma_t}\!\w{\tD}\w{T} , 
\ee
where ${}^{\Sigma_t}\!\w{\tD}\w{T}$ is the original definition of
the $\w{\tD}$-derivative of $\w{T}$ within the manifold $\Sigma_t$,
as introduced above. Actually Eq.~(\ref{e:TP:tD-4D}) allows us to
manipulate $\w{\tD}$-derivatives as 4-dimensional objects, as we have
done already for $\w{D}$-derivatives. 

\subsubsection{Conformal decomposition of $\w{K}$} \label{s:TP:conf_dec_K}

In addition to the conformal decomposition of the spatial metric
$\w{\gamma}$, the conformal 3+1 formalism is based on a conformal 
decomposition of $\Sigma_t$'s extrinsic curvature $\w{K}$:
\be \label{e:TP:conf_decomp_K}
   \encadre{ \w{K} =: \Psi^\zeta  \w{\tilde A} + \frac{1}{3} K \w{\gamma} }, 
\ee 
where $\w{\tilde A}$ captures the traceless part of $\w{K}$:
$\gamma^{ij} \tilde A_{ij} = \tilde\gamma^{ij} \tilde A_{ij} = 0$ 
and the exponent 
$\zeta$ is usually chosen to be $-2$ or $4$. The choice $\zeta=-2$
has been introduced by Lichnerowicz \cite{Lichn44} and 
is called the {\em conformal transverse-traceless} decomposition of
$\w{K}$ \cite{Cook00,York79}; it leads to an expression of the 
momentum constraint equation (\ref{e:FO:mom_constraint})
which is independent of $\Psi$ for
maximal slices ($K=0$) in vacuum spacetimes. It has been notably
used to get the Bowen-York semi-analytical initial data for black
hole spacetimes \cite{BowenY80}. The choice $\zeta=4$ is called the 
{\em physical transverse-traceless} decomposition of $\w{K}$ 
\cite{Cook00,OMurcY74} and leads to an expression of $\w{\tilde A}$
in terms of the time derivative of the conformal metric, the shift vector
and the lapse function which is independent of $\Psi$ 
[Eq.~(\ref{e:TP:A_dottgam_lbeta}) below]. 
This choice has been employed mostly in time evolution studies
\cite{ShibaN95,BaumgS99,BonazGGN04}. In the present article, we
do not choose a specific value for $\zeta$, so that the results are
valid for both of the cases above. 

Let us consider a coordinate system $(x^i)$ on each hypersurface
$\Sigma_t$ so that $(x^\alpha)=(t,x^i)$ constitute a smooth 
coordinate system on $\M$. We denote the shift vector of these 
coordinates by $\w{\beta}$. 
The coordinate time vector is
then $\tv = N \w{n} + \w{\beta}$ [Eq.~(\ref{e:FO:def_shift})]
and we define the time derivative
of the conformal metric $\w{\tilde\gamma}$ by
\be \label{e:TP:def_tgam_dot}
    \w{\dot{\tilde\gamma}} := \Lie{\tv} \w{\tilde\gamma} . 
\ee 
Written in terms of tensor components with respect to $(x^i)$,
this definition becomes
\be
   \dot{\tilde\gamma}_{ij} := \der{\tilde\gamma_{ij}}{t}
    = - \tilde\gamma_{ik} \tilde\gamma_{jl}  \der{\tilde\gamma^{kl}}{t} ,
\ee
where the second equality follows from 
$\tilde\gamma^{ik} \tilde\gamma_{kj} = \delta^i_{\ \, j}$. 
The trace of the relation (\ref{e:FO:gam_evol}) between the extrinsic
curvature $\w{K}$ and the time derivative of the metric $\w{\gamma}$
leads to the following evolution equation for the conformal factor $\Psi$
\be
\label{e:TP:Psidot}
    \der{}{t} \ln\Psi - \Lie{\w{\beta}} \ln\Psi =
        \frac{1}{6} \left( \w{\tD}\!\cdot\!\w{\beta} - N K \right) , 
\ee
whereas its traceless part gives a relation between $\w{\tilde A}$ and
$\w{\dot{\tilde\gamma}}$:
\be \label{e:TP:A_dottgam_lbeta}
    \encadre{ \w{\tilde A} = \frac{\Psi^{4-\zeta}}{2N}
        \left[ \Kil{\w{\tD}}{\underline{\w{\tilde\beta}}}
         - \frac{2}{3} (\w{\tD}\!\cdot\!\w{\beta})\, 
         \w{\tilde\gamma}
         - \w{\dot{\tilde\gamma}}            \right] },
\ee
where $\w{\tD}$ denotes the connection associated
with the conformal metric $\w{\tilde\gamma}$ and
$\underline{\w{\tilde\beta}}$ is the 1-form dual to the shift
vector via the conformal metric:
\be
    \underline{\w{\tilde\beta}} := \w{\tilde\gamma}(\w{\beta},.) 
        = \Psi^{-4} \underline{\w{\beta}},      \label{e:TP:def_utbeta}
\ee
or in index notation:
\be
    {\tilde\beta}_i := \tilde\gamma_{ij} \beta^j = \Psi^{-4} \beta_j . 
\ee
Inserting Eq.~(\ref{e:TP:A_dottgam_lbeta}) into Eq.~(\ref{e:TP:conf_decomp_K})
leads to the following expression of the extrinsic curvature
\be \label{e:TP:K_decomp}
    \w{K} = \frac{\Psi^4}{2N} \left[
    \Kil{\w{\tD}}{\underline{\w{\tilde\beta}}} 
    + \frac{2}{3} (NK - \w{\tD}\!\cdot\!\w{\beta})\, 
         \w{\tilde\gamma} - \w{\dot{\tilde\gamma}} \right]  ,    
\ee
which is independent of $\zeta$.

\subsubsection{Conformal geometry of the 2-surfaces $\Sp_t$}

The conformal metric $\w{\tilde\gamma}$ induces an intrinsic and
an extrinsic geometry of the 2-surfaces $\Sp_t\subset\Sigma_t$. First of all, 
the vector normal to $\Sp_t$ and with unit length with respect to
$\w{\tilde\gamma}$ is 
\be \label{e:TP:def_ts}
    \w{\tilde s} := \Psi^2 \w{s} . 
\ee
We denote by $\underline{\w{\tilde s}}$ the 1-form dual to it with respect
to the metric $\w{\tilde\gamma}$ \footnote{This notation does not follow
the underlining convention stated in Sec.~\ref{s:IN:index_intrinsic}
[cf. Eq.~(\ref{e:IN:underbar})],
namely $\underline{\w{\tilde s}}$ is not the dual to $\w{\tilde s}$
provided by the metric $\w{g}$ (=$\w{\gamma}$ on $\Sigma_t$).}
\be \label{e:TP:def_uts}
   \underline{\w{\tilde s}} := \w{\tilde\gamma}(\w{\tilde s}, . ) 
   = \Psi^{-2} \w{\gamma}(\w{s},.)
   = \Psi^{-2} \underline{\w{s}} , 
\ee
or in components,
\be \label{e:TP:def_uts_index}
    \tilde s_\alpha := \tilde\gamma_{\alpha\mu} \, \tilde{s}^\mu = \Psi^{-2} s_\alpha . 
\ee
Combining Eqs.~(\ref{e:IN:us_MDr}) and (\ref{e:TP:def_ts}), we get
\be \label{e:TP:uts_dr}
   \underline{\w{\tilde s}} = \tilde M \, \w{D} u ,  
\ee 
with 
\be \label{e:TP:tM_M}
    \tilde M := \Psi^{-2} M . 
\ee
Let us notice that the orthogonal projector onto the 2-surface $\Sp_t$
is the same for both metrics $\w{\gamma}$ and $\w{\tilde\gamma}$:
\be
  \vec{\w{q}} = \vec{\w{\gamma}} - \langle \underline{\w{s}}, . \rangle \, \w{s}  
  = \vec{\w{\gamma}} - \langle \underline{\w{\tilde s}}, . \rangle 
  \, \w{\tilde s}  . 
\ee 

As a hypersurface of $\Sigma_t$ endowed with the conformal metric 
$\w{\tilde\gamma}$, the first fundamental form of $\Sp_t$ is
\be \label{e:TP:def_tilde_q}
    \w{\tilde q} = \Psi^{-4}  \w{q} = \w{\tilde\gamma}
    - \underline{\w{\tilde s}} \otimes \underline{\w{\tilde s}},
\ee
and its second fundamental form
is defined by a formula similar to Eq.~(\ref{e:TP:def_H}),
with $\w{D}$ replaced by $\w{\tD}$, $\w{s}$ replaced by
$\w{\tilde s}$ and the scalar product [simply denoted by a dot in 
Eq.~(\ref{e:TP:def_H})] taken with $\w{\tilde\gamma}$:
\be
	\begin{array}{cccc}
	\w{\tilde H}: & \T_p(\M)\times\T_p(\M) & \longrightarrow & \mathbb{R} \\
		& (\w{u},\w{v}) & \longmapsto & 
                \w{\tilde \gamma}\left( \w{u}, 
                    \w{\tD}_{\vec{\w{q}}(\w{v})} \w{\tilde s} \right).
	\end{array} 
\ee
In index notation, one has
\be \label{e:TP:tH_Dts_q_index}
    \tilde H_{\alpha\beta} = \tilde\gamma_{\alpha\mu} \, q^\nu_{\ \, \beta} \, 
        \tD_\nu {\tilde s}^\mu =
        \tD_\nu {\tilde s}_\alpha \, q^\nu_{\ \, \beta}   .
\ee
Similarly to $\w{H}$, $\w{\tilde H}$ is symmetric and we have the following
properties:
\be  \label{e:TP:tH_qstar_tDts}
    \w{\tilde H} = \vec{\w{q}}^* \w{\tD} \w{\tilde s} .
\ee

The ``acceleration'' $\w{\tD}_{\w{\tilde s}} \, \underline{\w{\tilde s}}$
is given by a formula similar to Eq.~(\ref{e:TP:Ds_s_DlnM})
\be \label{e:TP:Dts_ts_DlntM}
    \encadre{\w{\tD}_{\w{\tilde s}} \, \underline{\w{\tilde s}} = 
        - \w{\tDS}\ln \tilde M } ,       
\ee
where $\w{\tDS}$ denotes the connection associated with the conformal metric
$\w{\tilde q}$ in $\Sp_t$. The $\w{\tDS}$-derivatives of a tensor field
can be expressed in terms of $\w{\tD}$ by a projection formula identical
to Eq.~(\ref{e:KI:2der}), except for $\w{\nabla}$ in the right-hand
side replaced by $\w{\tD}$. 
It is  actually easy to establish Eq.~(\ref{e:TP:Dts_ts_DlntM})
 from Eq.~(\ref{e:TP:Ds_s_DlnM}):
the $\w{D}$-derivative and $\w{\tD}$-derivative of the 1-form 
$\underline{\w{s}}$
are related by Eq.~(\ref{e:TP:connect_deriv}).
Substituting $\Psi^2 \tilde s_j$ for $s_j$ [Eq.~(\ref{e:TP:def_uts_index})], 
we get
\be \label{e:TP:Ds_tDts}
    D_i s_j = \Psi^2 \left(  \tD_i \tilde s_j
        - 2 \tilde s_i \tD_j \ln\Psi + 2 \tilde s^k \tD_k \ln\Psi \, 
        \tilde\gamma_{ij}\right) ,  
\ee
from which we obtain $s^k D_k s_i = \tilde s^k \tD_k \tilde s_i
- 2 \tD_i\ln\Psi + 2 \tilde s^k \tD_k \ln\Psi \, \tilde s_i$.
Substituting Eq.~(\ref{e:TP:Ds_s_DlnM}) for $s^k D_k s_i$ and replacing
$M$ by $\Psi^2 \tilde M$ then leads to Eq.~(\ref{e:TP:Dts_ts_DlntM}).

Expressing $q^\nu_{\ \, \beta} = \gamma^\nu_{\ \, \beta} - \tilde s^\nu \tilde s_\beta$
in Eq.~(\ref{e:TP:tH_Dts_q_index}) and using Eq.~(\ref{e:TP:Dts_ts_DlntM})
together with the symmetry of $\w{\tilde H}$ leads to 
\be \label{e:TP:tH_tDts_index}
    \tilde H_{\alpha\beta} = \tD_\alpha \tilde s_\beta
     +  \tilde s_\alpha \,\tDS_\beta \ln \tilde M  , 
\ee
or
\be
    \w{\tilde H} = \w{\tD} \underline{\w{\tilde s}}
        + \w{\tDS} \ln \tilde M \otimes \underline{\w{\tilde s}} .
\ee
Let us denote by $\tilde H$ the trace of $\w{\tilde H}$ with respect
to the conformal metric $\w{\tilde\gamma}$:
\be 
    \tilde H := \tilde\gamma^{ij} \tilde H_{ij} . 
\ee
We then have, similarly to Eq.~(\ref{e:TP:trH_divs}), 
\be
 \tilde H = \w{\tD}\!\cdot\!\w{\tilde s}
\ee

If we substitute $\w{D}\underline{\w{s}}$ in 
Eq.~(\ref{e:TP:H_qstar_Ds}) by its expression (\ref{e:TP:Ds_tDts})
in terms of $\w{\tD}\underline{\w{\tilde s}}$ and compare with 
Eq.~(\ref{e:TP:tH_qstar_tDts}), we get 
\be \label{e:TP:H_tH_gradPsi}
    \encadre{ \w{H} = \Psi^2 \left[ \w{\tilde H}
        + 2 (\w{\tD}_{\w{\tilde s}} \ln\Psi) \; \w{\tilde q} \right] } . 
\ee
The trace of this equation writes
\be  \label{e:TP:trace_H}
    H = \Psi^{-2} \left( \tilde H + 4 \w{\tD}_{\w{\tilde s}} \ln\Psi \right) . 
\ee

\subsubsection{Conformal 2+1 decomposition of the shift vector}

As in Sec.~\ref{s:TP:conf_dec_K}, we consider a coordinate system $(x^i)$
on $\Sigma_t$ and the associated shift vector $\w{\beta}$. 
In Sec.~\ref{s:IN:stacoord} we have already introduced the 2+1
orthogonal decomposition of $\w{\beta}$ with respect to the surface $\Sp_t$
[cf. Eq.~(\ref{e:IN:shift_2p1})]:
\be
     \w{\beta} = b\, \w{s} - \w{V} \qquad \mbox{with} \quad \w{V}\in\T_p(\Sp_t) .
\ee
Let us re-write this decomposition as 
\be \label{e:TP:beta_decomp}
    \w{\beta} = \tilde b\,  \w{\tilde s} - \w{V}, 
\ee
with 
\be
   \encadre{ \tilde b := \Psi^{-2} b 
    = \w{\tilde\gamma}(\w{\tilde s},\w{\beta}) 
    = \langle \underline{\w{\tilde s}},\w{\beta} \rangle}.
\ee
If $(t,x^i)$ constitutes a coordinate system stationary with
respect to $\Hor$, then $b = N$ on $\Hor$ [Eq.~(\ref{e:IN:b_N_station})],
so that 
\be \label{e:TP:b_NPsi2_H}
    \mbox{( $(x^\alpha)$ stationary w.r.t. $\Hor$\ )}
    \iff \tilde b \equalH N \Psi^{-2} ,
\ee
where $\equalH$ means that the equality holds only on $\Hor$. 

As a consequence of Eq.~(\ref{e:TP:beta_decomp}), 
the 1-form $\underline{\w{\tilde\beta}}$
defined by Eq.~(\ref{e:TP:def_utbeta}) has the
2+1 decomposition
\be
    \underline{\w{\tilde\beta}} = {\tilde b}\, \underline{\w{\tilde s}}
    - \underline{\w{\tilde V}} , 
\ee
with 
\be \label{e:TP:def_tilde_uV}
    \underline{\w{\tilde V}} := \w{\tilde\gamma}(\w{V},.) 
        = \Psi^{-4} \underline{\w{V}} .      
\ee
In index notation:
\be
    \tilde\beta_i := {\tilde b}\tilde s_i - \tilde V_i 
    \qquad \mbox{with}\quad \tilde V_i = \tilde\gamma_{ij} V^j 
        = \Psi^{-4} V_i . 
\ee
From this expression and Eq.~(\ref{e:TP:tH_tDts_index}),
we get
\bea
    \tD_i\tilde\beta_j + \tD_j\tilde\beta_i &=& b\left( 2\tilde H_{ij}
        - \tilde s_i \tDS_j \ln\tilde M - \tilde s_j \tDS_i \ln\tilde M \right)
        +  \tilde s_i \, \tD_j {\tilde b}+ \tilde s_j \, \tD_i {\tilde b} \nonumber\\
        & & - \tD_i \tilde V_j  - \tD_j \tilde V_i 
\eea
and
\be \label{e:TP:tdiv_beta_2p1}
    \tD_i \beta^i = {\tilde b}\tilde H + \tilde s^i \tD_i {\tilde b}- \tDS_a V^a
        - V^a \, \tDS_a \ln\tilde M . 
\ee
Injecting these last two relations into the expression
(\ref{e:TP:K_decomp}) of the extrinsic curvature, we get
\bea
    K_{ij} & = & \frac{\Psi^4}{2N} \Bigg[  
    b\left( 2\tilde H_{ij}
        - \tilde s_i \tDS_j \ln\tilde M - \tilde s_j \tDS_i \ln\tilde M \right)
        +  \tilde s_i \, \tD_j {\tilde b}+ \tilde s_j \, \tD_i {\tilde b} 
        - \tD_i \tilde V_j   \nonumber\\
     & &  - \tD_j \tilde V_i
     +\frac{2}{3} \left(NK -  {\tilde b}\tilde H - \tilde s^k \tD_k {\tilde b}+ \tDS_a V^a
        + V^a \, \tDS_a \ln\tilde M  \right) \tilde\gamma_{ij} \nonumber\\
     & &  - \dot{\tilde\gamma}_{ij}  \Bigg] \label{e:TP:K_dec2p1_conf_index}
\eea 
We deduce immediately from this expression the scalar $\w{K}(\w{s},\w{s})$
which appears in formulae of Sec.~\ref{s:TP:3p1decomp} and 
\ref{s:TP:2p1decomp}: 
\bea
    \w{K}(\w{s},\w{s}) & = & \frac{1}{N} \Bigg[ 
    \frac{1}{3} \left( 2\tilde s^k\tD_k {\tilde b}- 2 V^a \, \tDS_a \ln \tilde M
        + NK - {\tilde b}\tilde H + \tDS_a V^a \right) \nonumber \\
         & &   - \frac{1}{2} \dot{\tilde\gamma}_{kl} \tilde s^k \tilde s^l 
            \Bigg] . \label{e:TP:Kss}
\eea
We deduce also from Eq.~(\ref{e:TP:K_dec2p1_conf_index}) that
\bea
    K_{kl} s^k q^l_{\ \, i} & = & \frac{\Psi^2}{2N} \left(
    \tDS_i {\tilde b}- {\tilde b}\, \tDS_i \ln\tilde M - \tilde s^k \tDS_k \tilde V_l
     q^l_{\ \, i} + \tilde H_{ik} V^k - \dot{\tilde\gamma}_{kl} 
     s^k q^l_{\ \, i} \right) \nonumber \\
     & & \label{e:TP:Ksq}
\eea
and
\bea
   K_{kl} q^k_{\ \, i} q^l_{\ \, j} & = & 
     \frac{\Psi^4}{2N}  \Bigg[ 2 {\tilde b}\tilde H_{ij} - \tDS_i \tilde V_j
     - \tDS_j \tilde V_i 
     - \dot{\tilde\gamma}_{kl} \, q^k_{\ \, i} q^l_{\ \, j}  \nonumber \\
        & &
     + \frac{2}{3} \left( N K - {\tilde b}\tilde H
     - \tilde s^k \tD_k {\tilde b}+ \tDS_a V^a + V^a \tDS_a\ln\tilde M \right)
        \tilde q_{ij} \Bigg] .      \label{e:TP:Kqq_index}
\eea

\subsubsection{Conformal 2+1 decomposition of $\Hor$'s fields}

We are now in position to give expressions 
of the various fields related to $\Hor$'s null geometry
in terms of conformal 2+1 quantities. 
First of all, replacing $\w{K}(\w{s},\w{s})$ by formula (\ref{e:TP:Kss})
in Eq.~(\ref{e:TP:kappa_3p1}) leads to
\bea
    \encadre{
        \begin{array}{rcl}
    \kappa & = &\displaystyle \ell^\mu \nabla_\mu \ln N + \Psi^{-2} \tilde s^k \tD_k N
    + \frac{1}{2} \, \dot{\tilde\gamma}_{kl} \, \tilde s^k \tilde s^l 
            \\
    & & \displaystyle + \frac{1}{3} \left( {\tilde b}\tilde H - N K 
    + 2 V^a \, \tDS_a \ln \tilde M -  \tDS_a V^a - 2\tilde s^k \tD_k \tilde b
    \right) .
    \end{array}} 
\eea
For a coordinate system stationary with respect to $\Hor$,
one has ${\tilde b}\equalH N \Psi^{-2}$
[Eq.~(\ref{e:TP:b_NPsi2_H})] and $\el \equalH \w{t} + \w{V}$
[Eq.~(\ref{e:IN:el_t_V_station})], so that the above expression can 
be written
\be
     \encadre{
        \begin{array}{rcl}
        \kappa & \stackrel{\Hor,\mathrm{sc}}{=} & \displaystyle
   \der{}{t}\ln N + V^a \, \tDS_a \ln N + \Psi^{-2} \tilde s^k \tD_k N 
    + \frac{N}{3} \left( \Psi^{-2} \tilde H - K \right) \\[3mm]
    & & \displaystyle 
    + \frac{1}{2} \, \dot{\tilde\gamma}_{kl} \, \tilde s^k \tilde s^l
    +  \frac{2}{3} \left( 
    V^a \, \tDS_a \ln \tilde M -  \frac{1}{2} \tDS_a V^a - 
    \tilde s^k \tD_k {\tilde b}\right) ,
    \end{array} }
\ee
where $\stackrel{\Hor,\mathrm{sc}}{=}$ means that the 
equality holds only on $\Hor$ and for a coordinate system stationary
with respect to $\Hor$. 

Next, replacing $H$ by formula (\ref{e:TP:trace_H}) and
$\w{K}(\w{s},\w{s})$ by formula (\ref{e:TP:Kss})
in Eq.~(\ref{e:TP:theta_H_K}) leads to 
\be
    \encadre{
        \begin{array}{rcl}
    \theta & = &\displaystyle  N \Psi^{-2} \left( 4 \tilde s^k \tD_k\ln \Psi 
        + \frac{2}{3} \tilde H \right) 
        - \frac{1}{2} \, \dot{\tilde\gamma}_{kl} \, \tilde s^k \tilde s^l 
        \\[3mm]
    & &\displaystyle  + \frac{1}{3} \left[ (N\Psi^{-2}-\tilde b) \tilde H
    + 2 \tilde s^k \tD_k {\tilde b}- 2 V^a \, \tDS_a \ln \tilde M 
     + \tDS_a V^a - 2NK \right] .
     \end{array} }      \label{e:TP:theta_2p1conf}
\ee
For a coordinate system stationary with respect to $\Hor$, 
this expression simplifies somewhat, thanks to
the relation (\ref{e:TP:b_NPsi2_H}):
\be
    \encadre{
        \begin{array}{rcl}
        \theta & \stackrel{\Hor,\mathrm{sc}}{=} & \displaystyle
           N \Psi^{-2} \left( 4 \tilde s^k \tD_k\ln \Psi 
        + \frac{2}{3} \tilde H \right) 
     - \frac{1}{2} \, \dot{\tilde\gamma}_{kl} \, \tilde s^k \tilde s^l\\[3mm]
        & & \displaystyle + \frac{2}{3} \left(
     \tilde s^k \tD_k {\tilde b}-  V^a \, \tDS_a \ln \tilde M 
     + \frac{1}{2} \tDS_a V^a - NK \right) .
        \end{array} }
\ee

The expression of $\Hor$'s second fundamental form $\w{\Theta}$ in 
terms of the conformal 2+1 quantities is obtained by replacing
$\w{H}$ and $\vec{\w{q}}^*\w{K}$ in Eq.~(\ref{e:TP:Theta_H_K}) by
their expressions (\ref{e:TP:H_tH_gradPsi}) and (\ref{e:TP:Kqq_index}):
\be
    \encadre{
        \begin{array}{rcl}
    \w{\Theta} & = & \displaystyle \Psi^4 \Bigg\{ (N\Psi^{-2} - \tilde b) 
        \w{\tilde H}
        + \frac{1}{2}\Kil{\w{\tDS}}{\underline{\w{\tilde V}}}
        + \frac{1}{2} \vec{\w{q}}^* \w{\dot{\tilde\gamma}}
         + \Bigg[ 2N\Psi^{-2} \w{\tD}_{\w{\tilde s}} \ln\Psi \\[3mm]
     & &  \displaystyle  
        + \frac{1}{3} \left( \tilde b \tilde H +
        \w{\tD}_{\w{\tilde s}}\tilde b - NK - \w{\tDS}\!\cdot\!\w{V}
        - \w{\tDS}_{\w{V}}\, \ln\tilde M \right) \Bigg] \, \w{\tilde q}
        \Bigg\} , 
        \end{array} }       \label{e:TP:Theta_2p1conf}
\ee
or, in index notation,
\bea
    \Theta_{ab} & = & \Psi^4 \Bigg\{ (N\Psi^{-2} - \tilde b) 
        {\tilde H}_{ab}
        + \frac{1}{2} \left( \tDS_a \tilde V_b + \tDS_b \tilde V_a \right) 
        + \frac{1}{2} \dot{\tilde\gamma}_{kl} \, q^k_{\ \, a} q^l_{\ \, b}
         \nonumber \\
   & & + \Bigg[ 
       \frac{1}{3} \left( \tilde b \tilde H +
        \tilde s^k\tD_k \tilde b - NK - \tDS_a V^a 
        - V^a\tDS_a \ln\tilde M \right) \nonumber \\
    & & +     2N\Psi^{-2} \tilde s^k\tD_k \ln\Psi \Bigg] \, \tilde q_{ab}
        \Bigg\} . 
\eea
For a coordinate system stationary with respect to $\Hor$, 
Eq.~(\ref{e:TP:Theta_2p1conf}) simplifies to 
\be
    \encadre{
        \begin{array}{rcl}
    \w{\Theta} & \stackrel{\Hor,\mathrm{sc}}{=} & \displaystyle \Psi^4 
    \Bigg\{ \frac{1}{2}\Kil{\w{\tDS}}{\underline{\w{\tilde V}}}
        + \frac{1}{2} \vec{\w{q}}^* \w{\dot{\tilde\gamma}}
         + \Bigg[ 2N\Psi^{-2} \w{\tD}_{\w{\tilde s}} \ln\Psi \\[3mm]
     & &  \displaystyle  
        + \frac{1}{3} \left( \tilde b \tilde H +
        \w{\tD}_{\w{\tilde s}}\tilde b - NK - \w{\tDS}\!\cdot\!\w{V}
        - \w{\tDS}_{\w{V}}\, \ln\tilde M \right) \Bigg] \, \w{\tilde q}
        \Bigg\} , 
        \end{array} }     .
\ee
The shear tensor of $\Sp_t$ is deduced from Eqs.~(\ref{e:TP:theta_2p1conf})
and (\ref{e:TP:Theta_2p1conf}) by $\w{\sigma} = \w{\Theta} - 1/2 \, \theta
\w{q}$ [Eq.~(\ref{e:KI:Theta_split})]:
\be
    \encadre{
        \begin{array}{rcl}
    \w{\sigma} & = & \displaystyle \frac{\Psi^4}{2} \Bigg\{
     \Kil{\w{\tDS}}{\underline{\w{\tilde V}}}
    - (\w{\tDS}\!\cdot\!\w{V})\, \w{\tilde q}  
    +  \vec{\w{q}}^* \w{\dot{\tilde\gamma}}
    + \frac{1}{2} \w{\dot{\tilde\gamma}}(\w{\tilde s},\w{\tilde s})
    \, \w{\tilde q} \\[3mm]
    & & \displaystyle + 2(N\Psi^{-2} -\tilde b)\left( \w{\tilde H}
    - \frac{1}{2}\tilde H\,  \w{\tilde q} \right) \Bigg\} ,
        \end{array} }     \label{e:TP:shear_2p1conf}
\ee
or in index notation 
\bea
    \sigma_{ab} &=& \frac{\Psi^4}{2} \Bigg\{ 
    \tDS_a \tilde V_b + \tDS_b \tilde V_a - (\tDS_c V^c) \, \tilde q_{ab}
    + \dot{\tilde\gamma}_{kl} \left( q^k_{\ \, a} q^l_{\ \, b}
    + \frac{1}{2} \tilde s^k \tilde s^l \, \tilde q_{ab} \right)  
    \nonumber \\
    & & + 2(N\Psi^{-2} -\tilde b)\left( \tilde H_{ab} 
    - \frac{1}{2}\tilde H\,  \tilde q_{ab} \right) \Bigg\} . 
\eea

For a coordinate system stationary with respect to $\Hor$, 
this expression becomes very simple:
\be
    \encadre{
    \w{\sigma} \stackrel{\Hor,\mathrm{sc}}{=} \frac{\Psi^4}{2} \left[
     \Kil{\w{\tDS}}{\underline{\w{\tilde V}}}
    - (\w{\tDS}\!\cdot\!\w{V})\, \w{\tilde q}  
    +  \vec{\w{q}}^* \w{\dot{\tilde\gamma}}
    + \frac{1}{2} \w{\dot{\tilde\gamma}}(\w{\tilde s},\w{\tilde s})
    \, \w{\tilde q} \right] } .     
\ee
Note that 
$\Kil{\w{\tDS}}{\underline{\w{\tilde V}}}- (\w{\tDS}\!\cdot\!\w{V})\, \w{\tilde q}$
is the {\em conformal Killing operator} associated with the metric $\w{\tilde q}$
and applied to $\w{V}$.

Let us now give the 2+1 conformal decomposition of the transversal
deformation rate $\w{\Xi}$. From Eq.~(\ref{e:TP:thetak_H_K}),
its trace becomes 
\be
     \encadre{
        \begin{array}{rcl}
  \theta_{(\w{k})} & = & \displaystyle -\frac{1}{N} \Bigg\{ 
   2 \Psi^{-2} \left( \tilde s^k \tD_k\ln \Psi 
        + \frac{1}{3} \tilde H \right) + \frac{1}{4N} 
         \, \dot{\tilde\gamma}_{kl} \, \tilde s^k \tilde s^l 
      + \frac{1}{3} \Bigg[   \left( \frac{\tilde b}{N} - \Psi^{-2}
    \right)  \frac{\tilde H}{2}
          \\[3mm]
   & & \displaystyle +  K + \frac{1}{N} \left( V^a \, \tDS_a \ln \tilde M
    - \frac{1}{2} \tDS_a V^a   - \tilde s^k \tD_k {\tilde b}\right) \Bigg] 
    \Bigg\} , 
    \end{array} }
\ee
which for a coordinate system stationary with respect to $\Hor$
results in 
\be
    \encadre{
        \begin{array}{rcl}
        \theta_{(\w{k})} & \stackrel{\Hor,\mathrm{sc}}{=} & \displaystyle
       -\frac{1}{N} \Bigg\{ 
   2 \Psi^{-2} \left( \tilde s^k \tD_k\ln \Psi 
        + \frac{1}{3} \tilde H \right) + \frac{1}{4N} 
         \, \dot{\tilde\gamma}_{kl} \, \tilde s^k \tilde s^l \\[3mm]
         & & \displaystyle
   + \frac{1}{3} \left[ K + \frac{1}{N} \left( V^a \, \tDS_a \ln \tilde M
    - \frac{1}{2} \tDS_a V^a   - \tilde s^k \tD_k {\tilde b}\right) \right] 
    \Bigg\} . 
        \end{array} }    
\ee
Replacing
$\w{H}$ and $\vec{\w{q}}^*\w{K}$ in Eq.~(\ref{e:TP:Xi_H_K}) by
their expressions (\ref{e:TP:H_tH_gradPsi}) and (\ref{e:TP:Kqq_index})
yields
\be
    \encadre{
        \begin{array}{rcl}
    \w{\Xi} & = & \displaystyle -\frac{\Psi^4}{2N^2} 
    \Bigg\{ (N\Psi^{-2} + \tilde b) 
        \w{\tilde H}
        - \frac{1}{2}\left[ \Kil{\w{\tDS}}{\underline{\w{\tilde V}}}
        + \vec{\w{q}}^* \w{\dot{\tilde\gamma}} \right]
         + \Bigg[ 2\Psi^{-2} \w{\tD}_{\w{\tilde s}} \ln\Psi \\[3mm]
     & &  \displaystyle  
        + \frac{1}{3} \left( NK - \tilde b \tilde H -
        \w{\tD}_{\w{\tilde s}}\tilde b  + \w{\tDS}\!\cdot\!\w{V}
        + \w{\tDS}_{\w{V}}\, \ln\tilde M \right) \Bigg] \, \w{\tilde q}
        \Bigg\} , 
        \end{array} }       
\ee

The conformal 2+1 expression of \hajicek's form is
obtained by inserting Eq.~(\ref{e:TP:Ksq}) into Eq.~(\ref{e:TP:Omega_DN_qKs}):
\be
\encadre{   
        \begin{array}{rcl}
    \w{\Omega} & = & \displaystyle \w{\tDS}\ln N
    +\frac{\Psi^2}{2N} \Bigg[ \tilde b \; \w{\tDS}\ln\tilde M
    - \w{\tDS} \tilde b - \w{\tilde H}(\w{V},.)
    + \vec{\w{q}}^* \, \w{\tD}_{\w{\tilde s}}\, \underline{\w{\tilde V}}
    \\[3mm]
    & & \displaystyle +\vec{\w{q}}^*  \w{\dot{\tilde\gamma}}(\w{\tilde s},.)
    \Bigg] . 
        \end{array} }       
\ee
For a coordinate system stationary with respect to $\Hor$,
one has ${\tilde b}\equalH N \Psi^{-2}$
[Eq.~(\ref{e:TP:b_NPsi2_H})], so that the above expression 
results in
\be 
    \encadre{ \w{\Omega} \stackrel{\Hor,\mathrm{sc}}{=} 
    \frac{1}{2} \, \w{\tDS} \ln\left( \Psi^2 N \tilde M \right)
    + \frac{\Psi^2}{2N} \left[ 
    \vec{\w{q}}^* \, \w{\tD}_{\w{\tilde s}}\, \underline{\w{\tilde V}}
   +\vec{\w{q}}^*  \w{\dot{\tilde\gamma}}(\w{\tilde s},.)
    - \w{\tilde H}(\w{V},.) \right] } . 
\ee

\subsubsection{Conformal 2+1 expressions for $\w{\Theta}$, $\theta$
and $\w{\sigma}$ viewed as deformation rates of $\Sp_t$'s metric}

As already noticed in Remark~\ref{r:TP:already_2p1}, we have obtained 
2+1 expressions of $\w{\Theta}$
and $\theta$ in
Sec.~\ref{s:KI:deform_rate} [cf. Eqs.~(\ref{e:KI:Theta_Kil_V}) and
(\ref{e:KI:theta_q_Lie_t_q})]. These expressions involve the time derivative 
of $\Sp_t$'s metric $\w{q}$, whereas the expressions derived above involve
time derivative of the conformal 3-metric $\w{\tilde \gamma}$. 
Therefore it is worth performing a conformal decomposition of the 
equations of Sec.~\ref{s:KI:deform_rate}, since they will 
lead to expressions letting appear time derivatives different than to those
found above. 

Let us start from Eq.~(\ref{e:KI:Theta_Kil_V}) for $\w{\Theta}$. 
The first term in the right-hand side is conformaly decomposed as
\be \label{e:TP:qstar_Lie_t_q}
    \vec{\w{q}}^* \Lie{\tv} \w{q} = \vec{\w{q}}^* \Lie{\tv} (\Psi^4 
    \w{\tilde q}) = \Psi^4 \left[ \vec{\w{q}}^* \Lie{\tv} \w{\tilde q}
    + 4  (\Lie{\tv} \ln\Psi) \, \w{\tilde q} \right] , 
\ee
where we have used $\vec{\w{q}}^* \w{\tilde q} = \w{\tilde q}$. 
The second term in the right-hand side of  Eq.~(\ref{e:KI:Theta_Kil_V})
involves $\DS_a V_b + \DS_b V_a$. Now similarly to the relation 
(\ref{e:TP:connect_deriv}), 
\be
    \DS_a V_b = \tDS_a V_b - {}^2\! C^c_{ba} V_c , 
\ee
with 
\bea
    {}^2\!C^c_{\ \, ab} & := & \frac{1}{2} q^{cd} \left( \tDS_a q_{db}
        + \tDS_b q_{ad} - \tDS_d q_{ab} \right) \nonumber  \\
        & = & 2 \left( \tDS_a \ln \Psi \, q^c_{\ \, b}
            + \tDS_b \ln \Psi \, q^c_{\ \, a}
            - \tDS^c\ln \Psi \tilde q_{ab} \right) . 
\eea
Combining with $V_b = \Psi^4 \tilde V_b$ [Eq.~(\ref{e:TP:def_tilde_uV})],
we get 
\bea
  \DS_a V_b & = & \Psi^4 \Big[ \tDS_a \tilde V_b + 2 \tDS_a \ln \Psi \, \tilde V_b
  - 2 \tDS_b \ln \Psi \, \tilde V_a \nonumber \\
  & & + 2 (V^c \, \tDS_c\ln \Psi) \, \tilde q_{ab}
  \Big] . 
\eea
Hence
\be \label{e:TP:Kil_V_Kil_tV}
  \DS_a V_b + \DS_b V_a = \Psi^4 \left[ \tDS_a \tilde V_b +
   \tDS_b \tilde V_a  + 4 (V^c \, \tDS_c\ln \Psi) \, \tilde q_{ab} \right] . 
\ee
Finally the 2+1 split of the last term in Eq.~(\ref{e:KI:Theta_Kil_V})
is nothing than Eq.~(\ref{e:TP:H_tH_gradPsi}). Inserting this last relation,
as well as Eqs.~(\ref{e:TP:qstar_Lie_t_q}) and (\ref{e:TP:Kil_V_Kil_tV})
into Eq.~(\ref{e:KI:Theta_Kil_V}) leads to
\be
    \encadre{
        \begin{array}{rcl}
   \w{\Theta} & = & \displaystyle 
   \frac{\Psi^4}{2} \Bigg\{ \vec{\w{q}}^* \Lie{\tv} \w{\tilde q}
    + 4  (\Lie{\tv} \ln\Psi) \, \w{\tilde q} 
    + \Kil{\w{\tDS}}{\underline{\w{\tilde V}}} 
    + 4 (\w{\tDS}_{\w{V}} \ln\Psi)\, \w{\tilde q} \\[3mm]
    & & \displaystyle + 2 (N\Psi^{-2} -\tilde b)
    \left[ \w{\tilde H}
        + 2 (\w{\tD}_{\w{\tilde s}} \ln\Psi) \; \w{\tilde q} \right] \Bigg\} ,
        \end{array}}        \label{e:TP:Theta_Lie_t_q}
\ee
or in index notation
\bea
    \Theta_{ab} & = & \frac{\Psi^4}{2} \Bigg\{ \Liec{\tv} q_{\mu\nu}
        \; q^\mu_{\ \, a} q^\nu_{\ \, b}
        + 4 (\Liec{\tv} \ln\Psi) \, \tilde q_{ab}
        + \tDS_a \tilde V_b +
   \tDS_b \tilde V_a  \nonumber \\
   & & + 4 (V^c \, \tDS_c\ln \Psi) \, \tilde q_{ab} 
   + 2 (N\Psi^{-2} -\tilde b) \left[ \tilde H_{ab} + 2 (\tilde s^k \tD_k\ln\Psi)\, 
   \tilde q_{ab} \right] \Bigg\} . \nonumber \\
\eea
Contracting this equation with $q^{ab}$ leads immediately to an 
expression for the expansion scalar $\theta$:
\be
    \encadre{
        \begin{array}{rcl}
        \theta &=& \displaystyle
        \frac{1}{2}\tilde q^{\mu\nu} \Liec{\tv} \tilde q_{\mu\nu}
        + 4 \Liec{\tv} \ln\Psi + 
        \tDS_a V^a + 4 V^a \tDS_a\ln\Psi \\[3mm]
        & & \displaystyle (N\Psi^{-2} -\tilde b) (\tilde H
        + 4 \tilde s^k \tD_k\ln\Psi) . 
        \end{array} }   \label{e:TP:theta_q_Lie_t_q}
\ee
We then obtain the shear tensor $\w{\sigma}$ by forming
$\w{\Theta} - \theta/2\; \w{q}$ from Eqs.~(\ref{e:TP:Theta_Lie_t_q})
and (\ref{e:TP:theta_q_Lie_t_q}). All the terms involving $\ln\Psi$
cancel and there remains
\be
     \encadre{
        \begin{array}{rcl}
   \w{\sigma} & = & \displaystyle 
   \frac{\Psi^4}{2} \Bigg\{ \vec{\w{q}}^* \Lie{\tv} \w{\tilde q}
  - \frac{1}{2} \left( \tilde q^{\mu\nu} \Liec{\tv} \tilde q_{\mu\nu}\right)
   \w{\tilde q} + \Kil{\w{\tDS}}{\underline{\w{\tilde V}}} 
   - (\w{\tDS}\!\cdot\!\w{V}) \w{\tilde q} \\[3mm]
  & & \displaystyle + 2 (N\Psi^{-2} -\tilde b) \left( \w{\tilde H}
  - \frac{1}{2} \tilde H \, \w{\tilde q} \right)\Bigg\}  , 
        \end{array} }   \label{e:TP:shear_Lie_t_q}
\ee
or, in index notation
\bea
    \sigma_{ab} & = & \frac{\Psi^4}{2} \Bigg\{ \Liec{\tv} q_{\mu\nu}
        \; q^\mu_{\ \, a} q^\nu_{\ \, b}
    - \frac{1}{2} \left( \tilde q^{\mu\nu} \Liec{\tv} \tilde q_{\mu\nu}\right)
    \tilde q_{ab} + \tDS_a \tilde V_b + 
   \tDS_b \tilde V_a \nonumber\\
   & & - (\tDS_c V^c)\,  \tilde q_{ab} 
   + 2 (N\Psi^{-2} -\tilde b) \left( \tilde H_{ab} - \frac{1}{2} \tilde H \, 
   \tilde q_{ab} \right) \Bigg\} .
\eea

If one uses a coordinate system stationary with respect to $\Hor$, 
the above equations simplifies somewhat, thanks to the vanishing of
$N\Psi^{-2} -\tilde b$ [Eq.~(\ref{e:TP:b_NPsi2_H})]: 
\be
   \w{\Theta} \stackrel{\Hor,\mathrm{sc}}{=}
   \frac{\Psi^4}{2} \left[ \vec{\w{q}}^* \Lie{\tv} \w{\tilde q}
    + 4  (\Lie{\tv} \ln\Psi) \, \w{\tilde q} 
    + \Kil{\w{\tDS}}{\underline{\w{\tilde V}}} 
    + 4 (\w{\tDS}_{\w{V}} \ln\Psi)\, \w{\tilde q} \right] ,
\ee
\be
        \theta \stackrel{\Hor,\mathrm{sc}}{=}
        \frac{1}{2}\tilde q^{\mu\nu} \Liec{\tv} \tilde q_{\mu\nu}
        + 4 \Liec{\tv} \ln\Psi + 
        \tDS_a V^a + 4 V^a \tDS_a\ln\Psi , 
\ee
\be
   \w{\sigma} \stackrel{\Hor,\mathrm{sc}}{=}
   \frac{\Psi^4}{2} \left[ \vec{\w{q}}^* \Lie{\tv} \w{\tilde q}
  - \frac{1}{2} \left( \tilde q^{\mu\nu} \Liec{\tv} \tilde q_{\mu\nu}\right)
   \w{\tilde q} + \Kil{\w{\tDS}}{\underline{\w{\tilde V}}} 
   - (\w{\tDS}\!\cdot\!\w{V}) \, \w{\tilde q} \right] .
\ee
Moreover, in the case of a coordinate system $(t,x^i)$
adapted to $\Hor$, 
the term $\tilde q^{\mu\nu} \Liec{\tv} \tilde q_{\mu\nu}$ can be 
expressed in terms of the variation of determinant $\tilde q$ of
the conformal 2-metric components $\tilde q_{ab}$ in this coordinate system:
\be
     \tilde q^{\mu\nu} \Liec{\tv} \tilde q_{\mu\nu} =
    \Liec{\tv} \ln \tilde q. 
\ee

\begin{rem}
Equations (\ref{e:TP:Theta_Lie_t_q}), (\ref{e:TP:theta_q_Lie_t_q})
and (\ref{e:TP:shear_Lie_t_q}) constitute 2+1 expressions of 
respectively $\w{\Theta}$, $\theta$
and $\w{\sigma}$ in terms of $\Lie{\tv}\w{\tilde q}$ 
and (for $\w{\Theta}$ and $\theta$ only) $\Lie{\tv}\ln\Psi$. 
On the other side, Eqs.~(\ref{e:TP:Theta_2p1conf}), 
(\ref{e:TP:theta_2p1conf}) and (\ref{e:TP:shear_2p1conf}), 
provide the same quantities $\w{\Theta}$, $\theta$
and $\w{\sigma}$ in terms of $\Lie{\tv}\w{\tilde \gamma}$.
The equivalence between the two sets can be established in view of
the two identities:
\bea
    \vec{\w{q}}^* \Lie{\tv} \w{\tilde\gamma} & =& 
    \vec{\w{q}}^* \Lie{\tv} \w{\tilde q} , \\
   \Lie{\tv}\ln\Psi & = & \tilde b \, \w{\tD}_{\w{\tilde s}} \ln\Psi
    - \w{\tDS}_{\w{V}} \ln\Psi + \frac{1}{6} \Bigg( \tilde b \tilde H +
        \w{\tD}_{\w{\tilde s}}\tilde b - NK 
            \nonumber \\
       && - \w{\tDS}\!\cdot\!\w{V}  - \w{\tDS}_{\w{V}}\, \ln\tilde M \Bigg) .
\eea
The first identity is an immediate consequence of 
$\w{\tilde q} = \w{\tilde\gamma}- 
 \underline{\w{\tilde s}} \otimes \underline{\w{\tilde s}}$
[Eq.~(\ref{e:TP:def_tilde_q})] and 
$\vec{\w{q}}^* \, \underline{\w{\tilde s}} =0$,
whereas the second identity is nothing but the 2+1 split of the 
on the expression of $\Lie{\tv}\ln\Psi$ as given by 
Eq.~(\ref{e:TP:Psidot}) [cf. Eq.~(\ref{e:TP:tdiv_beta_2p1})]. 
\end{rem}

%% file: applinitdata.tex
%
%
\section{Applications to the initial data and slow evolution problems}
\label{s:BC}

Let us apply the results of previous sections 
to derive {\em inner} boundary conditions for
the partial differential equations arising from
the 3+1 decomposition of Einstein equations.
More specifically, we consider the problem of constructing numerically,
within the 3+1 formalism, a spacetime containing a black hole in 
quasi-equilibrium by employing some excision technique.
By {\em excision} is meant the removal of a 2-sphere $\Sp_t$ and its interior
from the initial Cauchy surface $\Sigma_t$. 
If we ask $\Sp_t$ to represent the apparent 
horizon \cite{Th87,Ma03,Da04} of a black hole in quasi-equilibrium\footnote{One can also
set the excised sphere $\Sp_t$ {\it inside} the horizon \cite{SeideS92,AlcubB01,AlcubBPST01},
instead of prescribing it to be the actual apparent horizon. In such
a case one must find numerically the apparent horizon (see \cite{Thorn96}, \cite{Thorn04} and 
references therein), or even the event 
horizon \cite{LibsoMSSW96,Diene03,CavenAM03}, and then the quasi-local techniques
in this article can be used in an {\it a posteriori} analysis, rather than as 
inner boundary conditions.}, the 
quasi-local tools presented previously
are very well suited to set the appropriate values,
or more generally constraints, to be satisfied by the 3+1 fields 
on this inner boundary.

Two most important physical problems where
these boundary conditions can be naturally applied are:
\begin{itemize}
\item[{\it i)}] The construction of initial data for binary black holes in 
quasi-circular orbits. By \emph{quasi-circular} is meant orbits for
which the decay due to gravitational radiation can be neglected. This is
a very good approximation for sufficiently separated systems and the
spacetime can then be considered as being endowed with a helical Killing
vector (see e.g. Refs.~\cite{GourgGB02} and \cite{FriedUS02} for a discussion). 
\item[{\it ii)}] The slow dynamical evolution of spacetimes containing
a black hole (more precisely, slow evolution of initial data 
containing a marginally trapped surface). For concreteness, the 
system of coupled elliptic equations in the {\it minimal no-radiation} 
approximation proposed in Ref. \cite{SchafG04}
(a gravitational analog of the magneto-hydrodynamics approximation in
electromagnetism), provides an appropriate framework for implementing
the isolated horizon prescriptions as inner boundary conditions.
Alternatively, the fully constrained evolution scheme presented in 
Ref.~\cite{BonazGGN04} (or approximations based on it) can also prove
to be useful, provided appropriate quasi-equilibrium 
initial free data are chosen on the initial slice\footnote{If the
initial data encode modes whose evolution propagate towards the
horizon and eventually fall into it, the enforcing of isolated
horizon conditions would lead to an ill-posed problem. Therefore 
initial data must be carefully chosen so as to minimize these
perturbations (e.g. making them smaller than the numerical noise).}.
\end{itemize}

\subsection{Conformal decomposition of the constraint equations}

The Hamiltonian and momentum constraint equations arising from
the 3+1 decomposition of Einstein equation have been presented
in Sec.~\ref{s:FO:3p1Einstein} [Eqs.~(\ref{e:FO:Ham_constraint})
and (\ref{e:FO:mom_constraint})]. Besides we have introduced in 
Sec.~\ref{s:TP:conf_decomp} a conformal decomposition of the 3-metric
$\w{\gamma}$ [Eq.~(\ref{e:TP:def_conf_metr})]
and the extrinsic curvature $\w{K}$ of the hypersurface
$\Sigma_t$ [Eq.~(\ref{e:TP:conf_decomp_K})]. Let us examine the 
impact of these decompositions on the constraint equations. 

\subsubsection{Lichnerowicz-York equation}

From the conformal decomposition $\w{\gamma} = \Psi^4 \, \w{\tilde \gamma}$
[Eq.~(\ref{e:TP:def_conf_metr})], the Ricci scalar ${}^3\!R$
associated with $\w{\gamma}$ is related to the Ricci scalar 
${}^3\!{\tilde R}$ related to  $\w{\tilde \gamma}$ by
${}^3\!R = \Psi^{-4} \; {}^3\!{\tilde R}- 8 \Psi^{-5} \tD_k \tD^k \Psi$,
so that the Hamiltonian constraint equation (\ref{e:FO:Ham_constraint})
becomes the {\em Lichnerowicz-York equation} for $\Psi$:
\be \label{e:BC:Lichne}
   \encadre{  \tD_k \tD^k \Psi - \frac{{}^3\!{\tilde R}}{8} \, \Psi
    + \frac{1}{8} \tilde A_{ij} \tilde A^{ij} \, \Psi^{2\zeta-3}
    + \left( 2\pi E - \frac{K^2}{12} \right) \Psi^5 = 0 }, 
\ee
where use has been made of the conformal decomposition 
(\ref{e:TP:conf_decomp_K}) of $\w{K}$ and $\tilde A^{ij}$
is related to $\tilde A_{ij}$ by means of the conformal metric:
\be \label{e:BC:def_Aij}
    \tilde A^{ij} := \tilde\gamma^{ik} \tilde\gamma^{jl} \tilde A_{kl} .
\ee

For a fixed conformal metric $\w{\tilde \gamma}$ and a fixed $\w{\tilde A}$, 
Eq.~(\ref{e:BC:Lichne}) is a non-linear equation for $\Psi$;
it has been first derived and analyzed by Lichnerowicz
\cite{Lichn44} in the special case $\zeta=-2$, $K=0$ and $E=0$ 
or $E={\rm const}$. The negative sign of the exponent $2\zeta - 3$ in
this case is crucial to guarantee the existence and uniqueness of the 
solution to this equation.
It has been extended to the case $K\not =0$
and discussed in great details by York \cite{York79,York04}.

\subsubsection{Conformal thin sandwich equations}
\label{s:BC:CTS}

We place ourselves in the framework of the {\em conformal thin sandwich} 
approach to the 3+1 initial data problem developed by York and Pfeiffer
\cite{York99,PfeifY03,York04,Cook00}. 
Regarding the problem of binary black hole on close circular orbits, 
this approach has led to the
most successful numerical solutions to date \cite{GrandGB02,CookP04,Ansor05}.
It is based on the transformation of the momentum constraint equation
(\ref{e:FO:mom_constraint}) into an elliptic equation for the
shift vector $\w{\beta}$\footnote{Likewise, 
York's original approach \cite{York79,York04} reduces the resolution of
the momentum constraint to an analogous vectorial elliptic equation;
see \cite{PfeifY03} and \cite{Pfeif04} for the discussion of the
relation between both approaches.}. Indeed inserting the expression 
(\ref{e:TP:K_decomp}) for $\w{K}$ into Eq.~(\ref{e:FO:mom_constraint})
yields
\be \label{e:BC:cts_shift}
    \encadre{
    \begin{array}{lcl}
    \displaystyle \tD_k \tD^k \beta^i +  \frac{1}{3} \tD^i \tD_k \beta^k 
    + {}^3\!{\tilde R}^i_{\ \, k} \beta^k & = & 
    \displaystyle 16\pi\Psi^4 N J^i + \frac{4}{3} N \tD^i K  
        - \tD_k \dot{\tilde\gamma}^{ij}\\
        & & \displaystyle+ 2N \Psi^{\zeta-4} \tilde A^{ik} 
            \tD_k\ln(N\Psi^{-6}) 
    \end{array} }, 
\ee
where ${}^3\!{\tilde R}^i_{\ \, j} := 
\tilde\gamma^{ik} \, {}^3\!{\tilde R}_{kj}$, ${}^3\!{\tilde R}_{ij}$
being the Ricci tensor of the connection $\w{\tD}$ compatible
with $\w{\tilde\gamma}$, and 
\be \label{e:BC:def_tgamuu_dot}
    \dot{\tilde\gamma}^{ij} := \der{\tilde\gamma^{ij}}{t}
        = - \tilde\gamma^{ik} \tilde\gamma^{jl}  \dot{\tilde\gamma}_{kl} , 
\ee
where $\dot{\tilde\gamma}_{kl}$ is the quantity introduced in 
Eq.~(\ref{e:TP:def_tgam_dot}). 
Notice that the quantity $\tilde A^{ij}$ which appears in
Eq.~(\ref{e:BC:cts_shift}) is considered as a function of the shift
vector according to 
\be
    A^{ij} = \frac{1}{2N\Psi^{\zeta-4}} \left(
        \tD^i \beta^j + \tD^j \beta^i - \frac{2}{3} \tD_k \beta^k
        \, \tilde\gamma^{ij} + \dot{\tilde\gamma}^{ij} \right) ,
\ee
which follows from Eq.~(\ref{e:BC:def_Aij}), (\ref{e:TP:A_dottgam_lbeta})
and (\ref{e:BC:def_tgamuu_dot}). 

When solving the initial data problem with the Hamiltonian constraint
under the form of the Lichnerowicz-York equation 
(\ref{e:BC:Lichne}) and the momentum constraint under the form of the
vector elliptic equation (\ref{e:BC:cts_shift}), one can choose freely
the conformal 3-metric
$\w{\tilde\gamma}$, its time derivative $\w{\dot{\tilde\gamma}}$ and
the trace of $\Sigma_t$'s extrinsic curvature, $K$.
Prescribing in addition the value of the time derivative
$\dot K = \dert{K}{t}$, leads to the {\em extended conformal thin sandwich}
formalism as presented in Ref.~\cite{PfeifY03}. Indeed prescribing
$\dot K$ leads to a ``constraint'' on the lapse function $N$, which 
can be derived by taking the trace of the dynamical Einstein 
equation (\ref{e:FO:evol_K_t}):
\be \label{e:BC:cts_lapse}
    \encadre{
    \begin{array}{lcl}
    \displaystyle \tD_k \tD^k N + 2 \tD_k\ln\Psi\, \tD^k N 
    & = &\displaystyle 
     \Psi^4 \Bigg\{ N \left[ 4\pi (E+S) 
    + \frac{K^2}{3} \right] \\
    & &\displaystyle  \ \qquad  - \dot K + \beta^k \tD_k K \Bigg\}
    + N \Psi^{2\zeta-4} \tilde A_{kl} \tilde A^{kl}  .
    \end{array} }
\ee
To summarize, in the extended conformal thin sandwich framework, 
the free data (modulo boundary values of the constraint parameters)
are the fields
$(\w{\tilde\gamma},\w{\dot{\tilde\gamma}},K,\dot K)$
on some spatial hypersurface $\Sigma_0$. 
The elliptic equations (\ref{e:BC:Lichne}), (\ref{e:BC:cts_shift})
and (\ref{e:BC:cts_lapse}) are to be solved for the 
conformal factor $\Psi$, the shift vector $\w{\beta}$ and the lapse
function $N$. One then gets a valid initial data set 
$(\Sigma_0,\w{\gamma},\w{K})$, i.e. a data set satisfying the
Hamiltonian and momentum constraints. Moreover the initial time
development of these initial data will be such that $\dot K$
takes the prescribed value. 
One should note that according to a recent study \cite{PfeifY05},
the uniqueness of a solution $(\Psi,N,\w{\beta})$ of the extended conformal
thin sandwich equations is not guaranteed: for some choice of 
free data $(\w{\tilde\gamma},\w{\dot{\tilde\gamma}},K,\dot K)$,
two distinct solutions $(\Psi,N,\w{\beta})$ have been found in a
spatial slice where no sphere has been excised. 

However, a unique solution
of the extended conformal thin sandwich has been found for problems
of direct astrophysical interest, like binary neutron stars 
\cite{BaumgCSST98,GourgGTMB01,MarroMW99,UryuE00} or binary black holes
\cite{GrandGB02,CookP04}. Moreover, this method of solving the
initial data problem has been recognized 
to have greater physical content than previous conformal formulations
\cite{Lichn44,OMurcY74,York79,Cook00}, because it allows a direct
control on the time derivative of the conformal metric. In particular, 
in a quasi-equilibrium situation, it is natural to choose
$\w{\dot{\tilde\gamma}}=0$ and $\dot K = 0$ 
to guarantee that the coordinates are adapted to the approximate
Killing vector reflecting the quasi-equilibrium
(see Refs. \cite{Cook02,ShibaUF04}).

In the rest of this Section, we translate the isolated
horizon geometrical conditions into boundary conditions for the
constrained parameters of the conformal thin sandwich formulation. 
We separate this analysis in NEH (Sec.~\ref{s:BC:BC_NEH})
and WIH boundary conditions (Sec.~\ref{s:BC:BCWIH}).

\subsection{Boundary conditions on a NEH} \label{s:BC:BC_NEH}

As seen in Sec. \ref{s:NE}, a NEH is characterized by the
vanishing of its second fundamental form $\w{\Theta}$ 
[see Eq.~(\ref{e:NE:Theta_null})]. According to the transformation rules
in Table \ref{t:KI:scaling} under rescalings of $\el$, the condition
$\w{\Theta}=0$ is independent of the 
specific choice for the null normal $\el$.
Using the decomposition (\ref{e:KI:Theta_split}) of $\w{\Theta}$, this
condition translates into the vanishing of the expansion $\theta$ and shear
$\w{\sigma}$  associated with $\el$.

\subsubsection{Vanishing of the expansion: $\theta=0$}

Imposing this  condition on a sphere $\Sp_t$ defines it as a marginally
outer trapped surface in $\Sigma_t$ (see
Sec. \ref{s:NE:trapped_app}). 
If in addition $\theta_{(\w{k})}\leq 0$, it corresponds 
to a {\it future marginally trapped surface}.
In this second case, we find from Eqs.~(\ref{e:TP:theta_H_K}) and  
(\ref{e:TP:thetak_H_K})
\bea
\label{e:BC:fut_marg_trapped}
\w{K}(\w{s},\w{s})-K=\frac{\theta}{2N} + N\theta_{(\w{k})} \leq 0 \ .
\eea
If $\Sp_t$ is the {\it outermost} marginally trapped surface, then
it is properly called an apparent horizon (see again
Sec. \ref{s:NE:trapped_app}). 
The condition  $\theta=0$  can be expressed in a variety of forms.
A convenient expression
follows from Eq. (\ref{e:TP:theta_H_K}) when substituting the 
conformal decomposition of the metric $\w{\gamma}$: 
\bea
\label{e:BC:app_hor_Kss}
\encadre{
4 \tilde{s}^i \tilde{D}_i \mathrm{ln} \Psi +
\tilde{D}_i\tilde{s}^i + \Psi^{-2} K_{ij} \tilde{s}^i\tilde{s}^j -
\Psi^2 K = 0
} \ .
\eea
If the conformal and $2+1$ decomposition for $\w{K}$ is included,
it follows from Eq. (\ref{e:TP:theta_2p1conf})
\be
\label{e:BC:app_hor_2+1}
    \encadre{
        \begin{array}{rcl}
    && \displaystyle  N \Psi^{-2} \left( 4 \tilde s^k \tD_k\ln \Psi 
        + \frac{2}{3} \tilde H \right) 
        - \frac{1}{2} \, \dot{\tilde\gamma}_{kl} \, \tilde s^k \tilde s^l 
        \\[3mm]
    & &\displaystyle  \ \  + \frac{1}{3} \left[ (N\Psi^{-2}-\tilde b) \tilde H
    + 2 \tilde s^k \tD_k {\tilde b}- 2 V^a \, \tDS_a \ln \tilde M 
     + \tDS_a V^a - 2NK \right] = 0.
     \end{array} 
}  
\ee
An alternative expression, that exploits
the relation between the expansion $\theta$ and the time evolution
of the volume element,  follows by substituting 
the value for $\Lie{\w{t}}\Psi$ provided by (\ref{e:TP:Psidot}) into Eq. 
(\ref{e:TP:theta_q_Lie_t_q})
\be
\label{e:BC:app_hor_Lie_t}
\encadre{
\begin{array}{rcl}
 & & 4\left( \beta^i\tilde{D}_i\Psi +  (N\Psi^{-2} -\tilde b) \tilde{s}^i+
V^i\right) \tilde{D}_i \Psi \\
&&\ \  + \Psi\left[\frac{2}{3}(\tilde{D}_i\beta^i - NK) + \tDS_a V^a +
 \frac{1}{2}\tilde q^{\mu\nu} \Liec{\tv} \tilde q_{\mu\nu}+ 
 (N\Psi^{-2} -\tilde b) \tilde{H} \right] = 0. 
\end{array}
}
\ee 
Eqs. (\ref{e:BC:app_hor_Kss}),  (\ref{e:BC:app_hor_2+1})
and (\ref {e:BC:app_hor_Lie_t}) 
can be seen as boundary conditions for the 
conformal factor $\Psi$ in the resolution of the Hamiltonian
constraint in a conformal decomposition, i.e. Eq. (\ref{e:BC:Lichne})
[see however the end of Sec. \ref{s:BC:other} for other possibilities].
They express the same geometrical condition in terms of 
different sets of fields. The appropriate form to be used 
must be chosen according to the details of the problem we want 
to solve. 
Finally,  we note that the boundary condition $\theta=0$
has been extensively studied in the 
literature (see \cite{Th87} for a numerical perspective and \cite{Ma03,Da04}
for an analytical one).

\subsubsection{Vanishing of the shear: $\sigma_{ab}=0$}
From Eq. (\ref{e:TP:shear_Lie_t_q}),
the vanishing of the shear $\w{\sigma}$ translates into
\bea
\label{e:BC:shearzero}
0&=&\underbrace{\left( \Lie{\w{t}} \tilde{q}_{ab} -
\frac{1}{2}\left(\Lie{\w{t}}\mathrm{ln}\tilde{q}\right)
\tilde{q}_{ab}\right)}_{\hbox{
\small \emph{I}: initial free data}}
+ \underbrace{\left(\tDS_a \tilde V_b + \tDS_b \tilde V_a 
    - (\tDS_c V^c)\,  \tilde q_{ab}\right)}_{\hbox{
\small \emph{II}: intrinsic geometry of }\Sp_t} \nn \\
&+&\underbrace{\left(N\Psi^{-2} - \tilde b\right)
\left(\tilde{H}_{ab} -\frac{1}{2} \tilde{q}_{ab}
\tilde{H}\right)} _{\hbox{\small \emph{III}: ``extrinsic'' geometry of }\Sp_t}
\eea
Defining, from Parts $I$ and $III$ in the previous equation, the 
symmetric traceless tensor 
\bea
\label{e:BC:source_V}
C_{ab} := - \left[ \left( \Lie{\w{t}} \tilde{q}_{ab} -
\frac{1}{2}\left(\Lie{\w{t}}\mathrm{ln}\tilde{q}\right)
\tilde{q}_{ab}\right) + \left(N\Psi^{-2} - \tilde b\right)
\left(\tilde{H}_{ab} -\frac{1}{2} \tilde{q}_{ab}
\tilde{H}\right) \right] \ \ , 
\eea
we can write Eq.~(\ref{e:BC:shearzero}) as
\bea
\tilde{q}_{bc} \tDS_aV^c + 
\tilde{q}_{ac} \tDS_b V^c
- \tilde{q}_{ab}  \tDS_c V^c
= C_{ab} \ \ .
\eea
If we contract with $\tDS^a$ and use the Ricci equation 
(\ref{e:IN:Ricci_ident}) (properly contracted), we get
\bea
\tilde{q}_{bc}\tDS^a\tDS_a V^c + {}^2\!\tilde{R}_{db}V^d =
\tDS^aC_{ab} \ \ .
\eea
Finally, defining $\tilde{C}_b^{\ a}:=\tilde{q}^{ca}C_{bc}$, we obtain the
following elliptic equation for $V^a$ on $\Sp_t$:
\bea
\label{e:BC:lapl_V}
\encadre{
^{2}\!\tilde{\Delta}V^a +  {}^2\!\tilde{R}^a_{\ b} V^b = \tDS^b
      \tilde{C}_b^{\ a} \ \ .
}
\eea
Once we have solved this equation on the sphere, we employ the
solution as a Dirichlet
boundary condition for the tangential part of the shift $\w{V}$.
Therefore, the vanishing of the shear (vanishing of two independent 
functions) can be completely attained by an appropriate choice of 
the (two-dimensional) vector $\w{V}$.
An important particular case occurs when we enforce the
vanishing of Parts $(I+II)$ and $III$ separately.
The vanishing of Part $III$ is motivated in the literature 
in two different manners. On the one hand, this term 
cancels if we demand the coordinate radius of the horizon to remain
fixed in a dynamical evolution [cf. Eq.~(\ref{e:IN:b_N_station})], 
something desirable from a numerical 
point of view. On the other hand, in order to make tractable the analytical
study on the well-posedness of the initial data problem with 
quasi-equilibrium boundary conditions, results in the literature
proceed by decoupling the momentum constraint from the 
Hamiltonian one. In particular, in this strategy, Part $III$ must
vanish on its own; if this is not the case,
the presence of the conformal factor $\Psi$ in the coefficient multiplying 
the {\it extrinsic geometry} part  would couple the
equation on $\Psi$ with the equation on $\w{\beta}$.

\noindent {\it Vanishing of Part $(I+II)$}

If Part $III$ is zero, the vanishing of $(I+II)$ is obtained by solving 
Eq. (\ref{e:BC:lapl_V}) with a  traceless symmetric tensor 
$C_{ab}$ in Eq. (\ref{e:BC:source_V}) completely characterised by
the traceless part of $ \Lie{\w{t}} \tilde{q}_{ab}$.
A specially important case corresponds to the choice $\dot{\tilde{\w{\gamma}}}=0$
motivated by bulk {\it quasi-equilibrium} considerations 
\cite{Cook02,ShibaUF04}.
In this case, the condition $(I+II) = 0$ reduces to
\bea
\label{e:BC:V_conf_Killing}
\encadre{
\tilde{q}_{bc} \tDS_aV^c + 
\tilde{q}_{ac} \tDS_b V^c
- \tilde{q}_{ab}  \tDS_c V^c = 0
} \ ,
\eea
which states that $\w{V}=-\vec{\w{q}}(\w{\beta})$ is a conformal Killing 
vector with respect to the metric
$\w{\tilde q}$, and hence to the (conformally related) metric $\w{q}$. 
These boundary conditions [generally $(I+II)=0$] can be found
in Refs.~\cite{CookP04,JaramGM04,DJK05} (see also Ref.~\cite{Eardl98} for 
a previous related work).
\begin{rem}
In the case of a stationary spacetime $(\M,\w{g})$, 
it is natural to choose coordinates
such that $\tv$ coincides with the Killing vector associated with 
stationarity. Then Part~$I$ in Eq.~(\ref{e:BC:shearzero}) 
vanishes identically. Moreover if we ask $\Hor$ to be
preserved by the spacetime symmetry, it must be transported
to itself by $\tv$, which implies that $\tv$ is tangent to $\Hor$.
Hence, from Eqs.~(\ref{e:IN:t_tangent_H}) and (\ref{e:IN:b_N_station}),
$b\equalH N$. This results
in the vanishing of Part $III$. 
Therefore for the choice of $\tv$ as a Killing vector in a stationary 
spacetime, the NEH condition reduces to the
vanishing of Part~II. As shown above, this implies that $\w{V}$ is 
a conformal Killing vector of $(\Sp_t,\w{q})$. 
In the case where $\Hor$ is a black hole event horizon, 
this result is linked to a first step 
in the demonstration of {\em Hawking's strong rigidity
theorem} \cite{Hawki72,Hawki73,HawkiE73}, 
which states that a stationary event horizon
which is not static must be in addition axisymmetric.
Indeed in the present case, $\el\equalH \tv + \w{V}$ 
[Eq.~(\ref{e:IN:el_tau_V}) with $b\equalH N$] and either $\w{V}=0$
or $\w{V}$ is a conformal symmetry of $(\Sp_t,\w{q})$. 
Via Hawking's theorem,
this conformal symmetry is then extended to a full symmetry (i.e. axisymmetry) 
of $(\M,\w{g})$. In particular $\el$ is then a Killing vector, hence the name
{\em rigidity}: $\Hor$'s null generators cannot move independentely 
of the spacetime symmetries. 
\end{rem}

\noindent {\it Vanishing of Part $III$}

Two manners of imposing the vanishing of the Part $III$ in
Eq. (\ref{e:BC:shearzero}) follow from the motivations presented after
Eq. ({\ref{e:BC:lapl_V}):
\begin{itemize}
\item[{\it a)}] If we choose a coordinate system stationary with respect
to the horizon (see Sec. \ref{s:IN:stacoord}), then Eq. (\ref{e:TP:b_NPsi2_H}) automatically implies
the vanishing of the coefficient $(N\Psi^{-2} - \tilde b)$. 
Therefore, the Dirichlet condition
for the radial part of the shift
\bea
\label{e:BC:radial_shift_dir}
\encadre{
\tilde{b}=N\Psi^{-2} 
} \ ,
\eea
together with $(I+II)=0$, guarantees the vanishing of the shear.
This is the choice in \cite{CookP04,JaramGM04}.

\item[{\it b)}] Even though the previous Dirichlet condition for 
the radial part of the shift $\tilde{b}$ is well motivated (since choosing a
stationary coordinate system with respect to $\Hor$ is convenient
from a numerical point of view), it presents the following
problem for the solution of the constraints.
If the value of $\tilde{b}$ is fixed on the boundary $\Sp_t$, we 
loose control on the value of its radial derivative
$\tilde{\w{s}}\cdot\w{\tD} \tilde{b}$ on $\Sp_t$. In particular, this means via 
Eq. (\ref{e:TP:Kss}),
that we cannot prescribe the sign of $\w{K}(\w{s},\w{s})$.
On the one hand, Eq. (\ref{e:BC:fut_marg_trapped}) then implies
that we cannot guarantee $\Sp_t$ to be
a {\it future} marginally trapped surface.
On the other hand and perhaps more importantly,
the positivity of the conformal factor $\Psi$ cannot be guaranteed 
when solving the Hamiltonian constraint,
since the
sign of $\w{K}(\w{s},\w{s})$ appearing  in the ``apparent horizon'' 
boundary condition (\ref{e:BC:app_hor_Kss}), must be controlled
in order to apply a maximum principle to Eq. (\ref{e:BC:Lichne}). 
This problem is discussed
in Ref. \cite{DJK05}. The solution proposed there for guaranteeing 
the vanishing of Part $III$ consists in choosing initial
free data $\tilde{\w{\gamma}}$ such that 
\bea
\label{e:BC:H_traceless}
\encadre{
\tilde{H}_{ab} -\frac{1}{2} \tilde{q}_{ab} \tilde{H}=0 
} \ , 
\eea
is satisfied.
This condition on the extrinsic curvature of the sphere $\Sp_t$,
i.e. on the shape of $\Sp_t$  inside $\Sigma_t$, is known as the
{\it umbilical} condition.
The boundary condition for the radial part of
the shift is obtained in Ref. \cite{DJK05} by imposing
\bea
\label{e:BC:Kss_neg}
\w{K}(\w{s},\w{s})= h_1 \ ,
\eea
where $h_1$ is a given function on $\Sp_t$ that can be considered
as a free data on $\Sigma_t$. Using Eq. (\ref{e:TP:Kss}),
this condition is expressed as a {\it mixed} condition on $\tilde{b}$
\bea
\label{e:BC:radial_shift_neu}
\encadre{
2\tilde s^k\tD_k \tilde{b} - \tilde{b} \tilde H = 
3 N h_1 - \tDS_k V^k - 2 V^k \, \tilde{D}_{\tilde{s}} \tilde{s}_k -N K
} \ .
\eea
Note
the change of sign convention in the tangential part of the shift
$\w{V}$ with respect
to \cite{DJK05}, where in addition a maximal slicing $K=0$ is assumed.

\end{itemize}

Finally we emphasize the fact that, in order to enforce the
NEH structure, it is enough to impose the appropriate boundary conditions
for the conformal factor and the tangential part of the shift:
Eq. (\ref{e:BC:app_hor_Kss}) for $\Psi$ and Eq. (\ref{e:BC:lapl_V})
for $\w{V}$. Besides, a boundary condition for the normal
part of the shift can be provided by making a choice relative to the coordinate system, 
Eq. (\ref{e:BC:radial_shift_dir}), or by fixing $\w{K}(\w{s},\w{s})$
on $\Sp_t$ through Eq. (\ref{e:BC:radial_shift_neu})
[i.e. fixing $\theta_{(\w{k})}$ in the maximal slicing case; cf. Eq. 
(\ref{e:BC:fut_marg_trapped})].
In brief, the NEH structure together with an (additional) appropriate choice for
$\tilde{b}$, permit to fix boundary conditions for $\Psi$ and
$\w{\beta}$. In other words, this first level in the isolated horizon
hierarchy
provides {\it enough} number of inner boundary conditions for addressing the
resolution of the constraint equations, as exploited by Cook \&
Pfeiffer \cite{CookP04} (see also Ref. \cite{Ansor05}). Incorporating
Eq. (\ref{e:BC:cts_lapse}) for the lapse in the construction of initial
data (see Sec. \ref{s:BC:CTS}) demands, in principle,
some additional geometrical structure on $\Hor$. This is considered in
next section.

\subsection{Boundary conditions on a WIH}
\label{s:BC:BCWIH}

In Sec. \ref{s:IH} we showed how the addition of a WIH structure on a NEH
permits to  fix the foliation $(\Sp_t)$ of the underlying
null hypersurface $\Hor$ in an intrinsic manner. This determination 
of the foliation proceeded
in two steps: firstly, by choosing a particular WIH class $[\el]$
(Sec. \ref{s:IH:prefWIH}) and, secondly, by choosing a foliation
$(\Sp_t)$ compatible with that class (Sec. \ref{s:IH:goodcuts}).
This procedure in two steps is necessary when adopting an approach 
strictly intrinsic to the null surface, since in this case there is no privileged
{\it starting} slice $\Sp_0$ in $\Hor$. 
In brief, simply fixing $\el$ does not determines the foliation.
This represents what we referred in the Introduction
(Sec. \ref{s:INTRO}) as an ``up-down''approach.

The situation changes completely
when adopting a 3+1 point of view. A  main feature in this case
is the actual {\it construction} of the spacetime starting
from an initial Cauchy slice $\Sigma_0$, which is then evolved by using Einstein equations.
In this setting, in which $\Sigma_0$ is given, fixing the lapse
determines the foliation of $\M$. Moreover,
fixing the evolution vector $\el$ on $\Hor$ {\it does} fix the lapse 
on $\Sp_0$ and consequently the foliation,  in contrast 
with the intrinsic approach to the geometry of $\Hor$ in the previous 
paragraph\footnote{As we have seen in Sec. \ref{s:IN:normal_l}  the converse is also
true on $\Hor$: fixing the foliation $(\Sp_t)$ not only fixes the
lapse (up to a ``time'' reparametrization) but also the null normal
$\el$. This is in contrast with the bulk case, where $(\Sigma_t)$
fixes $N$ but not the evolution vector $\w{t}$ due to the shift 
ambiguity. The difference relies on the existence of a single 
null direction in $\Hor$.}. This constitutes the ``down-up'' strategy 
mentioned in the Introduction.

Even though we adopt here the ``down-up'' approach, the organization 
of this section rather follows the
conceptual order dictated by the intrinsic geometry of $\Hor$.
Firstly we derive the implications of the choice of a 
WIH-compatible slicing. Then we apply 
the prescription in Sec. \ref{s:IH:prefWIH} for 
specifying a particular WIH and, 
finally, we revisit Sec. \ref{s:IH:goodcuts} and its
determination of the foliation once the WIH class is chosen.
As a result, we present different boundary values for the lapse
associated with each step.

\subsubsection{WIH-compatible slicing: $\kappa=\mathrm{const}$. 
Evolution equation for the lapse}
\label{s:BC:kappa_const}

Given an arbitrary but fixed WIH $(\Hor,[\el])$,
demanding the slicing defined by $N$ to be WIH-compatible (see
Sec. \ref{s:IH:WIHdefinition})
requires 
that the non-affinity coefficient $\kappa$ associated with the null vector 
$\w{\el}=N(\w{n}+\w{s})\in[\el]$ to be constant. 
Under the condition $\kappa=\mathrm{const}$ on the whole $\Hor$, the null 
normal $\el$ actually builds a WIH
structure\footnote{\label{fn:BC:kappa_const}Note 
that the condition $\kappa=\mathrm{const}$ does
not fix the WIH on $\Hor$, since a transformation (\ref{e:IH:fam_WIH}) changes
the WIH without affecting the constancy of $\kappa$. In this section 
we are assuming a given $[\el]$; in the next section
\ref{s:BC:prefWIH} the choice of a particular $[\el]$ is revisited.} or, in the
language of Sec. \ref{s:IH:IHhierarchy}, an $(A,B)$-horizon.
If, motivated by the 
discussion in Sec. \ref{s:IH:PhysParam} , 
we choose the representative $\el_0\in[\el]$ with non-affinity coefficient
given by the Kerr surface gravity, 
$\kappa_0=\kappa_{\Hor}(R_\Hor, J_\Hor)$, then it follows 
directly from Eq. (\ref{e:TP:kappa_3p1_index})
\be
\label{e:BC:evol_N}
\encadre{
\kappa_{\Hor}(R_\Hor, J_\Hor) = \LieH{\el} \ln N + s^i D_i N 
   - N K_{ij} s^i s^j
} \ .
\ee
This is an evolution equation for $N$ on $\Hor$ (see \cite{JaramGM04}).
As such, it can be employed to fix the values of $N$ along the 
horizon $\Hor$ once the lapse has been freely chosen on a initial
slice. This can be useful for fixing Dirichlet inner boundary conditions 
in the {\it slow} dynamical
evolution of a quasi-equilibrium black hole (e.g. the evolution 
of a black hole during a late ringing down phase). 
On the contrary, in the context of the construction of initial data,
Eq. (\ref{e:BC:evol_N}) by itself does not prescribe
a boundary condition for the lapse in Eq. (\ref{e:BC:cts_lapse}). 
This is precisely due to the presence of the $\LieH{\el} \ln N$ term,
which cannot be expressed in terms of the data on an initial slice. 
This is in agreement with the fact that imposing the slicing
to be WIH-compatible, through $\kappa=\kappa_{\Hor}$, 
does not determine the WIH class. A gauge choice
has to be made to fix, up to a constant, $\el$ and therefore the
family of slices $(\Sp_t)$ 
(cf. footnote \ref{fn:BC:kappa_const}, Secs. \ref{s:IH:prefWIH}
and \ref{s:IH:goodcuts}, as well as the rest of this section).

Consequently, if we are indeed interested in using Eq. (\ref{e:BC:evol_N})
for fixing a boundary condition for the lapse on the horizon, 
we are obliged to make a choice for the value of $\LieH{\el}N$ on 
$\Sp_0$.
In the quasi-equilibrium context, it is natural to  demand the lapse 
not to evolve: $\LieH{\el}N=0$. More generally, writing
$\LieH{\el}N=h_2$, with $h_2$ a function to be prescribed on $\Sp_0$,
Eq.~(\ref{e:BC:evol_N}) leads 
to the mixed boundary condition on $\Sp_0$
\bea
\label{e:BC:JGM_BC}
\left.\begin{array}{c}
\hbox{Eq. (\ref{e:BC:evol_N})} \\
+ \; \LieH{\el}N=h_2 
\end{array}\right\}
\Longrightarrow
\encadre{
\kappa_{\Hor}(R_\Hor, J_\Hor) = s^i D_i N 
   - N K_{ij} s^i s^j + h_2
} \ .
\eea
In this case, considered as a condition only on $\Sp_0$, the corresponding
$\el=N(\w{n}+\w{s})$  is associated {\it only} with an infinitesimal $(A)$-horizon.
Finally, another manner of looking at (\ref{e:BC:evol_N}) 
consists in freely prescribing the values for $N$ 
along the horizon $\Hor$ and consider Eq. (\ref{e:BC:evol_N}) as a constraint
on the rest of the fields, e.g. on the value of $K_{ij}s^is^j$ (see
example \ref{ex:BC:cond_mixed} below).

\subsubsection{Preferred WIH class: $\Lie{\el}\theta_{(\w{k})} =0$ }
\label{s:BC:prefWIH}

In Sec. \ref{s:IH:prefWIH} we prescribed a specific choice $[\el]$ of 
non-extremal WIH class among those that can be implemented on a
{\it generic} NEH.
This was achieved by imposing the derivative along $\el$ of the expansion 
associated with the ingoing null vector $\w{k}$, $\theta_{(\w{k})}$,
to vanish.  
In fact, such a condition could have been generalized to
\be
\label{e:BC:Lie_l_thetak_gauge}
\Lie{\el}\theta_{(\w{k})} = h_3 \ ,
\ee
with $\kappa_{(\el)}=\mathrm{const}$,
where the choice of $h_3$ corresponds to the choice of gauge in the WIH 
structure [different
choices for the function $h_3$ fix distinct values for the function $B$ 
in the transformation (\ref{e:IH:alpha}); the choice
$h_3=0$ corresponds to a  $(A,B,C)$-horizon, again in the
language of Sec. \ref{s:IH:IHhierarchy}].

In a 3+1 formulation where a given starting slice $\Sp_0$ is specified
and a WIH class is fixed, the 
choice of the only representative $\el\in [\el]$ 
characterized by $\kappa_{(\el)} =\kappa_{\Hor}(R_\Hor, J_\Hor)$
determines the slicing on $\Hor$. Therefore, a 
condition on the lapse must follow from
Eq.~(\ref{e:BC:Lie_l_thetak_gauge}).
If we substitute expressions  (\ref{e:TP:Omega_DN_qKs}) and  
(\ref{e:TP:theta_k_3p1}) in Eq. (\ref{e:IH:Lie_thetak}), we obtain indeed
\be
\label{e:BC:L_l_thetak_lapse}
\encadre{
\begin{array}{l}
^2D^\mu\DS_\mu N - 2 K_{\mu\nu} s^\nu \, \DS^\mu N \\
+\left(- \DS^\rho(q^\mu_{\ \, \rho} K_{\mu\nu} s^\nu) + q^{\mu\rho}
(K_{\mu\nu}s^\nu)(K_{\rho\sigma}s^\sigma)
-\frac{1}{2} {}^2\!R + \frac{1}{2} q^{\mu\nu}R_{\mu\nu} \right) N \nn \\  
+\frac{\kappa_{\Hor}(R_\Hor, J_\Hor)}{2} 
\left( D_\mu s^\mu - K_{\mu\nu} s^\mu s^\nu + K\right)= N h_3 \nn \ .
\end{array}
} 
\ee
This equation can
be used as a boundary condition for the lapse on
a cross-section $\Sp_0$ of $\Hor$. 
In this sense, it 
can be employed in combination with Eq. (\ref{e:BC:evol_N}): 
Eq. (\ref{e:BC:L_l_thetak_lapse}) fixes
the initial value of $N$ on $\Sp_0$ whereas Eq. (\ref{e:BC:evol_N}) dictates its ``time'' 
evolution. The freedom due to the presence of the $\LieH{\el}\ln N$
term in Eq. (\ref{e:BC:evol_N}) 
guarantees the compatibility between both equations.

Cook \cite{Cook02} has proposed 
a condition\footnote{See Refs. \cite{CookP04,Pfeif03} for a discussion on the degeneracy
occurring when using this boundary condition in conjunction with
other quasi-equilibrium conditions.} very similar to Eq.~(\ref{e:BC:Lie_l_thetak_gauge})
in order to fix the lapse on a initial slice $\Sp_0$. 
Using the null normal normalized as in Eq. (\ref{e:IN:def_hat_ell}) together with its dual (\ref{e:IN:hat_k}), 
i.e. $\hat{\el}:= (\w{n} + \w{s})/\sqrt{2}$ and $\hat{\w{k}}:= (\w{n} - \w{s})/\sqrt{2}$, 
and imposing
\bea
\label{e:BC:Cook_lapseBC}
\LieH{\hat{\el}}\theta_{(\hat{\w{k}})} = 0
\eea
on $\Sp_0$, leads to the condition on $N$ proposed by Cook \cite{Cook02}
and closely related to Eq. (\ref{e:BC:L_l_thetak_lapse}).
Using Eqs. (\ref{e:IN:el_nps}) and (\ref{e:NH:k_n_s}), together with the transformation law
for $\theta_{(\hat{\w{k}})}$ derived from Table \ref{t:KI:scaling}, it follows
\bea
\label{e:BC:relat_prefWIH_Cook}
\LieH{\el}\theta_{(\w{k})} = - (\LieH{\el} \ln N) \theta_{(\w{k})} +
\LieH{\hat{\el}}\theta_{(\hat{\w{k}})} ,
\eea
where in addition $\kappa_{(\el)}=\mathrm{const}$ must be satisfied.
Choosing as gauge condition in Eq. (\ref{e:BC:Lie_l_thetak_gauge})
$h_3=- (\LieH{\el} \ln N) \theta_{(\w{k})}$, both conditions 
(\ref{e:BC:Lie_l_thetak_gauge}) and (\ref{e:BC:Cook_lapseBC})
are the same.
Note also that in Eq. (\ref{e:BC:Cook_lapseBC}) we can always keep 
$\kappa_{(\el)}$ constant as  
long as we let $\LieH{\el} \ln N$ to be determined by
Eq. (\ref{e:BC:evol_N}). 

\subsubsection{Fixing the slicing: $\w{\DS}\cdot \w{\Omega} = h$.
Dirichlet boundary condition for the lapse}
\label{s:BC:fixslic}

In  Sec. \ref{s:IH:goodcuts} we concluded that, in the setting of an
``up-down'' approach, once a WIH class
has been fixed on $\Hor$ the choice of the exact part
$\w{\Omega}^{\mathrm{exact}}$ of 
the \hajicek\
form determines the foliation $(\Sp_t)$. We argued that, since its 
divergence-free part is 
fixed by relation (\ref{e:NE:Hdomega}), then 
the condition 
\bea
\label{e:BC:div_Omega}
\w{\DS}\cdot \w{\Omega} = h_4 \ ,
\eea
for some {\it gauge}  choice of $h_4$, actually  fix the foliation.
From the 3+1 perspective\footnote{We acknowledge 
B. Krishnan for the discussion in this section.}, this conclusion is a 
straightforward consequence
of Eq. (\ref{e:TP:Omega3+1}). Indeed, contracting (\ref{e:TP:Omega3+1}) with
$\DS^\mu$ and inserting (\ref{e:BC:div_Omega}), we obtain
\bea
\label{e:BC:lapl_N}
\encadre{
{}^2\Delta \mathrm{ln} N = \DS^\rho (q^\mu_{\ \, \rho} K_{\mu\nu} s^\nu) + h_4
} \ .
\eea
If we make now a gauge choice for $h_4$ (e.g. $h_4=0$ or the Pawlowski gauge 
as suggested in Sec. \ref{s:IH:goodcuts}), we dispose of an elliptic equation 
that fixes $N$ on $\Sp_t$ up to a constant value (or a function constant on $\Sp_t$,
if thinking in terms of $\Hor$). This fixes the foliation $(\Sp_t)$,
understanding the latter as the ensemble of leaves in $\Hor$
[distinct values of the integration ``constant'' only entail
different ``speeds'' to
go through the slicing $(\Sp_t)$].
In particular, the resulting lapse can be used as a Dirichlet boundary 
condition for the elliptic equation (\ref{e:BC:cts_lapse}).

As we can see, all boundary conditions derived at the
WIH level, and aimed at being imposed on a initial sphere $\Sp_0$,
involve the choice of some function $h_i$ that cannot be fixed 
in the context of the initial data problem.
This is the case of $h_2 = \LieH{\el}N$ in Eq. (\ref{e:BC:JGM_BC}),
which shows that the WIH structure, with its
``constant surface gravity'' characterization, cannot be captured
in terms of initial data. 
Regarding $h_3$ in Eq. (\ref{e:BC:L_l_thetak_lapse})
and $h_4$ in Eq. (\ref{e:BC:lapl_N}), they are directly related to the gauge
ambiguity in the free data of a WIH (more precisely to the {\it active}
and {\it passive} versions in Sec. \ref{s:NE:NEHfreedata}, respectively).
In sum, there exists an intrinsic ambiguity in the determination of the 2+1
slicing of a WIH. 
In consequence, we can conclude that the WIH on its own does not
permit to fully determine the boundary conditions of the
(extended) initial data problem and the prescription of an
additional condition (a function on $\Sp_0$) is unavoidable\footnote{Note that
this conclusion refers specifically to the initial data problem. A WIH
structure contains notions which are essentially dynamical 
(``second derivatives'' in time) and cannot be captured by the
initial data . The
WIH remains fully useful in the context of other problems. For instance, 
if we rather search an appropriate foliation for pursueing an
{\it a posteriori} analysis of
a full spacetime (not only a 3-slice) containing a NEH $\Hor$
(for instance the late time result of the numerical simulation
of a collapse), in a first step we could disregard not WIH-compatible foliations.
Then, after fixing a single 
spatial slice, the foliation on $\Hor$ can be completely
determined by using Eq. (\ref{e:BC:evol_N}).}.
Therefore the approach in Refs. \cite{CookP04,Ansor05}, where an
{\it effective} boundary condition on $\Sp_0$ is chosen for the lapse,
is fully justified from geometrical point of view.
Alternatively, we rather maintain here the geometrical expressions derived in
this section, and encode their effective character (as boundary conditions
on $\Sp_0$) through the free functions $h_i$ to be specified. Proceeding in this manner
{\it i)} the geometrical origin
of the ambiguity is made explicit, and {\it ii)} we can make use of 
the geometrical nature of the expressions to 
rearrange the correspondances between the 
different boundary conditions and the constrained parameter in the initial
data problem, in such a way that the WIH {\it effective} condition is not 
necessarily related to the lapse. We illustrate this second point in the following 
section.

\subsection{Other possibilities}
\label{s:BC:other}
In the previous two sections we have translated the NEH and WIH geometrical 
characterizations into boundary conditions on the constrained parameters 
of the initial data problem, with special emphasis in the conformal thin 
sandwich approach. In particular, we have first interpreted Eq. (\ref{e:BC:app_hor_Kss})
resulting from the vanishing of $\theta_{(\el)}$ as a boundary condition
for the conformal factor $\Psi$. The vanishing of the shear has translated
into the boundary condition (\ref{e:BC:lapl_V}) for the tangential part of 
the shift $\w{V}$, or simply into condition (\ref{e:BC:V_conf_Killing}) if an
additional condition on $\tilde{b}$ is enforced [either
Eq. (\ref{e:BC:radial_shift_dir}) or Eq. (\ref{e:BC:radial_shift_neu})
with the umbilical condition (\ref{e:BC:H_traceless})]. Finally, WIH conditions
in Sec. \ref{s:BC:BCWIH} are mainly interpreted as boundary conditions for the lapse.

However, a key feature of the present geometrical approach is the fact that
a given boundary condition is not necessarily associated with  
a single constrained parameter. It simply states a relation to be satisfied among 
different 3+1 fields and, consequently, can be in principle enforced as 
a boundary condition for different parameters. In this sense, the above-mentioned 
identification between boundary conditions and constrained parameters is
a well motivated choice but it is not the only one.
In order to facilitate the choice  of other possible combinations, we
recapitulate the boundary conditions presented along Sec. \ref{s:BC} in 
the Table \ref{t:BC:boundary_conditions}, where we make explicit the 
geometrical meaning of each of them. In particular, they are classified in NEH 
boundary conditions, in WIH-{\it motivated} conditions (since, as we have seen
in the previous section, they do not actually  construct a WIH) and a third 
set of boundary conditions, not necessarily related to a quasi-equilibrium
regime, that can also be used in general dynamical settings\footnote{The 
condition for fixing the slicing $\w{\DS}\cdot \w{\Omega} = h_4$ could also
be included in this category but, since we have mainly used the \hajicek{} form
in the quasi-equilibrium context, we keep it as a WIH-{\it motivated} condition.}.

\begin{table}
\begin{center}
\begin{tabular}{|c|c|c|l|}
\hline 
NEH & $\theta_{(\el)}=0$ & 
{\scriptsize
$4 \tilde{s}^i \tilde{D}_i \mathrm{ln} \Psi +
\tilde{D}_i\tilde{s}^i + \Psi^{-2} K_{ij} \tilde{s}^i\tilde{s}^j -
\Psi^2 K = 0$} & Eq. (\ref{e:BC:app_hor_Kss})  \\
\cline{2-4}
b. c.&$\w{\sigma} = 0$ &
{\scriptsize 
$^{2}\!\tilde{\Delta}V^a +  {}^2\!\tilde{R}^a_{\ b} V^b = \tDS^b
      \tilde{C}_b^{\ a}$ }
& {\small Eq. (\ref{e:BC:lapl_V})} \\
\hline
Non-eq. &  $r=\mathrm{const}$ &   
{\scriptsize $\tilde{b}=N\Psi^{-2}$} 
& {\small  Eq. (\ref{e:BC:radial_shift_dir}) }
\\
\cline{2-4}
b. c. & $K_{ij}s^is^j = h_1$ &
{\scriptsize
$2\tilde s^k\tD_k \tilde{b} - \tilde{b} \tilde H = 
3 N h_1 - \tDS_k V^k - 2 V^k \, \tilde{D}_{\tilde{s}} \tilde{s}_k -N K$}
& {\small Eq. (\ref{e:BC:radial_shift_neu}) }
\\
\hline
 &  $\LieH{\el}N=h_2$ & 
{\scriptsize
$\kappa_{\Hor}(R_\Hor, J_\Hor) = s^i D_i N 
   - N K_{ij} s^i s^j + h_2$
}  &{\small Eq. (\ref{e:BC:JGM_BC}) }
\\
\cline{2-4}
WIH &  &
{\scriptsize
$^2D^\mu\DS_\mu N - 2 K_{\mu\nu} s^\nu \, \DS^\mu N +
\left(- \DS^\rho(q^\mu_{\ \, \rho} K_{\mu\nu} s^\nu) +\right.$}
&
\\
b. c.&
$\LieH{\el}\theta_{(\w{k})}=h_3$ 
&
{\scriptsize
$\left. q^{\mu\rho}
(K_{\mu\nu}s^\nu)(K_{\rho\sigma}s^\sigma)
-\frac{1}{2} {}^2\!R + \frac{1}{2} q^{\mu\nu}R_{\mu\nu} \right) N +$}  
&
{\small Eq. (\ref{e:BC:L_l_thetak_lapse})}
\\
&&
{\scriptsize
$\frac{\kappa_{\Hor}(R_\Hor, J_\Hor)}{2} 
\left( D_\mu s^\mu - K_{\mu\nu} s^\mu s^\nu + K\right) = N h_3$}
&
\\        
\cline{2-4}
 &  $\w{\DS}\cdot \w{\Omega} = h_4$ & 
{\scriptsize
${}^2\Delta \mathrm{ln} N = \DS^\rho (q^\mu_{\ \, \rho} K_{\mu\nu} s^\nu) + h_4$
}  &
{\small Eq. (\ref{e:BC:lapl_N})}\\   
\hline
\end{tabular}
\end{center}
\vspace{2ex}
\caption[]{\label{t:BC:boundary_conditions}
Boundary conditions (b. c.) on $\Sp_0$ derived in Sec. \ref{s:BC}, together with
their geometrical content.}
\end{table}

As an illustration of a possible alternative combination of boundary conditions,
we provide a simple example (see Ref. \cite{JaramL05})
which represents, at the same time, a non-trivial implementation of the isolated 
horizon boundary conditions beyond the analytical stationary examples provided
in the rest of the article.
\begin{exmp}
\label{ex:BC:cond_mixed}
In the straightforward interpretation of Eqs. (\ref{e:BC:JGM_BC}) 
and (\ref{e:BC:lapl_N}) in Sec. \ref{s:BC:BCWIH}, they have been proposed as 
alternative boundary conditions for $N$, between which we must choose. However, 
we have also pointed out that Eq. (\ref{e:BC:JGM_BC}) can also be
understood as fixing the value of $K_{ij}s^is^j$. In that case we can use
it to determine the free function $h_1$ in boundary condition (\ref{e:BC:radial_shift_neu})
for $\tilde{b}$. Therefore a particular combination of boundary conditions in
Table \ref{t:BC:boundary_conditions} is given by: vanishing expansion 
for $\Psi$, conformal Killing condition for $\w{V}$, condition
(\ref{e:BC:radial_shift_neu}) for $\tilde{b}$ with $K_{ij}s^is^j$ fixed by
Eq.  (\ref{e:BC:JGM_BC}) with $h_2=0$
\bea
\label{e:BC:k_ss_JGM}
K_{ij}s^is^j =\frac{1}{N}\left(s^i D_i N - \kappa_{\Hor}(R_\Hor, J_\Hor)\right) \ \ ,
\eea
and, finally, condition (\ref{e:BC:lapl_N}) for the lapse.
Figure \ref{f:BC:kss_example} shows the maximum and minimum values 
of $K_{ij}s^is^j$ during the iteration of a numerical implementation
of these boundary conditions,  where we have chosen
$\dot{\tilde{\gamma}}_{ij} = K 
= 0$, $\tilde{\gamma}$ a flat metric, $\w{V} =\mathrm{const} \cdot \w{\partial_\varphi}$
(a symmetry on $\Sp_0$) and $h_4=0$ in Eq. (\ref{e:BC:lapl_N}), together with an integration constant 
$C=\ln 0.2$ for $\ln N$.
Since this implementation is performed in maximal slicing, in particular
the constructed quasi-equilibrium horizon is a 
{\it future} marginally trapped surface.
\begin{figure}
\centerline{\includegraphics[width=0.8\textwidth]{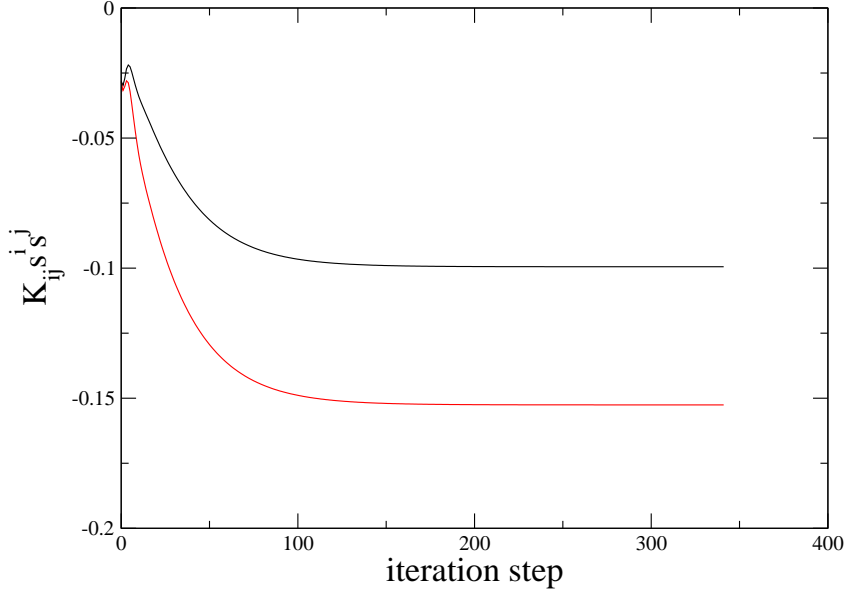}}
\caption[]{\label{f:BC:kss_example} 
Value of $K_{ij}s^is^j$ along the iteration of the simultaneous numerical implementation
of Eqs.(\ref{e:BC:JGM_BC}) and (\ref{e:BC:lapl_N}).}
\end{figure}
\end{exmp}
In brief, keeping boundary conditions in geometrical form we gain in flexibility for 
combining them in different manners. See Ref. \cite{JaramL05} for other
possibilities, in particular the enforcing of the vanishing of $\theta$
as a condition on the normal part of the shift, $\tilde{b}$, instead of a condition on $\Psi$.
In conjunction with Eq. (\ref{e:BC:lapl_V}) for $\w{V}$, this means
that the NEH condition $\w{\Theta}=0$ can be completely fulfilled by an appropriate
choice of the shift $\w{\beta}$.

%% file: concl.tex
%
%
\section{Conclusion}
\label{s:CO}

In this article, we have developed an approach to null hypersurfaces
based on the 3+1 formalism of general relativity, the main motivation
being the application of the isolated horizon formalism to numerical 
relativity. Although the geometry of a null hypersurface $\Hor$ 
can be elegantly studied from a purely intrinsic point of view, 
i.e. without referring to objects defined outside $\Hor$, the present
3+1 strategy proves to be useful, at least for two reasons. 
First of all, any 3+1 spacelike
slicing $(\Sigma_t)$ provides
a natural normalization of the null normal $\el$ to $\Hor$, 
along with a projector $\w{\Pi}$ onto $\Hor$ --- fixing the ambiguities
inherent to null hypersurfaces. Secondly, this permits to express
explicitly $\Hor$'s intrinsic quantities in terms of fields of 
direct interest for numerical relativity, 
like the extrinsic curvature $\w{K}$ of the hypersurface
$\Sigma_t$ and its the 3-metric $\w{\gamma}$
[or its conformal representation $(\Psi,\w{\tilde\gamma})$]. In addition, 
we have adopted a fully 4-dimensional point of view, by introducing an auxiliary
null foliation $(\Hor_u)$ in a neighborhood of $\Hor$.  
Not only this facilitates the link between the null geometry and 
the 3+1 description, but also reduces the actual computations to 
standard 4-dimensional tensorial calculus (e.g. involving the spacetime
connection $\w{\nabla}$) and 4-dimensional exterior calculus 
(e.g. the differential $\dd\uel$ and $\dd\uk$
of 1-forms associated with 
the null normal $\el$ and the ingoing null vector $\w{k}$, 
in connection with Frobenius theorem related to the submanifolds
$\Hor$ and $\Sp_t=\Hor\cap\Sigma_t$, or the decomposition of the curvature 
tensor following from Cartan's structure equations).

Thanks to the projector $\w{\Pi}$, we have performed a 4-dimensional
extension of $\Hor$'s second fundamental form $\w{\Theta}$,
and of the \hajicek\ 1-form $\w{\Omega}$. Besides, we have introduced 
as a basic object the transversal deformation rate $\w{\Xi}$.
By performing various projections of the Einstein equation, 
we have recovered, in addition to the null Raychaudhuri equation,
the Damour-Navier-Stokes equation, and have derived an evolution 
equation for $\w{\Xi}$. Independently of the Einstein equation, 
the so-called tidal force
equation (which involves the Weyl tensor)
is recovered. All these equations constitute a set of 
evolution equations along the null generators of $\Hor$. They
hold for any null hypersurface. 

Following \hajicek\  and Ashtekar et al., we have then considered 
non-expan\-ding null hypersurfaces (more specifically 
\emph{non-expanding horizons}) 
as a first step in the modelisation of a black
hole horizon in quasi-equilibrium. At this stage, a new geometrical
structure enters into the scene, namely the connection 
$\w{\hat\nabla}$ on $\Hor$ compatible with the degenerate metric $\w{q}$
and induced by the spacetime connection $\w{\nabla}$. This can be
achieved thanks to the vanishing of the second fundamental form $\w{\Theta}$
for a non-expanding null hypersurface. 
The couple $(\w{q},\w{\hat\nabla})$ then completely characterizes
the geometry of a non-expanding horizon. Once the null normal $\el$
is fixed by some 3+1 slicing $(\Sigma_t)$, 
this geometry is encoded in the fields 
$(\w{q},\w{\Omega},\kappa,\w{\Xi})$ evaluated in a spatial cross-section
$\Sp_t = \Hor \cap \Sigma_t$. The change in time of these quantities
is obtained by specializing the evolution equations to the case
$\w{\Theta}=0$. A number of possible constraints then follow for
characterizing  a horizon 
in quasi-equilibrium, beyond 
being simply non-expanding, leading to a hierarchy of structures
on $\Hor$. In particular, an intermediate notion 
of quasi-equilibrium is provided by the \emph{weakly isolated horizon} 
structure introduced by Ashtekar et al. and defined by 
requiring (i) a time-independent $\w{\Omega}$ and (ii) $\kappa$ 
to be constant
over $\Hor$. This permits a quasi-local
expression of physical parameters, like mass and angular momentum
and gives constraints on the 3+1 slicing, even if it does not
further constrain the geometry of $\Hor$. On the contrary, the
\emph{isolated horizon} structure, that requires 
all fields $(\w{q},\w{\Omega},\kappa,\w{\Xi})$ to be time-independent,
represents the maximal degree of equilibrium imposed in a
quasi-local manner. It really restricts $\Hor$ among all possible
non-expanding horizons. It also provides tools for extracting 
information in the neighborhood of $\Hor$. 

Thanks to explicit formul\ae\  relating $\w{q}$, $\w{\Theta}$, $\w{\Omega}$,
$\kappa$ and $\w{\Xi}$ to 3+1 fields, including the lapse function $N$
and shift vector $\w{\beta}$, we have then translated the isolated horizon
hierarchy into inner boundary conditions onto an excised sphere 
in the spatial hypersurface $\Sigma_t$. This permits us to study the
problem of initial data and spacetime evolution in a constrained scheme,
making the links with existing results in 
for numerical relativity. This connection with the 3+1 Cauchy problem
illustrates the ``down-up'' strategy mentioned in the Introduction, i.e.
the description of the null geometry on $\Hor$ by {\em constructing} it from
initial data on a spacelike 2-surface. 
This provides an alternative point of view
to the ``up-down'' picture, generally considered in the isolated horizon literature, 
and in which data on spatial slices are determined a posteriori
from a {\em given} 3-geometry on $\Hor$ as a whole.
In conclusion, the tools discussed in this article
are aimed to provide a useful setting for studying black holes in 
realistic astrophysical scenarios involving regimes close to the 
steady state.

\begin{ack}
We thank V\'\i ctor Aldaya, Lars Andersson, Marcus Ansorg, 
Abhay Ashtekar, Carlos Barcel\'o, 
Silvano Bonazzola, Brandon Carter, Gregory B. Cook, 
Sergio Dain, Thibault Damour, John Friedman,  
Achamveedu Gopakumar, 
Philippe Grandcl\'ement,  Mikolaj Korzynski,
Badri Krishnan, Jos\'e Mar\'\i a Mart\'\i n Garc\'\i a,
Jean-Pierre Lasota, 
Jurek Lewandow\-ski, Fran\c cois Limousin, Richard Matzner, 
Guil\-lermo Mena Marug\'an, J\'er\^ome Novak, Tomasz  Paw\-lowski,
Harald Pfeiffer, Jos\'e M. M. Senovilla, Koji Uryu, and Ra\"ul Vera
for very instructive discussions. 
J.L.J. is supported by a Marie
Curie Intra-European contract MEIF-CT-2003-500885 
within the 6th European Community Framework Programme.
\end{ack}

%% file: lieflow.tex
%
%
\section{Flow of time: various Lie derivatives along $\el$}
\label{s:LF}

The choice (\ref{e:NH:l_norm1}) for $\el$, as the tangent vector of 
$\Hor$'s null generators associated with the parameter $t$, means that
$\el$ can be considered as the ``advance-in-time'' vector associated with 
$t$. This is also manifest in the relation
$\langle \dd t, \el \rangle = 1$ [Eq.~(\ref{e:NH:l_norm2})]
or $\el = \tv + \w{V} + (N-b)\w{s}$ [Eq.~(\ref{e:IN:el_tau_V})], 
which shows that $\el$ is equal to the coordinate time vector $\w{t}$
plus some vector tangent to $\Sigma_t$ [namely $\w{V} + (N-b)\w{s}$].
Therefore in order to describe the ``time evolution'' of the objects
related to $\Hor$, it is natural to introduce the Lie derivative
along $\el$. 
However, it turns out that various kinds of such Lie derivatives
can be defined. 
First of all, there is the Lie derivative along the vector field
$\el$ within the spacetime manifold $\M$, which is denoted by 
$\Lie{\el}$. But since $\el\in\T(\Hor)$, there is also the Lie derivative
along the vector field $\el$ within the manifold $\Hor$, which 
we denote by $\LieH{\el}$. Finally, since $\el$ Lie drags 
the 2-surfaces $\Sp_t$ (cf. Sec.~\ref{s:IN:normal_l} and 
Fig.~\ref{f:IN:Lie_St}), one may define within the manifold $\Sp_t$ a
Lie derivative ``along $\el$'', which we denote $\LieS{\el}$.
We present here the precise definitions of these Lie derivatives 
and the relationships between them.


\subsection{Lie derivative along $\el$ within $\Hor$: $\LieH{\el}$}

Since $\el$ is a vector field on $\Hor$, one may naturally construct
the Lie derivative along $\el$ of any tensor field $\w{T}$ on $\Hor$.
This results in another tensor field on $\Hor$, which we denote
by $\LieH{\el}\w{T}$. 
Now, we may extend $\w{T}$ into a tensor field on $\M$ 
thanks to the push-forward
mapping $\Phi_*$ for vectors [cf. Eq.~(\ref{e:NH:push_forward})]
and the operator $\w{\Pi}^*$ for linear forms [cf. Eq.~(\ref{e:NH:def_P_star})].
It is then legitimate to ask for the relation between 
the Lie derivative $\Lie{\el}$ of this 4-dimensional extension within 
the manifold $\M$, and $\LieH{\el}\w{T}$.

Firstly we notice that both Lie derivatives
coincide onto vectors\footnote{More precisely we should write 
Eq.~(\ref{e:NE:LieH_Lie_vector}) as 
$\Phi_*\LieH{\el}\w{v} = \Lie{\Phi_*\el}\Phi_*\w{v}$, 
where $\Phi_*$ is the push-forward operator associated with the embedding
of $\Hor$ in $\M$ (cf. Sec.~\ref{s:NH:def_hyp}), but according to our
4-dimensional point of view, we do not distinguish between $\w{v}$
and $\Phi_*\w{v}$.}: 
\be \label{e:NE:LieH_Lie_vector}
    \encadre{ \forall\w{v}\in\T(\Hor),\quad
        \LieH{\el}\w{v} = \Lie{\el}\w{v} } . 
\ee
This follows from the very definition of the Lie derivative
(cf. Fig.~\ref{f:KI:Lie_vect}).
Let us now consider an arbitrary 1-form on $\Hor$: 
$\w{\varpi}\in\T^*(\Hor)$. 
Then invoking the Leibnitz rule on contractions and using
the property (\ref{e:NE:LieH_Lie_vector})
\bea
  \forall\w{v}\in\T(\Hor),\quad
  \langle \LieH{\el}\w{\varpi},\w{v}\rangle
   & = & \LieH{\el}\langle\w{\varpi},\w{v}\rangle
  - \langle \w{\varpi}, \LieH{\el}\w{v} \rangle \nonumber \\
  & = & \Lie{\el}\langle\w{\varpi},\w{v}\rangle
  - \langle \w{\varpi}, \Lie{\el}\w{v} \rangle \nonumber \\
  & = & \Lie{\el}\langle\w{\Pi}^*\w{\varpi},\w{v}\rangle
  - \langle \w{\Pi}^*\w{\varpi}, \Lie{\el}\w{v} \rangle \nonumber \\
  & = & \langle \Lie{\el} (\w{\Pi}^*\w{\varpi}), \w{v} \rangle .
\eea 
We conclude that the 1-forms 
$ \LieH{\el}\w{\varpi}$ and $\Lie{\el} (\w{\Pi}^*\w{\varpi})$
coincide on $\T(\Hor)$. Therefore their extensions to $\T(\M)$
provided by the projector $\w{\Pi}$ also coincide, and we can write
\be \label{e:NE:LieH_Lie_form}
     \forall\w{\w{\varpi}}\in\T^*(\Hor),\quad
       \w{\Pi}^* \, \LieH{\el}\w{\varpi} = \w{\Pi}^*\Lie{\el}
       (\w{\Pi}^*\w{\w{\varpi}})  . 
\ee
By taking tensorial products, we above analysis can be extended 
straightforwardly
to any field $\w{A}$ of multilinear forms acting on $\T_p(\Hor)$, 
so that we get
\be  \label{e:LF:LieH_Lie}
    \encadre{ \forall\w{\w{A}}\in\T^*(\Hor)^{\otimes n},\quad
       \w{\Pi}^* \, \LieH{\el}\w{A} = \w{\Pi}^*\Lie{\el}
       (\w{\Pi}^*\w{\w{A}}) }  . 
\ee


\subsection{Lie derivative along $\el$ within $\Sp_t$: $\LieS{\el}$}

\begin{figure}
\centerline{\includegraphics[width=0.7\textwidth]{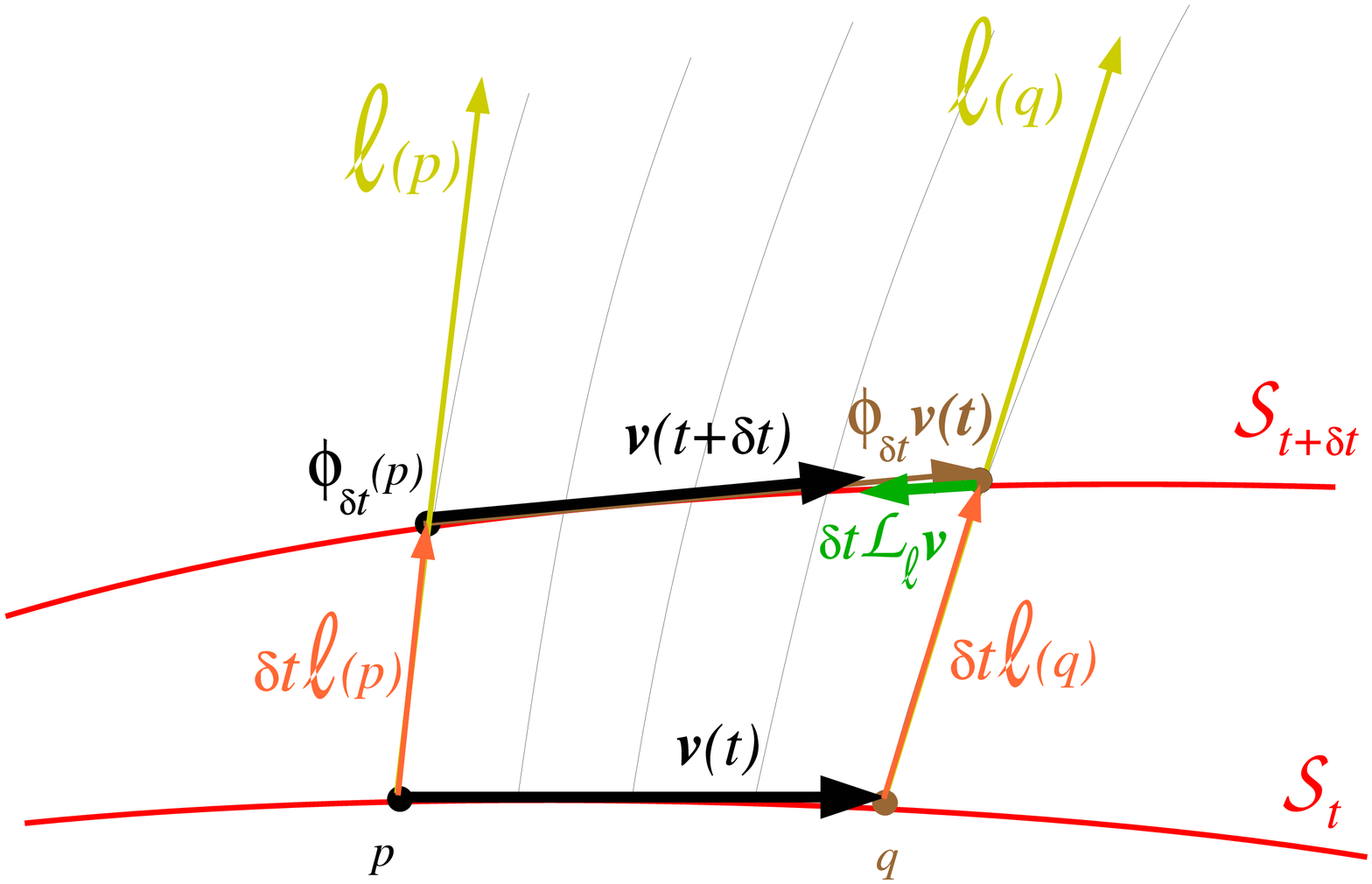}}
\caption[]{\label{f:KI:Lie_vect} 
Geometrical construction showing that $\Lie{\el} \w{v} \in\T(\Sp_t)$
for any vector $\w{v}$ tangent to the 2-surface $\Sp_t$: on $\Sp_t$,
a vector can be identified by a infinitesimal displacement between 
two points, $p$ and $q$ say. These points are transported onto the
neighbouring surface $\Sp_{t+\delta t}$ along the field lines of the
vector field $\el$ (thin lines on the figure) 
by the diffeomorphism $\phi_{\delta t}$
associated with $\el$: the displacement between $p$ and $\phi_{\delta t}(p)$
is the vector $\delta t\, \el$. The couple of points 
$(\phi_{\delta t}(p),\phi_{\delta t}(q))$ defines the vector 
$\phi_{\delta t} \w{v}(t)$ tangent to $\Sp_{t+\delta t}$. The Lie derivative
of $\w{v}$ along $\el$ is then defined by the difference between the
value of the vector field $\w{v}$ at the point $\phi_{\delta t}(p)$,
i.e. $\w{v}(t+\delta t)$, and the vector transported from $\Sp_t$ along
$\el$'s field lines, i.e. $\phi_{\delta t} \w{v}(t)$ :
$\Lie{\el}{\w{v}}(t+\delta t) = \lim_{\delta t\rightarrow 0} [ \w{v}(t+\delta t) -
\phi_{\delta t} \w{v}(t)]/\delta t$. Since both vectors $ \w{v}(t+\delta t)$
and $\phi_{\delta t} \w{v}(t)$ are in $\T(\Sp_{t+\delta t})$, it is then
obvious that $\Lie{\el}{\w{v}}(t+\delta t) \in\T(\Sp_{t+\delta t})$.}
\end{figure}

We have seen in Sec.~\ref{s:IN:normal_l} that $\el$
Lie drags the 2-surfaces $\Sp_t$: $\Sp_{t+\delta t}$
is obtained from the neighbouring surface $\Sp_t$ by an infinitesimal
displacement $\delta t \el$ of each point of $\Sp_t$. 
As stressed by Damour \cite{Damou79}, an immediate 
consequence of this is that the Lie derivative along $\el$ of any
vector tangent to $\Sp_t$ is a vector which is also tangent to
$\Sp_t$:
\be \label{e:KI:Lie_v}
    \forall \w{v}\in\T(\Sp_t),\quad \Lie{\el} \w{v} \in\T(\Sp_t) .
\ee
This is obvious from the geometrical definition of a Lie derivative
(see Fig.~\ref{f:KI:Lie_vect}). It can also be established ``blindly'':
consider $\w{v}\in\T(\Sp_t)$; then $\el\cdot\w{v}=\w{k}\cdot\w{v}=0$, 
so that 
\be \label{e:KI:l_Lie_v}
    \el \cdot \Lie{\el} \w{v} = \el \cdot (\w{\nabla}_{\el} \w{v}
        - \w{\nabla}_{\w{v}} \el) = 
        \underbrace{\el \cdot \w{\nabla}_{\el} \w{v}}_{=-
        \w{v} \cdot \w{\nabla}_{\el} \el}
            - \underbrace{\el \cdot  \w{\nabla}_{\w{v}} \, \el}_{=0}
            = - \w{v} \cdot (\kappa \el) = 0 . 
\ee
and
\be
    \w{k} \cdot \Lie{\el} \w{v} =  \w{k} \cdot \w{\nabla}_{\el} \w{v}
        - \w{k}\cdot \w{\nabla}_{\w{v}} \el
         = - \w{v} \cdot \w{\nabla}_{\el} \w{k}
          + \el \cdot \w{\nabla}_{\w{v}} \w{k}
          = \dd \uk (\w{v},\el) . 
\ee
With the expression (\ref{e:KI:dk}) of the exterior derivative of $\uk$
and the fact that $\langle \uel, \w{v}\rangle = 0$ and 
$\langle \uel, \el\rangle = 0$, we get immediately
\be \label{e:KI:k_Lie_v}
    \w{k} \cdot \Lie{\el} \w{v} = 0 . 
\ee
Equations (\ref{e:KI:l_Lie_v}) and (\ref{e:KI:k_Lie_v}), by stating
that $\Lie{\el} \w{v}$ is orthogonal to both $\el$ and $\w{k}$, 
show that $\Lie{\el} \w{v}$ is tangent to  $\Sp_t$ [cf. 
Eq.~(\ref{e:NH:span_perp_S})], and therefore establish (\ref{e:KI:Lie_v}).
The property (\ref{e:KI:Lie_v}) means that, although 
$\el\not\in \T(\Sp_t)$, $\Lie{\el}$ can be viewed
as an internal operator on the space $\T(\Sp_t)$ of vector fields tangent to
$\Sp_t$. We will denote it as $\LieS{\el}$ to stress this feature and
rewrite Eq.~(\ref{e:KI:Lie_v}) as 
\be \label{e:KI:LieS_v}
    \encadre{ 
        \forall \w{v}\in\T(\Sp_t),\quad \LieS{\el} \w{v}
        := \Lie{\el} \w{v} \in\T(\Sp_t) } .
\ee

The definition of $\LieS{\el}$ can be extended to 1-forms on $\Sp_t$
by demanding that the Leibnitz rule holds
for the contraction of a 1-form and a vector field: if
$\w{\varpi}\in\T^*(\Sp_t)$ is a 1-form on $\Sp_t$, 
we define the {\em Lie derivative 
$\LieS{\el}\w{\varpi}$ of
$\w{\varpi}$ along $\el$} as the 1-form whose action on vectors is
\be \label{e:KI:def_Lie_S}
  \forall \w{v}\in\T(\Sp_t),\quad 
  \langle \LieS{\el}\w{\varpi}, \w{v} \rangle :=
  \Lie{\el} \langle \w{\varpi}, \w{v} \rangle 
  - \langle \w{\varpi}, \LieS{\el} \w{v} \rangle .
\ee
Note that the right-hand side of this equation is well defined
since $\LieS{\el} \w{v} \in\T(\Sp_t)$, so that we can apply 
the 1-form $\w{\varpi}$ to it. 
We can extend the definition of the Lie derivative $\LieS{\el}$
to bilinear forms on $\Sp_t$, and more generally to 
multilinear forms, by means of Leibnitz rule:
\be
    \LieS{\el}(\w{\varpi}_1 \otimes \w{\varpi}_2) =
    \LieS{\el} \w{\varpi}_1 \otimes \w{\varpi}_2
    + \w{\varpi}_1 \otimes \LieS{\el} \w{\varpi}_2 . 
\ee
Taking into account the property (\ref{e:KI:LieS_v}), which also holds
for any tensorial product of vectors, we finally conclude that
the Lie derivative operator $\LieS{\el}$ 
is defined for any tensor field on $\Sp_t$: it is internal
to $\Sp_t$ in the sense that it transforms a tensor field on $\Sp_t$
into another tensor field on $\Sp_t$. This 2-dimensional
operator has been introduced by Damour \cite{Damou79,Damou82} 
and called by  him the {\em ``convective derivative''}. 

Now, any 1-form $\w{\varpi}\in \T^*(\Sp_t)$
can also be seen as a 1-form on $\M$ thanks to the orthogonal
projector $\vec{\w{q}}$ on $\T(\Sp_t)$ : it is the 1-form
$\vec{\w{q}}^*\w{\varpi}$ defined by Eq.~(\ref{e:KI:def_q_star})
\footnote{In this Appendix, we make a distinction between 
$\w{\varpi}\in \T^*(\Sp_t)$ and $\vec{\w{q}}^*\w{\varpi}\in \T^*(\M)$,
whereas in the remaining of the article we use the same symbol to 
denote both applications, considering $\w{\varpi}$ as the pull-back
of $\vec{\w{q}}^*\w{\varpi}$ by the embedding of $\Sp_t$ in $\M$.}.
Let us then investigate the relation between the ``4-dimensional'' 
Lie derivative $\Lie{\el} \vec{\w{q}}^*\w{\varpi}$ and the
``2-dimensional'' one, $\LieS{\el}\w{\varpi}$. 
The first thing to notice is that 
the 1-forms $\Lie{\el} \vec{\w{q}}^*\w{\varpi}$ and 
$\vec{\w{q}}^* \; \LieS{\el} \w{\varpi}$ coincide when restricted
to $\T(\Sp_t)$. Indeed 
\bea
    \forall \w{v}\in\T(\Sp_t),\quad 
    \langle \Lie{\el} \vec{\w{q}}^*\w{\varpi} , \w{v} \rangle & =&
        \Lie{\el} \langle \vec{\w{q}}^*\w{\varpi} , \w{v} \rangle 
        - \langle \vec{\w{q}}^*\w{\varpi} , \Lie{\el} \w{v} \rangle \nonumber \\
& =& \Lie{\el} \langle \w{\varpi} , \vec{\w{q}}(\w{v}) \rangle 
        - \langle \w{\varpi} , \vec{\w{q}}(\Lie{\el} \w{v}) \rangle \nonumber \\    
            & = & \Lie{\el} \langle \w{\varpi} , \w{v} \rangle 
        - \langle \w{\varpi} , \Lie{\el} \w{v} \rangle  \nonumber \\ 
        & = & \langle \LieS{\el}\w{\varpi}, \w{v} \rangle ,  
\eea
where the third equality follows from property (\ref{e:KI:Lie_v}) and
the fourth one from the definition (\ref{e:KI:def_Lie_S}). Hence
\be
    \forall \w{\varpi}\in\T^*(\Sp_t),\quad
     \LieS{\el}\w{\varpi} = 
     \left. \left( \Lie{\el} \vec{\w{q}}^*\w{\varpi}  \right) 
     \right| _{\T(\Sp_t)} .  \label{e:KI:Lqpi} 
\ee
Moreover, 
\bea
    \forall \w{\varpi}\in\T^*(\Sp_t),\quad
        \langle \Lie{\el} \vec{\w{q}}^*\w{\varpi} , \el \rangle & =&
        \Lie{\el} 
        \underbrace{\langle \vec{\w{q}}^*\w{\varpi} , \el \rangle}_{=0} 
        - \langle \vec{\w{q}}^*\w{\varpi} , 
            \underbrace{\Lie{\el} \el}_{=0} \rangle \nonumber \\
        \langle \Lie{\el} \vec{\w{q}}^*\w{\varpi} , \el \rangle & =& 0 . 
                \label{e:KI:Lqpi_l}     
\eea
Since $\langle \vec{\w{q}}^*\; \LieS{\el}\w{\varpi},\el \rangle
= 0$ (for $\vec{\w{q}}(\el)=0$), 
we can combine Eqs.~(\ref{e:KI:Lqpi}) and (\ref{e:KI:Lqpi_l}) in 
\be \label{e:KI:LqH}
    \forall \w{\varpi}\in\T^*(\Sp_t),\quad
    \left. \left( \vec{\w{q}}^*\; \LieS{\el}\w{\varpi} 
        \right) 
        \right| _{\T(\Hor)} =
     \left. \left( \Lie{\el} \vec{\w{q}}^*\w{\varpi}  \right) 
        \right| _{\T(\Hor)} .    
\ee
But regarding the direction transverse to $\Hor$ one has
\bea
    \forall \w{\varpi}\in\T^*(\Sp_t),\quad
        \langle \Lie{\el} \vec{\w{q}}^*\w{\varpi} , \w{k} \rangle & =&
        \Lie{\el} 
        \underbrace{\langle \vec{\w{q}}^*\w{\varpi} , \w{k} \rangle}_{=0} 
        - \langle \vec{\w{q}}^*\w{\varpi} , 
            \underbrace{\Lie{\el} \w{k}}_{=[\el,\w{k}]} \rangle \nonumber \\
        \langle \Lie{\el} \vec{\w{q}}^*\w{\varpi} , \w{k} \rangle & =& 
        \langle \w{\varpi}, \vec{\w{q}}([\w{k},\el]) \rangle , 
                            \label{e:KI:Lqpi_k}    
\eea
where $[\w{k},\el]$ denotes the commutator of vectors $\w{k}$ and $\el$.
The right-hand side of Eq.~(\ref{e:KI:Lqpi_k}) is in general different from zero.
Indeed, a simple calculation using Eq.~(\ref{e:KI:dk}) shows that
\be
    [k,\ell]^\alpha = k^\mu (\nabla_\mu \ell^\alpha + \nabla^\alpha \ell_\mu)
        - \frac{1}{2N^2} \nabla_{\el} \ln\left(\frac{N}{M}\right)
            \, \ell^\alpha , 
\ee
so that 
\be
   q^\alpha_{\ \, \mu} [k,\ell]^\mu = k^\mu q^{\alpha\nu}
        (\nabla_\mu \ell_\nu + \nabla_\nu \ell_\mu) . 
\ee
For instance a sufficient condition for the right-hand side of 
Eq.~(\ref{e:KI:Lqpi_k}) to vanish, and then 
$ \Lie{\el} \vec{\w{q}}^*\w{\varpi}$ to coincide
with $\vec{\w{q}}^* \LieS{\el} \w{\varpi}$,
consists in demanding $\el$ to be a Killing vector 
of spacetime: $\nabla_\alpha \ell_\beta + \nabla_\beta \ell_\alpha=0$. 

Another writing of Eq.~(\ref{e:KI:Lqpi}) is
\be \label{e:KI:qLq}
    \forall \w{\varpi}\in\T^*(\Sp_t),\quad
    \vec{\w{q}}^*\; \LieS{\el}\w{\varpi} = 
    \vec{\w{q}}^* \Lie{\el} \vec{\w{q}}^*\w{\varpi} ,
\ee
where each side of the equality is a 1-form on $\T(\M)$ and the 
operators $\vec{\w{q}}^*$ added with respect to Eq.~(\ref{e:KI:Lqpi})
effectively restrict the non-trivial action of these 1-forms 
to the subspace $\T(\Sp_t)$
of $\T(\M)$. 

By taking tensorial products, the above analysis can be extended
easily to any multilinear form $\w{A}$ acting on $\T(\Sp_t)$. In
particular Eq.~(\ref{e:KI:qLq}) can be generalized to 
\be \label{e:KI:qLqT}
   \encadre{ \forall \w{A}\in\T^*(\Sp_t)^{\otimes n},\quad
   \vec{\w{q}}^*\; \LieS{\el}\w{A} = 
    \vec{\w{q}}^* \Lie{\el} \vec{\w{q}}^*\w{A} } . 
\ee
Note the similarity between this relation and Eq.~(\ref{e:LF:LieH_Lie})
for $\LieH{\el}$. 

%% file: cartan.tex
%
%
\section{Cartan's structure equations} \label{s:CA}

Many studies about null hypersurfaces and isolated horizons
make use of the Newman-Penrose framework, which is based on 
the {\em complex} null tetrad introduced in Sec.~\ref{s:IN:NP}. 
An alternative approach is Cartan's formalism which is based on 
a {\em real} tetrad and exterior calculus (see e.g. Chap.~14 of
MTW \cite{MisneTW73} or Chap.~V.B of Ref.~\cite{ChoquDD77}
for an introduction). Cartan's formalism
is at least as powerful as the Newman-Penrose one, although it 
remains true that the latter is well adapted to null surfaces. 


\subsection{Tetrad and connection 1-forms}

In the present context, it is natural to consider the following bases
for, respectively, $\T(\M)$ (vector fields) and $\T^*(\M)$ (1-forms)
\be
    \w{e}_\alpha = (\el, \w{k},\w{e}_2,\w{e}_3)
    \quad\mbox{and}\quad
    \w{e}^\alpha = (-\uk, -\uel,\w{e}^2,\w{e}^3) , \label{e:CA:tetrad}
\ee
where $\w{e}_2$ and $\w{e}_3$ are two vector fields tangent to the 2-surface
$\Sp_t$ which constitute an orthonormal basis of $\T(\Sp_t)$
(with respect to the induced Riemannian metric $\w{q}$ of $\Sp_t$)  and
$\w{e}^2$ and $\w{e}^3$ are the two 1-forms in $\T^*(\M)$ such 
that the basis $(\w{e}^\alpha)$ of $\T^*(\M)$ is the dual of the 
basis $(\w{e}_\alpha)$ of $\T(\M)$, i.e. it satisfies
\be \label{e:CA:dual_tetrad}
    \langle \w{e}^\alpha, \w{e}_\beta \rangle = \delta^\alpha_{\ \, \beta},
\ee
where $ \delta^\alpha_{\ \, \beta}$ denotes the Kronecker symbol.
The vector basis $(\w{e}_\alpha)$ is usually called a {\em tetrad},
or {\em moving frame} or {\em rep\`ere mobile}. Note that
the ordering $(\w{e}_0=\el,\; \w{e}_1=\w{k})$ and
$(\w{e}^0 =  -\uk,\;  \w{e}^1=-\uel)$ has been chosen to ensure
Eq.~(\ref{e:CA:dual_tetrad}) for $\alpha,\beta\in\{0,1\}$, by
virtue of the fact that $\el$ and $\w{k}$ are null vectors and
satisfy $\el\cdot\w{k}=-1$ [Eq.~(\ref{e:NH:l_k_m1})]. 
Note that the tetrad $(\el, \w{k},\w{e}_2,\w{e}_3)$ is the same as 
that used to construct the complex Newman-Penrose null tetrad in
Sec.~\ref{s:IN:NP}.

Thanks to the properties (\ref{e:NH:l_k_m1}) and (\ref{e:NH:span_perp_S}),
the metric tensor components with respect to the chosen tetrad are
\be \label{e:CA:gab}
    g_{\alpha\beta} = \w{g}(\we_\alpha,\we_\beta)  
    = \left( \begin{array}{cccc}
    0 & -1 & \ \, 0 & \ \, 0 \\
    -1 & 0 & \ \, 0 & \ \, 0 \\
    0 & 0 & \ \,  1 & \ \, 0 \\
    0 & 0 & \ \, 0 & \ \, 1 
    \end{array} \right) .   
\ee

The {\em connection 1-forms} of the spacetime connection $\w{\nabla}$
with respect to the tetrad  $(\w{e}_\alpha)$ are the sixteen 1-forms
$\w{\omega}^\beta_{\ \, \alpha}$ defined by 
\be
    \forall\w{v}\in\T(\M),\quad 
        \w{\nabla}_{\w{v}} \, \w{e}_\alpha
        = \langle \w{\omega}^\mu_{\ \, \alpha}, \w{v}\rangle \, 
            \w{e}_\mu . 
\ee
The expansions of the connection 1-forms on the basis 
$(\w{e}^\alpha)$ of $\T^*(\M)$ define
the {\em connection coefficients}\footnote{Note that we are following
MTW convention \cite{MisneTW73} for the ordering of the indices 
$\alpha \gamma$ of the connection coefficients, which is the 
reverse of Hawking \& Ellis' one \cite{HawkiE73}.}
 $\Gamma^\beta_{\ \, \alpha \gamma}$ 
of $\w{\nabla}$ with respect to the tetrad $(\w{e}_\alpha)$:
\be
    \w{\omega}^\beta_{\ \, \alpha} = \Gamma^\beta_{\ \, \alpha \mu} \, 
		\w{e}^\mu \qquad\mbox{or}\qquad
        \Gamma^\beta_{\ \, \alpha \gamma} = \langle \we^\beta,
            \w{\nabla}_{\we_\gamma} \we_\alpha \rangle.  
                                            \label{e:CA:connect_coef}
\ee
By direct computations using the formulas of Sec.~\ref{s:KI}, we get 
\bea
    & & \w{\omega}^0_{\ \, 0} = - \w{\omega}^1_{\ \, 1} = 
        \w{\omega} - N^{-2}\w{\nabla}_{\el} \sigma \, \uel 
                                                \label{e:CA:omega00} \\
    & & \w{\omega}^1_{\ \, 0} = \w{\omega}^0_{\ \, 1} = 0   \\
    & & \w{\omega}^a_{\ \, 0} = \w{\omega}^1_{\ \, a} = 
    (\Omega_a - \w{\nabla}_{\w{e}_a}\rho)\, \uel
        + \Theta_{ab} \, \w{e}^b  \\
    & & \w{\omega}^a_{\ \, 1} = \w{\omega}^0_{\ \, a}
        = - \Omega_a \, \uk - N^{-2} \w{\nabla}_{\w{e}_a}\sigma \, \uel
            + \Xi_{ab} \, \w{e}^b \\
    & & \w{\omega}^b_{\ \, a} = - \w{\omega}^a_{\ \, b} = 
    - \Gamma^b_{\ \, a0} \, \uk
    - \Gamma^b_{\ \, a1} \, \uel
    + \Gamma^b_{\ \, ac} \, \w{e}^c , \label{e:CA:omegaba} 
\eea
where $\rho$ is related to the lapse $N$ and the metric factor $M$
by $\rho=\ln(MN)$ [Eq.~(\ref{e:IN:def_M})] and 
we have introduced the abbreviation
\be
    \sigma := \frac{1}{2}\ln\left( \frac{N}{M} \right) .
\ee
$\Omega_a$, $\Theta_{ab}$ and $\Xi_{ab}$ denote
the components of respectively the \hajicek\ 1-form $\w{\Omega}$, 
the deformation rate $\w{\Theta}$ and transversal deformation rate
$\w{\Xi}$ with respect to the basis $(\w{e}^a)=(\w{e}^2,\w{e}^3)$
of $\T^*(\Sp_t)$:
\be
    \w{\Omega} = \Omega_a \, \w{e}^a , \quad
    \w{\Theta} = \Theta_{ab} \, \w{e}^a \otimes \w{e}^b  \quad \mbox{and} \quad 
    \w{\Xi} = \Xi_{ab} \, \w{e}^a \otimes \w{e}^b . \label{e:CA:expans_omthxi}
\ee
Note that the above expressions are not restricted to $\T(\Sp_t)$ but
do constitute 4-dimensional writings of the 1-form $\w{\Omega}$
and the bilinear forms $\w{\Theta}$ and $\w{\Xi}$, since all these forms
vanish on the vectors $\w{e}_0=\el$ and $\w{e}_1=\w{k}$
[cf. Eqs.~(\ref{e:NH:Theta_l_k_0}), (\ref{e:NH:Omega_l_k_0}),
and (\ref{e:KI:Xi_qstar_gradk})]. 
Note also that since the basis $(\w{e}_a)$ is orthonormal, one
has $\Theta^a_{\ \, b} = \Theta_{ab}$ and $\Xi^a_{\ \, b} = \Xi_{ab}$. 

The symmetries (or antisymmetries) of the 1-forms $\w{\omega}^\beta_{\ \, \alpha}$
when changing the indices $\alpha$ and $\beta$, as expressed in 
Eqs.~(\ref{e:CA:omega00})-(\ref{e:CA:omegaba}), are due to the constancy
of the components $g_{\alpha\beta}$ of the metric tensor $\w{g}$ 
in the basis $\w{e}^\alpha\otimes\w{e}^\beta$
[cf. Eq.~(\ref{e:CA:gab})]. Indeed this
constancy, altogether with the metric compatibility relation 
$\dd g_{\alpha\beta} = \w{\omega}_{\alpha\beta} + \w{\omega}_{\beta\alpha}$
(cf. e.g. Eq.~(14.31b) of MTW \cite{MisneTW73}), implies
$\w{\omega}_{\alpha\beta} = - \w{\omega}_{\beta\alpha}$, where
$\w{\omega}_{\alpha\beta} := g_{\alpha\mu} \w{\omega}^\mu_{\ \, \beta}$. 
In particular $\w{\omega}^2_{\ \, 2} = \w{\omega}^3_{\ \, 3} = 0$. 


\subsection{Cartan's first structure equation}

Cartan's first structure equation states that the exterior derivative 
of each 1-form $\w{e}^\alpha$ is a 2-form which is expressible 
as a sum of exterior products involving the connection 
1-forms\footnote{See Sec.~\ref{s:IN:exterior} for our conventions 
regarding exterior calculus.}: 
\be \label{e:CA:1st_struct}
    \dd \w{e}^\alpha = \w{e}^\mu \wedge \w{\omega}^\alpha_{\ \, \mu} .
\ee
These relations actually express the vanishing of the torsion of the
spacetime connection $\w{\nabla}$. 

For $\alpha=0$, Eq.~(\ref{e:CA:1st_struct}) results in 
\bea
    \dd\w{e}^0 & = & \w{e}^0 \wedge \w{\omega}^0_{\ \, 0}
     + \w{e}^1 \wedge \w{\omega}^0_{\ \, 1}
     + \w{e}^a \wedge \w{\omega}^0_{\ \, a} \nonumber \\
     -\dd \uk & = & -\uk\wedge (\w{\omega} 
     - N^{-2}\w{\nabla}_{\el} \sigma \, \uel)
     + \w{e}^a \wedge
     ( - \Omega_a \, \uk - N^{-2} \w{\nabla}_{\w{e}_a}\sigma \, \uel
            + \Xi_{ab} \, \w{e}^b ) \nonumber \\
    \dd\uk & = & \uk\wedge\w{\omega} - N^{-2}\w{\nabla}_{\el} \sigma \, 
    \uk\wedge\uel + \underbrace{\Omega_a \w{e}^a}_{=\w{\Omega}}\wedge\uk
    + N^{-2} \w{\nabla}_{\w{e}_a}\sigma \, \w{e}^a \wedge \uel 
    \nonumber \\
    & & - \underbrace{\Xi_{ab} \, \w{e}^a\wedge\w{e}^b}_{=0} \nonumber\\
     & = & \uk\wedge(\w{\Omega} - \kappa \uk)
        - \uk\wedge\w{\Omega} + N^{-2} \left(  - \w{\nabla}_{\el} \sigma \, \uk
        + \w{\nabla}_{\w{e}_a}\sigma \, \w{e}^a\right) \wedge \uel \nonumber \\
     & = & N^{-2} \left( - \w{\nabla}_{\w{k}} \sigma \, \uel
    - \w{\nabla}_{\el} \sigma \, \uk
        + \w{\nabla}_{\w{e}_a}\sigma \, \w{e}^a\right) \wedge \uel  \nonumber \\
    \dd\uk & = & N^{-2} \dd \sigma \wedge \uel  , \label{e:CA:Frobenius_k}
\eea
where we have used the symmetry of $\Xi_{ab}$, as well as the expression 
(\ref{e:KI:Omega_omega_k}) of $\w{\omega}$ in terms of $\w{\Omega}$
and $\uk$. 
Equation (\ref{e:CA:Frobenius_k}) is nothing but 
the Frobenius relation (\ref{e:KI:dk}). 

For $\alpha=1$, Eq.~(\ref{e:CA:1st_struct}) results in 
\bea
    \dd\w{e}^1 & = & \w{e}^0 \wedge \w{\omega}^1_{\ \, 0}
     + \w{e}^1 \wedge \w{\omega}^1_{\ \, 1}
     + \w{e}^a \wedge \w{\omega}^1_{\ \, a} \nonumber \\
     -\dd\uel & = & -\uel\wedge(-\w{\omega} 
     + N^{-2}\w{\nabla}_{\el} \sigma \, \uel)
     + \w{e}^a \wedge \left[ (\Omega_a - \w{\nabla}_{\w{e}_a}\rho)\, \uel
        + \Theta_{ab} \, \w{e}^b \right] \nonumber \\
    \dd\uel &=& (\w{\omega}-\w{\Omega} + \w{e}^a\w{\nabla}_{\w{e}_a} \rho)
        \wedge \uel 
        - \underbrace{\Theta_{ab} \, \w{e}^a\wedge\w{e}^b}_{=0} 
        = ( -\kappa \, \uk + \w{e}^a\w{\nabla}_{\w{e}_a} \rho) 
           \wedge \uel  \nonumber\\
    &=& (- \w{\nabla}_{\el}\,\rho\, \uk 
       - \w{\nabla}_{\w{k}}\,\rho\, \uel + \w{\nabla}_{\w{e}_a} \rho\, \w{e}^a)
           \wedge \uel  \nonumber\\
    \dd\uel & = & \dd\rho \wedge \uel ,         \label{e:CA:Frobenius_l}
\eea
where we have used the symmetry of $\Theta_{ab}$, as well as 
Eqs.~(\ref{e:KI:Omega_omega_k}) and (\ref{e:NH:def_kappa}). 
Again, we recover a previously derived Frobenius relation, namely 
Eq.~(\ref{e:NH:Frobenius_l}).

Finally, for $\alpha=a=2$ or $3$, Cartan's first structure equation 
(\ref{e:CA:1st_struct}) results in 
\bea
    \dd\w{e}^a & = & \w{e}^0 \wedge \w{\omega}^a_{\ \, 0}
     + \w{e}^1 \wedge \w{\omega}^a_{\ \, 1}
     + \w{e}^b \wedge \w{\omega}^a_{\ \, b} \nonumber \\
     & = & -\uk\wedge \left[ (\Omega_a - \w{\nabla}_{\w{e}_a}\rho)\, \uel
        + \Theta_{ab} \, \w{e}^b \right] 
        - \uel\wedge( - \Omega_a \, \uk - N^{-2} \w{\nabla}_{\w{e}_a}\sigma \, \uel
            + \Xi_{ab} \, \w{e}^b ) \nonumber \\
    & &    + \w{e}^b \wedge \left( - \Gamma^a_{\ \, b0} \, \uk
    - \Gamma^a_{\ \, b1} \, \uel
    + \Gamma^a_{\ \, bc} \, \w{e}^c \right)  \nonumber \\
   \dd\w{e}^a & = &  \left( 2\Omega_a - \w{\nabla}_{\w{e}_a}\rho\right)
     \uel\wedge\uk
   + \left( \Theta_{ab} - \Gamma^a_{\ \, b0} \right) \w{e}^b \wedge \uk
   + \left( \Xi_{ab} - \Gamma^a_{\ \, b1} \right) \w{e}^b \wedge \uel
    \nonumber \\
    & & +  \Gamma^a_{\ \, bc} \, \w{e}^b \wedge \w{e}^c .
                                                \label{e:CA:1st_struct_a}
\eea


\subsection{Cartan's second structure equation}

Cartan's second structure equation relates the exterior derivative of
the connection 1-forms $\w{\omega}^\alpha_{\ \, \beta}$ to the
connection curvature:
\be \label{e:CA:2nd_struct}
    \dd\w{\omega}^\alpha_{\ \, \beta} = \w{{\mathcal R}}^\alpha_{\ \, \beta}
        - \w{\omega}^\alpha_{\ \,\mu} \wedge \w{\omega}^\mu_{\ \, \beta} , 
\ee
where the $\w{{\mathcal R}}^\alpha_{\ \, \beta}$ are the sixteen 
{\em curvature 2-forms} associated with the connection $\w{\nabla}$
and the tetrad $(\w{e}_\alpha)$.
They are defined in terms of the spacetime Riemann curvature tensor 
(cf. Sec.~\ref{s:IN:curvat}) by 
\be \label{e:CA:def_curv2form}
    \forall (\w{u},\w{v})\in\T(\M)\times\T(\M),\quad
    \w{{\mathcal R}}^\alpha_{\ \, \beta}(\w{u},\w{v}) :=
    \mathrm{\bf Riem}(\w{e}^\alpha,\w{e}_\beta,\w{u},\w{v}) . 
\ee
Note that due to the symmetry property (\ref{e:IN:Riemann_antisym12})
of the Riemann tensor, there are actually only 6, and not 16,
independent curvature 2-forms. 
From Eq.~(\ref{e:CA:def_curv2form}), the curvature 2-forms can
be expressed in terms of the components 
$R^\alpha_{\ \, \beta\gamma\delta}$ of the Riemann tensor 
with respect to the bases $(\w{e}_\alpha)$ and 
$(\w{e}^\alpha)$ [cf. Eq.~(\ref{e:IN:def_Riemann})] as
\be
   \w{{\mathcal R}}^\alpha_{\ \, \beta} = R^\alpha_{\ \, \beta\mu\nu}
   \, \w{e}^\mu \otimes \w{e}^\nu
   = \frac{1}{2} R^\alpha_{\ \, \beta\mu\nu}
   \, \w{e}^\mu \wedge \w{e}^\nu ,
\ee
where the second equality follows from the antisymmetry of the
Riemann tensor with respect to its last two indices; it clearly exhibits
that $\w{{\mathcal R}}^\alpha_{\ \, \beta}$ is a 2-form. 
Conversely, one may express the Riemann tensor in terms of the
curvature 2-forms as
\be
    \mathrm{\bf Riem} = \w{e}_\mu \otimes \w{e}^\nu 
        \otimes \w{{\mathcal R}}^\mu_{\ \, \nu} .
\ee

For $\alpha=\beta=0$, Cartan's second structure equation 
(\ref{e:CA:2nd_struct}) results in
\bea
    & & \dd\w{\omega}^0_{\ \, 0} = \w{{\mathcal R}}^0_{\ \, 0}
        - \w{\omega}^0_{\ \, 1} \wedge \w{\omega}^1_{\ \, 0}
        - \w{\omega}^0_{\ \, a} \wedge \w{\omega}^a_{\ \, 0} \nonumber \\
    & & \dd \left( \w{\omega} - N^{-2}\w{\nabla}_{\el} \sigma \, \uel \right)
    = \w{{\mathcal R}}^0_{\ \, 0} 
    - \left( - \Omega_a \, \uk - N^{-2} \w{\nabla}_{\w{e}_a}\sigma \, \uel
            + \Xi_{ab} \, \w{e}^b \right) \wedge \nonumber \\
   & & \qquad\qquad\qquad\qquad\qquad\qquad\quad 
    \left[ (\Omega_a - \w{\nabla}_{\w{e}_a}\rho)\, \uel
        + \Theta_{ab} \, \w{e}^b \right] ,      \label{e:CA:2nd_struct_00_prov}
\eea
where, according to the definition (\ref{e:CA:def_curv2form})
and to the symmetry property (\ref{e:IN:Riemann_antisym12}) of the Riemann
tensor, 
\be
    \w{{\mathcal R}}^0_{\ \, 0} =
    \mathrm{\bf Riem}(\w{e}^0,\w{e}_0, . , .) 
    = \mathrm{\bf Riem}(-\uk,\el, . , .) = 
    \mathrm{\bf Riem}(\uel,\w{k}, . , .) .
\ee
Expanding Eq.~(\ref{e:CA:2nd_struct_00_prov}) 
[cf. Eq.~(\ref{e:NH:ext_deriv})] and using Eq.~(\ref{e:CA:Frobenius_l})
leads to 
\bea
    \dd \w{\omega} & = & \mathrm{\bf Riem}(\uel,\w{k}, . , .)
    - \Omega_b \Theta^b_{\ \, a} \, \w{e}^a\wedge\uk
    + \Theta_{ac}\Xi^c_{\ \, b} \, \w{e}^a\wedge\w{e}^b \nonumber \\
    & & + \bigg\{\  \dd(N^{-2}\w{\nabla}_{\el} \sigma)
        + N^{-2}\w{\nabla}_{\el} \sigma \, \dd\rho
        + \Omega^a(\Omega_a - \w{\nabla}_{\w{e}_a}\rho)\, \uk \nonumber \\
    & & \qquad - \left[ N^{-2} \w{\nabla}_{\w{e}_b} \sigma \Theta^b_{\ \, a} 
     + (\Omega_b - \w{\nabla}_{\w{e}_b}\rho) \Xi^b_{\ \, a} \right]
     \w{e}^a \ \bigg\} \wedge \uel . \label{e:CA:2nd_struct_00}
\eea

For $\alpha=0$ and $\beta=1$ (or $\alpha=1$ and $\beta=0$), 
Cartan's second structure equation 
(\ref{e:CA:2nd_struct}) results in the trivial equation $0=0$. 
For $\alpha=0$ and $\beta=a$ it gives
\bea
    && \dd\wo^0_{\ \, a}  = \w{{\mathcal R}}^0_{\ \, a}
        - \w{\omega}^0_{\ \, 0} \wedge \w{\omega}^0_{\ \, a}
        - \w{\omega}^0_{\ \, 1} \wedge \w{\omega}^1_{\ \, a}
        - \w{\omega}^0_{\ \, b} \wedge \w{\omega}^b_{\ \, a} \nonumber \\
   && \dd\left( - \Omega_a \, \uk - N^{-2} \w{\nabla}_{\w{e}_a}\sigma \, \uel
            + \Xi_{ab} \, \w{e}^b \right) 
            = \w{{\mathcal R}}^0_{\ \, a} \nonumber \\
    && \qquad - \left( \w{\omega} 
                - N^{-2}\w{\nabla}_{\el} \sigma \, \uel \right) 
                \wedge 
    \left( - \Omega_a \, \uk - N^{-2} \w{\nabla}_{\w{e}_a}\sigma \, \uel
            + \Xi_{ab} \, \w{e}^b \right) \nonumber \\
    && \quad + 
    \left( \Omega_b \, \uk + N^{-2} \w{\nabla}_{\w{e}_b}\sigma \, \uel
            - \Xi_{bc} \, \w{e}^c \right)
            \wedge 
    \left( - \Gamma^b_{\ \, a0} \, \uk
    - \Gamma^b_{\ \, a1} \, \uel
    + \Gamma^b_{\ \, ad} \, \w{e}^d  \right) . \nonumber \\
\eea
Expanding this expression and using Eqs.~(\ref{e:CA:Frobenius_k}) 
and (\ref{e:CA:Frobenius_l}), as well as 
$\w{{\mathcal R}}^0_{\ \, a} = - \mathrm{\bf Riem}(\uk,\w{e}_a, . , .)$,
leads to 
\bea
    \dd(\Xi_{ab}\we^b) & = &
         - \mathrm{\bf Riem}(\uk,\w{e}_a, . , .) + \bigg\{
          N^{-2}\left( \Omega_a\dd\sigma 
   + \w{\nabla}_{\we_a}\sigma \,
   \dd\rho + \w{\nabla}_{\we_a}\sigma \, \w{\omega} \right)  \nonumber \\
  & &  + \dd(N^{-2} \w{\nabla}_{\we_a}\sigma) 
  + \bigg[ N^{-2}\left( \Omega_a \w{\nabla}_{\el}\sigma
    + \w{\nabla}_{\we_b}\sigma \Gamma^b_{\ \, a0} \right)
  - \Omega_b \Gamma^b_{\ \, a1} \bigg] \uk \nonumber \\
  & & + \bigg[ \Xi_{bc} \Gamma^c_{\ \, a1} 
   - N^{-2} \left( \Xi_{ab} \w{\nabla}_{\el}\sigma 
   + \Gamma^c_{\ \, ab} \w{\nabla}_{\we_c}\sigma \right) \bigg] \we^b
          \bigg\} \wedge \uel      \nonumber \\
   & &  + \left[ \dd\Omega_a + \Omega_a \w{\omega}
    + \left( \Xi_{bc} \Gamma^c_{\ \, a0} - \Omega_c \Gamma^c_{\ \, ab}
    \right) \we^b \right] \wedge \uk 
    - \Xi_{ab} \, \wo \wedge \we^b \nonumber \\
   & & - \Xi_{bd} \Gamma^d_{\ \, ac} \, \we^b \wedge\we^c .
                                        \label{e:CA:2nd_struct_0a}
\eea

For $\alpha=1$ and $\beta=1$, Cartan's second structure equation
yields the same result as for $\alpha=0$ and $\beta=0$
(since $\wo^1_{\ \, 1} = - \wo^0_{\ \, 0}$ and 
$\w{{\mathcal R}}^1_{\ \, 1} = - \w{{\mathcal R}}^0_{\ \, 0}$), namely
Eq.~(\ref{e:CA:2nd_struct_00}). 
For $\alpha=1$ and $\beta=a$, it writes
\bea
    & & \dd \wo^1_{\ \, a} = \w{{\mathcal R}}^1_{\ \, a} 
        - \w{\omega}^1_{\ \, 0} \wedge \w{\omega}^0_{\ \, a}
        - \w{\omega}^1_{\ \, 1} \wedge \w{\omega}^1_{\ \, a}
        - \w{\omega}^1_{\ \, b} \wedge \w{\omega}^b_{\ \, a} \nonumber \\
    && \dd\left[ (\Omega_a - \w{\nabla}_{\w{e}_a}\rho)\, \uel
        + \Theta_{ab} \, \w{e}^b \right] = \w{{\mathcal R}}^1_{\ \, a} \nonumber\\
    & & \qquad + \left( \w{\omega} - N^{-2}\w{\nabla}_{\el} \sigma \, \uel 
        \right) \wedge \left[ (\Omega_a - \w{\nabla}_{\w{e}_a}\rho)\, \uel
        + \Theta_{ab} \, \w{e}^b \right] \nonumber \\
    && \qquad - \left[ (\Omega_b - \w{\nabla}_{\w{e}_b}\rho)\, \uel
        + \Theta_{bc} \, \w{e}^c \right]
        \wedge \left( - \Gamma^b_{\ \, a0} \, \uk
    - \Gamma^b_{\ \, a1} \, \uel
    + \Gamma^b_{\ \, ad} \, \w{e}^d  \right) .  
\eea
Expanding this expression and using Eqs.~(\ref{e:CA:Frobenius_k}) 
and (\ref{e:CA:Frobenius_l}), as well as 
$\w{{\mathcal R}}^1_{\ \, a} = - \mathrm{\bf Riem}(\uel,\w{e}_a, . , .)$,
leads to 
\bea
    \dd(\Theta_{ab}\we^b) & = &
         - \mathrm{\bf Riem}(\uel,\w{e}_a, . , .) + \bigg\{
    \dd(\w{\nabla}_{\we_a}\rho - \Omega_a)
    + (\Omega_a - \w{\nabla}_{\w{e}_a}\rho) (\wo-\dd\rho) \nonumber \\
    && + \Gamma^b_{\ \, a0} (\w{\nabla}_{\we_b}\rho - \Omega_b) \uk
    + \bigg[ N^{-2} \w{\nabla}_{\el}\sigma \Theta_{ab}
        + \Gamma^c_{\ \, ab}(\Omega_c - \w{\nabla}_{\w{e}_c}\rho)
        \nonumber \\
     && + \Gamma^c_{\ \, a1}\Theta_{bc}    \bigg] \we^b
            \bigg\} \wedge \uel 
            + \Theta_{bc} \Gamma^c_{\ \, a0} \, \we^b\wedge\uk
            + \Theta_{ab}\, \wo\wedge\we^b \nonumber \\
    & & - \Theta_{bd}\Gamma^d_{\ \, ac}\, \we^b\wedge\we^c . 
                                \label{e:CA:2nd_struct_1a}
\eea
As an application of this relation, we can express the 
Lie derivative of the second fundamental form along $\el$
restricted to the 2-plane $\T(\Sp_t)$, i.e. the quantity
$\vec{\w{q}}^* \Lie{\el}\w{\Theta}$. Indeed, by means of 
the expansion (\ref{e:CA:expans_omthxi}), 
let us write $\Lie{\el}\w{\Theta}$ as
\bea
    \Lie{\el}\w{\Theta} &=& \we^a \otimes \Lie{\el}(\Theta_{ab}\we^b)
    + \Theta_{ab} \Lie{\el}\we^a\otimes \we^b \nonumber \\
        &=&   \we^a \otimes 
        \big[ \el\cdot\dd(\Theta_{ab}\we^b) +
        \dd\underbrace{\langle\Theta_{ab}\we^b,\el\rangle}_{=0}\big] 
        + \Theta_{ab} \big( \el\cdot\dd\we^a + 
        \dd\underbrace{\langle\we^a,\el\rangle}_{=0}
        \big) \otimes \we^b \nonumber \\
    &=&    \we^a \otimes \left[ \el\cdot\dd(\Theta_{ab}\we^b) \right]
                 \nonumber \\
   && + \Theta_{ab} \left[ (2\Omega_a - \w{\nabla}_{\we_a} \rho) \uel
    + (\Theta^a_{\ \, c} - \Gamma^a_{\ \, c0}) \we^c \right]
   \otimes \we^b    , 
\eea
where we have used Cartan's identity (\ref{e:IN:Cartan_id}) to get
the second line and Cartan's first structure equation (\ref{e:CA:1st_struct_a})
to get the third one. Then
\be
   \Lie{\el}\w{\Theta}(\we_a,\we_b) = \dd(\Theta_{ac}\we^c)(\el,\we_b)
   + \Theta_{bc} (\Theta^c_{\ \, a} - \Gamma^c_{\ \, a0}) .  
                                            \label{e:CA:LieTheta_ab}
\ee
Applying the 2-form (\ref{e:CA:2nd_struct_1a}) to the couple of vectors
$(\el,\we_b)$ results in
\be
    \dd(\Theta_{ac}\we^c)(\el,\we_b) =
     - \mathrm{\bf Riem}(\uel,\w{e}_a, \el , \we_b)
     + \Theta_{bc} \Gamma^c_{\ \, a0} + \kappa \Theta_{ab} . 
                                            \label{e:CA:dThetaec_leb}
\ee 
Combining Eqs.~(\ref{e:CA:LieTheta_ab}) and (\ref{e:CA:dThetaec_leb})
yields
\be
   \Lie{\el}\w{\Theta}(\we_a,\we_b) = \kappa \Theta_{ab}
   + \Theta_{ac} \Theta^c_{\ \, b} 
   - \mathrm{\bf Riem}(\uel,\w{e}_a, \el , \we_b) .
\ee
Hence
\be
  \vec{\w{q}}^* \Lie{\el}\w{\Theta} = \kappa \w{\Theta}
    + \w{\Theta}\cdot\vec{\w{\Theta}} 
    - \vec{\w{q}}^*  \mathrm{\bf Riem}(\uel, . , \el , .) , 
                                        \label{e:CA:qLieTheta}
\ee
i.e. we recover Eq.~(\ref{e:DN:qLieTheta}).

For $\alpha=a$ and $\beta=0$, Cartan's second structure equation
yields the same result as Eq.~(\ref{e:CA:2nd_struct_1a}),
since $\wo^a_{\ \, 0} =  \wo^1_{\ \, a}$ and 
$\w{{\mathcal R}}^a_{\ \, 0} = \w{{\mathcal R}}^1_{\ \, a}$.
For $\alpha=a$ and $\beta=1$, it
yields the same result as for $\alpha=0$ and $\beta=a$
(since $\wo^a_{\ \, 1} =  \wo^0_{\ \, a}$ and 
$\w{{\mathcal R}}^a_{\ \, 1} = \w{{\mathcal R}}^0_{\ \, a}$),
namely Eq.~(\ref{e:CA:2nd_struct_0a}).
Finally, for $\alpha=a$ and $\beta=b$,
Cartan's second structure equation writes
\bea
    & & \dd \wo^a_{\ \, b} = \w{{\mathcal R}}^a_{\ \, b} 
        - \w{\omega}^a_{\ \, 0} \wedge \w{\omega}^0_{\ \, b}
        - \w{\omega}^a_{\ \, 1} \wedge \w{\omega}^1_{\ \, b}
        - \w{\omega}^a_{\ \, c} \wedge \w{\omega}^c_{\ \, b} \nonumber \\
    & & \dd \left( - \Gamma^a_{\ \, b0} \, \uk
    - \Gamma^a_{\ \, b1} \, \uel
    + \Gamma^a_{\ \, bc} \, \w{e}^c \right) = \Omega^a_{\ \, b} \nonumber \\
   & & - \left[ (\Omega_a - \w{\nabla}_{\w{e}_a}\rho)\, \uel
        + \Theta_{ac} \, \w{e}^c \right] \wedge
        \left( - \Omega_b \, \uk - N^{-2} \w{\nabla}_{\w{e}_b}\sigma \, \uel
            + \Xi_{bd} \, \w{e}^d \right) \nonumber \\
    & & - \left(  - \Omega_a \, \uk - N^{-2} \w{\nabla}_{\w{e}_a}\sigma \, \uel
            + \Xi_{ac} \, \w{e}^c\right) \wedge
        \left[ (\Omega_b - \w{\nabla}_{\w{e}_b}\rho)\, \uel
        + \Theta_{bd} \, \w{e}^d \right]  \nonumber \\
   & &     - \left( - \Gamma^a_{\ \, c0} \, \uk
    - \Gamma^a_{\ \, c1} \, \uel
    + \Gamma^a_{\ \, cd} \, \w{e}^d \right)
    \wedge 
    \left(  - \Gamma^c_{\ \, b0} \, \uk
    - \Gamma^c_{\ \, b1} \, \uel
    + \Gamma^c_{\ \, bf} \, \w{e}^f \right) ,  
\eea
which leads to 
\bea
    \dd (\Gamma^a_{\ \, bc} \we^c ) & = & 
    \mathrm{\bf Riem}(\we^a,\we_b, . , .) 
    + \bigg\{ \Gamma^a_{\ \, b0} N^{-2} \dd\sigma + \dd \Gamma^a_{\ \, b1} 
    +\Gamma^a_{\ \, b1} \dd\rho \nonumber \\ 
 & & + \left[ (\w{\nabla}_{\w{e}_a}\rho -\Omega_a)\Omega_b
     + (\Omega_b - \w{\nabla}_{\w{e}_b}\rho)\Omega_a
     - \Gamma^a_{\ \, c0} \Gamma^c_{\ \, b1}
     + \Gamma^a_{\ \, c1} \Gamma^c_{\ \, b0} \right] \uk \nonumber \\
 & &   + \bigg[ \Xi_{bc} (\Omega_a - \w{\nabla}_{\w{e}_a}\rho)
     + N^{-2} \left( \Theta^a_{\ \, c}   \w{\nabla}_{\w{e}_b}\sigma
     - \Theta_{bc} \w{\nabla}_{\w{e}_a}\sigma \right) \nonumber \\
 & &    - \Xi^a_{\ \, c} (\Omega_b - \w{\nabla}_{\w{e}_b}\rho) 
 - \Gamma^a_{\ \, d1} \Gamma^d_{\ \, bc}
     + \Gamma^a_{\ \, dc} \Gamma^d_{\ \, b1} \bigg] \, \we^c \bigg\} 
     \wedge \uel  \nonumber \\
 & &    +  \left[ \dd\Gamma^a_{\ \, b0} 
 + \left( \Theta^a_{\ \, c} \Omega_b - \Omega^a \Theta_{bc}
        - \Gamma^a_{\ \, d0} \Gamma^d_{\ \, bc}
     + \Gamma^a_{\ \, dc} \Gamma^d_{\ \, b0} \right) \we^c \right] \wedge \uk
     \nonumber \\
 & & + \left( \Xi^a_{\ \, d} \Theta_{bc} - \Theta^a_{\ \, c} \Xi_{bd} 
 + \Gamma^a_{\ \, fd} \Gamma^f_{\ \, bc} \right) \we^c \wedge \we^d .
                                                    \label{e:CA:2nd_struct_ab}
\eea


\subsection{Ricci tensor}

Having computed the tetrad components of the Riemann tensor 
via Cartan's second structure equation, we can evaluate the tetrad
components of the Ricci by means of the definition (\ref{e:IN:def_Ricci})
of the latter:
\be \label{e:CA:Ricci_ea_eb}
    \w{R}(\we_\alpha,\we_\beta) = \mathrm{\bf Riem}(\we^\mu, \we_\alpha,
        \we_\mu,\we_\beta) . 
\ee
For $\alpha=\beta=0$, we get
\bea
    \w{R}(\el,\el) & = & \mathrm{\bf Riem}(\we^\mu, \el,
        \we_\mu,\el) \nonumber \\
        & = & - \underbrace{\mathrm{\bf Riem}(\uk, \el, \el, \el)}_{=0}
        - \underbrace{\mathrm{\bf Riem}(\uel, \el, \w{k}, \el)}_{=0}
        + \mathrm{\bf Riem}(\we^a, \el, \we_a, \el) \nonumber \\
        & = & - \mathrm{\bf Riem}( \uel, \we_a, \we_a, \el) , 
\eea
where we have used the symmetry property (\ref{e:IN:Riemann_antisym12})
of the Riemann tensor. 
Let us substitute Eq.~(\ref{e:CA:2nd_struct_1a})
for $\mathrm{\bf Riem}( \uel, \we_a, ., .)$. 
We notice that the long term $\{\ldots\}\wedge\uel$ vanishes when applied
to $(\we_a, \el)$, as the term $\we^b\wedge\we^c$.
Moreover since
$\langle\wo,\el\rangle = \kappa$ [Eq.~(\ref{e:NH:chi_eigen_l_k})],
$\Theta_{ab}\delta^b_{\ \, a} = \theta$ [Eq.~(\ref{e:KI:def_theta})]
and $\Theta_{ab} \Gamma^b_{\ \, a0} = 0$ (by symmetry of $\Theta_{ab}$
and antisymmetry of $\Gamma^b_{\ \, a0}$ with respect to the indices 
$a$ and $b$), we get
\be
    \w{R}(\el,\el) = \dd(\Theta_{ab}\we^b)(\we_a, \el)
      + \kappa \theta . \label{e:CA:Rll_inter}
\ee
Let us express the exterior derivative of the 1-form $\Theta_{ab}\we^b$
by means of formula (\ref{e:IN:d1form_cov_der}):
\bea
    \dd(\Theta_{ab}\we^b)(\we_a, \el) & = &
     \langle \w{\nabla}_{\we_a} \, (\Theta_{ab}\we^b) ,\; \el \rangle
        - \langle \w{\nabla}_{\el} \, (\Theta_{ab}\we^b),\; \we_a \rangle 
            \nonumber \\
    & = & -\langle \Theta_{ab}\we^b,\; \w{\nabla}_{\we_a} \el \rangle
        - \w{\nabla}_{\el} \Theta_{ab} \langle \we^b,\we_a\rangle 
        + \Theta_{ab} \langle \we^b, \w{\nabla}_{\el} \we_a \rangle   
            \nonumber \\
    & = & \Theta_{ab} \left(
    \langle \we^b, \w{\nabla}_{\el} \we_a \rangle
     - \langle \we^b, \w{\nabla}_{\we_a} \el \rangle \right) 
      - \w{\nabla}_{\el} \Theta_{ab} \, \delta^b_{\ \, a}  \nonumber \\
    & = &   \Theta_{ab} \left(  \langle \wo^b_{\ \, a}, \el \rangle
        - \langle \wo^b_{\ \, 0}, {\we_a} \rangle \right)
        - \w{\nabla}_{\el}  \theta \nonumber \\
    & = & \Theta_{ab} \left( \Gamma^b_{\ \, a0} 
            - \Theta^b_{\ \,a} \right) - \w{\nabla}_{\el}  \theta  
            = - \Theta_{ab} \Theta^{ab} - \w{\nabla}_{\el}  \theta ,
\eea
where we have used the property $\Theta_{ab}  \Gamma^b_{\ \, a0} = 0$
noticed above. 
Inserting this relation into Eq.~(\ref{e:CA:Rll_inter}) yields
\be
 \w{R}(\el,\el) = - \w{\nabla}_{\el}  \theta 
                    + \kappa \theta - \Theta_{ab} \Theta^{ab} 
                                        \label{e:CA:Raychaud} .
\ee
We thus recover the null Raychaudhuri equation
(\ref{e:DN:Raychaud_prov}).

For $\alpha=0$ and $\beta=a$, Eq.~(\ref{e:CA:Ricci_ea_eb}) gives
\be
   \w{R}(\el,\we_a) =  \mathrm{\bf Riem}( \uel, \w{k},\el ,\we_a)
    - \mathrm{\bf Riem}( \uel, \we_b,\we_b ,\we_a) . 
\ee
Substituting Eq.~(\ref{e:CA:2nd_struct_00}) for 
$\mathrm{\bf Riem}( \uel, \w{k},. ,.)$ and Eq.~(\ref{e:CA:2nd_struct_1a})
for $\mathrm{\bf Riem}( \uel, \we_b,. ,.)$, we get
\be
    \w{R}(\el,\we_a) = \dd\wo(\el,\we_a) +
     \dd(\Theta^b_{\ \, c}\we^c)(\we_b,\we_a)
     + \theta  \Omega_a +\Gamma^c_{\ \, bc} \Theta^b_{\ \, a} , 
                                        \label{e:CA:Ricci_l_ea_prov}
\ee
where we have used once again the symmetry property 
$\Theta_{bc}\Gamma^b_{\ \, ca}=0$, as well as 
$\Theta_{bc}=\Theta^b_{\ \, c}$. Thanks to Cartan's identity
(\ref{e:IN:Cartan_id}), the first term in the right-hand side of 
Eq.~(\ref{e:CA:Ricci_l_ea_prov}) writes
\be
    \dd\wo(\el,\we_a) = \langle \Lie{\el} \wo 
    - \dd\underbrace{\langle\wo,\el\rangle}_{=\kappa},\ \we_a\rangle 
    = \langle \Lie{\el} \wo, \we_a \rangle 
        - \langle  \dd\kappa, \we_a\rangle . \label{e:CA:dom_l_ea}
\ee
The second term in the right-hand side of 
Eq.~(\ref{e:CA:Ricci_l_ea_prov}) is expressed by means of 
formula (\ref{e:IN:d1form_cov_der}):
\bea
   \dd(\Theta^b_{\ \, c}\we^c)(\we_b,\we_a) & = & 
   \langle \w{\nabla}_{\we_b}( \Theta^b_{\ \, c}\we^c), \we_a \rangle
   - \langle \w{\nabla}_{\we_a}( \Theta^b_{\ \, c}\we^c), \we_b \rangle
    \nonumber \\
    & = & \w{\nabla}_{\we_b} \Theta^b_{\ \, c} \langle \we^c, \we_a \rangle
    -  \Theta^b_{\ \, c} 
    \langle  \we^c, \w{\nabla}_{\we_b} \we_a \rangle \nonumber \\ 
     & & - \w{\nabla}_{\we_a} \Theta^b_{\ \, c} \langle \we^c, \we_b \rangle
     +  \Theta^b_{\ \, c} 
    \langle  \we^c, \w{\nabla}_{\we_a} \we_b \rangle   \nonumber \\
  & = & \w{\nabla}_{\we_b} \Theta^b_{\ \, a} -  \Theta^b_{\ \, c}
    \Gamma^c_{\ \, ab} - \w{\nabla}_{\we_a} 
    \underbrace{\Theta^b_{\ \, b}}_{=\theta}
    + \underbrace{\Theta^b_{\ \, c} \Gamma^c_{\ \, ba}}_{=0}  \nonumber \\
  & = & \langle \dd \Theta^b_{\ \, a}, \we_b\rangle 
    -  \Gamma^c_{\ \, ab} \Theta^b_{\ \, c}
     -  \langle \dd\theta, \we_a\rangle . 
                                    \label{e:CA:dThetae_eb_ea}
\eea
Inserting expressions (\ref{e:CA:dom_l_ea}) and (\ref{e:CA:dThetae_eb_ea})
into Eq.~(\ref{e:CA:Ricci_l_ea_prov}) leads to
\bea
  \w{R}(\el,\we_a) & = & \langle \Lie{\el} \wo, \we_a \rangle 
    + \theta \Omega_a 
        - \langle  \dd\kappa, \we_a\rangle -  \langle \dd\theta, \we_a\rangle
        \nonumber \\
    & &  + \langle \dd \Theta^b_{\ \, a}, \we_b\rangle   
      +\Gamma^c_{\ \, bc} \Theta^b_{\ \, a}
      -  \Gamma^c_{\ \, ab} \Theta^b_{\ \, c}
     \  . \label{e:CA:Ricci_l_ea_prov2} 
\eea
We recognize in the last term the component on $\we^a$ of the 
covariant divergence of $\vec{\w{\Theta}}$ with respect to the
connection $\w{\DS}$ induced by $\w{\nabla}$ in the 2-surface $\Sp_t$:
\be
  \left( \w{\DS}\cdot \vec{\w{\Theta}} \right) _a =
    \langle \dd \Theta^b_{\ \, a}, \we_b\rangle   
      +\Gamma^c_{\ \, bc} \Theta^b_{\ \, a}
      -  \Gamma^c_{\ \, ab} \Theta^b_{\ \, c} . 
\ee
Besides 
\bea
    \langle \Lie{\el} \wo, \we_a \rangle  & = &
    \langle \Lie{\el}(\w{\Omega} - \kappa\uk), \we_a \rangle 
    = \langle \Lie{\el}\w{\Omega} - \Lie{\el}\kappa \; \uk 
    - \kappa \Lie{\el}{\uk},
        \we_a \rangle \nonumber \\
   & = &\langle \Lie{\el}\w{\Omega} - \kappa \el\cdot \dd\uk, \we_a \rangle
    = \langle \Lie{\el}\w{\Omega} - \kappa N^{-2} 
    \el\cdot(\dd\sigma\wedge \el),\;  \we_a \rangle \nonumber \\
    & = & \langle \Lie{\el}\w{\Omega}, \we_a\rangle , 
\eea
so that Eq.~(\ref{e:CA:Ricci_l_ea_prov2}) can be written 
\be
      \w{R}(\el,\we_a) = \left\langle \Lie{\el} \w{\Omega} 
      + \theta \, \w{\Omega}-  \dd(\kappa+\theta)
      + \w{\DS}\cdot \vec{\w{\Theta}},\  \we_a \right\rangle , 
                                                \label{e:CA:DNS}
\ee
which is nothing but the Damour-Navier-Stokes equation under the
form (\ref{e:DN:Rlq_DNS}). 

Finally, let us consider the components of the Ricci tensor relative
to $\T(\Sp_t)$, i.e. $\w{R}(\we_a,\we_b)$. From Eq.~(\ref{e:CA:Ricci_ea_eb}),
we get
\bea    
    \w{R}(\we_a,\we_b) & = & - \mathrm{\bf Riem}(\uk, \we_a, \el,\we_b)
    - \mathrm{\bf Riem}(\uel, \we_a, \w{k},\we_b) \nonumber \\
   &&  + \mathrm{\bf Riem}(\we^c, \we_a, \we_c,\we_b) . 
                                            \label{e:CA:Ricci_ab_prov}
\eea
The first term in the right-hand side can be expressed, thanks to
Eq.~(\ref{e:CA:2nd_struct_0a})
\bea
    - \mathrm{\bf Riem}(\uk, \we_a, \el,\we_b) & = &
        \dd(\Xi_{ac}\we^c)(\el,\we_b)  - \w{\nabla}_{\we_b} \Omega_a
            - \Omega_a\Omega_b - \Xi_{bc}\Gamma^c_{\ \, a0} \nonumber \\
            & & + \Gamma^c_{\ \, ab}\Omega_c  + \kappa \Xi_{ab} \ , 
                        \label{e:CA:Riem_kalb}
\eea
whereas Eq.~(\ref{e:CA:2nd_struct_1a}) yields
\bea
   - \mathrm{\bf Riem}(\uel, \we_a, \w{k},\we_b) &=& \dd(\Theta_{ac}\we^c)
   (\w{k},\we_b) - \w{\nabla}_{\we_a}\w{\nabla}_{\we_b} \rho
   + \Gamma^c_{\ \, ab}  \w{\nabla}_{\we_c} \rho \nonumber \\
   & & - \w{\nabla}_{\we_a} \rho \, \w{\nabla}_{\we_b} \rho
   + \Omega_a \w{\nabla}_{\we_b} \rho + \Omega_b \w{\nabla}_{\we_a} \rho 
   + \w{\nabla}_{\we_b} \Omega_a \nonumber \\
   & & - \Gamma^c_{\ \, ab} \Omega_c - \Omega_a \Omega_b 
   - N^{-2} \w{\nabla}_{\el}\sigma\, \Theta_{ab} 
   - \Gamma^c_{\ \, a1} \Theta_{bc} \ ,  \label{e:CA:Riem_lakb}
\eea
and Eq.~(\ref{e:CA:2nd_struct_ab}) gives
\bea
    \mathrm{\bf Riem}(\we^c, \we_a, \we_c,\we_b) & = & 
    \dd(\Gamma^c_{\ \, ad}\we^d)(\we_c,\we_b)
    - \Theta_{ac} \Xi^c_{\ \, b} - \Xi_{ac} \Theta^c_{\ \, b}
    + \theta_{(\w{k})} \Theta_{ab} \nonumber\\
    && + \theta \Xi_{ab} - \Gamma^c_{\ \, db} \Gamma^d_{\ \, ac}
    + \Gamma^c_{\ \, dc} \Gamma^d_{\ \, ab} . \label{e:CA:Riem_cacb}
\eea
Collecting together Eqs.~(\ref{e:CA:Riem_kalb}), (\ref{e:CA:Riem_lakb})
and (\ref{e:CA:Riem_cacb}) enables us to write Eq.~(\ref{e:CA:Ricci_ab_prov})
as
\bea
    \w{R}(\we_a,\we_b) & = & \dd(\Xi_{ac}\we^c)(\el,\we_b)  
     - \Gamma^c_{\ \, a0} \Xi_{bc} 
       + (\kappa + \theta) \Xi_{ab}
       + \dd(\Theta_{ac}\we^c)(\w{k},\we_b) \nonumber \\
    & &   - \Gamma^c_{\ \, a1} \Theta_{bc} 
    + (\theta_{(\w{k})} - N^{-2} \w{\nabla}_{\el}\sigma) \Theta_{ab}
    - \Theta_{ac} \Xi^c_{\ \, b} - \Xi_{ac} \Theta^c_{\ \, b} \nonumber \\
    & & - \w{\nabla}_{\we_a}\w{\nabla}_{\we_b} \rho
   + \Gamma^c_{\ \, ab}  \w{\nabla}_{\we_c} \rho 
    - \w{\nabla}_{\we_a} \rho \, \w{\nabla}_{\we_b} \rho
   + \Omega_a \w{\nabla}_{\we_b} \rho + \Omega_b \w{\nabla}_{\we_a} \rho
   \nonumber \\ 
   & &  -2 \Omega_a \Omega_b 
   + \dd(\Gamma^c_{\ \, ad}\we^d)(\we_c,\we_b)
   - \Gamma^c_{\ \, db} \Gamma^d_{\ \, ac}
    + \Gamma^c_{\ \, dc} \Gamma^d_{\ \, ab} .   \label{e:CA:Ricci_ab_prov2}
\eea
Let us first notice that the last three terms of this expression 
are nothing but the Ricci tensor ${}^2\!\w{R}$ 
of the connection $\w{\DS}$ associated
with the induced metric in the 2-surface $\Sp_t$, applied to the
couple of vectors $(\we_a,\we_b)$. Indeed, using the moving frame 
$(\we_{a})$ in $\Sp_t$, the connection 1-forms ${}^2\!\wo^b_a$
of $\w{\DS}$ are given
by a formula similar to Eq.~(\ref{e:CA:connect_coef}), except that
the range of the summation index $\mu$ is now restricted to 
$\{2,3\}$:
\be
    {}^2\!\wo^b_{\ \, a} = \Gamma^b_{\ \, a c} \we^c . 
\ee
Expressing the curvature of $\w{\DS}$ by means of Cartan's second structure 
equation, we get then
\bea
    {}^2\!\w{R}(\we_a,\we_b) & = & \left( \dd ({}^2\!\wo^c_{\ \, a})
    -  {}^2\!\wo^c_{\ \, d} \wedge {}^2\!\wo^d_{\ \, a} \right)
     (\we_c,\we_b) \nonumber \\
     & = &  \dd(\Gamma^c_{\ \, ad}\we^d)(\we_c,\we_b)
   - \Gamma^c_{\ \, db} \Gamma^d_{\ \, ac}
    + \Gamma^c_{\ \, dc} \Gamma^d_{\ \, ab} .   \label{e:CA:2Rab}
\eea
Besides, we may express the term $\dd(\Xi_{ac}\we^c)(\el,\we_b)$ [resp.
$\dd(\Theta_{ac}\we^c)(\w{k},\we_b)$] which appears in the right-hand side
of Eq.~(\ref{e:CA:Ricci_ab_prov2}) in terms of the Lie derivative
$\Lie{\el}\w{\Xi}$ (resp. $\Lie{\w{k}}\w{\Theta}$). 
Indeed a computation similar to that which led to Eq.~(\ref{e:CA:LieTheta_ab})
for $\Lie{\el}\w{\Theta}$ gives 
\be
   \Lie{\el}\w{\Xi}(\we_a,\we_b) = \dd(\Xi_{ac}\we^c)(\el,\we_b)
   + \Xi_{bc} (\Theta^c_{\ \, a} - \Gamma^c_{\ \, a0}) .  
\ee
and
\be
   \Lie{\w{k}}\w{\Theta}(\we_a,\we_b) = \dd(\Theta_{ac}\we^c)(\w{k},\we_b)
   + \Theta_{bc} (\Xi^c_{\ \, a} - \Gamma^c_{\ \, a1}) .      
\ee
Inserting the above two equations, as well as Eq.~(\ref{e:CA:2Rab}),
into Eq.~(\ref{e:CA:Ricci_ab_prov2}) leads to
\bea
    \w{R}(\we_a,\we_b) & = & \Lie{\el}\w{\Xi}(\we_a,\we_b) 
    +  \Lie{\w{k}}\w{\Theta}(\we_a,\we_b)
    - 2\Theta_{ac} \Xi^c_{\ \, b} - 2\Xi_{ac} \Theta^c_{\ \, b} \nonumber \\
   & & + (\kappa + \theta) \Xi_{ab} 
        + (\theta_{(\w{k})} - N^{-2} \w{\nabla}_{\el}\sigma) \Theta_{ab}
         -2 \Omega_a \Omega_b 
          - \w{\DS}_{\we_a}\w{\DS}_{\we_b} \rho \nonumber \\
    & &
    - \w{\DS}_{\we_a} \rho \, \w{\DS}_{\we_b} \rho
   + \Omega_a \w{\DS}_{\we_b} \rho + \Omega_b \w{\DS}_{\we_a} \rho
   +  {}^2\!\w{R}(\we_a,\we_b).         \label{e:CA:Ricci_ab_final}
\eea

%% file: symplectic.tex
\section{Physical parameters and Hamiltonian techniques}
\label{s:SP}

In Sec. \ref{s:IH:PhysParam} we introduced quasi-local notions for the physical parameters
associated with the horizon, when no matter is present on $\Hor$. In that section we only 
presented the final results, since the actual derivations involved the
use of Hamiltonian techniques not discussed in this article.
In this Appendix we aim at providing some intuition on the actual use of these 
tools. Instead of using a formal presentation (see \cite{AshteFK00,AshteBL01,KLP04}) 
we introduce the basic concepts by illustrating them with examples extracted from the
isolated horizon literature, always in absence of matter on the
horizon.

As indicated in Sec. \ref{s:IH:PhysParam}, the physical parameters
are identified with quantities conserved under certain
transformations on the space $\Gamma$ of solutions of Einstein 
equation. More specifically, these envisaged solutions in $\Gamma$
contain a ``fixed'' WIH  
$(\Hor, [\el])$ as inner boundary. The transformations on
$\Gamma$ relevant for the definition of the conserved quantities,
are associated with the WIH-symmetries of this inner boundary.
An appropriate characterization of this {\it phase space} $\Gamma$,
where each {\it point} is a Lorentzian manifold $(\M,\w{g})$, must be therefore introduced.
This is accomplished by setting a well-posed variational problem
associated with  spacetimes $(\M, \w{g})$ containing a ``given'' WIH. 
A rigorous presentation
of this whole subject requires a careful definition of the involved objects
(domain of variation of the dynamical fields, variation
of the fields at the boundaries of this domain, etcetera).
As mentioned above, we simply aim here at underlying the most relevant steps, 
through the use of (simplified) examples,
referring the reader to the original references for the detailed
formulations. As in the rest of the article,
we restrict ourselves to the non-extremal case $\kappa\neq 0$.

\subsection { Well-posedness of the variational problem}

In our study of black hole spacetimes, we are interested in asymptotically flat
solutions to Einstein equation with a WIH as inner boundary.
In the variational formulation of spacetime dynamics,
the presence of boundaries in the manifold  generally demands the 
introduction of boundary terms in the action so as to compensate 
the variation of this action with respect to the dynamical fields. This is relevant for
the differentiability of the action with respect to the dynamical fields
as well as for guaranteeing that the derived equations of motion are actually 
the ones corresponding to the studied problem.

\begin{exmp}
\label{ex:IH:var_probl} 
This example shows how the WIH condition (\ref{e:IH:Lw}) guarantees 
the well-posedness of a first order action principle.
Details can be found in \cite{AshteBF99,AshteBL01,Kr02}.
An  appropriate first order action for our problem can be written
in terms of the cotetrad $(\w{e}^I)$  
and a real 1-form connection ${\w{A}^I}_J$, where the capital 
Latin letters correspond to Lorentz indices
which are raised and lowered with the Minkowski metric.
We can write the action as
\bea
\label{s:IH:var_ppl}
S(\w{e},\w{A}) \sim -\int_M \w{\Sigma}_{IJ}\wedge \w{F}^{IJ} + 
\int_{\tau_\infty}\w{\Sigma}_{IJ}\wedge \w{A}^{IJ} \ ,
\eea
where
\bea
\w{\Sigma}_{IJ} &=&\frac{1}{2} \epsilon_{IJKL} \w{e}^K\wedge \w{e}^L \nn \\
{\w{F}^J}_I &=& \dd {\w{A}^J}_I + {\w{A}^K}_I\wedge {\w{A}^J}_K , 
\eea
$\epsilon_{IJKL}$ is the alternating symbol in 4 dimensions  and $\tau_\infty$ 
denotes spatial infinity. In order to determine the equations of motion,
the action is varied with respect
to the fields $\w{e}^I$ and ${\w{A}^I}_J$ in a region $M$ of $\M$ delimited by 
two Cauchy surfaces $\Sigma_-$ and $\Sigma_+$ (on which the variation 
of the fields, schematically denoted by $\delta(\w{e},\w{A})$, vanishes),
by spatial infinity $\tau_\infty$ and the inner boundary 
$(\Hor, [\el])$ (see Fig. \ref{f:IH:variational}). 
\begin{figure}
\centerline{\includegraphics[width=0.7\textwidth]{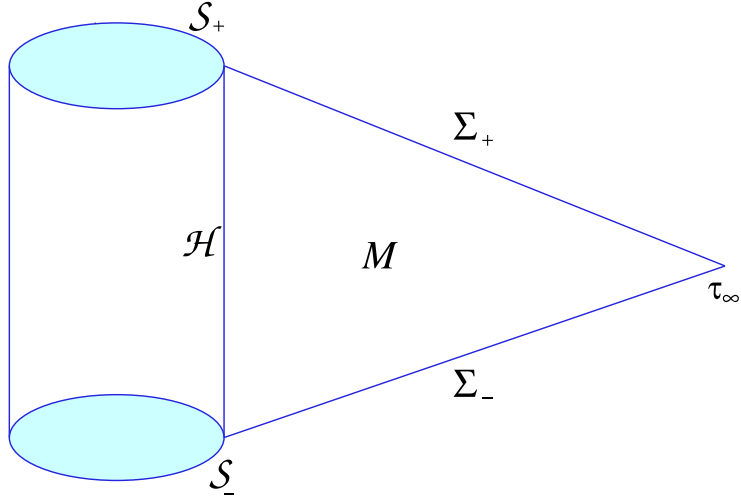}}
\caption[]{\label{f:IH:variational} 
Domain of variation in ${\mathcal M}$ bounded
by two Cauchy surfaces $\Sigma_-$ and $\Sigma_+$,
spatial infinity $\tau_\infty$ and the WIH $({\mathcal H}, [\el])$.
$\Sp_-$ and $\Sp_+$ denote the cross-sections resulting from
the intersections of the spatial slices $\Sigma_-$ and $\Sigma_+$
with the WIH.}
\end{figure}
In the variation of the action,  the bulk term gives rise to 
a boundary term at infinity, but it is exactly cancelled by the variation
of the boundary term at infinity in Eq. (\ref{s:IH:var_ppl}).
The resulting variation can be expressed as
\bea
\delta S(\w{e},\w{A})=\int \mathrm{(Equations \  of \ Motion)}\cdot \delta(\w{e},\w{A}) 
-\frac{1}{8\pi G} \int_{\mathcal H} \delta \w{\omega}\wedge
\w{{}^2\epsilon} \ \ .
\eea
The problem is well-posed and
Einstein equations are recovered as an extremal 
value of this action ( $\delta S(\w{e},\w{A})=0$), 
if the boundary integral at $\Hor$ vanishes. 
This is the crucial point we want to make in this example:
the WIH condition $\LieH{\el}\w{\omega}=0$, together with the NEH conditions and 
$\delta(\w{e},\w{A})=0$ on $\Sigma_-$ and $\Sigma_+$, suffices to guarantee
the vanishing of this boundary term (see \cite{AshteBF99} for the details).
\end{exmp}

\subsubsection{Phase space and canonical transformations}

Once the variational problem is well-posed, 
we must determine the phase space $\Gamma$.
We firstly introduce some general concepts and 
notation (see, e.g. Refs. \cite{AM78,Ar89,GS84}). 

The {\it phase space} is constituted by a pair
$(\Gamma, \w{J})$, where $\Gamma$ is an (infinite-dimensional)
manifold in which each
point represents a solution to the equations of motion and
$\w{J}$ is a closed two-form on $\Gamma$ known as the {\it symplectic form}: 
$\dd^\Gamma \w{J} = 0$, with $\dd^\Gamma$ 
the exterior differential in $\Gamma$.   
We can associate with each function $F:\Gamma\rightarrow \mathbb{R}$, a {\it Hamiltonian}
vector field $\delta_F$ in $\T(\Gamma)$ (the space of vector fields on 
$\Gamma$ in the notation introduced in Sec. \ref{s:notations}) as follows
\bea
i_{\delta_F} \w{J} = \dd^\Gamma F \ ,
\eea
where $i_{\delta_F} \w{J}:=\w{J}(\delta_F,\cdot)$.
Given two functions $F$ and $G$ with support on $\Gamma$, 
their {\it Poisson bracket} is defined as 
\bea
\{F,G\}=\w{J}(\delta_F, \delta_G) \ \ .
\eea
The pair $(\Gamma, \w{J})$ is also known as a {\it symplectic manifold}
({\it pre-symplectic} if the kernel of  $\w{J}$ is 
non-trivial).
Even though the phase spaces we are interested in are intrinsically  
infinite-dimensional, we shall skip all the subtleties 
related to infinite-dimensional symplectic spaces (see for instance
\cite{MR99}).

An infinitesimal transformation on the space $\Gamma$,
generated by the vector field $\delta_{\w{W}}$, is a {\it canonical transformation}
(also called {\it symplecto-morphism}) if it preserves the 
symplectic form $\w{J}$
\bea
{}^\Gamma\!{\mathcal L}_{\delta_{\w{W}}} \w{J} = 0 \ .
\eea
Using the closed character of $\w{J}$, this is locally 
equivalent to the existence of a Hamiltonian function $H_{\w{W}}$, i.e.
\bea
\label{e:IH:sympl_exact}
\encadre{
\left.\begin{array}{l}
\hbox{$\delta_{\w{W}}$ preserves (locally)} \\
\hbox{the Poisson brackets}
\end{array}
\right\}
\Longleftrightarrow \exists H_{\w{W}} \,\hbox{ such that } \,i_{\delta_W}\w{J} = \dd^\Gamma H_{\w{W}}  \ ,
}
\eea
Applying this expression to a generic vector field $\delta$ in 
$\T(\Gamma)$, yields
\bea
\label{e:IH:sympl_exact_applied}
\w{J}(\delta_{\w{W}},\delta) = \delta H_{\w{W}} \ ,
\eea
where the notation $\delta H_W:=\dd^\Gamma H_{\w{W}} (\delta)$ is designed to mimic
the {\it intuitive} physical notation. 
The evolution of a function $G$ along the flow of the vector field $\delta_W$  on $\Gamma$ 
({\it Hamilton equations}), can be evaluated as
\bea
\delta_{\w{W}} G = \{H_{\w{W}}, G\} \ .
\eea 
In particular, due to the anti-symmetry of $\w{J}$, $H_{\w{W}}$ remains
constant along the $\delta_{\w{W}}$ trajectories.
With these elements, the general strategy to associate physical parameters 
with the horizon will proceed 
via the following steps:
\begin{enumerate}
\item Construction of the appropriate phase space $\Gamma$ for our problem.
\item Extension of a given WIH-symmetry of $(\Hor,[\el])$ to an infinitesimal 
diffeomorphism $\w{W}$  {\it on  each space-time $\M$ of $\Gamma$}, 
giving rise to a family of vector fields $\{\w{W}\}_\Gamma$. 
Definition of a canonical transformation 
$\delta_{\w{W}}$ on $\Gamma$ out of the family $\{\w{W}\}_\Gamma$.
\item Identification of the physical parameter with the associated
conserved quantity $H_{\w{W}}$.
\end{enumerate}
We illustrate these steps by continuing Example \ref{ex:IH:var_probl}
(see again \cite{AshteFK00,AshteBL01}). 

\begin{exmp} {\em Phase space and canonical transformations}.
\label{ex:IH:phase_space} 

\noindent (1) {\it Phase space}. 

The phase space $\Gamma$ where we describe the dynamics defined by
the action (\ref{s:IH:var_ppl}), can be parametrized by the pairs
$(\w{e}^I, \w{A}^I_J)$ which satisfy Einstein equations
and contain an inner boundary given by a ``fixed'' WIH 
$({\mathcal H}, [\el])$. The so-called {\it conserved current 
method} \cite{CW87} provides a standard
manner to derive the relevant symplectic form from a given action. 
In our case this results in \cite{AshteBL01}
\bea
\w{J}(\delta_1,\delta_2) &=& -\frac{1}{16\pi G} \int_\Sigma
\left( \delta_1\w{\Sigma}^{IJ}\wedge  \delta_2 \w{A}_{IJ} - \delta_2\w{\Sigma}^{IJ}
\wedge  \delta_1 \w{A}_{IJ} \right) \nonumber \\
&+&\frac{1}{8\pi G} \int_{\Sp_t}\left(\delta_1 \w{{}^2\epsilon}\;\delta_2\psi - 
\delta_2 \w{{}^2\epsilon}\;\delta_1\psi \right)
\label{e:IH:symplgamma} \ \ ,
\eea
where $\psi$ is a function on $\Hor$ such that $\LieH{\el}\psi:= \kappa_{(\el)}$ 
and $\delta_1, \delta_2$ are arbitrary vector fields on $\T(\Gamma)$.

\noindent (2) {\it Canonical transformations induced by space-time transformations}.

On {\it each} point of $\Gamma$, i.e. on each spacetime $\M$ represented 
by a pair $(\w{e}^I, \w{A}^I_J)$, we consider an infinitesimal spacetime diffeomorphism 
$\w{W}(\w{e}, \w{A})$ (we make explicit the dependence of this vector field 
on the particular spacetime)
whose restriction to $\Hor$ is a WIH-symmetry. 
This family of spacetime vector fields $\{\w{W}(\w{e},\w{A})\}$ 
permits us to define an infinitesimal transformation
$\delta_{\w{W}}$ {\it on} $\Gamma$, which is 
defined in a point-wise manner (on each point $\M$ of $\Gamma$) by its action
on the coordinates $(\w{e}^I, \w{A}^I_J)$ of $\Gamma$
\be
\label{e:IH:def_deltaW}
\left.\left(\delta_{\w{W}} \w{e}^I\right)\right|_\M := {\mathcal L}_{\w{W}(\w{e}, \w{A})}\w{e}^I \ \ \ \ \ \ \
\left.\left(\delta_{\w{W}} \w{A}^I_J\right)\right|_\M := {\mathcal L}_{\w{W}(\w{e}, \w{A})}\w{A}^I_J \ ,
\ee
where right-hand side terms are evaluated on each spacetime $\M$,
associated respectively with the pair $(\w{e}^I,\w{A}^I_J)$.
In sum, starting from a family of WIH-symmetries $\{\w{W}(\w{e},\w{A})|_{\Hor}\}$
an infinitesimal transformation $\delta_{\w{W}}$ has been induced 
in the phase space $\Gamma$ (see Fig. \ref{f:IH:mtogamma}). 
\begin{figure}
\centerline{\includegraphics[width=1.0\textwidth]{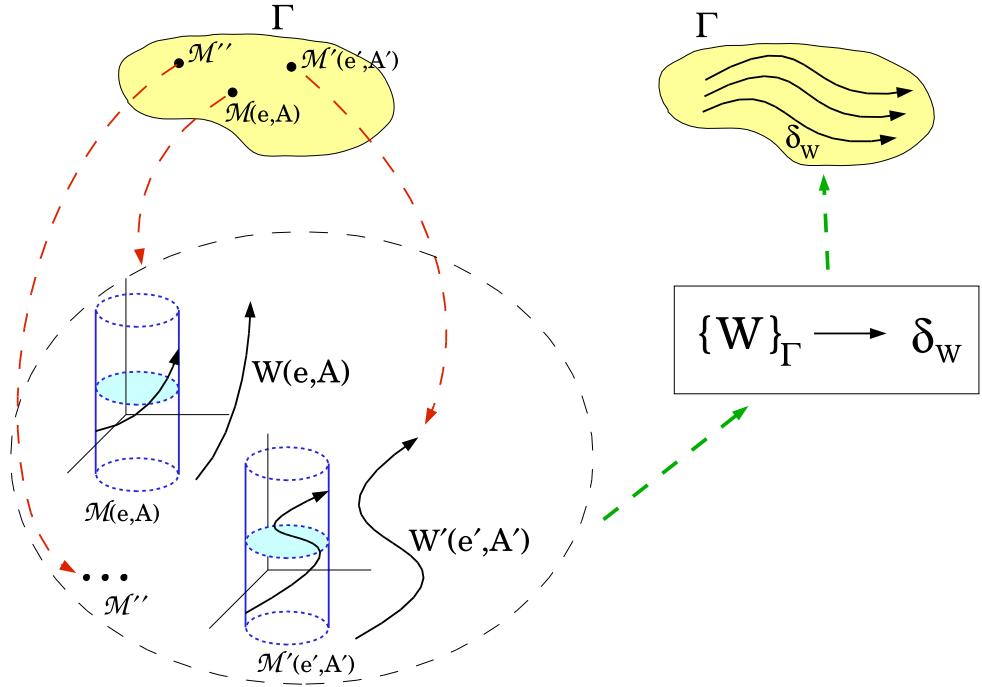}}
\caption[]{\label{f:IH:mtogamma} 
Illustration of the construction of a transformation $\delta_{\w{W}}$
on $\Gamma$ from the family $\{\w{W}\}_\Gamma$ on 
diffeomorphisms of spacetimes $\M$. On each point of $\Gamma$, a spacetime
$\M(\w{e},\w{A})$, we consider a diffeomorphism $\w{W}(\w{e},\w{A})$ whose
restriction to $\Hor$ is a WIH-symmetry. The ensemble of these
spacetime diffeomorphisms $\{\w{W}(\w{e},\w{A})\}_\Gamma$ generates 
a transformation $\delta_{\w{W}}$ on $\Gamma$ through Eq. (\ref{e:IH:def_deltaW}).
}
\end{figure}
The question now is to find out if such a
transformation $\delta_{\w{W}}$ is a canonical one. According to 
(\ref{e:IH:sympl_exact}), one must contract the
vector field $\delta_{\w{W}}$ with the symplectic form (\ref{e:IH:symplgamma})
and check if the resulting 1-form on $\Gamma$ is (locally) exact. Following 
(\ref{e:IH:sympl_exact_applied}), this contraction is applied on an 
arbitrary vector field $\delta$
\bea
\w{J}(\delta, \delta_{\w{W}})&=&
\frac{-1}{8\pi G}\int_{\Sp_t} 
\delta\left[\langle\w{q}^*\w{W},\w{\omega}\rangle \w{{}^2\epsilon}\right]
- \langle\delta \w{q}^*\w{W},\w{\omega}\rangle\w{{}^2\epsilon} +
\kappa_{(\w{W})}\delta \w{{}^2\epsilon} \nn \\
&+&\frac{1}{16\pi G}\int_{\Sigma_\infty} \w{A}^{IJ}\wedge 
\langle	\w{W},\Sigma_{IJ}\rangle + \langle\w{W},\w{A}^{IJ}\rangle\delta\Sigma_{IJ} \ ,
\label{e:IH:omegadelta}
\eea
where $\kappa_{(\w{W})}=\langle \w{W}-\w{q}^*\w{W},\w{\omega}\rangle$.

\noindent (3) {\it Conserved quantities and horizon physical parameters}.

The term at infinity is related to the standard ADM quantities (whenever
$\w{W}$ is a symmetry of the asymptotic metric at infinity). Consequently, 
it is associated with the exact variation of a function on $\Gamma$,
the corresponding ADM parameter.
Likewise, in order to associate a conserved quantity with the horizon $\Hor$
itself, the integral on $\Sp_t$ must be written  
as the exact variation of a function on $\Gamma$. 
We study this problem for the specific and physically relevant cases
of the angular momentum and the energy.
\end{exmp}

\subsection{Applications of examples (\ref{ex:IH:var_probl})  and (\ref{ex:IH:phase_space})} 

We offer some more details on the discussion developed 
in Secs. \ref{s:IH:ang_mom} and \ref{s:IH:mass}.

\subsubsection{ Angular momentum}

We restrict $\Gamma$ to its
subspace $\Gamma_{\w{\phi}}$ of spacetimes
containing a class II WIH. 
On the inner boundary $\Hor$ of each spacetime $\M$, the {\it same} rigid
azimuthal WIH-symmetry $\w{\phi}$ is {\it fixed}
(in fact $\w{\phi}$ is an isometry of the cross-section $\Sp_t$).
More specifically, we consider on every
spacetime in $\Gamma_{\w{\phi}}$ a vector field $\w{\phi}$ 
which is a $SO(2)$ axial isometry of
the induced metric $\w{q}$ on $\Hor$ with $2 \pi$ affine length.

This WIH-symmetry on $\Hor$ is then extended to a vector field $\w{\varphi}$
on each spacetime $\M$. Since we are interested in studying the
angular momentum related to the horizon itself, this extension 
$\w{\varphi}$ is enforced to vanish outside some compact neighbourhood
of the horizon. Evaluating expression (\ref{e:IH:omegadelta}) in this case, 
results in \cite{AshteBL01}
\bea
\w{J}(\delta,\delta_{\w{\varphi}}) =\delta \left(
\frac{-1}{8\pi G}\int_{\Sp_t} \w{\phi}\cdot\w{\omega}\ \w{{}^2\epsilon} \right)
\ .
\eea
The transformation $\delta_{\w{\varphi}}$
induces directly a locally canonical transformation on $\Gamma_{\w{\phi}}$.
Making $J_{\mathcal H}:=
H_{\w{\varphi}}$,  the conserved quantity $J_{\Hor}$
is identified as the angular momentum associated with the horizon
and Eq. (\ref{e:IH:angmom}) follows.
We also point out that this expression is conserved under the canonical
transformation $\delta_{\w{\varphi}}$ in $\Gamma$,
even if $\w{\phi}$ is not a WIH-symmetry. However, as mentioned in 
Sec. \ref{s:IH:ang_mom}, in the absence of a symmetry
the physical status of this expression is unclear.

\subsubsection{Mass}

As discussed in Sec. \ref{s:IH:mass}, the definition of the mass is related to 
the choice of an
evolution vector $\w{t}$ on each spacetime $\M$, which plays now the 
role of the vector $\w{W}$ in Example \ref{ex:IH:phase_space}.
We fix expression (\ref{e:IH:t_boundary_cond}) as the inner boundary condition
for $\w{t}$.
Regarding the outer boundary condition at infinity, we make $\w{t}$
to approach an observer $\w{t}_\infty$ inertial with respect
to the flat metric.

The first law of black hole mechanics (\ref{e:IH:first_law}) follows from 
{\it imposing} $\delta_{\w{t}}$ to
be a locally canonical transformation on $\Gamma_{\w{\phi}}$.
Expression (\ref{e:IH:omegadelta})  in this case
simplifies to 
\bea
\label{e:IH:sympl_delta_t}
\w{J}(\delta, \delta_{\w{t}})=\delta E^t_{\mathrm{ADM}} -\left(
\frac{\kappa_{(t)}}{8\pi G}\delta a_{\mathcal H}
+ \Omega_{(t)} \delta J_{\mathcal H} \right) \ \ ,
\eea
where $E^t_{\mathrm{ADM}}$ corresponds to the ADM energy 
and $a_{\mathcal H}=\int_{{\mathcal S}}\w{{}^2\epsilon}$ is the area 
of $\Sp_t$.
On each spacetime $\M$ in $\Gamma$, $\kappa_{(\w{t})}$
and $\Omega_{(\w{t})}$ are constant. However, the actual values of these
constants change from one spacetime to another:  $\kappa_{(\w{t})}$ and 
$\Omega_{(\w{t})}$ are functions on $\Gamma$.
As a necessary condition for $\w{J}(\cdot, \delta_t)$ to be an exact 
variation on $\Gamma$ so as to make $\delta_{\w{t}}$ a canonical transformation
via Eq. (\ref{e:IH:sympl_exact_applied}), 
the form $\w{J}(\cdot, \delta_t)$ must be closed.
Consequently,  functions $\kappa_{(\w{t})}$ and $\Omega_{(\w{t})}$
depend on $\Gamma$ {\it only} through an explicit dependence on $a_{\mathcal H}$
and $J_{\mathcal H}$, satisfying
\bea
\label{e:IH:integr_cond}
\frac{\partial \kappa_{(t)}(a_{\mathcal H}, J_{\mathcal H})}{\partial  J_{\mathcal H}}
=8\pi G \frac{\partial \Omega_{(t)}(a_{\mathcal H}, J_{\mathcal H})}{\partial  a_{\mathcal H}} \ .
\eea 
If $\delta_{\w{t}}$ is indeed  a canonical infinitesimal
transformation, we can write the second term in the right-hand side of Eq.(\ref{e:IH:sympl_delta_t}) 
as an exact variation $\delta E^t_{\mathcal H}$, and Eq. (\ref{e:IH:first_law}) follows.
As mentioned in Sec. \ref{s:IH:mass}, this does not fix the functional
forms of  $\kappa_{(\w{t})}(a_{\mathcal H}, J_{\mathcal H})$,
$\Omega_{(\w{t})}(a_{\mathcal H}, J_{\mathcal H})$ and 
$E^t_{\mathcal H}$. 
Finally, these dependences of the physical parameters in $a_{\mathcal H}$ and
$J_{\mathcal H}$, which are the same for all 
spacetimes in $\Gamma_{\w{\phi}}$,
are fixed by making them to coincide with those of the Kerr family
(a subspace of  $\Gamma_{\w{\phi}}$), as explained
in Sec.  \ref{s:IH:mass}.

%% file: kerr.tex
%
%
\section{Illustration with the event horizon of a Kerr black hole}
\label{s:KE}

In Examples~\ref{ex:NH:EF}, \ref{ex:FO:EF}, \ref{ex:IN:EF}, 
\ref{ex:KI:kin_EF} and \ref{ex:DN:EF}, we have considered for simplicity
a non-rotating static black hole (Schwarzschild spacetime). 
It is of course interesting to investigate rotating stationary black holes
(Kerr spacetime) as well. In particular, the \hajicek\  1-form
which has been found to vanish for a Schwarzschild horizon
[Eq.~(\ref{e:KI:OmegaH_EF})] is no longer zero for a Kerr horizon.
We discuss here the event horizon $\Hor$ of a Kerr black hole from
the 3+1 decomposition introduced in Sec.~\ref{s:TP}, by considering 
a foliation $(\Sigma_t)$ based on Kerr coordinates.

\subsection{Kerr coordinates}

In standard textbooks,
the Kerr solution is presented in {\em Boyer-Lindquist coordinates}
$(t_{\rm BL},r,\theta,\varphi_{\rm BL})$. However, being a generalization
of Schwarzschild coordinates to the rotating case, these coordinates
are singular on the event horizon $\Hor$, as discussed in 
Example~\ref{ex:NH:EF}. We consider instead {\em Kerr coordinates}, 
which are regular on $\Hor$. These are the coordinates
in which Kerr originally exhibited his solution \cite{Kerr63}; they are
a generalization of Eddington-Finkelstein coordinates to the rotating
case. Denoting them by $(V,r,\theta,\varphi)$, they are such that the curves
$V={\rm const}$, $\theta={\rm const}$ and $\varphi={\rm const}$
are ingoing null geodesics (they form a so-called {\em principal 
null congruence}), as in the Eddington-Finkelstein 
case\footnote{Note, however, a difference with Schwarzschild spacetime
in Eddington-Finkelstein coordinates: in the rotating Kerr case, the 
hypersurfaces $V={\rm const}$ are no longer null 
(see e.g. \cite{FletcL03}).}. As in the Schwarzschild case, we will 
use the coordinate
\be
    t := V - r 
\ee
instead of $V$ [cf. Eq.~(\ref{ex:NH:t_EF})]. 
The coordinates $(t,r,\theta,\varphi)$ are then simply a spheroidal version
of the well-known {\em Kerr-Schild coordinates} $(t,x,y,z)$: $t$ is 
the same coordinate and $(x,y,z)$ are related to $(r,\theta,\varphi)$
by
\bea
    & & x = (r\cos\varphi - a\sin\varphi)\sin\theta,\quad
    y = (r\sin\varphi + a\cos\varphi)\sin\theta,\quad \nonumber  \\
    & & z = r\cos\theta ,   \label{e:KE:Kerr_Schild_coord}
\eea
where $a$ is the angular momentum parameter of the Kerr solution, i.e. 
the quotient of the total angular momentum $J$ by the total mass $m$,
$a := J / m$. 
The relation with the Boyer Lindquist coordinates 
$(t_{\rm BL},r,\theta,\varphi_{\rm BL})$
is as follows:
\be \label{e:KE:coord_K_BL}
    dt = dt_{\rm BL} + \frac{dr}{\frac{r^2+a^2}{2mr}-1}
     \quad\mbox{and}\quad
    d\varphi = d\varphi_{\rm BL} + \frac{a\, dr}{r^2 - 2mr + a^2} \ \ . 
\ee

The metric components with respect to the ``3+1''
Kerr coordinates $(t,r,\theta,\varphi)$ are given by 
\bea
   g_{\mu\nu} dx^\mu dx^\nu  &= &  - \left( 1 - \frac{2mr}{\rho^2} \right) dt^2
   + \frac{4mr}{\rho^2}\, dt\, dr 
   - \frac{4amr}{\rho^2} \sin^2\theta \, dt\, d\varphi \nonumber \\
   & & + \left( 1 + \frac{2mr}{\rho^2} \right)  dr^2
   - 2 a \sin^2\theta \left( 1+ \frac{2mr}{\rho^2} \right) dr\, d\varphi
   \nonumber \\
   & & + \rho^2 d\theta^2 + \left( r^2 + a^2 + 
   \frac{2 a^2 m r\sin^2\theta}{\rho^2}\right)\sin^2\theta
   d\varphi^2 ,             \label{e:KE:metric_comp_Kerr}
\eea
with 
\be
    \rho^2 := r^2 + a^2 \cos^2\theta .  \label{e:KE:rho_def}
\ee
The event horizon $\Hor$ is located at 
\be
    r = r_{\Hor} := m + \sqrt{m^2-a^2} . \label{e:KE:r_H_def}
\ee
Since $r_{\Hor}$ does not depend upon $\theta$ nor $\varphi$, 
the Kerr coordinates are
{\em adapted to} $\Hor$, according to the terminology introduced in 
Sec.~\ref{s:IN:stacoord}.
Note that the metric components given by Eq.~(\ref{e:KE:metric_comp_Kerr})
are all regular at $r=r_{\Hor}$. On the contrary, most of them are
singular at $\rho=0$, which, via Eq.~(\ref{e:KE:rho_def}),
corresponds to $r=0$ and $\theta=\pi/2$, and, via 
Eq.~(\ref{e:KE:Kerr_Schild_coord}), to the ring $x^2+y^2=a^2$ in the
plane $z=0$. This is the ring singularity of Kerr spacetime. 
Note also that in the limit $a\rightarrow 0$, then $\rho\rightarrow r$ and
the line element (\ref{e:KE:metric_comp_Kerr}) reduces
to the line element (\ref{e:NH:metric_edd_fink}) in Eddington-Finkelstein
coordinates. The metric (\ref{e:KE:metric_comp_Kerr}) is clearly 
stationary and axisymmetric and the two vectors
\be \label{e:KE:Killing_vect}
    \w{\xi}_0 := \left( \der{}{t} \right) _{r,\theta,\varphi}
    \qquad \mbox{and} \qquad
    \w{\xi}_1 := \left( \der{}{\varphi} \right) _{t,r,\theta}
\ee
are two Killing vectors, $\w{\xi}_0$ being associated with the
stationarity and $\w{\xi}_1$ with the axial symmetry of the Kerr
spacetime.

\begin{rem}
The two Killing vectors $\w{\xi}_0$ and $\w{\xi}_1$ are identical to
the ``standard'' two Killing vectors which are formed from the
Boyer-Lindquist coordinates:
\be
    \w{\xi}_0 = \left( \der{}{t_{\rm BL}} \right) _{r,\theta,\varphi_{\rm BL}}
    \qquad \mbox{and} \qquad
    \w{\xi}_1 = \left( \der{}{\varphi_{\rm BL}} \right) _{t_{\rm BL},r,\theta} .
\ee
This properties follows easily from the transformation law
(\ref{e:KE:coord_K_BL}) between the two sets of coordinates.
Consequently, the metric coefficients $g_{tt} = \w{\xi}_0 \cdot \w{\xi}_0$,
$g_{t\varphi} = \w{\xi}_0 \cdot \w{\xi}_1$ and 
$g_{\varphi\varphi} = \w{\xi}_1 \cdot \w{\xi}_1$, which can be read
on Eq.~(\ref{e:KE:metric_comp_Kerr}) are the same than those for
Boyer-Lindquist coordinates, as it can be checked by comparing with
e.g. Eq.~(33.2) in MTW \cite{MisneTW73}.
\end{rem}

\subsection{3+1 quantities}

Let us consider the foliation of Kerr spacetime by the hypersurfaces
$\Sigma_t$ of constant Kerr time $t$.
From the line element (\ref{e:KE:metric_comp_Kerr}), we read
the corresponding lapse function
\be \label{e:KE:lapse}
    N = \frac{\rho}{\sqrt{\rho^2 + 2mr}},
\ee 
the shift vector
\be
    \beta^i = \left( \frac{2mr}{\rho^2+2mr},0,0\right)
    \quad\mbox{and}\quad
    \beta_i = \left( \frac{2mr}{\rho^2},0,-\frac{2amr}{\rho^2}\sin^2\theta
        \right)
\ee
and the 3-metric
\be \label{e:KE:gam_cov}
    \gamma_{ij} = \left( \begin{array}{ccc}
    1 + \frac{2mr}{\rho^2} & 0\  & 
         - a \left( 1+ \frac{2mr}{\rho^2}\right) \sin^2\theta \\
    0 & \rho^2 & 0 \\
    - a \left( 1+ \frac{2mr}{\rho^2}\right) \sin^2\theta \   & 0 &
        \frac{A}{\rho^2}\sin^2\theta       
    \end{array} \right) ,
\ee
\be
    \gamma^{ij} = \left( \begin{array}{ccc}
    \frac{A}{\rho^2(\rho^2+2mr)}\ & 0 & \frac{a}{\rho^2} \\
    0 & \rho^{-2} & 0 \\
    \frac{a}{\rho^2} & 0 & \frac{1}{\rho^2\sin^2\theta} 
    \end{array} \right) , 
\ee
with 
\bea
    A & := & (r^2+a^2)^2 - (r^2 -2mr + a^2) a^2 \sin^2\theta \nonumber \\
     & = & \rho^2 (r^2+a^2) + 2 a^2 m r \sin^2\theta .
\eea
The unit timelike normal to $\Sigma_t$ is deduced from the values
of the lapse function and the shift vector via  
Eq.~(\ref{e:FO:def_shift}), which results in
\bea
    n^\alpha & = & \left( \frac{1}{\rho} \sqrt{\rho^2 + 2mr},
    - \frac{2mr}{\rho  \sqrt{\rho^2 + 2mr}},0,0 \right) \label{e:KE:n_comp} \\
    n_\alpha & = & \left( - \frac{\rho}{\sqrt{\rho^2 + 2mr}},0,0,0 \right) . 
\eea
Finally the extrinsic curvature tensor of $\Sigma_t$ is obtained from
Eq.~(\ref{e:FO:gam_evol}) with $\partial \gamma_{ij}/\partial t = 0$:
\bea
    K_{ij} & = & \left( \begin{array}{ccc}
        \frac{2m(a^2\cos^2\theta -r^2) (\rho^2+mr)}{\rho^5
        \sqrt{\rho^2+2mr}} \ & 
        \frac{2a^2m r \sin\theta\cos\theta}{\rho^3\sqrt{\rho^2+2mr}} \ &
      \frac{a m(r^2- a^2\cos^2\theta)\sin^2\theta
      \sqrt{\rho^2+2mr}}{\rho^5} \\[5mm]
      {\rm sym.} & \frac{2mr^2}{\rho\sqrt{\rho^2+2mr}} &
      - \frac{2a^3 m r \sin^3\theta\cos\theta}{\rho^3\sqrt{\rho^2+2mr}} \\[5mm]
      {\rm sym.} & {\rm sym.} & \displaystyle
      { \frac{2mr \sin^2\theta}{\rho\sqrt{\rho^2+2mr}} \times \atop
       \left[ r 
      + \frac{a^2 m (a^2\cos^2\theta -r^2) \sin^2\theta}{\rho^4} 
      \right] }
    \end{array} \right) . \nonumber \\
        \label{e:KE:Kij}
\eea
As a check of this formula, we may compare it with Eqs.~(A2.33)-(A2.38)
of Ref.~\cite{Thornb93}. 

\subsection{Unit normal to $\Sp_t$ and null normal to $\Hor$}

The 2-surface $\Sp_t\subset\Sigma_t$ is defined by 
$r={\rm const}=r_{\Hor}$. Its outward unit normal $\w{s}$
lying in $\Sigma_t$ is obtained from $s_i = (\alpha,0,0)$, with
$\alpha$ such that $\gamma^{ij} s_i s_j = 1$. We get
\bea
   s_i & = & \left( \rho \sqrt{ \frac{\rho^2+2mr}{A} },\ 0,\ 0 \right), 
    \label{e:KE:us_comp} \\
   s^i & = & \left( \frac{1}{\rho} \sqrt{ \frac{A}{\rho^2+2mr} },\  0,\ 
   \frac{a}{\rho} \sqrt{ \frac{\rho^2+2mr}{A} } \right) . \label{e:KE:s_comp}
\eea
As a check, we verify that $\w{n}$ and $\w{s}$ given by Eqs.~(\ref{e:KE:n_comp})
and (\ref{e:KE:s_comp}) coincide with the first two vectors of
the orthonormal basis $(\vec{\w{E}}_\alpha)$ introduced by King et al.
\cite{KingLK75} [cf. their Eq.~(2.4), noticing that their coordinate
vectors are $\left( \partial/\partial V \right) _{r,\theta,\varphi}
= \left( \partial/\partial t \right) _{r,\theta,\varphi}$ and 
$\left( \partial/\partial r \right) _{V,\theta,\varphi} 
= \left( \partial/\partial r \right) _{t,\theta,\varphi} 
- \left( \partial/\partial t \right) _{r,\theta,\varphi}$].

We then get the null normal to $\Hor$ associated
with the Kerr slicing, $\el$, by inserting  expressions (\ref{e:KE:lapse}), 
(\ref{e:KE:n_comp}) and (\ref{e:KE:s_comp}) into
$\el=N(\w{n}+\w{s})$ [Eq.~(\ref{e:IN:el_nps})]:
\be \label{e:KE:ell_comp}
    \ell^\alpha = \left( 1,\ \frac{\sqrt{A}-2mr}{\rho^2+2mr},\ 0,\
        \frac{a}{\sqrt{A}} \right) . 
\ee
The value on the horizon is obtained by 
noticing that $A\equalH (2m r_{\Hor})^2$; we get
$\ell^\alpha \equalH (1,0,0,a/(2mr_{\Hor}))$, i.e., 
from Eq.~(\ref{e:KE:Killing_vect}),
\be \label{e:KE:el_xi0_xi1}
   \encadre{ \el \equalH \w{\xi}_0 + \Omega_{\Hor} \, \w{\xi}_1 }, 
\ee
with\footnote{The constant $\Omega_{\Hor}$, which constitutes an
equivalent expression for $\Omega_{\Hor}$ in
Eq. (\ref{e:IH:physparam}),  should not be confused with
the \hajicek\ 1-form $\w{\Omega}$.} 
\be
    \encadre{\Omega_{\Hor} := \frac{a}{2mr_{\Hor}} 
    = \frac{a}{2m(m+\sqrt{m^2-a^2})}} .
\ee
Eq.~(\ref{e:KE:el_xi0_xi1}) shows that on the horizon, the null normal
$\el$ is a linear combination of the two Killing vectors $\w{\xi}_0$
and $\w{\xi}_1$ with constant coefficients (compare with the inner
boundary (\ref{e:IH:t_boundary_cond}) for the evolution vector
$\w{t}$ in Sec. \ref{s:IH:mass}). It is therefore a Killing
vector itself. This implies
\be \label{e:KE:el_Killing}
    \Lie{\el} \w{A} \equalH 0 ,
\ee
for any tensor field $\w{A}$ which respects the stationarity and 
axisymmetry of the Kerr spacetime. Another phrasing of this is saying
that $\Hor$ is a {\em Killing horizon} \cite{Carte69}.
Comparing Eq.~(\ref{e:KE:el_xi0_xi1}) with Eq.~(\ref{e:IN:el_t_V_station})
(taking into account that $\w{t}=\w{\xi}_0$), we get the surface velocity
of $\Hor$ with respect to Kerr coordinates:
\be
   \encadre{ \w{V} = \Omega_{\Hor} \, \w{\xi}_1 
    =  \Omega_{\Hor} \, \left( \der{}{\varphi} \right) _{t,r,\theta}}. 
\ee
Hence the quantity $\Omega_{\Hor}$ can be viewed as the angular 
velocity of $\Hor$ with respect to the coordinates $(t,r,\theta,\varphi)$.
The fact that $\Omega_{\Hor}$ is a constant over $\Hor$ reflects the
{\em rigidity theorem} of stationary black holes (see e.g. Theorem 4.2
of Ref.~\cite{Carte73}; more generally, in the WIH setting of
Sec. \ref{s:IH}, the constancy of $\Omega_{\Hor}$
guarantees $\w{t}$ to be a WIH-symmetry on $\Hor$).  

\subsection{3+1 evaluation of the surface gravity $\kappa$}

We will need the orthogonal projector $\vec{\w{q}}$ on $\Sp_t$. 
Its components with respect to the coordinates $(r,\theta,\varphi)$ are
given by the formula $q^i_{\ \, j} = \delta^i_{\ \, j} - s^i s_j$;
from Eqs.~(\ref{e:KE:us_comp})-(\ref{e:KE:s_comp}), we get
($i$ = row index, $j$ = column index)
\be \label{e:KE:qproj}
    q^i_{\ \, j} = \left( \begin{array}{ccc}
        0 & 0\quad & 0 \\
        0 & 1\quad & 0 \\
        \displaystyle 
        - \frac{a}{A} \left( \rho^2 + 2mr \right)\  &0\quad  & 1
        \end{array} \right) . 
\ee
We will also need the 1-form $\w{K}(\w{s},.)$. From 
Eqs.~(\ref{e:KE:Kij}) and (\ref{e:KE:s_comp}), we get
\bea
    K_{rj} s^j & = & \frac{m(r^2-a^2\cos^2\theta)}{\rho^4(\rho^2+2mr)\sqrt{A}}
    \left[ \rho^2 a^2\sin^2\theta - 2(\rho^2+mr)(r^2+a^2)\right] 
        \nonumber \\
    K_{\theta j} s^j & = & \frac{2a^2 mr\sin\theta\cos\theta}{(\rho^2 
        + 2mr) \sqrt{A}} \nonumber \\
    K_{\varphi j} s^j &= & \frac{am\sin^2\theta}{\rho^4\sqrt{A}}
        \left[ r^2(3r^2+a^2\cos^2\theta) + a^2(r^2-a^2\cos^2\theta)\right] . 
                    \label{e:KE:Ks} 
\eea
 
Let us start by evaluating the non-affinity parameter $\kappa$ from the 3+1
expression (\ref{e:TP:kappa_3p1}). The first part of this relation 
is computed from Eqs.~(\ref{e:KE:ell_comp}) and (\ref{e:KE:lapse}):
\be
    \ell^\mu \nabla_\mu \ln N = \frac{m}{\rho^2} 
    \frac{r^2-a^2\cos^2\theta}{(\rho^2+2mr)^2} (\sqrt{A}-2mr) . 
\ee
Since $A\equalH (2m r_{\Hor})^2$, this implies
\be \label{e:KE:ellnablogN}
    \ell^\mu \nabla_\mu \ln N \equalH 0 , 
\ee
in agreement with Eq.~(\ref{e:KE:el_Killing}).
The second term in the right-hand side of Eq.~(\ref{e:TP:kappa_3p1}) 
is computed from Eqs.~(\ref{e:KE:s_comp}) and (\ref{e:KE:lapse})
\be
    s^i D_i N = \frac{m}{\rho^2} 
    \frac{(r^2-a^2\cos^2\theta)\sqrt{A}}{(\rho^2+2mr)^2} , 
\ee
resulting in the following value on the horizon
\be \label{e:KE:sgradN}
    s^i D_i N \equalH 
    \frac{2m^2 r_{\Hor} (r_{\Hor}^2 - a^2\cos^2\theta)}{ \rho_{\Hor}^2
    ( \rho_{\Hor}^2 +2m r_{\Hor})^2} , 
\ee
with $\rho_{\Hor}^2 := r_{\Hor}^2+a^2\cos^2\theta 
= 2mr_{\Hor} - a^2\sin^2\theta$. 
Finally from Eqs.~(\ref{e:KE:lapse}), (\ref{e:KE:Ks}) and (\ref{e:KE:s_comp}), 
we evaluate the last term which enters in formula (\ref{e:TP:kappa_3p1}), 
namely
$N K_{ij} s^i s^j$. The obtained expression is rather complicated; however 
combining its value on the horizon with the results
(\ref{e:KE:ellnablogN}) and (\ref{e:KE:sgradN}) yields a very 
simple expression for the non-affinity parameter:
\be \label{e:KE:kappa}
    \encadre{\kappa = \frac{r_{\Hor}-m}{2m r_{\Hor}}
        = \frac{\sqrt{m^2-a^2}}{2m(m+\sqrt{m^2-a^2})} }. 
\ee
Note that $\kappa$ does not depend on $\theta$, in agreement with 
the fact that $\Hor$, endowed with the null normal $\el$ given by 
Eq. (\ref{e:KE:el_xi0_xi1}), is an isolated horizon [zeroth law of black
hole mechanics, cf. Eq.~(\ref{e:IH:WIHzerolaw})]. Actually we recover for
$\kappa$ the classical value of the surface gravity of a Kerr
black hole (see e.g. Eq.~(12.5.4) of Wald \cite{Wald84}). 

\subsection{3+1 evaluation of the \hajicek\ 1-form $\w{\Omega}$}

Let us now compute the \hajicek\ 1-form from 
the 3+1 formula (\ref{e:TP:Omega3+1}). From Eqs.~(\ref{e:KE:lapse}),
(\ref{e:KE:Ks}) and (\ref{e:KE:qproj}), we get
\bea
    \Omega_\theta & = & - \frac{2 a^2 m r \sin\theta\cos\theta}{\rho^2+2mr}
        \left( \frac{1}{\sqrt{A}} + \frac{1}{\rho^2} \right) \\
    \Omega_\varphi & = & - \frac{am\sin^2\theta}{\rho^4\sqrt{A}}
        \left[ r^2(3r^2+a^2\cos^2\theta) + a^2(r^2-a^2\cos^2\theta)\right] , 
\eea
from which we deduce the following values on the horizon:
\bea
    \Omega_\theta &\equalH& - \frac{a^2 \sin\theta \cos\theta}{2m r_{\Hor}
        - a^2 \sin^2\theta} \\
    \Omega_\varphi &\equalH& \frac{a}{r_{\Hor}} \sin^2\theta
    \frac{(2m^2-3m r_{\Hor} + r_{\Hor}^2)\cos^2\theta
        - r_{\Hor}(r_{\Hor}+m)}{(r_{\Hor}+ (2m-r_{\Hor})\cos^2\theta)^2} .
                \label{e:KE:Omega_phi_hor}
\eea
As a check, let us recover the total angular momentum $J_{\Hor}=am$
from the integral (\ref{e:IH:angmom}) which involves $\w{\Omega}$. 
The symmetry generator $\w{\phi}$ which appears in the integral
is of course in the present case the Killing vector 
$\w{\xi}_1 = \left( \partial/\partial{\varphi} \right) _{t,r,\theta}$,
so that formula (\ref{e:IH:angmom}) results in ($G=1$) 
\be
    J_{\Hor} = -\frac{1}{8\pi} \int_{\Sp_t} \Omega_\varphi \, 
    \w{{}^2\!\epsilon} .
\ee
Let us express the integral in terms of the coordinates $(\theta,\varphi)$
which span $\Sp_t$:
\be \label{e:KE:J_prov}
    J_{\Hor}  = -\frac{1}{8\pi} \int_0^\pi \int_0^{2\pi}
            \Omega_\varphi \, \sqrt{q} \, d\theta\, d\varphi ,
\ee
where $q = \det q_{ab}$, with the 2-metric components $q_{ab}$
read from Eq.~(\ref{e:KE:gam_cov}):
\be \label{e:KE:qab}
    q_{ab} = \left( \begin{array}{cc}
        \rho^2 & 0 \\
        0\  & \displaystyle \frac{A}{\rho^2}\sin^2\theta 
        \end{array} \right) . 
\ee
Hence $\sqrt{q}=\sqrt{A}\sin\theta$, so that 
$\sqrt{q}\equalH 2mr_{\Hor} \sin\theta$ and the integral (\ref{e:KE:J_prov})
becomes
\be
    J_{\Hor}  = -\frac{2mr_{\Hor}}{8\pi} \int_0^\pi \int_0^{2\pi}
       \Omega_\varphi \, \sin\theta \, d\theta\, d\varphi .
\ee
Substituting Eq.~(\ref{e:KE:Omega_phi_hor}) for $\Omega_\varphi$,
we get
\be
    J_{\Hor} = - \frac{am}{4}
    \int_0^\pi \frac{(2\lambda-1)(\lambda-1)\cos^2\theta
        - \lambda(2\lambda+1)}{(\lambda + (1-\lambda)\cos^2\theta)^2}
            \, \sin^3\theta \, d\theta , 
\ee
where we have set $\lambda := r_{\Hor}/(2m)$. 
It turns out that the above integral is independent of $\lambda$
and is simply equal to $-4$, hence
\be
    \encadre{J_{\Hor} = am}, 
\ee
as it should be for a Kerr black hole. 

\subsection{3+1 evaluation of $\w{\Theta}$ and $\w{\Xi}$}

In order to apply the formul\ae\ derived in Sec.~\ref{s:TP:Th_Xi_H},
let us first compute the second fundamental form $\w{H}$ of
the 2-surface $\Sp_t$ (as a hypersurface of $\Sigma_t$).
From the relation $H_{ij} = D_k s_l \, q^k_{\ \, i} q^l_{\ \, j}$
[Eq.~(\ref{e:TP:H_qstar_Ds})] and expressions (\ref{e:KE:us_comp})
and (\ref{e:KE:qproj}), we get the following values on the horizon:
\bea
    H_{\theta\theta} &\equalH& \frac{2m r_{\Hor}^2}{\sqrt{
    (2mr_{\Hor}- a^2\sin^2\theta)(4mr_{\Hor}- a^2\sin^2\theta})}, \label{e:KE:Hthth} \\
    H_{\theta\varphi} &\equalH& - \frac{2 a^3 m r_{\Hor}
        \sin^3\theta\cos\theta}{
   (2mr_{\Hor}- a^2\sin^2\theta)^{3/2} (4mr_{\Hor}- a^2\sin^2\theta)^{1/2}}, \\
   H_{\varphi\varphi} &\equalH& \frac{mr_{\Hor}^3 \sin^2\theta}{4
   (2mr_{\Hor}- a^2\sin^2\theta)^{5/2} (4mr_{\Hor}- a^2\sin^2\theta)^{1/2}}
   \times \nonumber \\
   & & \left[ 4m^3 + 9m r_{\Hor}^2 + 3 r_{\Hor}^3 - 4a^2(r_{\Hor}+3m)
   \cos 2\theta 
   + a^4 \frac{m-r_{\Hor}}{r_{\Hor}^2} \cos 4\theta \right] . 
   \nonumber \\
    \label{e:KE:Hpp}
\eea
We are then in position to evaluate $\Hor$'s second fundamental form
$\w{\Theta}$ via Eq.~(\ref{e:TP:Theta_H_K}): 
$\w{\Theta} = N(\w{H} -\vec{\w{q}}^* \w{K})$. Using the value of
$\w{K}$ and $\vec{\w{q}}$ given by Eqs.~(\ref{e:KE:Kij})
and (\ref{e:KE:qproj}), we get 
\be 
    \encadre{\w{\Theta} \equalH 0} .
\ee
Hence we recover the fact that the event horizon of a Kerr black
hole is a non-expanding horizon. 

Regarding the transversal deformation rate $\w{\Xi}$, we use
the formula $\w{\Xi}=-1/(2N)\, (\w{H} -\vec{\w{q}}^* \w{K})$
[Eq.~(\ref{e:TP:Xi_H_K})]. Using expressions (\ref{e:KE:Hthth})-(\ref{e:KE:Hpp}),
(\ref{e:KE:Kij}), (\ref{e:KE:qproj}) and (\ref{e:KE:lapse}), we get
\bea
    \Xi_{\theta\theta} &\equalH& - \frac{2mr_{\Hor}^2}{2mr_{\Hor}-a^2\sin^2\theta}, \\
    \Xi_{\theta\varphi} &\equalH& \frac{2 a^3 m r \sin^3\theta\cos\theta}{
    (2mr_{\Hor}-a^2\sin^2\theta)^2} ,\\
    \Xi_{\varphi\varphi} &\equalH& - \frac{2m r_{\Hor}\sin^2\theta}{
    2mr_{\Hor}-a^2\sin^2\theta} \left[ r_{\Hor} 
    + \frac{a^2 m (a^2\cos^2\theta- r_{\Hor}^2)\sin^2\theta}{
    (2mr_{\Hor}-a^2\sin^2\theta)^2} \right].
\eea
Contracting with $q^{ab}$ [obtained as the inverse of the matrix (\ref{e:KE:qab})],
we get the transversal expansion scalar
\be
    \theta_{(\w{k})} \equalH - \frac{2m+ 3r_{\Hor} - \frac{a^2}{m}\left[
            1  + \left(\frac{m}{r_{\Hor}}-1\right) \cos^2\theta \right]}{2(2mr_{\Hor}
            -a^2\sin^2\theta)} .
\ee
As a check, one can easily verify that in the non-rotating limit
($a=0$, $r_{\Hor}=2m$), the values of $\w{\Omega}$, $\w{\Xi}$ and
$\theta_{(\w{k})}$ derived above reduce to that obtained in 
Example~\ref{ex:KI:kin_EF} for the Eddington-Finkelstein slicing of
the Schwarzschild horizon [cf. Eqs.~(\ref{e:KI:XiH_EF})
and (\ref{e:KI:OmegaH_EF})].

%% file: notsummary.tex
%
%
\section{Symbol summary}
\label{s:SY}

The various metrics and associated connections (intrinsic geometries)
used in this article are collected in Table~\ref{t:SY:metric_connect}, whereas
the symbols used to describe the extrinsic geometries of the various
submanifolds are collected in Table~\ref{t:SY:extrinsic}. 

\begin{table}
\begin{center}
\begin{tabular}{ccccc}
\hline 
\hline 
Manifold & Metric & Signature & Compatible  & Ricci \\
 & & & connection & tensor \\
\hline 
$\M$ & $\w{g}$ & $(-,+,+,+)$ & $\w{\nabla}$ & $\w{R}$  \\
$\Hor$ & $\w{q}$ & $(0,+,+)$ & $\times$ & $\times$ \\
$\Sigma_t$ &  $\w{\gamma}$ & $(+,+,+)$ & $\w{D}$ & ${}^3\!\w{R}$ \\
$\Sigma_t$ &  $\w{\tilde \gamma}$ & $(+,+,+)$ & $\w{\tilde D}$ & ${}^3\!\w{\tilde R}$ \\
$\Sp_t$ &  $\w{q}$ & $(+,+)$ & $\w{\DS}$ & ${}^2\!\w{R}$  \\
$\Sp_t$ &  $\w{\tilde q}$ & $(+,+)$ & $\w{\tDS}$ & ${}^2\!\w{\tilde R}$  \\
\hline
$\Hor$ NEH & $\w{q}$ & $(0,+,+)$ & $\w{\hat\nabla}$ & not used \\
\hline
\hline
\end{tabular}
\end{center}
\vspace{2ex}
\caption[]{\label{t:SY:metric_connect}
Metric tensors and associated connections used in
this article. The symbol '$\times$' in the second line means that
there is not a unique connection on $\Hor$ compatible with $\w{q}$,
for the latter is degenerate. The last line regards the particular case
of a non-expanding horizon.}
\end{table}

\begin{table}
\begin{center}
\begin{tabular}{cccc}
\hline 
\hline 
Submanifold & Projector &  Second fundamental
     & Normal \\
 of $\M$   &  onto it & form(s) / embedding manifold & vector(s) \\
\hline 
$\Hor$  & $\w{\Pi}$ & $\w{\Theta}$ / $(\M,\w{g},\el)$ & $\el$ \\
$\Sigma_t$ &  $\vec{\w{\gamma}}$ & $\w{K}$ / $(\M,\w{g})$ & $\w{n}$ \\
$\Sp_t$ & $\vec{\w{q}}$ &  $\w{H}$ / $(\Sigma_t,\w{\gamma})$ & $\w{s}$ \\
$\Sp_t$ & $\vec{\w{q}}$ &  $\w{\tilde H}$ / $(\Sigma_t,\w{\tilde \gamma})$  & $\w{\tilde s}$  \\
$\Sp_t$ & $\vec{\w{q}}$ &  $(\w{\Theta},\w{\Xi})$ / $(\M,\w{g},\el)$ & $(\el,\w{k})$ \\
$\Sp_t$ & $\vec{\w{q}}$ &  $(\vec{\w{q}}^* \w{K},\w{H})$ / $(\M,\w{g})$ & $(\w{n},\w{s})$ \\
\hline
$\Hor$ NEH & $\w{\Pi}$ & $0$ / $(\M,\w{g})$ & $\el$ \\
\hline
\hline
\end{tabular}
\end{center}
\vspace{2ex}
\caption[]{\label{t:SY:extrinsic}
Extrinsic geometry of various submanifolds of $\M$. Note that the projectors
$\vec{\w{\gamma}}$ and $\vec{\w{q}}$ are orthogonal ones with respect
to the ambient metric $\w{g}$, whereas $\w{\Pi}$ is not. The notation 
``$\w{\Theta}$ / $(\M,\w{g},\el)$'' means that the second fundamental form
$\w{\Theta}$ depends not only on the ambient geometry $(\M,\w{g})$, but
also on the choice of the null normal $\el$.
The pairs
$(\w{\Theta},\w{\Xi})$ and $(\vec{\w{q}}^* \w{K},\w{H})$ are required 
for the embedding of the 2-surfaces $\Sp_t$ in $\M$ because 
the correct object to describe such an two-dimensional embedding is not a 
single bilinear form
but a type $\left({1 \atop 2}\right)$ tensor 
[see Eqs.~(\ref{e:KI:CarterK_Theta_Xi}) and (\ref{e:TP:CarterK_qK_H})].
The last line regards the particular case of a non-expanding horizon.}
\end{table}

%% file: ref.tex
%
%